%% file: panda_tdr_FSC_pub.tex
\documentclass[pdftex]{panda_book}
\markpanda{Strong interaction studies with antiprotons}{Technical Design Report - FSC}
\begin{document}
%
%
\include{panda_tdr_FSC_tit_pub}
%
%
\include{panda_tdr_FSC_exs}
\pagenumbering{arabic}
\include{panda_tdr_FSC_int_pub}

\include{panda_tdr_FSC_req}

\include{panda_tdr_FSC_mech_pub}
\include{panda_tdr_FSC_photo}

\include{panda_tdr_FSC_cal}
\include{panda_tdr_FSC_sim}

\include{panda_tdr_FSC_perf_pub}
\include{panda_tdr_FSC_org_pub}
\include{panda_tdr_FSC_end}
%
\end{document}

%% file: panda_tdr_FSC_tit_pub.tex
\pagenumbering{roman}
\onecolumn
%
%
\thispagestyle{empty}
\vspace*{0.5cm}
\begin{center}

{\huge\bf Technical Design Report  \\ \  for the \Panda \ \\ Forward Spectrometer Calorimeter \\
\ \\ {\sf\small (Anti\underline{P}roton \underline{AN}nihilations at \underline{DA}rmstadt)}}\\
\vskip 1cm
{\large \Panda{} Collaboration}
%
\vskip 0.5cm
\fbox{\today}
\end{center}
\vskip 1cm
%
%
\vskip 1cm
\begin{center}
\includegraphics[width=0.8\dwidth]{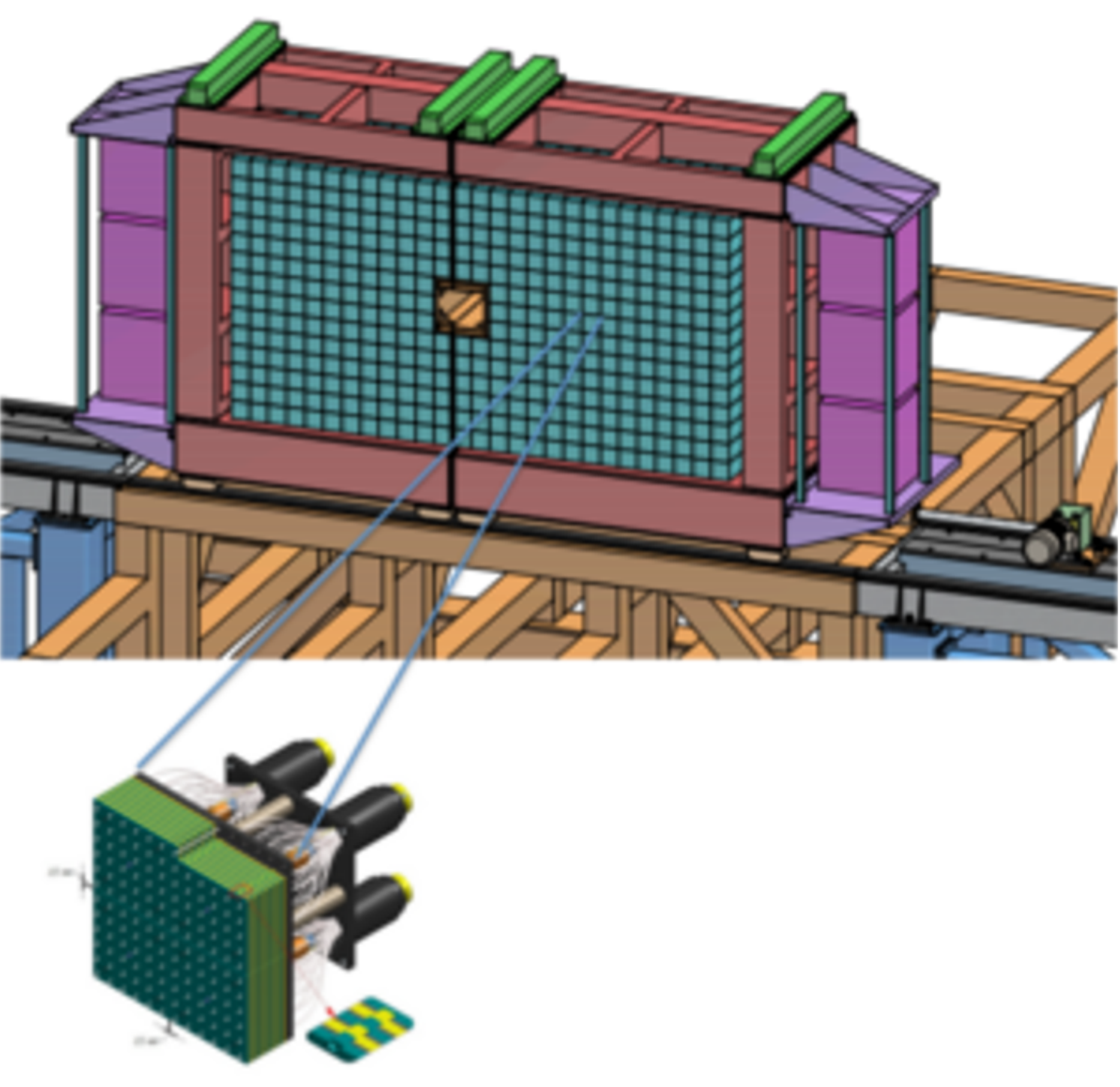}
\end{center}
\vfill

\newpage

\vspace{16cm}

\vfill
{\bf Cover}:
The figure shows the \Panda Forward Spectrometer Calorimeter placed on the Forward Spectrometer support.
The region of a single module is zoomed, allowing a view into the calorimeter structure with the sandwich of
scintillator and lead tiles and the bunches of traversing WLS fibres, funneled to photo detectors, one for each of the four cells of one module. 
The ``LEGO''-type locks used to firmly join the tiles are zoomed even more.
This document is devoted to the electromagnetic calorimeter of the Forward Spectrometer and describes the design considerations, the technical layout, the expected performance, and the production readiness.


%
%
\newpage
\begin{center}
\vspace*{3mm }
{\LARGE The \Panda{} Collaboration}
\vskip 7mm
\input{./main/authors}
\end{center}
%
%
\vfill
\hrulefill\\
\begin{tabbing}
Editors:
\hspace{3cm} \= Anatoliy Derevschikov  \hspace{0.5cm}  \= Email: \verb$derevshchikov@ihep.ru$\\ \ \\
\> Herbert L\"ohner  \hspace{0.5cm}  \> Email: \verb$loehner@kvi.nl$ \\ \ \\
\> Dmitriy Morozov   \hspace{0.5cm}  \> Email: \verb$morozov@ihep.ru$ \\ \ \\
\> Rainer Novotny  \hspace{0.5cm}  \> Email: \verb$Rainer.Novotny@exp2.physik.uni-giessen.de$ \\ \ \\
\> Herbert Orth  \hspace{0.5cm}  \> Email: \verb$H.Orth@gsi.de$ \\ \ \\
\> Andrey Ryazantsev  \hspace{0.5cm}  \> Email: \verb$ryazants@ihep.ru$ \\ \ \\
\>  Pavel Semenov  \hspace{0.5cm}  \> Email: \verb$Pavel.Semenov@ihep.ru$ \\ \ \\
\> Alexander Vasiliev  \hspace{0.5cm}  \> Email: \verb$Alexander.Vasiliev@ihep.ru$ \\ \ \\

Technical Coordinator: \> Lars Schmitt \hspace{0.5cm} \> Email: \verb$l.Schmitt@gsi.de$ \\ \ \\
Spokesperson: \>  James Ritman \hspace{0.5cm} \> Email: \verb$j.ritman@fz-juelich.de$ \\
Deputy:  \> Diego Bettoni  \hspace{0.5cm} \> Email: \verb$bettoni@fe.infn.it$ \\
\end{tabbing}
\hrulefill\\
\vfill
%
%
\cleardoublepage
\input{./main/preamble}
%
%
\cleardoublepage
\tableofcontents
%
%

%% file: main/authors.tex
%
%
\institem{Aligarth Muslim University, Physics Department,{ \bf Aligarth}, India}
\authitem{B.~Singh}
\lastitem
\institem{Universit\"at Basel,{ \bf Basel}, Switzerland}
\authitem{W.~Erni},
\authitem{I.~Keshelashvili},
\authitem{B.~Krusche},
\authitem{M.~Steinacher},
\authitem{N.~Walford}
\lastitem
\institem{Institute of High Energy Physics, Chinese Academy of Sciences,{ \bf Beijing}, China}
\authitem{B.~Liu},
\authitem{H.~Liu},
\authitem{Z.~Liu},
\authitem{X.~Shen},
\authitem{C.~Wang},
\authitem{J.~Zhao}
% ...
\lastitem

%% file: main/preamble.tex
%
\begin{center}
\vspace*{2cm}
{\Large\bf Preface}\addcontentsline{toc}{chapter}{Preface}
\vskip 2cm
\begin{minipage}[t]{8cm}
\sloppy\large
This document is a Technical Design Report (TDR) of the Forward Spectrometer 
of the \Panda{} Electromagnetic Calorimeter (EMC)  and describes the design considerations, the technical layout, the expected performance, and the production readiness of the Forward Spectrometer EMC.
The \Panda{}
Electromagnetic Calorimeter consists of the Target Spectrometer EMC and the Forward 
Spectrometer EMC. While the two EMC components operate as a united calorimeter, they are based on
completely different detection techniques. The Target Spectrometer EMC is covered by a separate
TDR, which was submitted earlier, and this document may refer to it.

\end{minipage}
\end{center}
\vspace*{2cm}
\centerline{
}
\vfill
\clearpage
\vspace*{18cm}
\hrulefill\\
\vspace*{2cm}\\
\begin{minipage}[t]{10cm}
\sloppy
The use of registered names, trademarks, \etc in this publication does not
imply, even in the absence of specific statement, that such names are exempt
from the relevant laws and regulations and therefore free for general use.
\end{minipage}
\vfill
%

%% file: panda_tdr_FSC_exs.tex
%
%
\cleardoublepage
\chapter{Executive Summary}
\label{sec:exs}

This Technical Design Report (TDR) illustrates the technical layout and the expected performance
of the Forward Spectrometer Calorimeter (FSC) in the \PANDA spectrometer. This document is divided into 
nine chapters. 
We start with a motivation for the \PANDA experiment and outline important aspects for the design of the experimental setup. After explaining in \Refchap{sec:req} the motivation for choosing a sampling calorimeter of the shashlyk type instead of a crystal calorimeter,  we describe the construction of shashlyk modules and the readout chain with photo multiplier in \Refchap{chap:mech}. There we also outline the complete calorimeter block as it is integrated into the \PANDA Forward Spectrometer. The readout electronics is presented in  \Refchap{sec:roel} and the important aspects of calibration and monitoring are given in \Refchap{chap:cal}. Simulations of important  properties  of the calorimeter are summarised in \Refchap{sec:sim}. The results of test-beam studies of different prototypes are presented and compared with simulations in  \Refchap{sec:perf}. The TDR concludes with details about the management of the whole project. 
\section{The \PANDA experiment}
	\PANDA is a next-generation hadron physics detector planned to be operated at the
Facility for Antiproton and Ion Research (FAIR) at Darmstadt,
Germany. It will use antiprotons generated with a 30 GeV proton beam
from the Synchrotron SIS 100 interacting with a Nickel production
target. The antiprotons are collected and cooled in the collector ring
(CR) and then accumulated in the High Energy Storage Ring (HESR) where they are
further cooled to precisely study collisions with an internal proton target (or nuclear targets) at momenta between 
1.5 GeV/$c$ and 15 GeV/$c$ .

With the \PANDA spectrometer it is planned to carry out a rich and diversified hadron-physics 
research program. The experiment is designed to fully exploit the extraordinary physics potential 
arising from the availability of high-intensity cooled antiproton beams. The aim of the versatile 
experimental program is to significantly improve our knowledge of the strong interaction and 
of the mechanisms leading to hadron structure and masses. 
Thanks to the expected tremendous boost in statistical accuracy and precision of measured data, 
significant progress beyond the present understanding of the field will be made.


%
For precision spectroscopy of exotic hadrons and charmonium states, the full detector acceptance is 
required to perform a sensitive partial-wave analysis. Thus, the main goal of the \PANDA electromagnetic 
calorimetry is to detect photons in almost the full solid angle. 
The shashlyk-type electromagnetic calorimeter described in this TDR  is part of the Forward 
Spectrometer of \PANDA. It will cover about 0.74\percent of the solid angle in the forward direction, which is essential 
for the  \PANDA performance, since due to the relativistic boost the secondary particle density in the forward 
direction will largely surpass the isotropic distribution.
Since final states with many photons in a wide energy range are very likely,  a low photon threshold is 
a mandatory requirement for the electromagnetic calorimeters of \PANDA, in order to guarantee excellent photon 
recognition and resolution. 



\subsection {General setup}

To achieve almost full acceptance and good momentum resolution over a large momentum range, the Target Spectrometer is housed in 
a solenoid magnet supporting the identification of tracks with high transverse momenta.  The Forward Spectrometer employs 
a dipole magnet for tracking the reaction products going forward. The superconducting solenoid magnet
provides a maximum field strength of 2~T and has a coil opening of 1.89 m and a coil length of 2.75 m. The Target
Spectrometer is arranged in a barrel part covering angles between 22\degrees and 140\degrees, a forward endcap part covering the 
forward-angle range down to 5\degrees in the vertical and 10\degrees in the horizontal plane, and a backward endcap part 
covering the region between about 145\degrees and 170\degrees. 

The dipole magnet has a field integral of up to 2 Tm with an aperture of 1.4 m in width and 0.7 m in height. 
The Forward Spectrometer covers the very forward angles. Both spectrometer parts are equipped with tracking, 
charged-particle identification, electromagnetic calorimetry and muon identification. In order to operate the experiment 
at high rates and to deal with different physics-event topologies in parallel, a self-triggering readout scheme was adopted.

\subsection {Tracking detectors}

The micro-vertex detector consists of a four-layer barrel detector and six detector wheels in the forward direction 
made from radiation hard silicon pixel and strip sensors. For the tracking in the solenoid field, low-mass straw tubes arranged
in straight and skewed configurations are foreseen. The straws have a diameter of 1 cm and a length of 1.5 m.
Tracks at small polar angles (5\degrees $< \theta <$ 22\degrees) are measured by large 
planar GEM detectors. Three stations are placed between 1.1 m and 1.9 m downstream of the interaction point. 
Further downstream, in the Forward Spectrometer, straw-tube chambers  
with a tube diameter of 1 cm will be employed.     

\subsection {Particle identification}

Charged-particle identification is required over a large momentum range from 200 MeV/$c$ up to
almost 10 GeV/$c$. 
The velocity of charged particles is determined primarily by Cherenkov detectors. In the Target Spectrometer
two DIRC detectors based on the Detection of Internally Reflected Cherenkov light are being developed, one
consisting of long rectangular quartz bars for the barrel region (Barrel DIRC), the other one (Disk DIRC) being shaped as a disc for the forward endcap.
 
For the particle identification in the forward region a Ring Imaging Cherenkov Detector (RICH) is 
planned, combined with a Time Of Flight (TOF)-system downstream of the RICH. TOF can be 
exploited in \PANDA, although no dedicated start detector is available. 

Using a scintillator tile hodoscope (SciTil) covering the barrel section in front of the electromagnetic calorimeter (EMC) and
a scintillator wall after the dipole magnet, relative timing of charged particles with
very good time resolution of about 100 ps can be achieved.  The energy loss within the trackers 
will be employed as well for particle identification below 1 GeV/$c$ since the individual charge is obtained by analog
readout or a time-over-threshold measurement.

The detection system is complemented by a Muon Detection system based on drift tubes located inside the segmented
magnet yoke, between the spectrometer magnets, and at the downstream end of the spectrometer. Muon detection is implemented 
as a range system with interleaved absorbing material and detectors to better distinguish muons from pions
in the low-momentum range of \Panda. 

\subsection {Calorimetry}

In the Target Spectrometer high-precision electromagnetic calorimetry is required over a large
range from a few MeV up to several GeV in energy deposition. Lead-tungstate (PWO) is chosen as scintillating crystal 
for the calorimeters in the Target
Spectrometer because of its high density, fast response, and good light yield, enabling high energy resolution 
and a compact calorimeter configuration.
The concept of \PANDA places the Target Spectrometer EMC inside the super-conducting coil of the solenoid.
Therefore, the basic requirements of the appropriate scintillator material are compactness to minimise the
radial thickness of the calorimeter layer, fast response to cope with high interaction rates, sufficient 
energy resolution and efficiency over the wide dynamic range of photon energies given by the physics program,
and finally an adequate radiation hardness. 

To achieve the required very low energy threshold, the light yield has to be maximised. Therefore, improved
lead-tungstate (PWOII) crystals are employed with a light output twice as high as used in CMS at LHC at CERN. 
Operating these crystals at -25\degrees C increases the light output by another factor of four. In addition,
large-area Avalanche PhotoDiodes (APDs) are used for the readout of scintillation light, providing high quantum 
efficiency and an active area four times larger than used in CMS. 

The crystal calorimeter is complemented in the Forward Spectrometer with a shashlyk-type sampling calorimeter
consisting of 378 modules, each composed of four independent cells of 55$\times$55  {\ensuremath{\mm^2}\xspace} size, covering in total an area of about 3 m $\times$ 1.5 m.   
This document presents the details of the technical design of the shashlyk electromagnetic
calorimeter of the \Panda Forward Spectrometer.   

\section {The \PANDA Forward Spectrometer Calorimeter}

 %
\subsection {Introduction}

 The \PANDA fixed-target experimental setup requires the ability to measure single photons,
\piz as well as $\eta$ mesons (with a mass of 550 MeV) in a wide energy range with superior energy and 
position resolutions. A fine-sampling calorimeter, covering a large area of  
over 4 {\ensuremath{\m^2}\xspace} downstream of the interaction point in \PANDA, was chosen since it meets the requirements and can be built at 
modest costs as well. The energy range in the \PANDA experiment will reach from a few MeV 
up to 15 GeV photon energies. The investigation of the optimal parameters of such a fine-sampling 
calorimeter in this wide energy region was essential for this TDR.

\subsection {Design of the modules}

Prototypes of fine-sampling electromagnetic calorimeter modules were constructed at IHEP Protvino. The design
was based on the electromagnetic calorimeter for the KOPIO experiment at Brookhaven National Laboratory, USA, 
with additional modifications to adapt 
the calorimeter for the wide energy range. In particular, the total depth was increased up to 20 radiation lengths. The modules were assembled from 380 alternating layers of lead and 
scintillator plates. The thickness of lead plates was 275 $\mu$m, and the
thickness of scintillator plates 
was 1.5 mm.  Lead plates were doped with 3\percent of antimony to improve their rigidity.
Scintillator plates were made of doped polystyrene.  
The effective radiation length was 34 mm, and the effective Moli\`ere radius was 59 mm. 
The scintillators were manufactured at the scintillator workshop of IHEP Protvino 
with the use of injection moulding technology.

\subsection {Experimental setup for prototype studies and calibration} 

Two calorimeter prototypes (Type-1 and Type-2 modules) were assembled and tested at the U-70 beams at IHEP. 
The Type-1 prototype consisted of
nine modules assembled into a 3$\times$3 matrix installed on a remotely controlled (x,y) moving support, which positioned 
the prototype across the beam with a precision of 0.4 mm. The module size was 110$\times$110$\times$675  {\ensuremath{\mm^3}\xspace}. 
The Type-2 prototype consisted of 64 cells assembled into an 8$\times$8 matrix installed on
the same moving support. The cell size was 55$\times$55$\times$675  {\ensuremath{\mm^3}\xspace}
and four cells are combined into on module. The beam tests were
carried out in 2006 and 2008.

The Type-2 and improved Type-3 prototypes were tested at Mainz with electrons 
of energy between 100 and 700 MeV. These tests were carried out in 2012 and 
2014. The experimental setup as well as the performance results are described 
in detail in \Refchap{sec:perf}.

\subsection {Performance}

After some dedicated calibration runs at the Protvino accelerator, when each module was exposed to the 19 GeV/$c$ beam, 
the Type-1 and Type-2 prototypes
were positioned, so that the beam hits the central module. It was exposed to beams at momenta 1, 2, 3.5, 5, 7, 10, 14 and 
19 GeV/$c$. 
The energy resolution is obtained from a Gaussian fit to the peak in the energy over momentum distribution at  E/p=1.
The stochastic term for both prototypes (Type-1 and Type-2) 
was found to be around 3\percent, which is less than the anticipated 4\percent stated by 
the \PANDA requirements.

The position resolution has been determined by comparing the exact impact coordinate of the beam particle, 
measured by the last drift chamber, and the centre-of-gravity of the electromagnetic shower developed 
in the calorimeter prototype. 
As a result, the stochastic term of about 
15 mm was found for Type-1 and 8 mm for Type-2.

%
The Type-2 prototype fulfils the basic requirements on the energy and position resolution 
for the electromagnetic calorimeter in the \PANDA Forward Spectrometer.
However, the non-uniformity of the energy response across the cell in the Type-2 prototype was as big
as 30\percent. This led to
the design and assembly of the Type-3 prototype in order to significantly decrease the non-uniformity of the
energy response.  

%

The Type-3 modules represent a significant improvement of the overall performance with respect to
energy, position and time resolution necessary for shower reconstruction. The significantly better
reproducibility and homogeneity of the modules and individual cells should guarantee to reach 
the necessary performance even down to photon energies as low as 10-20 MeV. 
The implementation
of additional reflector material and WLS fibres of better quality and light collection have almost
doubled the recorded light yield of $\sim$2.8$\pm$0.3 photo electrons per MeV deposited energy. 
The Type-3 prototype finally fulfils all the basic 
requirements for the electromagnetic calorimeter in the 
\PANDA Forward Spectrometer and will be chosen for mass production. 


\subsection {Photosensors}

The \PANDA FSC has to register energy deposition in a high dynamic range with low 
noise at high rate of forward photons. Since the sensors are
outside of the magnetic field, the most appropriate and robust device is a photomultiplier tube (PMT). We selected
the PMT Hamamatsu R7899. The concept of the Cockcroft-Walton high-voltage base, the 
performance of this photosensor with respect to efficiency, linearity, 
count rate capability, dynamic range and noise level are presented 
in \Refchap{chap:mech}.

\subsection {The readout electronics}

The input characteristics of the front-end electronics
should match the electrical properties of the FSC modules. 
Both \PANDA calorimeters, the Target Spectrometer EMC, based on lead-tungstate crystals,
 and the shashlyk Forward Spectrometer Calorimeter, have a number of common features 
like the output signal parameters
and the signal treatment. Thus, the structure of the readout electronics 
will be similar for the FSC and the Target Spectrometer EMC. The main 
difference, however, is caused by the higher rates and shorter signals 
from the FSC which requires specially designed digitiser modules.

The readout includes Digitiser, Data Concentrator (DCON) and Compute Node
modules for on-line computing. The digitiser module contains Sampling ADCs
for continuous digitisation of the detector signals and a Field
Programmable Gate Array for on-line data processing. 
Using a serial optical-link connection, the reduced data are transferred
to the DCON module, located outside the \PANDA detector. 

The DCON module collects data from several digitisers, performs data 
pre-processing and sends them
to the Compute Node
for on-line reconstruction of the physics signatures, like shower detection
and particle identification. 

\subsection {Calibration and Monitoring}

The FSC should have about 3$\percent$  energy resolution (stochastic term) and 3.5 mm
position resolution (at the centre of the cell). To fully utilise such a good performance
one needs to have a monitoring system to measure variations of the PMT gain
at the percent level or better in order to compensate the gain changes. 

Each FSC cell needs to be periodically calibrated. A pre-calibration with vertical 
cosmic-ray muons is a fast and reliable method to adjust the PMT gains, especially for 
the initial settings. 

The fine calibration of the FSC exploiting neutral pion
decays, will apply algorithms which are well known in high-energy physics.
The second algorithm exploits the E/p (energy/momentum) ratio for electrons from decays of
miscellaneous particles. In this method we transfer the energy scale from the forward
tracker to the FSC by measuring the E/p ratio for isolated electrons (E from the
calorimeter, p from the tracker). 
The fine calibration by these two methods  can be simultaneously performed for the entire 
FSC within a couple days.    

Two types of light monitoring system are going to be implemented for the 
\PANDA FSC. Each type has specific features and will be used for different goals.
The front-side monitoring system consists of a set of LEDs, one LED for 
each FSC module, to provide light for each module (i.e. four cells)
independently. Each LED is installed below the front cover of the module and illuminates
all the fibre loops of the module. 
This simple and inexpensive system can be useful for detector
commissioning and maintenance.

A more complex and precise light-monitoring system to monitor PMT gains
will be installed at the back side of the module. 
A fixed fraction of a light pulse 
is transported to each module by means of quartz fibres. Each fibre connects to
the optical connector of the module and is divided into four parts inside the module 
to inject light into each PMT. 
An essential element of the light monitoring system is a stable
reference photodetector with a good sensitivity at short wavelengths.


\subsection {Simulations}

The software tools,
reconstruction algorithms, and the digitisation procedures are described.
Simulations focused  on the threshold dependence of energy and spatial 
resolution for reconstructed photons and electrons, the influence 
of the material budget in front of the FSC, and the electron-hadron separation.
Because most of the physics channels have very low production cross section,
typically between pb and nb, a background rejection power up to 10${^9}$ 
has to be achieved. This requires an electromagnetic calorimeter which 
allows an accurate photon reconstruction in the energy range from 10-20 MeV
to 15 GeV and an effective and clean electron-hadron separation.


\subsection {Test-beam studies}

The first beam test was performed at the U70 accelerator in Protvino using a mixed secondary beam (electrons, muons,  $\pi^-$ and $K^-$)  
with a momentum tagging system for a matrix of the Type-1 shashlyk modules. At electron energies between 1 and 19 GeV, 
the energy resolution was compatible with the results from KOPIO and consistent with GEANT3 Monte-Carlo simulations. 
However, the position resolution of 6 mm in the centre of a module was insufficient for the required $\pi^0$ identification.
Therefore, the Type-2 shashlyk module has been produced by subdividing the Type-1 module into 4 cells which are read out independently. 
Due to smaller lateral cell size, an improved spatial resolution can be expected. 
The sampling ratio, the overall thickness, and number of layers have been kept the same as for Type-1. 
The lead plates, however, were kept in common for the four optically isolated cells. 
Two subarrays composed of 4$\times$4 or 3$\times$3 modules with a granularity of 8$\times$8 or 6$\times$6 individual cells 
have been prepared for test experiments at the IHEP Protvino facility and at the tagged-photon facility of MAMI in Mainz, Germany,
extending the response function for photons down to 50~MeV. To summarise the results briefly, three statements can be made: 
(a) The position resolution has indeed improved by almost a factor 2, 
(b) the single-cell time resolution is about 100 ps at 1 GeV, 
(c) however the strong position dependence of the relative energy resolution as much as 30$\percent$ makes Type-2 modules  absolutely unacceptable for \Panda. 

A variety of reasons was found to explain these findings, resulting in the improved Type-3 shashlyk module 
which incorporated many improvements in light collection and mechanical stability. A matrix of 4$\times$4 cells was tested with the tagged-photon beam at Mainz. To summarise the results:
The Type-3 shashlyk modules represent a significant improvement of the overall performance with respect to energy, position and time information necessary for shower reconstruction. The significantly better reproducibility and homogeneity of the modules and individual cells should guarantee to reach the necessary performance even at photon energies as low as 10-20~MeV. The implementation of additional reflector material and WLS fibres of better quality and light collection have almost doubled the recorded light yield which now amounts to $\sim$2.8$\pm$0.3 photo electrons per MeV deposited energy. 
The obtained experimental results can be well understood and reproduced by a standard simulation based on GEANT4.

\subsection {Conclusion} 

Measurements {\bf at high energies} of the energy and position resolutions of 
two electromagnetic 
calorimeter prototypes exploiting the
fine-sampling technique for the \PANDA
experiment at FAIR have been carried out at the IHEP Protvino test beam facility using the 70 GeV accelerator.
Studies were made in the electron beam energy range from 1 to 19 GeV. The energy tagging has allowed us to measure 
the stochastic term in the 
energy resolution as about 3\percent for both prototypes.
Taking into account the effect of the light 
transmission in the scintillator tiles
and WLS fibres, photon statistics as well as the noise of the entire electronic chain, resulted in a good agreement 
between the measured energy 
resolutions and the GEANT Monte Carlo simulations. 

However, the non-uniformity of the energy response in the Type-2 prototype was as large as  30\percent, and this observation led to
the design and assembly of the Type-3 prototype in order to drastically improve the non-uniformity of the energy response. 
The Type-2 and 
Type-3  prototypes were tested in 2012 and 2014 {\bf at low energies} 
at MAMI in Mainz with
electrons of energy between 100 and 700 MeV. 

It has turned out that  the performance parameters of the  Type-3 
prototype with cell sizes of 5.5$\times$5.5 {\ensuremath{\cm^2}\xspace}, with four pins between the scintillator and lead plates,
with scintillator plates wrapped by Tyvek, and with the KURARAY optical fibres 
fulfil all the basic requirements for 
the electromagnetic calorimeter of the
\PANDA Forward Spectrometer, including uniformity of the energy response. The overall detector design
appears to be well appropriate for the \PANDA application also with respect to the lateral and longitudinal
dimensions to provide the required granularity and to minimise shower leakage up to the highest photon energies. 

The necessary mass production of such 
high-quality finely segmented 
modules has been achieved at IHEP Protvino. The general layout of the mechanical structure is completed including 
the estimates of the integration into the \PANDA detector. Such a calorimeter
with 1512 readout channels can be produced at IHEP Protvino within three-four years. 
Experimental
data together with the elaborate design concepts and simulations for the  finely segmented shashlyk electromagnetic 
calorimeter of the Forward Spectrometer
show that the ambitious physics program of \PANDA can be fully explored based on the measurement of electromagnetic probes, 
such as photons, 
electrons and positrons, or the reconstruction of the invariant mass of neutral mesons.

%


%

%% file: panda_tdr_FSC_int_pub.tex
%
%
\cleardoublepage
\chapter{Overview of the \PANDA Experiment}
\label{sec:int}

\section{FAIR, HESR, and goals of the \PANDA experiment}
	The \PANDA experiment at the Facility for Antiproton and Ion Research (FAIR) 
 will use the antiproton beam from the High Energy Storage Ring (HESR) colliding with 
an internal proton target and a general purpose spectrometer to carry out a rich and 
diversified hadron physics program. FAIR is an international research centre located
at the site of GSI in Darmstadt, Germany, and funded by 16 countries. FAIR is a very versatile
particle accelerator complex providing a variety of experimental facilities for a large number of international teams.

\begin{figure*}
\begin{center}
\includegraphics[width=0.99\dwidth]{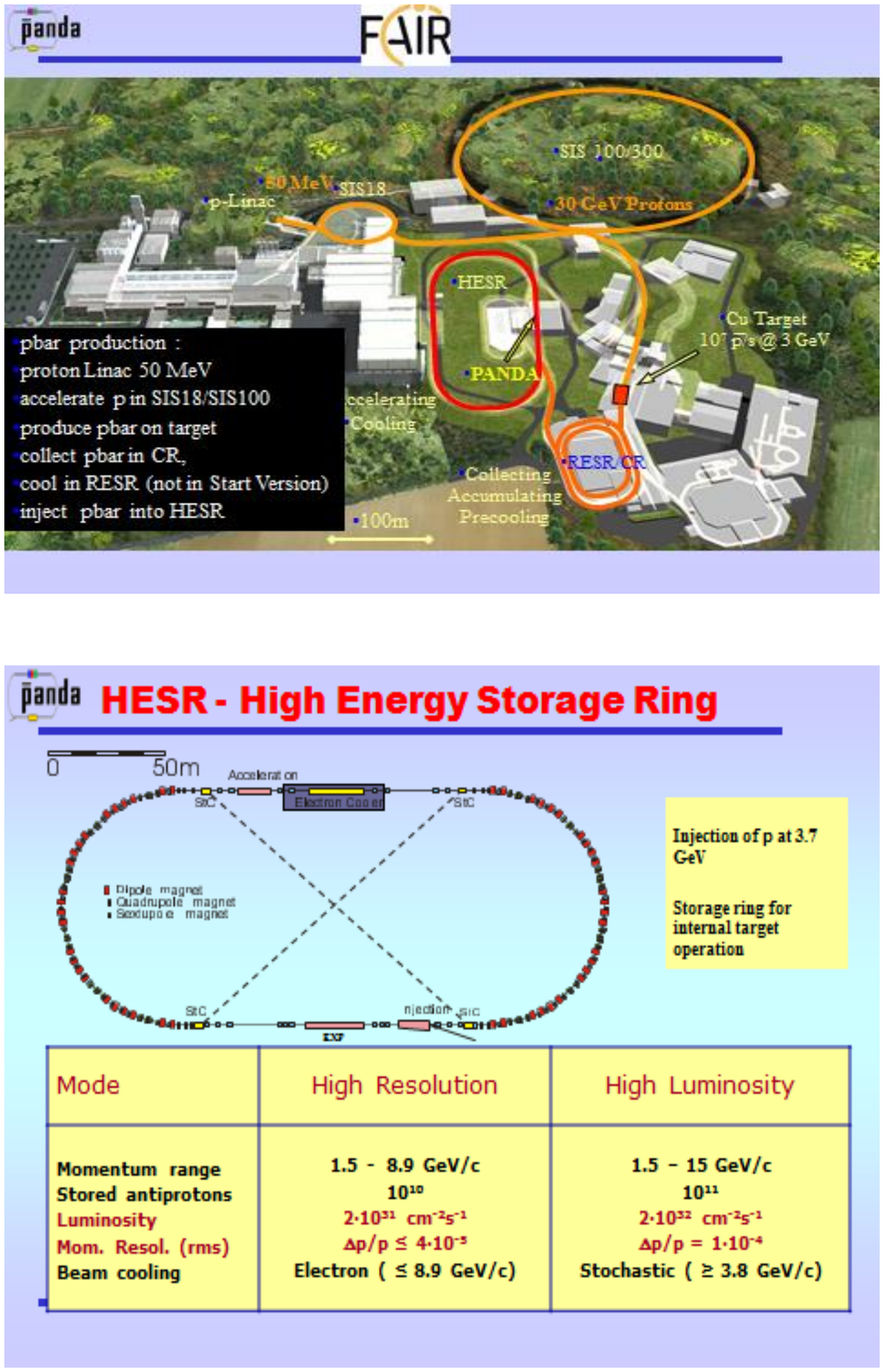}
\caption[\FAIR accelerator complex and HESR schematic view]
{\FAIR accelerator complex: top, global overview; bottom, HESR schematic view.}
\label{fig:FAIR_and_HESR}
\end{center}
\end{figure*}

{\bf FAIR accelerator complex}. The FAIR accelerator complex is shown in 
\Reffig{fig:FAIR_and_HESR} and is described in detail in 
\cite{fair_baseline}.
The Modularised Start Version (MSV) is described in 
\cite{fair_start_version}. The starting point of all future particle beams at FAIR
is the existing GSI research centre with its
universal linear accelerator (UNILAC) and the Schwerionen-Synchrotron SIS-18.
In the following, we concentrate on the production of antiprotons.

{\bf Antiproton production}. The existing GSI accelerator will be upgraded by a proton linear accelerator 
(p-LINAC), feeding SIS-18
by multi-turn injection with a 70 MeV proton beam of 35 mA current. Roughly 2$\cdot$10$^{12}$ protons
will be accumulated and accelerated to a kinetic energy of 2 GeV
\cite{proton_linac}. Subsequently, the protons are transferred to the SIS-100, a superconducting fast cycling
synchrotron with a bending power of 100 Tm and a circumference of 1083.6 m. Several injections from 
the SIS-18 are needed to accumulate roughly 2$\cdot$10$^{13}$ protons before these are further 
accelerated
to a final energy of 29 GeV. An additional acceleration to 90 GeV by the SIS-300 (not in the MSV), 
a superconducting synchrotron with a bending power of 300 Tm, will later be possible but not used
for antiproton production \cite{antiproton_complex}.

The high-energy protons from the SIS-100 will be directed in bunches of 50 ns on 
a nickel target for antiproton production every 10 seconds. The repetition
rate is limited by the cycle length of the subsequent Collector Ring (CR). The remaining
proton accelerator time is shared among other experiments running in parallel to the antiproton production
beam line.   

{\bf Antiproton extraction}. As a result of protons interacting with a solid-state 
target, a large diversity of secondary particles is being produced. By the very nature of the many random processes the momentum
and angular distributions of those particles are very wide. For a high antiproton 
collection efficiency a combination of a magnetic horn and a momentum separation
station is foreseen on the transfer path to the CR. Antiprotons will be accepted with
a momentum of 3.8 GeV/$c$ {$ \pm 3 \percent $}, while the transverse emittance
is expected to be cut to 240 mm mrad by that transfer beam line setup 
\cite{antiproton_complex}.

{\bf CR}. The collector ring (CR) provides full acceptance of those separated antiprotons.
The major task is the collection of transported antiprotons, the cooling of the large
phase space and the de-bunching of the beam within one cycle \cite{collector_ring}.
The momentum of the antiprotons at injection into the CR and at extraction from the CR
is the same, namely 3.8 GeV/$c$. However, the momentum bite  {$\delta$}p/p which is 
{$ 3 \percent$}  at injection, will be decreased down to {$0.1 \percent$} at extraction.
Similarly the transverse emittance will be decreased from 240 mm mrad at injection
down to 5 mm mrad at extraction. In the first years of physics runs, the pre-cooled   
beam will be directly fed into the HESR beam line. A later upgrade involves the
construction of the Recuperated Experimental Storage Ring (RESR) \cite{resr}.

{\bf RESR}. The need for a high-intensity beam requires an accumulation of 
antiprotons coming from the CR. The RESR
will be located in the same hall as the CR. It is designed to accumulate within
3 hours up to 10$^{11}$ antiprotons at a momentum of 3.8 GeV/$c$ \cite{tdr_resr}.
During accumulation the beam emittance is further reduced by stochastic cooling.

{\bf HESR}. Although the high-energy storage ring HESR 
(see \Reffig{fig:FAIR_and_HESR}) will not exclusively host the \PANDA experiment, 
it is specially designed for its demands. Antiprotons at FAIR will reach the HESR through the following stages:
proton source, 70 MeV proton linac, proton acceleration in SIS18/SIS100, \pbar production on a target, 
\pbar collection and pre-cooling in the CR, \pbar cooling in the RESR (not in the FAIR MSV), \pbar injection into HESR.

An important feature of the HESR is the combination of phase-space cooled
beams and dense internal targets, comprising challenging beam parameters in 
two operation modes. The first one is high-luminosity mode with up to 10$^{11}$ particles in the ring
(peak luminosity up to 2$\cdot$10$^{32}$ cm$^{-2}$ s$^{-1}$). 
The second operation mode is the high-resolution 
mode with up to 10$^{10}$ particles in the ring (peak luminosity up to 
2$\cdot$10$^{31}$ cm$^{-2}$ s$^{-1}$) and a momentum spread down to a 
few times 10$^{-5}$. Powerful stochastic and electron cooling 
systems are necessary to meet the 
experimental requirements. 
The racetrack-shaped storage ring consists of two 180\degrees arcs and two 155 m long 
straight sections with a maximum of possible symmetry in beam optics. The complete  lattice consists of 44 dipole magnets for bending and 84 quadrupole magnets for focussing.
The total circumference sums up to 575 m. 

On the eastern straight section, injection kicker magnets are placed as well as the \PANDA
experiment with the internal target system.  Among the main components of the HESR are
multi-harmonic RF-cavities which allow to accelerate or decelerate the antiproton
beam in the momentum range of 1.5 GeV/$c$ to 15 GeV/$c$ for the high-luminosity mode and 
1.5 GeV/$c$ to 8.9 GeV/$c$ for the high-resolution mode. 
\PANDA contains a solenoid and a dipole magnet and
both have to be compensated by respective anti-fields in the magnetic chicane. The
compensating solenoid magnet will most likely be placed upstream of \PANDA, whereas
two dipole magnets, upstream and downstream, compensate the dipole field at
\PANDA with a total bending angle of 40 mrad \cite{hesr}. 

The other straight section will host the beam cooling system. It consists of a powerful
stochastic cooling system with its pick-ups on the \PANDA side and an electron cooler for
lowest transverse emittances of beam particles. In addition, the KOALA \cite{koala} 
and SPARC \cite{sparc} experiments are planned to be placed on the beam cooler side 
of the HESR. However, up to now the exact locations of those experiments are not yet
fixed.

{\bf The goals of \PANDA}. The \PANDA experiment is being designed to fully exploit 
the extraordinary physics potential arising from the availability of high-intensity 
cooled antiproton beams.  The aim of the versatile experimental program is to 
answer burning questions in the field of Quantum Chromo Dynamics (QCD). In contrast
to collider experiments as CMS, ATLAS or LHCb at CERN with proton-proton collisions,
or CDF and D0 at Fermilab, USA, with antiproton-proton collisions, the emphasis is not put
on highest centre-of-mass (c.m.) energies for studies at the energy frontier.
\PANDA was designed for very high precision measurements in the lower c.m. energy
regime between 2.3 and 5.5 GeV. The high precision is achieved both by a state-of-the-art
detector design and by a very high interaction rate of a high precision 
antiproton beam with an internal stream of hydrogen atoms as target material.  Thanks to the expected tremendous boost 
in statistical accuracy and precision of measured data, sufficient progress beyond 
the present understanding of the field will be made.
The annihilation of antiprotons and protons gives direct access to a broad range of particles, which is
not restricted to vector-meson production like in electron-positron collisions.

	The study of QCD bound states is of fundamental importance for a better, 
quantitative understanding of
 QCD. Particle mass spectra can be computed within the framework of non-relativistic
potential models, effective field theories, and Lattice QCD. Precision measurements 
are needed to distinguish 
between the different approaches and identify the relevant degrees of freedom. 
The studies to be carried out in \PANDA include the spectroscopy of charmonium 
states and other heavy hadrons, the search for exotic states such as gluon-rich hadrons 
(hybrids and glueballs), multi-quark and molecular states, heavy hadrons in matter, and
hypernuclei.  In addition to the spectroscopic studies, \PANDA will
be able to investigate the structure of the nucleon using
electromagnetic processes, such as Wide Angle Compton Scattering
(WACS) and the process $\pbarp \to \ee$, which will allow the
determination of the electromagnetic form factors of the proton in the
time-like region over an extended q$^2$ region.

Another interesting topic is the study of hyperon and vector meson polarisation in a wide range
of production angles and at different energies. These studies allow to investigate the nature 
of the strong interaction in general and the origin of polarisation phenomena in particular.
\section{The physics case}
One of the most challenging and fascinating goals of modern nuclear and hadronic physics is
the achievement of a quantitative understanding of the strong interaction. 
Lattice QCD predicts the
existence of a whole spectrum of bound states of gluons, glueballs, and in addition
gluonic excitations of hadrons, the hybrids. Antiproton-proton annihilation via two- or 
three-gluon processes have been shown at LEAR experiments to be a copious source of
gluons, so especially gluonic degrees of freedom may be favourably studied compared
to other processes where the presence of quark and antiquark suppresses the production of 
glueballs. 
LEAR experiments and the E760/E835 experiments at Fermilab demonstrated the unique advantage of
the combination of an intense high-resolution 
antiproton beam with a state-of-the-art 4{$\pi$}-experiment in the field of hadron
spectroscopy. 
In these experiments not only the most prominent glueball
candidate f$_0$(1500) was discovered, they also dominate the relevant PDG results 
with their precision, which is due to cooled antiproton 
beams achieved by stochastic and electron cooling.

Significant progress has been achieved in recent years thanks to
considerable advances in experimental techniques and theoretical understanding. 
New experimental
results have stimulated a very intense theoretical activity and a
refinement of the theoretical tools. Still there remain a number of fundamental
questions which so far can only be answered qualitatively. Phenomena such as the
confinement of quarks, the existence of glueballs and hybrids, and the
origin of the masses of hadrons in the context of 
chiral-symmetry breaking are long-standing puzzles and represent the
intellectual challenge in our attempt to understand the nature of the
strong interaction and of hadronic matter.

Experimentally, the structure of hadrons  can be studied with
various probes such as electrons, pions, kaons, protons or
antiprotons. In antiproton-proton annihilations, however, particles with gluonic
degrees of freedom as well as particle-antiparticle pairs are
copiously produced, allowing spectroscopic studies with very high
statistical accuracy and unprecedented precision due to the direct production 
of resonances with various spin-parity quantum numbers. 
Therefore, antiproton annihilations are an excellent tool
to address the open fundamental problems in hadron physics.

The physics scope of \PANDA is subdivided into several pillars: heavy hadron
spectroscopy, including charmonium studies and the search for exotic forms of matter
(glueballs, hybrids), heavy hadrons in matter, open-charm physics, hypernuclei,
nucleon structure, and hyperon polarisation. 
\subsection{Charmonium spectroscopy}

	Ever since its discovery in 1974 the charmonium system has been a
 powerful laboratory for improving the understanding of the strong interaction. The
 high mass of the $c$ quark (1.5 GeV) makes it plausible to attempt a
 description of the dynamical properties of the \ccbar system in terms
 of non-relativistic potential models, in which the functional form of
 the potential is chosen to reproduce the asymptotic properties of the
 strong interaction. The free parameters in these models are to be
 determined from a comparison with the experimental data. Now, more
 than forty years after the \jpsi discovery, charmonium physics
 continues to be an exciting and interesting field of research. The
 recent discoveries of new states (e.g. X(3872)), and the
 exploitation of the $B$ factories as rich sources of charmonium states
 have given rise to renewed interest in heavy quarkonia, and
 stimulated a lot of experimental and theoretical activities.

The gross features of the charmonium spectrum are reasonably well
 described by potential models, but these obviously cannot cover the
 whole story: relativistic corrections are important and coupled-channel 
effects are significant and can
 considerably affect the properties of the \ccbar states. To explain the
 finer features of the charmonium system, model calculations and
 predictions are made within various complementary theoretical
 frameworks. Substantial progress in an effective field theoretical
 approach, called Non-Relativistic QCD, has been achieved in
 recent years. Complementary to this analytical approach, we may expect
 significant progress in lattice gauge theory calculations, which have
 become increasingly more capable in dealing quantitatively with
 non-perturbative dynamics in all its aspects, starting from the first
 principles of QCD.

Experimentally the charmonium system has been studied mainly in \ee and
{\ensuremath{\bar{p}p}\xspace} experiments. In \ee annihilations direct
charmonium formation is possible only for states with the quantum
numbers of the photon, $J^{PC} = 1^{--}$, namely J/$\psi$, $\psi\prm$
and $\psi$(3770) resonances.  Precise measurements of masses and
widths of these states can be obtained from the energy of the electron
and positron beams, which are known with good accuracy. All other
states can be reached by means of other production mechanisms, such as
photon-photon fusion, initial-state radiation, $B$-meson decay and
double charmonium production.

	On the other hand, all \ccbar states can be directly formed in
 {\ensuremath{\bar{p}p}\xspace} annihilations through the coherent
 annihilation of the three quarks in the proton with the three
 antiquarks in the antiproton. This mechanism, originally proposed by
 P. Dalpiaz in 1979, could be successfully exploited a few years later
 at CERN and FNAL thanks to the development of stochastic
 cooling. With this method the masses and widths of all charmonium
 states can be measured with excellent accuracy, determined by the
 very precise knowledge of the initial {\ensuremath{\bar{p}p}\xspace}
 state, and not limited by the resolution of the detector. The search for
 new hadronic states with heavy quarks is mandatory, in order to investigate,
 which characteristic features of QCD are realised in the  nature of
 elementary particles.  There are many candidates for charmonium-like
 particles above the \DDbar threshold (the open-charm threshold), which have
 been observed during the past few years at BELLE \cite{belle}, BaBaR \cite {babar}, 
CDF \cite{cdf},  BESIII \cite {bes3} and LHCb \cite {lhcb}. 
Among the observed (XYZ)-resonances are states like 
X(3872), X(3940), Y(3940), Z(3930), X(4160),
Y(4008), Y(4664), Z(4430), Z(4058), Z(4258), X(4630),Y(4260),
Y(4320),Y(4140), where  X(3872) and Z(4430) are the most reliable states at present.
Recently, three spectacular charged charmonium-like Z-states were observed by at least two 
independent experiments: Zc(3900) \cite{newZ1, newZ2}, Zc(4020) \cite{newZ3, newZ4}, and Zc(4430)$^{\pm}$ \cite{newZ5, newZ6}. 
These states have quantum numbers J$^P$=1$^+$ and were observed by BESIII, BELLE or LHCb.

Since charmonia decay via leptonic channels, which are otherwise rare at these 
energies in the hadronic environment, \PANDA can often profit from clean 
signal/background ratios. For the newly discovered X, Y, Z states, which often
possess very narrow widths, a precise determination of the excitation curve is
necessary to distinguish between the different theoretical interpretations. This can
be done decisively better with \PANDA in formation mode due to the strong phase-space cooling
of the antiproton beam in HESR, compared to the production of these states in the
decay chain of heavier particles.

\subsection{Search for gluonic excitations (glueballs and hybrids)}

One of the main challenges of hadron physics is the search for gluonic
excitations, i.e. hadrons in which the gluons can act as constituent partons. 
These gluonic hadrons fall into two main categories:
glueballs, i.e. states of pure glue, and hybrids, which consist
of a \qqbar pair and excited glue. The additional degrees of freedom
carried by gluons allow these hybrids and glueballs to have $J^{PC}$
exotic quantum numbers: in this case mixing effects with nearby \qqbar
states are excluded and, consequently, their experimental identification
becomes easier. The properties of glueballs and hybrids are determined by the
long-distance features of QCD and their study will yield fundamental
insight into the structure of the QCD vacuum. Antiproton-proton
annihilations provide a very favourable environment to search for
gluonic hadrons.

%
%

\subsection {Study of hadrons in nuclear matter}

The influence of matter on hadrons will be investigated with
various target materials.  
Medium modifications of hadrons embedded in hadronic matter are studied in order 
to understand the origin of hadron masses in the context of spontaneous chiral symmetry 
breaking in QCD and its partial restoration in a hadronic environment.  So far, 
experiments have been
focused on the light-quark sector. The high-intensity \pbar beam with momentum of up
to 15 GeV/$c$ will allow an extension of this program to the charm
sector both for hadrons with hidden and with open charm. The in-medium
masses of these states are expected to be affected primarily by the gluon condensate.

Another sensitive study which advantageously can be carried out in \PANDA is the 
measurement
of \jpsi and $D$ meson production cross sections in \pbar annihilations
on a series of nuclear targets. The comparison of the resonant \jpsi
yield obtained from \pbar annihilation on protons and different
nuclear targets allows to deduce the \jpsi-nucleus dissociation
cross section, a fundamental parameter to understand \jpsi
suppression in relativistic heavy ion collisions interpreted as a
signal for quark-gluon plasma formation.

In contrast to other experiments, \PANDA has a special design consisting of 
a dipole magnet in the Forward Spectrometer that bends the accelerator beam 
away from the zero-degree region. Thus, basically the full solid angle for decay 
products is covered, providing an enormous advantage for threshold 
experiments and for the determination of the quantum numbers of quarkonia 
with the help of amplitude analysis. For the implantation of strange or 
charmed baryons in nuclear matter the existence of a particle-antiparticle pair creates  
the unique advantage that the detection of either one
of them provides an excellent trigger to study the reaction of the partner 
inside nuclear matter.

\subsection {Open-charm spectroscopy}

The energy range around the open-charm threshold allows direct observation 
of open-charm pairs. With the HESR running at full luminosity and at 
\pbar momenta larger than
6.4 GeV/$c$ a large number of $D$ meson pairs will be produced. The high
yield (e.g. 100 charm pairs per second) around the $\psi$(4040) and the
well defined production kinematics of $D$ meson pairs will allow to
carry out a significant charm-meson spectroscopy program which would
include, for example, the rich $D$ and \Ds meson spectra. Also here 
energy scan experiments, especially at production thresholds, will be 
performed. 

A very typical excited open-charm hadron state is the recently observed 
D$^*$$_s$$_0$ (2317). There exists a great uncertainty as to whether it can 
be described more adequately as a two-meson molecule with a leading four-quark 
Fock state or as a conventional meson with a leading two-quark Fock
state. \PANDA will measure the resonance shape of that narrow state which 
is directly connected with the inner structure of the resonance.

\subsection {Hypernuclear physics}

Hypernuclei are systems in which up or down quarks are replaced by
strange quarks. In this way a new quantum number, strangeness, is
introduced into the nucleus. Although single and double
{\ensuremath{\Lambda}\xspace}-hypernuclei were discovered many decades
ago, only 6  double {\ensuremath{\Lambda}\xspace}-hypernuclei
were observed up to now.  \PANDA will search for new states in a later stage. 
To this end, target materials in the form of wires will be inserted into \PANDA. The availability 
of \pbar beams at FAIR will
allow efficient production of hypernuclei with more than one strange
hadron, making \PANDA competitive with planned dedicated
facilities. This will open new perspectives for nuclear-structure
spectroscopy and for studying the forces between hyperons and nucleons.

\subsection{Electromagnetic form factor in the time-like region} 

The electromagnetic probe is an excellent tool to investigate the
structure of the nucleon. The \PANDA experiment offers the unique
opportunity to make a precise determination of the electromagnetic
form factors in the time-like region with unprecedented accuracy. The
form factors (FF) measured in electron scattering are  internally
connected with those measured in the annihilation process.  Moreover,
they are observables that can test our understanding of the nucleon
structure in the regime of non-perturbative QCD as well as at higher
energies where perturbative QCD applies.

The interaction of the
electron with the nucleon is described by the exchange of one photon
with space-like four-momentum transfer q$^2$.  The
lepton vertex is described completely within QED and on the nucleon
vertex, the structure of the nucleon is parametrised by two real
scalar functions depending on one variable q$^2$
only. These real functions are the Dirac form factor and the Pauli
form factor. The form factors are analytical functions of the four-momentum 
transfer q$^2$ ranging from
q$^2$ = $-\infty$ to $+\infty$.  While in
electron scattering the form factors can be accessed in the range of
negative (space-like) q$^2$, the annihilation
process allows to access positive (time-like) q$^2$. 
Unitarity of the matrix element requires that space-like
form factors are real functions of q$^2$ while for
time-like q$^2$ they are complex functions.

The \PANDA experiment offers a unique opportunity to determine the moduli
of the complex form factors in the time-like domain by measuring the
angular distribution of the process in a q$^2$
range from about 5 (GeV/$c$)$^2$ up to 14 (GeV/$c$)$^2$. A determination of the
magnetic form factor up to a q$^2$ of 22 (GeV/$c$)$^2$
will be possible by measuring the total cross section.

\subsection {Polarisation of hyperons and vector mesons}

        Spin is one of the fundamental quantum properties of particles, and understanding the spin of particles
 can help to penetrate deeper into their structure and the
 interaction dynamics. Presently there are no plans to combine 
\PANDA experiments with
 polarised beams or polarised targets. Nevertheless, the availability of
 4$\pi$-geometry and a detector system capable of identifying both charged and
 neutral particles allow to determine the polarisation of hyperons and vector
 mesons by measurements of angular distributions of their decay products.
 Measurements carried out in hadron-hadron, hadron-nucleus,
 nucleus-nucleus and lepton-nucleus collisions revealed large, in the 
order of some ten percent, polarisation of secondary
 particles with spin 1/2 and spin 1. 

Significant polarisation was observed
 not only in inclusive, but also in exclusive reactions. Polarisation
 effects are especially significant at moderate energies and diminish 
rapidly as the energy is increasing. 
 Spin effects can be associated with fundamental problems of the strong
 interaction, such as spontaneous chiral-symmetry breaking, the appearance
 of masses of hadrons and quarks, the formation of dynamical quarks,
 and the emergence of their large anomalous chromo-magnetic moments, as
 well as quark confinement.

\section{The \PANDA detector}

\begin{figure*}
\begin{center}
\includegraphics[width=0.8\dwidth]{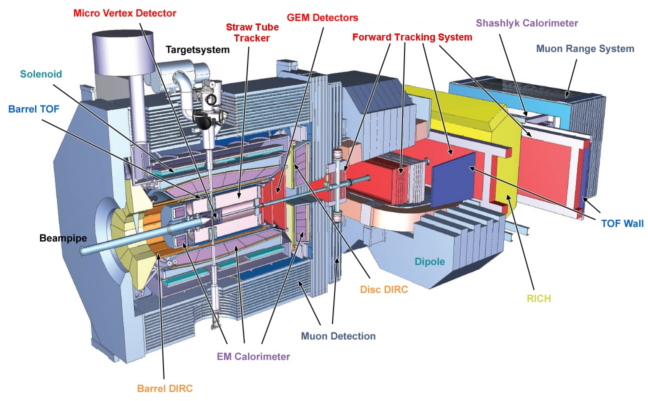}
\caption[\PANDA detector overview]
{Layout of the \Panda detector consisting of a Target Spectrometer, surrounding the interaction region, and a Forward Spectrometer to detect particles emitted into the forward region. The beam traverses \Panda from left to right.}
\label{fig:panda_det}
\end{center}
\end{figure*}

The main objectives of the design of the \PANDA experiment are the coverage of the
full geometrical acceptance, high resolution for tracking, particle
identification and calorimetry, high rate capabilities, and a versatile
readout providing efficient event selection. To obtain a good momentum resolution, the
detector (see \Reffig{fig:panda_det}) is split into a Target Spectrometer housed in a
superconducting solenoid magnet surrounding the interaction point and a Forward Spectrometer 
employing a dipole magnet for the momentum analysis of charged small-angle tracks. A silicon vertex detector surrounds the
interaction point.   Tracking and identification of charged
particles, electromagnetic calorimetry and muon
identification are available in both spectrometer parts to enable the detection of the complete mass range
of final-state particles relevant for the \PANDA physics objectives.

An almost full acceptance and good momentum resolution over a
large momentum range is achieved by combining a solenoid magnet for deflecting tracks with high transverse
momenta (in the Target Spectrometer) and a dipole magnet for the momentum analysis of 
forward-going reaction products (in the Forward Spectrometer). 

The solenoid magnet is designed to analyse particles emitted between 5\degrees and
140\degrees in the vertical and between 10\degrees and 140\degrees in the horizontal plane. The field is 
generated by a barrel-shaped superconducting magnet enclosed by 
ferromagnetic flux return yokes. The superconducting solenoid magnet 
provides a longitudinal field  of up to 2~T for beam momenta above the
injection momentum of 3.8 GeV/$c$. Below that momentum, the field strength has to be reduced 
to 1~T in order to ensure a stable lattice setting of the HESR. 
The solenoid has a coil opening of 1.89 m and a coil length of 2.75 m.

The  Target Spectrometer is arranged in a barrel part for angles larger
than 22\degrees and an endcap part for the forward range down to
5\degrees in the vertical and 10\degrees in the horizontal plane.

The dipole magnet with a window frame has a 1 m gap and more than 2 m aperture. 
In the current planning, the magnet yoke will occupy 
about 2.5 m in beam direction starting from 3.5 m downstream of the 
target.  The maximum bending power of the magnet 
will be 2 Tm and the resulting deflection of the antiproton beam at the 
maximum momentum of 15 GeV/$c$ will be 2.2\degrees. The design acceptance for 
charged particles covers a dynamic range of a factor 15 with the detectors 
downstream of the magnet. For particles with lower momenta, detectors will 
be placed inside the yoke opening. The beam deflection will be compensated 
by two correcting dipole magnets, placed in the beam line around the \PANDA detector.

The deflection of particle trajectories in the field of the dipole magnet 
will be measured with a set of straw tubes of 1 cm diameter, two stations 
placed in front, two within and two behind the dipole magnet. This arrangement will 
allow to track particles with highest momenta as well as very low momentum 
particles whose tracks will curl up inside the magnetic field. A wall of 
slabs made of plastic scintillator, read out on both ends by fast photo tubes, 
will serve as time-of-flight stop counter placed at about 7 m from 
the target. In addition, similar detectors will be placed inside the dipole 
magnet opening, to detect low-momentum particles which do not exit the 
dipole magnet. 

For the detection of photons and electrons a shashlyk-type 
calorimeter with high resolution and efficiency will be employed. The 
detection is based on lead-scintillator sandwiches read out with wavelength 
shifting fibres passing through the block and coupled to photomultipliers. 
To cover the forward acceptance, 28 rows and 54 columns are required with 
a cell size of 55$\times$55 mm$^2$, i.e. 1512 cells in total, which will be placed at 
a distance of 7 to 8 m from the target. For the very forward part of the 
muon spectrum a further range tracking system consisting of interleaved 
absorber layers and rectangular aluminium drift-tubes is being designed, 
similar to the Muon Detection system of the Target Spectrometer, but laid out for 
higher momenta. The system allows discrimination of pions from muons, 
detection of pion decays and, with moderate resolution, also energy 
determination of neutrons and antineutrons.

\subsection {Targets}


     In order to reach the designed peak luminosity of 2$\cdot$10$^{32}$ cm$^{-2}$s$^{-1}$ a target 
thickness of about 4$\cdot$10$^{15}$ hydrogen atoms per cm$^2$ is required assuming 10$^{11}$ 
stored antiprotons in the HESR ring. 
These conditions pose a real challenge for an internal target inside a storage ring. At present, 
two different, complementary techniques for the internal target \cite{targets_tdr} are being developed: the cluster-jet target 
and the pellet target. Both techniques are capable of providing sufficient densities for hydrogen at the 
interaction point, but exhibit different properties concerning their effect on the beam quality and the 
definition of the interaction point. 
A cluster-jet target with thickness up to 2$\cdot$10$^{15}$ atoms/cm$^{2}$ is expected 
to be available for \PANDA experiments. 
This target has the advantage that the density can be varied continuously, allowing 
to provide a constant event rate. The weaker definition of the interaction point compared 
to the pellet target, however, is a disadvantage. 
For a pellet target we may assume an average thickness of 4$\cdot$10$^{15}$ hydrogen atoms per 
cm$^2$. 
Further developments are needed to minimise instantaneous jumps in event rate due to variations in the pellet size,  
followed by sharp dips in the event rate in the intervals between subsequent pellets. In addition, other internal targets of 
heavier gases, like deuterium, nitrogen or argon can be made available. For non-gaseous nuclear targets 
the situation is different in particular in case of the planned hyper-nuclear experiment. In these studies 
the whole upstream endcap and part of the inner detector geometry will be modified.

\subsection {Tracking system}

 The design of the micro-vertex detector (MVD) \cite{mvd_tdr} for the Target Spectrometer 
is optimised  for the detection 
of secondary vertices from D-mesons and hyperon decays and a maximum acceptance close 
to the interaction 
point. The MVD will also strongly improve the transverse momentum resolution. The concept 
of the MVD is based 
on radiation hard silicon pixel detectors with fast individual pixel readout circuits 
and silicon strip 
detectors. The layout foresees a four-layer barrel detector with an inner radius 
of 2.5 cm and an outer 
radius of 15 cm. It is about 40 cm long. The two innermost layers will consist 
of hybrid pixel detectors while 
the outer two layers 
are considered to consist of double-sided silicon strip detectors. Six detector wheels 
arranged 
perpendicular to the beam will achieve the best acceptance for the forward part of 
the particle spectrum. 
Here again, the inner four disks are  made of hybrid pixel detectors and the following 
two are a combination 
of strip detectors on the outer radius and pixel detectors closer to the beam pipe. 
Additional silicon strip disk layers are considered further downstream to 
achieve a better acceptance for hyperon cascades. The present design of the pixel 
detectors comprises silicon sensors, 
which are 100 $\mu$m thick corresponding to 0.1\percent of a radiation length.

     The charged particle tracking devices must handle the high particle fluxes that are anticipated for 
a luminosity of up to several 10$^{32}$ cm$^{-2}$s$^{-1}$. The momentum resolution {$\delta$p}/p has to be on 
the percent level. The detectors should have good detection efficiency for secondary vertices which can 
occur outside the inner vertex detector (e.g. from K$^0$$_S$). This is achieved by the combination of the MVD 
close to the interaction point with two outer systems. One is covering a large area and is designed as 
a barrel around the MVD. This will be a stack of straw tubes and is called straw tubes tracker (STT) \cite{straw_tube_tdr}. 
The forward angles will be covered using three sets of GEM trackers. The STT consists of aluminised 
Mylar straw tubes which are self-supporting by the operation at 1 bar overpressure. The straws are 
arranged in planar layers which are mounted in a hexagonal shape around the MVD. In total there are 
27 layers of which the eight central ones are tilted to achieve an acceptable resolution of 3 mm also 
in z-direction (i.e. parallel to the beam), while the resolution in the transverse directions is anticipated to be 150 $\mu$m.  
The gap to the surrounding detectors is filled with further individual straws. In total there are 
4600 straws around the beam pipe at radial distances between 15 cm and 42 cm with an overall length 
of 150 cm. All straws have a diameter of 1 cm.

\subsection {Particle Identification}

Charged-particle identification is required over a large momentum range from 200 MeV/$c$ up to
almost 10 GeV/$c$. To this end, various physics processes are employed.
The main fraction of charged particles is identified by several Cherenkov detectors. In the Target Spectrometer
two DIRC detectors based on the Detection of Internally Reflected Cherenkov light are being developed, one
consisting of long rectangular quartz bars for the barrel region (BarrelDIRC), the other one (DiscDIRC) being shaped as a disc for the forward endcap.

Time of flight can be partially exploited in \PANDA, although no dedicated start detector is available. 
Using a scintillator tile hodoscope(SciTil) covering the barrel section in front of the electromagnetic calorimeter (EMC) and
a scintillator wall after the dipole magnet, relative timing of charged particles with
very good time resolution of about 100 ps can be achieved.  The energy loss within the trackers will be employed 
as well for particle identification below 1 GeV/$c$ since the individual charge is obtained by analog
readout or a time-over-threshold measurement. 
 
Muons are an important probe for e.g. \jpsi decays, semi-leptonic 
D-meson decays, and the Drell-Yan process. The majority of background particles 
are pions and muons as their decay daughter particles. However, at the low momenta of the
\PANDA experiments the signature is less clean than in high-energy physics experiments. 
Nevertheless, in order to provide a proper separation of primary muons from pions and 
decay muons, a range tracking system will be implemented in the yoke of the 
solenoid magnet. Here, a fine segmentation of the yoke as absorber with 
interleaved tracking detectors allows to distinguish energy-loss 
processes of muons and pions from kinks due to pion decays. Only in this way a 
high separation of primary muons from the background can be achieved. 
In the barrel region the yoke is segmented in a first layer of 6 cm iron 
followed by 12 layers of 3 cm thickness. The gaps for the detectors are 
3 cm wide. This amount of material is sufficient for the absorption of pions 
in the momentum and angular range of \PANDA. 

\subsection{Scintillator tile barrel (time-of-flight)}

For slow particles at large polar angles, particle identification will be supported 
by a time-of-flight (TOF) detector positioned just outside the Barrel DIRC, where 
it can also be used to detect photon conversions in the DIRC radiator. The SciTil 
detector is based on very fast scintillator tiles of 28.5$\times$28.5 mm$^2$ size, individually 
read out by two silicon photo multipliers (SiPM) attached to each end of a tile. 
The full system consists of 5,760 tiles in the barrel part.  Material budget and 
the dimension of this system are optimised such that a value of less than 2\percent of 
one radiation length, including readout and mechanics, and less than 2 cm 
radial thickness will be reached, respectively. The expected time resolution 
of 100 ps will allow to achieve precision timing of tracks for event building 
and fast software triggers. The detector also provides well-timed input with 
a good spatial resolution for on-line pattern recognition.

\subsection {Electromagnetic calorimetry}

In the Target Spectrometer high-precision electromagnetic calorimetry is required over a large
range from a few MeV up to several GeV in energy deposition. Lead-tungstate (PWO) is chosen as scintillating crystal for the calorimeters in the Target
Spectrometer because of its high density, fast response, and good light yield, enabling high energy resolution and a compact calorimeter configuration.
The concept of \PANDA places the Target Spectrometer EMC inside the superconducting coil of the solenoid.
Therefore, the basic requirements of the appropriate scintillator material are compactness to minimise the
radial thickness of the calorimeter layer, fast response to cope with high interaction rates, sufficient 
energy resolution and efficiency over the wide dynamic range of photon energies given by the physics program,
and finally an adequate radiation hardness. In order to fulfil these requirements, even a compact geometrical 
design must provide a high granularity leading to a large quantity of crystal elements. The largest sub-detector
is the barrel calorimeter with 11360 crystals of 200 mm length. In the backward direction 592 crystals 
provide hermeticity at worse resolution due to the presence of material needed for signal readout and supply lines of other \PANDA sub-detectors. 
As compared to the barrel calorimeter, the 3600 crystals in the forward endcap face a much higher range of particle rates up to 500 kHz per crystal. 

To achieve the required very low energy threshold, the light yield has to be maximised. Therefore, improved
lead-tungstate (PWOII) crystals are employed with a light output twice as high as used in CMS at LHC at CERN. 
Operating these crystals at -25\degrees C increases the light output by another factor of four. In addition,
large-area Avalanche PhotoDiodes (APDs) are used for the readout of scintillation light, providing high quantum efficiency and an active area four times larger than used in CMS. 

The crystal calorimeter is complemented in the Forward Spectrometer with a shashlyk-type sampling calorimeter 
consisting of 378 modules each made from 4 cells of 55$\times$55  {\ensuremath{\mm^2}\xspace} size covering about 3 m $\times$ 1.5 m.   
This document presents the details of the technical design of this
calorimeter.

\subsection {Data acquisition}

The challenge of \PANDA experiments is the high physics background created 
by decays of conventional hadronic states. 
Moreover, the expected background particles reveal an event topology very 
similar to that of decaying 
exotic-matter states. This ambiguity makes it extremely challenging to 
select and measure only decays of interest and 
excludes a conventional triggered readout. 
Only an advanced on-line event reconstruction with optimised resolution 
will allow us to uniquely identify the exotic-matter candidate events. 
Since the production probability for the exotic states is extremely low, 
the antiproton annihilation rate has to be as high as 20~MHz. 
Correspondingly, data are produced at a rate of about 200~Gb/s, which 
impossibly can be stored for the off-line analysis. 
In order to provide an adequate data-acquisition scheme, the trigger-less 
detector-readout technique shall be employed.

The newly developed approach 
foresees a complete on-line reconstruction of a measured event before 
it is classified as a background or a signal event. 
The trigger-less readout is realised with self-triggered intelligent 
front-end electronics (FEE), a very precise time-distribution 
system and on-line data processing algorithms. 
An intelligent front-end detects 
particle hits in the detectors and processes the raw data. The resulting 
data are sent to the data-concentrator modules which 
assign incoming data to certain time periods, called bursts, and send 
them for on-line processing to compute nodes (CN). 
An event-building network guarantees that the data from all \PANDA 
sub-detectors, which belong to the same burst, arrive at a single CN. 
Another key functionality of the DC is the fast and precise distribution 
of synchronisation signals from a single source to all front-end modules.
The \PANDA concept provides a high degree of flexibility in the 
choice of trigger algorithms.

\section{Conclusion}

\PANDA will be the first experiment for detecting both charged and neutral particles from antiproton on proton annihilation with
a truly 4{$\pi$}-detector exploring the energy regime of charm with high precision.
The in-flight mode at \PANDA allows for the production of resonances in the production 
mode, where a particle is produced together with one or more additional particles, or
in the direct formation of the resonance. The comparison of both methods helps 
classifying the resonances and identifying those resonances with exotic quantum 
numbers, i.e. quantum numbers forbidden for ordinary quark-antiquark mesons. 
In general, no restrictions for quantum numbers of resonances produced in \PANDA 
exist, thus all energetically allowed states can be populated.

The \PANDA experiment together with the high-quality internal antiproton beam at the
HESR will be a powerful tool to address fundamental questions of hadron physics in the
charmed and multi-strange hadron sector. Despite the impressive wealth of recent new
observations in the hidden charm meson sector, \PANDA will be able to deliver decisive
contributions to this field, due to complementarity of the antiproton-proton entrance
channel and the capabilities of the detector system under construction.

%
%
\newpage
\bibliographystyle{panda_tdr_lit}


%


%% file: panda_tdr_FSC_req.tex
%
\cleardoublepage
\chapter{Design Considerations for the Forward Spectrometer Calorimeter}
\label{sec:req}
\section{Engineering constraints}

The reconstruction efficiency of exclusive channels in $\pbarp$ annihilation has a strong  
dependence on the geometrical  acceptance.
The angular coverage of \PANDA should be maximised to achieve close to 99\percent $4\pi$ solid angle in the 
centre-of-mass system. The \PANDA Forward Spectrometer electromagnetic Calorimeter (FSC) is optimised to detect photons and 
electrons within the energy range between 10 MeV and 15 GeV. It covers 
about only 0.74\percent of the full solid angle, however the acceptance of the FSC for the inclusive photons production is almost 
8\percent.   The dimensions of the FSC are determined by its position at
$\approx$7 meters from the interaction point and by the size of the opening in the forward endcap
of the Target Spectrometer EMC. The FSC is situated between the Forward TOF and the forward Muon Range System, 
thus the total depth of the FSC is limited by the overall length of the Forward Spectrometer.

\section{Calorimeter concept}

Electromagnetic calorimeters are based on measuring the total energy deposition of photons or electrons
in the detector medium. The energy deposited by the secondary electromagnetic shower particles 
is detected either as Cherenkov radiation, caused by electrons and positrons like in a lead-glass calorimeter, 
or as scintillation light emitted by an active homogeneous medium. Alternatively, sampling calorimeters, 
constructed from alternating layers of organic scintillator and heavy absorber materials, 
have been used in high-energy physics since tens of years. 

The sampling fraction of such calorimeters, i.e. the thickness ratio of active 
(scintillator) and passive (absorber) layers, determines the lateral size of the electromagnetic
showers, expressed by the Moli\`ere radius: the larger the fraction of the absorber material, 
the narrower the shower is. The thickness of the scintillator plates affects the light 
yield and, in turn, 
determines the stochastic term of the energy resolution. The thickness of the absorber plates,
expressed in terms of radiation lengths, determines its sampling term. The interaction probability
of the secondary shower particles is smaller in thinner absorber plates. Hence, to achieve a better
energy resolution, thicker scintillator and thinner absorber plates are needed. 
Therefore, the choice of the calorimeter sampling is based on a compromise between 
the lateral shower size and the required energy resolution, defined by the physics requirements.

In a shashlyk-type sampling calorimeter the scintillation light is absorbed, re-emitted and transported to a photo detector by wavelength 
shifting (WLS) optical fibres running through the calorimeter modules longitudinally 
(i.e. along the beam direction). In the past, the typical stochastic term of the energy 
resolution was about 10\percent for large electromagnetic calorimeters of sampling type.

The energy resolution can be significantly improved in fine-sampling electromagnetic calorimeters in
the energy range from 50 MeV to well above 1 GeV. 
The best energy resolution ever achieved by sampling calorimeters was recently demonstrated by 
improved
electromagnetic calorimeter modules with a very fine sampling, developed for the KOPIO experiment
\cite{KOPIOweb,KOP08,KOP04}
at BNL, USA .
The stochastic term of the energy resolution of these modules, measured with photons of 
50-1000 MeV energy was about 3\percent. A similar high-performance electromagnetic calorimeter is now being
considered for \PANDA.

\section{Advantages and disadvantages of the shashlyk detector}
\label{sec:req:advandis}
The shashlyk technology, selected for the electromagnetic calorimeter in the \PANDA Forward Spectrometer,
has been proven by several experiments 
\cite{PHENIX,LHCb,KOPIOweb} for its high performance at a comparatively low price. 
The main disadvantage of the shashlyk detector is the non-uniformity in the transverse light output caused by the 
inhomogeneity of this type of calorimeter. 
This effect would cause an inhomogeneous response for photons or electrons with perpendicular incidence with 
respect to the front face of  the calorimeter. However, such a situation is negligible for the \PANDA FSC since 
the interaction point is 8 meters away from the FSC and particles enter the FSC with an angle of incidence significantly 
different from perpendicular incidence (>1.5 degree).

Another disadvantage is the large volume of the calorimeter, especially for the efficient detection of
high-energy photons and electrons. The large dimensions are not so crucial for the electromagnetic calorimeter 
in the \PANDA Forward Spectrometer in comparison with the Target
Spectrometer, where space inside the solenoid is rather limited. 

\section{General considerations}

In this chapter we will address the parameters required to optimise the performance of the FSC.
In order to achieve the required background suppression, a good separation of electrons and pions over a wide momentum 
range up to  $\approx$15 $\gevc$ is mandatory, and the PID information from the individual detector components  has to be exploited.

The \PANDA experiment aims at a variety of physics topics related to the very nature of the strong binding force
at the long-distance scale. Although the specific decay chains of various final states turn out to be different, 
most channels share one important  feature, namely many photons and/or electrons/positrons in the final state. 
Examples are hidden-charm decays of charmonium hybrids with neutral recoils and low-mass isoscalar  
S-waves (appearing in $\piz$ $\piz$), radiative charm decays, and the nucleon structure physics. 
These goals put special emphasis on the electromagnetic calorimeter. Its basic performance parameters 
have to be tuned to accomplish an efficient detection of the channels of interest.

Photons in the final state can originate from various sources. The most abundant sources are  decay photons of
copiously produced
$\piz$  and $\eta$ mesons. Important probes for the mechanism of the strong interaction are radiative charmonium  
decays (like $\chicone \to \jpsi\gamma$), 
which are suppressed by the charm production yield, or direct photons from rare electromagnetic processes.
In order to distinguish radiatively decaying charmonium and direct photons from background,
it is of utmost  importance to very efficiently detect and identify $\piz$ and $\eta$ mesons and to
reduce as much as possible inefficiencies in photon detection caused by solid angle limitations or energy threshold requirements.

\PANDA will not employ a threshold Cherenkov detector to discriminate pions from electrons and positrons. 
Therefore, the FSC has to add complementary information to the basic energy-momentum ratio (E/p). The lateral shower shape 
information is needed to discriminate e$^\pm$  from background. 
Hadronic showers (caused by K$_L$, n, charged hadrons) in an electromagnetic calorimeter 
differ significantly from electromagnetic showers due to the difference in the energy loss per interaction and the elementary statistical properties of 
these processes. The quality of this discrimination does (in first order) not depend on the actual choice of the geometry of
detector elements or the readout, as long as the lateral dimension of the detector cell is matched to the Moli\`ere radius. 
Therefore, this requirement does not place a strong restriction on the calorimeter design. Nevertheless, the final design 
process must incorporate an optimisation of the electron-pion separation power.

\subsection{Coverage and energy threshold considerations}

Apart from the energy resolution, the minimum photon energy E$_{thres}$ that can be detected
with the FSC is a crucial issue since it determines the very acceptance of  low-energy photons.
From the physics point of view, the low-energy threshold of the FSC should be the same as in the PWO calorimeter of the Target Spectrometer,
which is 10~MeV for a photon- or electron-induced cluster of cells, while the threshold for individual detector cells can be
as low as 3~MeV with correspondingly low noise levels of 1~MeV (see \cite{EMC} and Table 3.1 therein).
Although  a photon detection threshold of E$_{thres}$ = 10~MeV would be ideal,
technical limitations like noise or a reasonable coverage of a low-energy shower may 
increase this value, but at least  E$_{thres}$ = 20~MeV should be
achieved in order to meet the physics goals of \PANDA. More details about simulations are given in \Refchap{sec:sim}.

In the \Panda experiment, most of the foreseen physics channels  
have such a low cross section (in the nb region), that one needs  a background 
rejection power up to $10^9$. This strong suppression requires an adequate accuracy for photon 
and electron reconstruction within the energy range from 10 MeV to 15 GeV.  
In addition and in particular for the exclusive channels and for complex decay trees, optimal energy resolution is achieved through fitting
the kinematical parameters of detected tracks and photon candidates
by constraining their total momentum and energy sum to the initial $\ppbar$ system (kinematic fit).
We use commonly applied algorithms for the identification of exclusive channels 
such as decay trees with $\jpsi$ in 
final states: 
\begin{equation}
 \pbarp \to C \to \jpsi + X
  \label{eq:ctojpsi}
\end{equation}
where $X$ can be e.g. $\gamma$, $\piz$, $\eta$.

The identification algorithms start with the $\jpsi$ recognition.  
The first step of the reconstruction of $\jpsi$ decays is searching 
for  two candidates with opposite charge, identified as electrons.
The particle identification applies the likelihood selection algorithm. 
In this procedure, one of the 
candidates should have a likelihood value $L > 0.2$ and the other
a value $L > 0.85$. Both candidates are combined and accepted as $\jpsi$ 
candidate, if their invariant mass 
is found within the interval [2.98, 3.16] $\gevcc$. The reconstructed  tracks of
these two charged particles are kinematically and geometrically fitted to a 
common vertex and their invariant mass is constrained to the nominal $\jpsi$ mass.
Subsequently,  $C$ candidates are formed by combining the accepted  $\jpsi$ and $X$ 
candidates, whose invariant mass lies
within a suitable range. In the last step, the corresponding tracks  of 
$\jpsi$ and X candidates of the final state are kinematically fitted by 
constraining their momentum and energy sum to the initial $\ppbar$ 
system, and the invariant mass of lepton-pair candidates to the $\jpsi$ mass. 

\begin{figure}
\begin{center}
\includegraphics[width=\swidth]{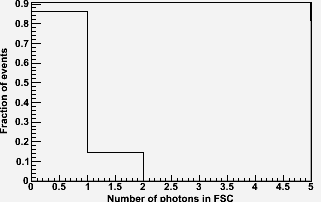}
\includegraphics[width=\swidth]{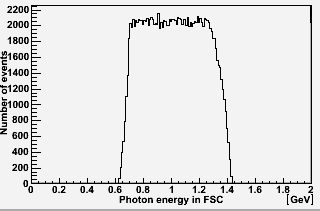}
\caption[ $\psiprime$ decay to $\jpsi\eta$: energy and fraction of photons in FSC]
{Simulation of photons from $\psiprime$ decay to $\jpsi\eta$, showing the fraction of events with zero or one photon in the FSC (top) and the photon energy detected in the FSC (bottom).}
\label{fig:req:jpsi}
\end{center}
\end{figure}

\begin{figure}
\begin{center}
\includegraphics[width=\swidth]{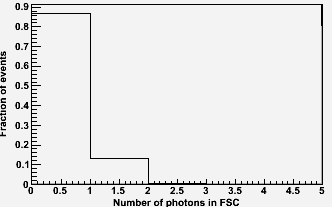}
\includegraphics[width=\swidth]{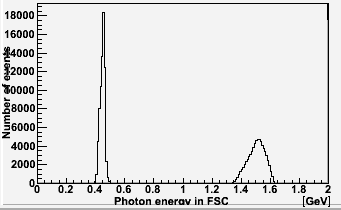}
\caption[$\psiprime$ decay to $\chi_{c2}\gamma$: energy and fraction of photons in FSC]
{Simulation of photons from $\psiprime$ decay to $\chi_{c2}\gamma$,  showing the fraction of events with zero or one photon in the FSC (top) and the photon energy detected in the FSC (bottom).}
\label{fig:req:chi2c}
\end{center}
\end{figure}

One of the main physics goals in the experimental program of \PANDA is charmonium 
spectroscopy. However, the high hadronic background in $\ppbar$ annihilations 
creates tough conditions  for   the decay  reconstruction. To avoid this 
problem,  it is  necessary to select those exclusive decays of charmonium, 
which are less  distorted by background.
As an example, for the two exclusive charmonium decay channels
\begin{equation}
\psiprime(3686) \to \jpsi\eta \to \ee\gamma\gamma
\label{eq:psiptojpsi}
\end{equation}
\begin{equation}
\psiprime(3686) \to \chi_{c2}\gamma \to \jpsi\gamma\gamma \to \ee\gamma\gamma 
\label{eq:psiptochi2}
\end{equation}
the addition of the FSC
will provide a significant ($\approx$ 15\percent) increase of two-photon events as compared to solely using the Target Spectrometer 
(\Reffig{fig:req:jpsi} and \Reffig{fig:req:chi2c}).

The actual partitioning between the forward
endcap EMC and the FSC is optimised to allow high-momentum tracks to enter 
the spectrometer dipole unhindered. This is a crucial requirement to provide 
clean Dalitz-plot analyses in the search for new forms of  matter, like glueballs or hybrids. 
Thus, the FSC enriches our ability to find new bound states of QCD above the open-charm threshold. 
Furthermore, the solid-angle coverage provided by the FSC is 
extremely useful to study the proton form factor in the time-like 
kinematical region. In addition, the presence of the FSC
allows for a more efficient detection 
of hyperons emitted forward and decaying electromagnetically. 
Polarisation measurements of hyperons and vector mesons with \PANDA can 
help to clarify the strong interaction dynamics in the confinement 
region. 

\subsection{Dynamical energy range}

\begin{figure}
\begin{center}
\includegraphics[width=\swidth]{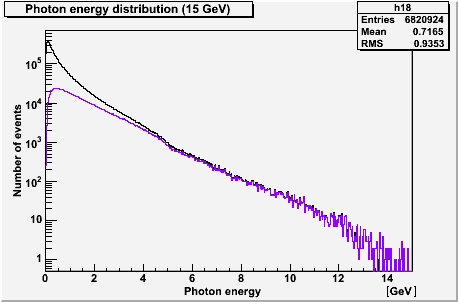}
\caption[DPM simulation at 15 \gevc: photon spectrum in FSC acceptance]
{Results of DPM simulations for 15 \gevc  antiproton annihilations: Energy distribution of all photons (black curve)  compared to the photon energy distribution in the FSC acceptance (red curve).}
\label{fig:req:allphotons15}
\end{center}
\end{figure}

\begin{figure}
\begin{center}
\includegraphics[width=\swidth]{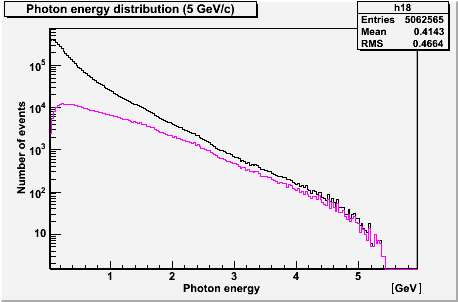}
\caption[DPM simulation at 5 \gevc: photon spectrum in FSC acceptance]
{Results of DPM simulations for 5 \gevc  antiproton annihilations: Energy distribution of all photons (black curve)  compared to the photon energy distribution in the FSC aperture (red curve).}
\label{fig:req:allphotons5}
\end{center}
\end{figure}

Figure~\ref{fig:req:allphotons15} and \Reffig{fig:req:allphotons5} show the photon energy distributions, obtained from DPM generator calculations, 
for the antiproton momentum of 15 \gevc and 5 \gevc, respectively. Shown are the distributions for all photons and those emitted in the acceptance of he FSC.
The highest energies are found in the forward direction. Since low-energy capabilities 
are mandatory for all \PANDA calorimeters, the dynamic range is mainly driven by the highest photon energy expected in the respective kinematic region.
The dynamical range for the \Panda FSC should  cover 10(20) MeV - 15 GeV for photons or charged particles. One can deduce the dynamical range requirements of the photo detector and readout chain from this information. Since one cell of the detector registers at most 75\percent of the electromagnetic shower, the photo detectors of a single cell should register signals in the range of 3 MeV - 12 GeV.   

\section{Resolution requirements}
\subsection{Energy resolution}

The best possible energy resolution is required to ensure the exclusiveness of events.
A precisely measured energy/momentum ratio (E/p) is an important asset to positively identify electrons and positrons against pions. 
Various constraints determine  the appropriate  energy resolution:
\begin{itemize}
\item Precise measurement of electron and positron energies for
 \begin{itemize}
 \item very accurate E/p determination, and
 \item optimum  $\jpsi$  mass resolution
 \end{itemize}
\item Efficient recognition of light mesons (e.g. $\piz$ and $\eta$) to reduce potential combinatorial background.
\end{itemize}
Another consequence of insufficient energy resolution is the bad determination of $\piz$ and $\eta$ meson masses. 
At low energies the resolution is dominated by the  $1/\sqrt{\rm E}$  dependence of the energy resolution, while at high 
energies the resolution 
is dominated by the constant term. Experiments like Crystal Barrel, CLEO, BaBar and  BESIII, with
similarities in the topology and composition of  final states, have proven, that a  $\piz$ width of  less than  
8~MeV and an $\eta$ width of less than  30~MeV is necessary for a reasonable  final-state decomposition. 
Assuming an energy dependence of the energy resolution of  the form
\begin{equation}
  \frac{\sigma_{\rm E}}{\rm E} =
\frac{b}{\sqrt{\rm{E/GeV}}} \oplus c
\label{eq:psiptochi22}
\end{equation}
where the $\oplus$ operator indicates the quadratic sum of operands,
leads to the requirement for the constant term $c \approx$ 1\percent  and the stochastic term $b \approx$ (2-3)\percent. 
This balance of values also ensures a  $\jpsi$ resolution which is well matched with the resolution of the
typical light recoil mesons.

By varying the sampling ratios and cell sizes of the shashlyk FSC
one can select the optimal resolution and size of the detector. Since
the space is not strictly limited for the \PANDA FSC, we were able to select the
sampling ratio for the best calorimeter performance. In the previous studies
carried out at IHEP Protvino  for the KOPIO experiment, the optimal lead-scintillator ratio
corresponding to a stochastic term of 3\percent/$\sqrt{\rm E}$ was found \cite{KOP04,
KOPIOweb,KOP08}. The equipment at the IHEP scintillator department was 
adjusted for that ratio. 

\begin{figure}[ht]
\centering
\includegraphics[width=1.1\swidth]{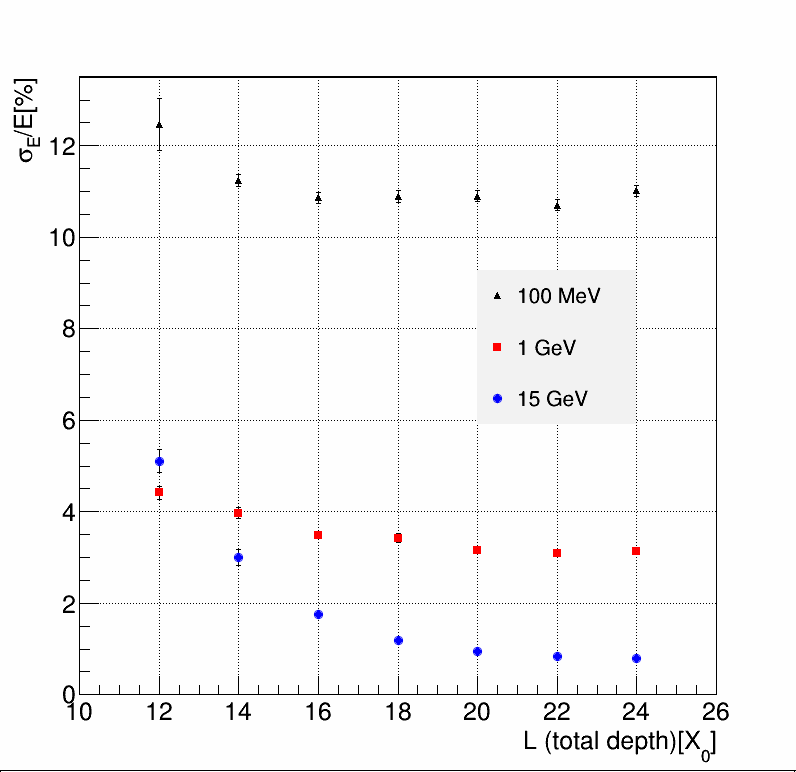}
\caption[Dependence of energy resolution on FSC depth]
{Dependence of the energy resolution on the depth of the FSC for three energies of incident electrons.}
\label{fig:radlen}
\end{figure}

During the optimisation process discussed above, we kept the depth of the FSC constant at 20 radiation lengths (X$_0$), which is
de facto standard for electromagnetic calorimeters working at high energies. To ensure that this is the optimal depth for the 
\PANDA FSC,  a MC simulation with PandaRoot (see~\Refchap{sec:sim}) was carried out. As can be seen from \Reffig{fig:radlen}, the depth of 20 X$_0$
is still close to optimal for \Panda energies, as the energy resolution is not deteriorated for 15 GeV electrons. A shashlyk  calorimeter for lower energies could be built with shorter depth, if needed. 
    
\subsection{Single-cell threshold}
\begin{figure}
\begin{center}
\includegraphics[width=\swidth]{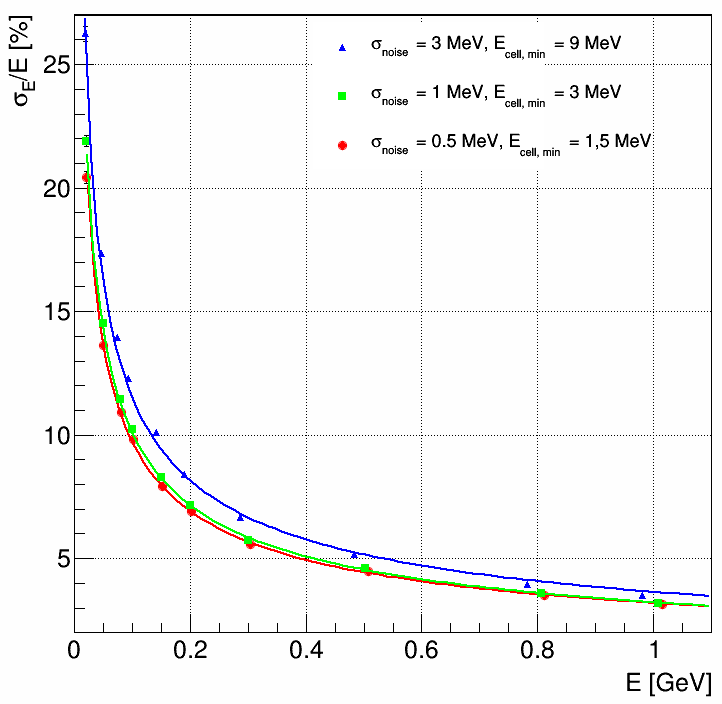}
\caption[Dependence of energy resolutions on reconstruction thresholds]
{Comparison of the energy resolutions as function of the photon energy for three different single-cell reconstruction thresholds. 
The most realistic scenario with a noise term of $\sigma_{noise} \, = \,1 \, \mev$
and a single-cell threshold of $E_{1}=3\,\mev$ is illustrated by green rectangles, a worse case
($\sigma_{noise} \, = \, 3\, \mev$, $E_{1} \, = \, 9 \, \mev$) by blue triangles, and a better case ($\sigma_{noise} \, = \, 0.5 \, \mev$, 
$E_{1} \,= \, 1.5 \, \mev$) by red circles.}
\label{fig:req:Ethreshold}
\end{center}
\end{figure}

The energy threshold and the energy resolution place a requirement on the minimum detected
energy E$_1$ in a single FSC cell. As a consequence, this threshold  puts a limit on the single-cell noise, since the energy cut (of several  MeV) should be 
high enough to exclude a random assignment of photons. This requirement can be relaxed by demanding a higher single-cell energy 
to identify a bump, i.e. a local maximum in the energy deposition (e.g. 10 MeV). Starting with
those bumps as a central cell, additional cell energies are only collected in the vicinity of the central cell. With typically 
10 neighbours and not more than 10 particles hitting the FSC, we expect at most one random hit per event for a single-cell cut of E$_1$ = 3$\sigma_{noise}$. Figure~\ref{fig:req:Ethreshold}  shows that a single-cell
threshold of  E$_1$ = 3~MeV is needed to obtain the required energy resolution. In addition, the detection of  
low-energy photons in more than the central cell
can only be achieved for E$_1$ $\le$ 3~MeV. From these considerations we deduce a limit for the 
total noise of  $\sigma_{Enoise}$ = 1 MeV.

\subsection{The FSC cell size and spatial resolution}
\label{sec:req:resolutions}
The spatial resolution is mainly governed by the granularity of the detector-cell structure. The reconstruction of the point 
of impact is achieved by a weighted average of  hits in adjacent cells. In addition, to identify overlapping photons 
(e.g. due to $\piz$ decays with small opening angles) it is mandatory to efficiently split multiple-hit clusters into 
individual photons. This procedure requires, that  the central hits of the involved  photons are separated by at least two cell 
widths to assure two local maxima in energy deposition. The highest $\piz$ energy for the highest incident 
antiproton momentum of 15 \gevc will be 15 GeV. The minimum opening angle for the decay of $\piz$ into two gamma 
quanta will be 18 mrad. For a distance between the \PANDA target and the FSC of 8 m, we thus obtain 144 mm between 
two photons at the FSC front face. Consequently, the FSC cell size is required to be smaller than 72 mm.

The needed spatial resolution is mainly governed by the required width of the $\piz$ invariant-mass peak 
in order to assure proper final-state decomposition. Since the main parameters of the FSC should be the same 
as for the Target Spectrometer EMC, the FSC spatial resolution will have to achieve a $\piz$ mass resolution of about 
5 MeV/$c^2$ for 15 GeV $\piz$ mesons (see \cite{EMC} and Fig. 3.6 therein).
The worst spatial resolution is obtained
at the centre of the cell, and the best one is found near the boundary between two cells. Let us take the worst case, 
a $\piz$ decay into two photons, when both photons hit the FSC close to the centre of the two cells. 
In this case for 5 MeV/$c^2$ mass resolution, the angular resolution will be about 3.4\percent (the energy resolution 
at 15 GeV is about 1.6\percent (see test-beam results in \Refchap{sec:perf}). To achieve this angular resolution 
at a distance of 8 m from the target, the spatial resolution at the centre of each cell will have to be equal to 
$18\cdot10^{-3}\cdot0.034\cdot8000$ mm/$\sqrt{2} \approx 3.5$ mm.

Thus, we deduce the following requirements for the FSC cell size and spatial resolution:
\begin{itemize}
\item the cell size must be less than 72 mm;
\item the spatial resolution at the centre of the cell should be less than 3.5 mm.   
\end{itemize}

\section{Count rates}

\begin{figure}
\begin{center}
\includegraphics[width=1.07\swidth]{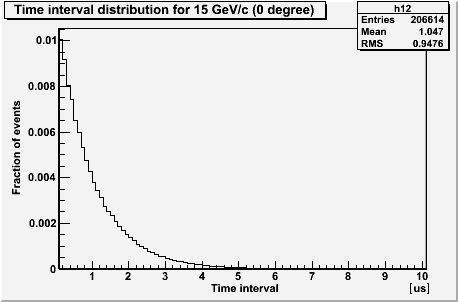}
\caption[Hit rate near beam pipe]
{Distribution of time between hits for energies E $>$ 1~MeV simulated with the DPM background generator for 
the calorimeter region near the beam pipe.}
\label{fig:req:hitrate_pipe}
\end{center}
\end{figure}

\begin{figure}
\begin{center}
\includegraphics[width=\swidth]{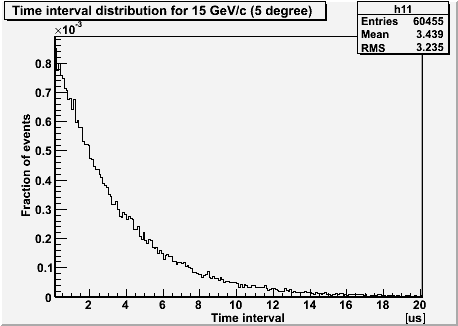}
\caption[Hit rate near FSC outer edge]
{Distribution of time between hits for energies E $>$ 1~MeV simulated with the DPM background generator for 
the calorimeter region at the FSC edge.}
\label{fig:req:hitrate_edge}
\end{center}
\end{figure}

Figure~\ref{fig:req:hitrate_pipe} and \Reffig{fig:req:hitrate_edge} show  the hit 
rates  for energies E $>$ 1~MeV simulated with the DPM background generator for 
two different  calorimeter regions: near the beam pipe and near the vertical edge of the detector.
The mean rates per cell  are  $\approx$1 MHz near the beam pipe and $\approx$300 kHz  at  the detector edge. 

\section{Radiation hardness}
Since
the FSC is positioned relatively close to the interaction region, and a large fraction of particles
produced at the target will be boosted forward into the FSC acceptance, the FSC will receive a considerable load of particles and, therefore, 
all the FSC components must withstand an adequate 
radiation dose. The estimated maximum absorbed dose reaches
nearly 0.25 Gy/h in the calorimeter body closest to the 
beam axis.
Assuming an overall lifetime of \PANDA of 10 years
and a duty cycle of 50\percent, the integral absorbed 
dose of the FSC could reach up to 10 kGy at a position near the beam pipe.
The most sensitive components of the FSC are scintillator
plates and WLS fibres. Thus, both the scintillator and the WLS
fibres must satisfy the radiation requirements mentioned above.

\section{Calibration and monitoring prerequisites}
The energy calibration of the FSC cells has to be performed at a precision much better than the energy resolution, 
thus at the sub-percent level, for the off-line data analysis. Since the \PANDA data acquisition system (DAQ) relies
on on-line trigger decisions performed on compute  nodes, the FSC calibration constants have to be available
for the on-line trigger decisions with sufficient precision, i.e. at least at the percent level, in real time. 
Accordingly, the monitoring system must detect any gain variations at this level in the readout chain, starting 
with the light transmission in the calorimeter elements and ending at the digitising modules.

\begin{table}[h]
\sf
\caption[Compilation of FSC requirements]
{Requirements for the \PANDA FSC. Rates and doses are based on a luminosity of L = 2$\cdot$10$^{32}$ cm$^{-1}$s$^{-1}$.}
\begin{tabular}{|p{0.45\swidth}|p{0.45\swidth}|}
\hline
Common properties & Required performance value\\
\hline
energy resolution $\sigma_{\rm E}$/{\rm E}             & $\approx$ (2-3)\% /$\sqrt{\rm{E/GeV}}$ $\oplus$  1\%  \\
energy threshold (photons) E$_{thres}$      & 10 MeV (20 MeV tolerable) \\
energy threshold (single cell) E$_1$       & 3 MeV \\
noise (energy equiv.) $\sigma_{Enoise}$ & 1 MeV \\
angular coverage                           & 0\degrees - 5\degrees \\
energy range from E$_{thres}$ to            & 15 GeV \\
spatial resolution                         & 3.5 mm \\
load per cell                              & $\approx$ 1 MHz \\
radiation hardness (maximum integrated dose) & 10 kGy   \\
\hline
\end{tabular}
\label{tab:req:reqs} 
\end{table}

\section{Conclusion}
The full list of requirements for the \PANDA FSC  is compiled in \Reftbl{tab:req:reqs}.
Based on these requirements, the performance of the readout electronics for individual detector channels 
has been designed. The readout electronics 
should work reliably up to a load of a few MHz and provide a linear response from signal levels of 3 MeV up to 12 GeV
($\approx$75\percent of the electromagnetic shower deposited in the central cell of a cluster).
In this technical design report of the FSC, we will demonstrate that the designed 
calorimeter  will  fulfil all the listed requirements on energy resolution, spatial resolution, 
energy threshold, timing, and radiation hardness.

%
%
\newpage
\bibliographystyle{panda_tdr_lit}

%% file: panda_tdr_FSC_mech_pub.tex
%
\cleardoublepage
\chapter{The Design of the FSC}
\label{chap:mech}

The concept of the FSC design is based on the existing experience with the KOPIO project at BNL, USA. 
The structure of the detector is composed of lead plates as passive absorber and
organic scintillator tiles as active elements. The development is performed in three iterations 
with different and continuously improving performance and better adaptation to the
\PANDA needs.
The main mechanical properties of the \Panda FSC are compiled in \Reftbl{tab:fsc_prop}.

\begin{table*}[h!]
\begin{center}
\begin{tabular}{lcc}
\hline
Title & Value  & Units \\
\hline
Overall detector width (x direction) & 4.9 &  m \\
Overall detector height (y direction) & 2.2 &  m \\
Overall detector depth (z direction) &  1150 &  mm \\
Overall detector weight (see \Reftbl{tab:weight_list}) &  14.7 &  tons \\
Number of channels (including beam pipe zone) & 1512  &  pcs \\
Number of modules  (including beam pipe zone) &  378 &  pcs \\
Module weight & 21 &  kg \\
Module cross section &  110$\times$110 & mm$^2$ \\
Cell cross section    &  55$\times$55 & mm$^2$ \\
Scintillator tile thickness & 1.5  & mm \\
Lead plate thickness &   0.275 &  mm \\
Beam pipe zone & 9 = (3$\times$3) & modules \\
\hline
\end{tabular}
\caption[Compilation of mechanical properties of the FSC]
{Main mechanical properties of the FSC.}
\label{tab:fsc_prop}
\end{center}
\end{table*}

\section{The FSC components, geometry and dimensions}

%
%

The FSC is located behind the dipole magnet of the \Panda Forward Spectrometer, just downstream of the RICH detector,
and is designed in planar geometry, covering the most forward angular range 
up to 5\degrees in the vertical and 10\degrees in the horizontal direction. 
The exact position and the dimensions of the FSC are defined by the size of the central hole 
in the forward endcap EMC of the Target Spectrometer \cite{temc_tdr}.
The active volume of the FSC consists of 54$\times$28 cells. 
The overall dimensions of the detector frame are 3.6 m in width and 2.2 m in height.

\begin{figure}
\begin{center}
\includegraphics[width=\swidth]{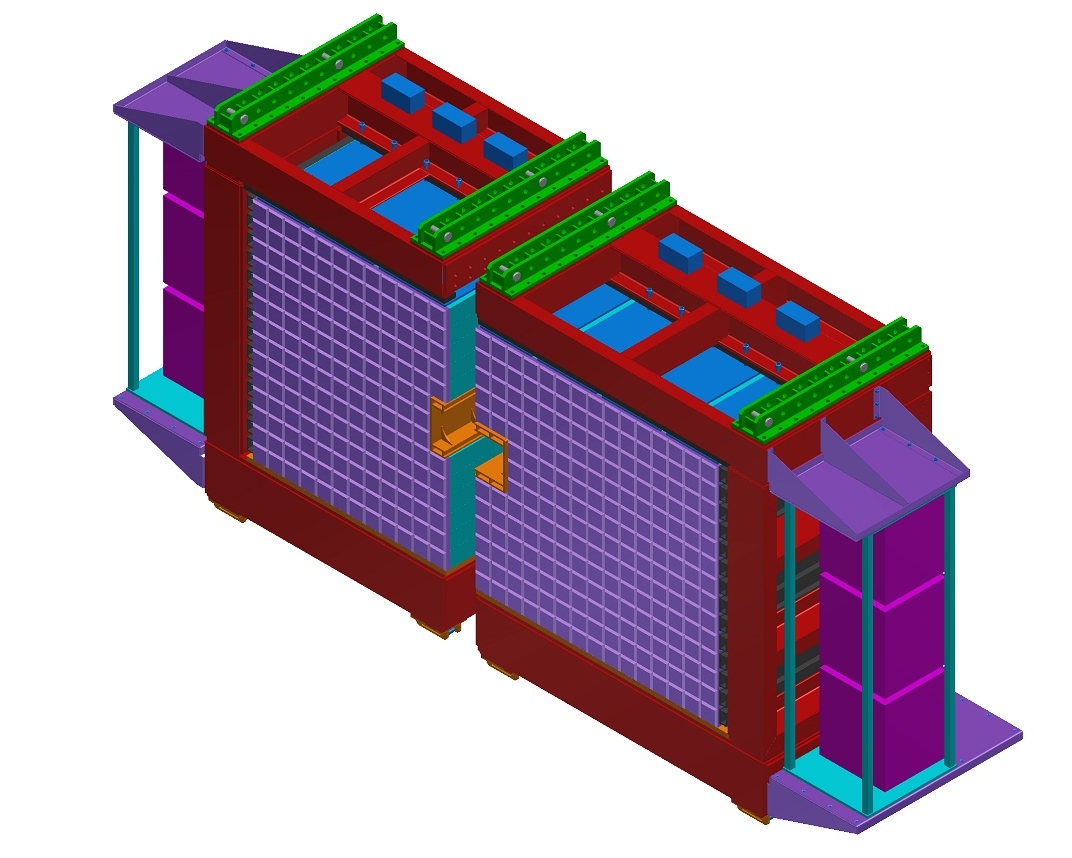}
\caption[FSC global view]
{Sketch of the global view of the FSC as seen from the front side.}
\label{fig:mech:det_over}
\end{center}
\end{figure}

\begin{figure}
\begin{center}
\includegraphics[width=\swidth]{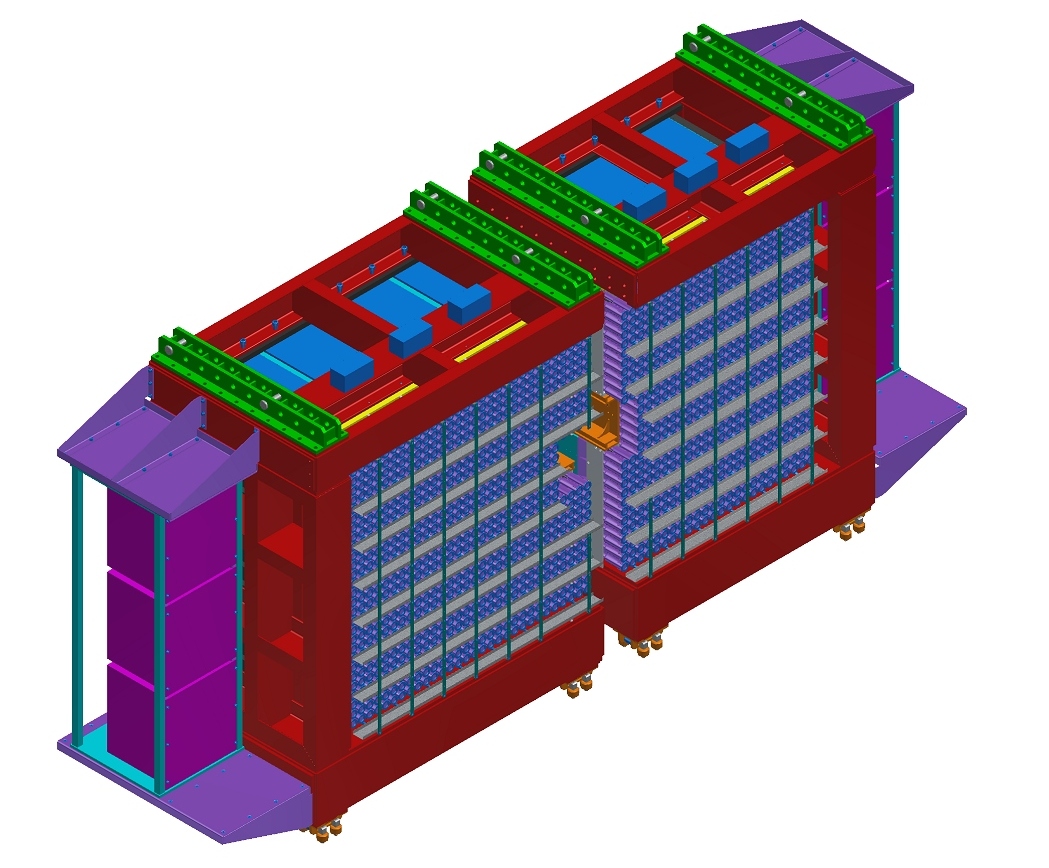}
\caption[FSC back-side view]
{Sketch of the global view of the FSC as seen from the back side.}
\label{fig:mech:det_backover}
\end{center}
\end{figure}

The main parts of the FSC are 378 modules, 
the support frame, the moving system, the electronics and monitoring system, and cable trays. 
According to the design requirements, the calorimeter setup consists of two sections 
which are tied together with bolts and positioning pins during the FSC operation. 
Each section is divided into two zones, separated by a 30 mm thick aluminium-alloy back-plate. 
In front of the back-plate and attached to it in the axial direction, are the modules which house 
the scintillator- and lead-plate arrangement of the shashlyk calorimeter. Photo detectors, electronics 
and cable trays are located behind the back-plate.  A global view of the FSC setup is sketched in
\Reffig{fig:mech:det_over} and \Reffig{fig:mech:det_backover} as seen from the front side and 
the back side, respectively.

\begin{figure*}[h]
\begin{center}
\includegraphics[width=\dwidth]{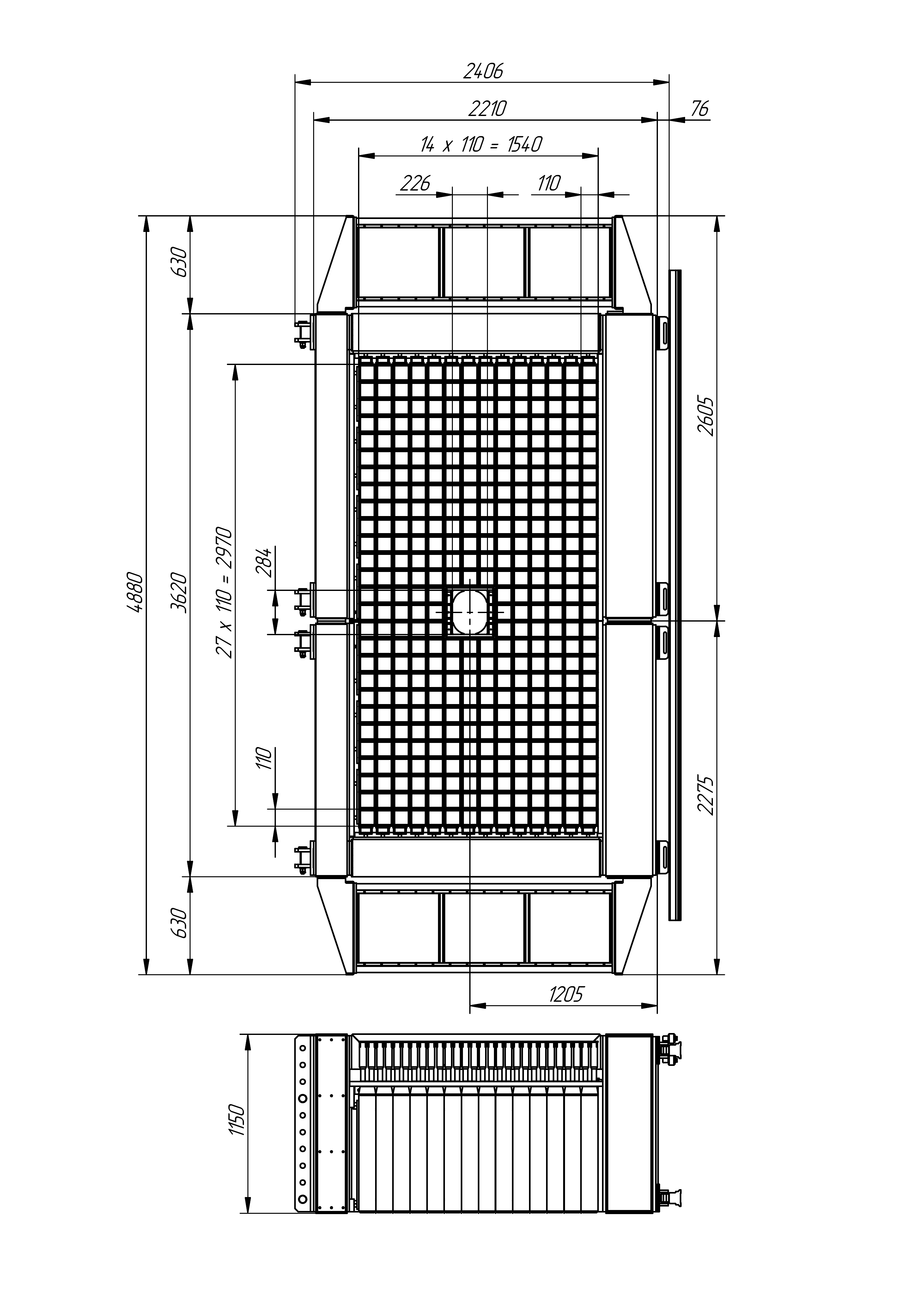}
\caption[Drawing with dimensions of closed FSC]
{Dimensions of the closed FSC frame in the beam position.}
\label{fig:mech:det_dims}
\end{center}
\end{figure*}

The FSC detector covers an active area of 
1540$\times$2970 mm$^2$ and consists of two sections to facilitate the installation 
and maintenance procedures. The depth of the detector along the beam line is 
1150 mm which results from the characteristic module length (1033 mm) and additional space for 
the front-side monitoring system (20 mm) and the back-side cable trays (97 mm).
The main outer dimensions of the calorimeter in the closed position,  i.e., when both sections are tied together,  
are given in the technical drawing in
\Reffig{fig:mech:det_dims}. The hole in the detector provides enough room for the HESR beam pipe. 
The size of the hole is 3$\times$3 modules. During the design of the FSC frame the diameter of the beam
pipe changed several times. The beam pipe hole in the FSC was also changed accordingly. The previous design
contained a hole of 2$\times$2 modules. Some of the pictures in this TDR may refer to the old design of the 
beam hole.


\section{Design of the individual module}

The general design of the module is described in this section. Later, in the sections below,
several modifications of the design, which were used to build various prototypes, will be presented.
Each module is a sandwich of 380 layers of lead absorber plates and plastic scintillator tiles (1.5 mm thick), 
corresponding 
to a total thickness of 19.6 radiation lengths. The individual detector module is composed of four 
optically isolated parts, called cells,  
with a cross section of 5.5$\times$5.5 cm$^2$  each. The four cells are combined and aligned by common 
absorber plates made 
of 0.275 mm thick lead sheets providing additional fixing holes for the individual scintillator tiles.  
Four scintillator tiles correspond to one lead plate, which has 144 holes (1.3 mm diameter) with a pitch of  9.3 mm. 

\begin{figure*}
\begin{center}
\includegraphics[width=0.8\dwidth]{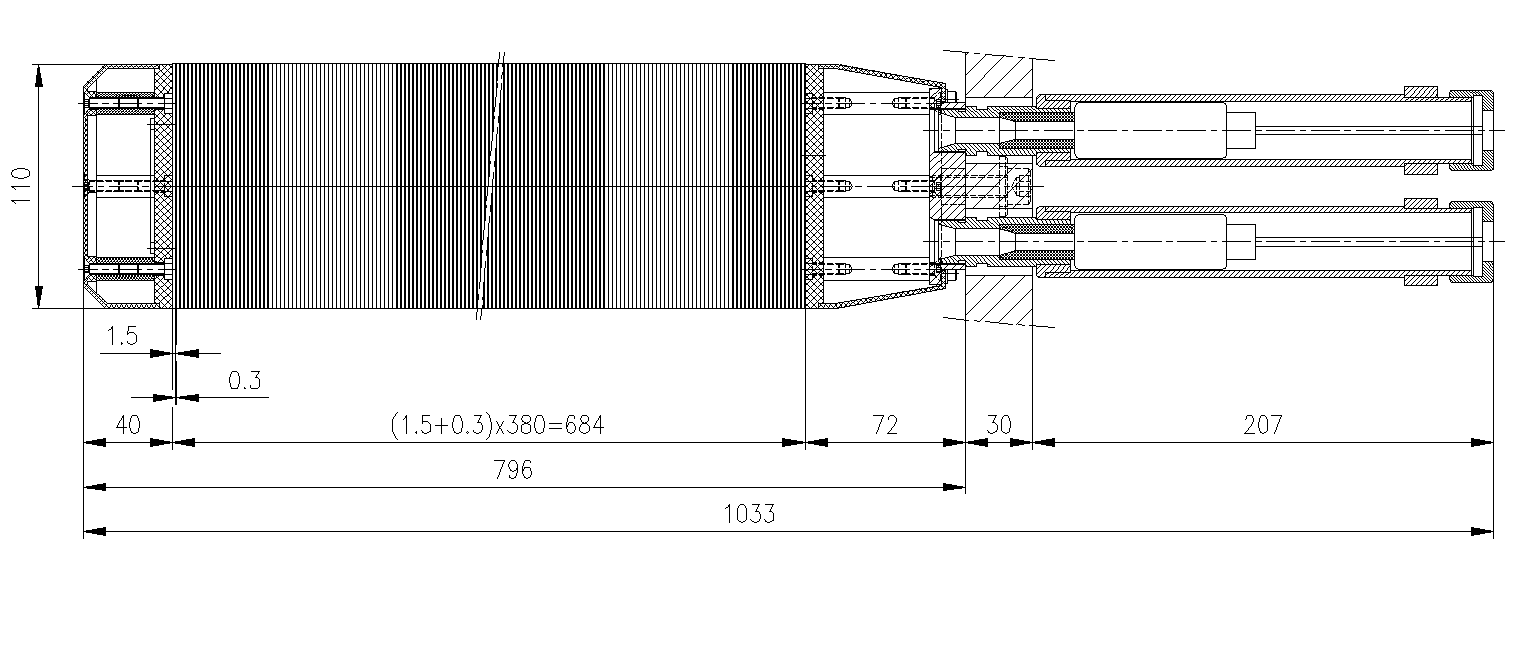}
\caption[Type-2 module dimensions]
{The global dimensions of a shashlyk FSC module of Type-2.}
\label{fig:mech:mod_dims}
\end{center}
\end{figure*}

The characteristic module length including photo detectors and HV bases is 1033 mm. The longitudinal 
cross section and global dimensions of the FSC module are shown in \Reffig{fig:mech:mod_dims}. It is
worthwhile to mention that the drawing in this figure represents a module without Tyvek between lead and 
scintillator tiles (Type-2, will be defined in \Refsec{sec:types_of_proto}). A subsequent prototype (Type-3) is 76 mm longer because of the additional width of the Tyvek sheets. To match the 
final FSC detector design requirements, the length of the CW-base can be shortened.

\begin{figure}[h]
\begin{center}
\includegraphics[width=0.8\swidth]{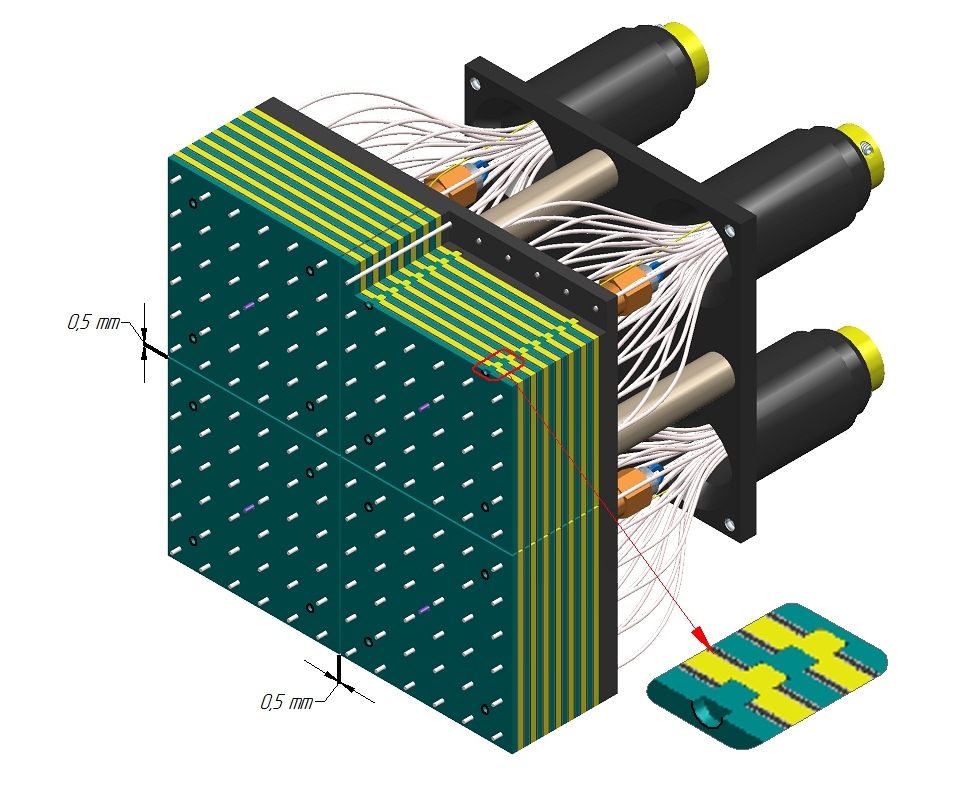}
\caption[LEGO-type locking of lead and scintillator tiles]
{3D view of the stack of scintillator tiles locked by LEGO-type pins and holes.}
\label{fig:lego}
\end{center}
\end{figure}

Important parts of the mechanical design of the module are the LEGO-type 
locks for the scintillator tiles. Four pins per tile fix 
the relative position of the scintillators and provide 0.3 mm gaps which are sufficiently wide 
to place the 0.275 mm thick lead plates without optical contact 
between lead and scintillator (\Reffig{fig:lego}). 
The module is wrapped with reflective material. Internal edges of the 
tiles are covered by white paint to provide optical isolation and to increase the light output.

\begin{figure}
\begin{center}
\includegraphics[width=0.8\swidth]{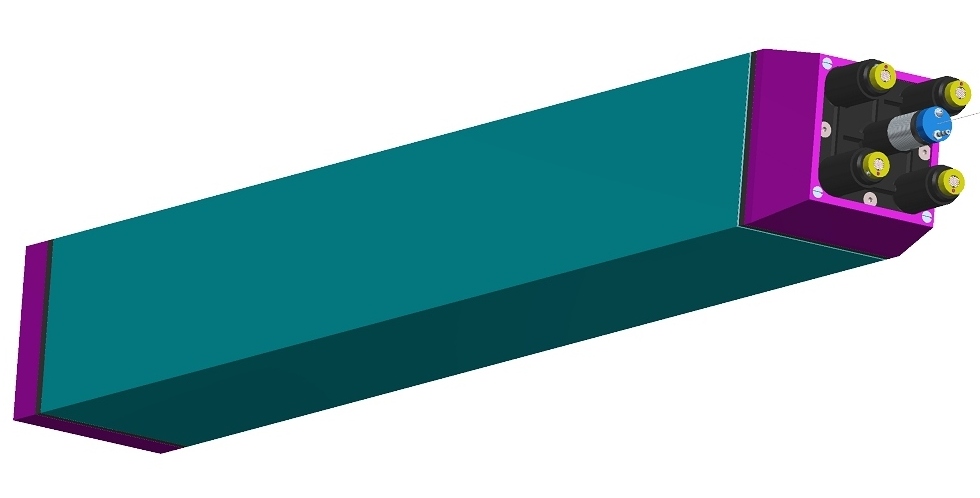}
\caption[Assembled FSC module]
{A global view of the assembled FSC module without PMT compartment.}
\label{fig:mech:mod_back}
\end{center}
\end{figure}

Figure \ref{fig:mech:mod_back} shows a 3D view of the assembled FSC module. The sandwich structure of 
alternating scintillator tiles and lead plates, axially traversed by WLS fibres, 
is pulled together between two 8 mm thick 
pressure plates attached to the front and the back side, respectively. The set is tightened by 
steel strings, 
running from the front to the back side, with reinforcement screws. 
Optical fibres are assembled into four bunches on the back side of the module, i.e. the photo detector 
side. The ends of
the fibres are glued, cut and polished to make four clear light-output ports. The photo detector is
attached to this port with optical-quality silicone cookies. 
Connectors for the light monitoring system are placed in the centre of the back-side support plate. 
A back-side cover protects outgoing optical fibres from mechanical damage. On the front side of the 
module the WLS fibres form loops and are protected by the front cover.

\begin{figure}
\begin{center}
\includegraphics[width=0.8\swidth]{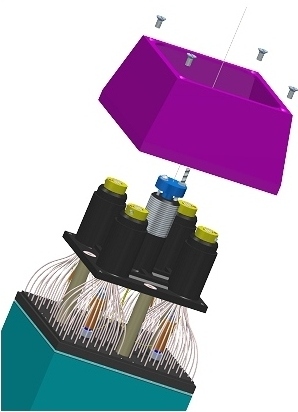}
\caption[Back-side view of FSC module]
{A back-side exploded view of the FSC module.}
\label{fig:mech:mod_back_cover}
\end{center}
\end{figure}

All custom parts of the \Panda FSC modules like scintillator tiles, lead plates, front and back covers, 
and
pressure plates can be produced at the IHEP Protvino scintillator department using existing manufacturing 
equipment 
and methods which were established for the KOPIO shashlyk calorimeter production.
Figure \ref{fig:mech:mod_back_cover} shows an exploded view of the back side of a module revealing the 
components:
the back-side pressure plate, four tightening screws, the back-side support plate fastened with four 
bolts, connectors, and the back-side cover with fastening screws.

\begin{figure}
\begin{center}
\includegraphics[width=0.8\swidth]{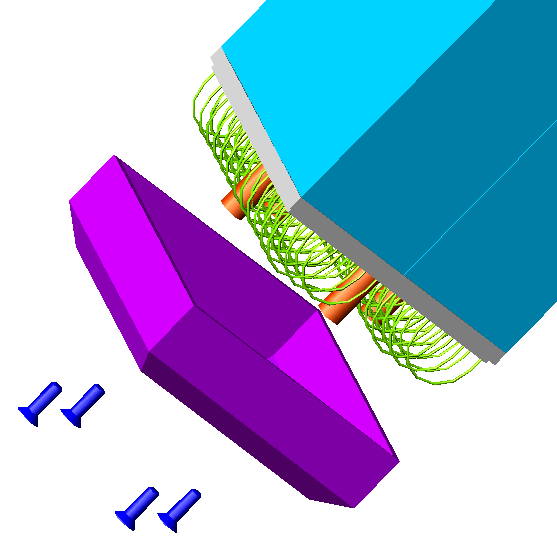}
\caption[Front-side view of FSC module]
{A front-side view of the FSC module showing the front cover, the WLS fibre loops, and the pressure plate.}
\label{fig:mech:mod_front_cover}
\end{center}
\end{figure}

\begin{figure}
\begin{center}
\includegraphics[width=0.8\swidth]{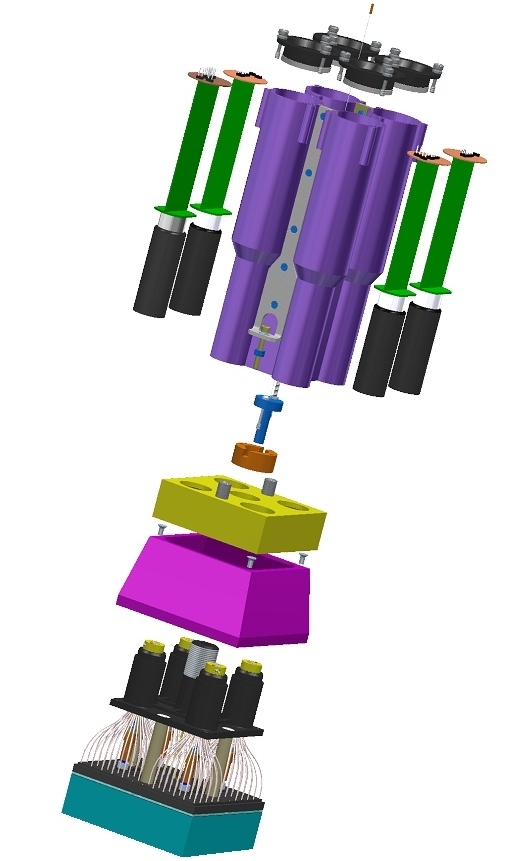}
\caption[Exploded back-side view of FSC module]
{Exploded view of a fully assembled module.}
\label{fig:mech:mod_back_exploded}
\end{center}
\end{figure}

Figure \ref{fig:mech:mod_front_cover} shows an exploded view of the front side of a module revealing 
the components:
the front-side pressure plate, four posts and the front cover with fastening screws. 
Figure \ref{fig:mech:mod_back_exploded} shows an exploded view of the fully assembled module with photo 
detectors, HV bases with protection tubes, details of the monitoring system connector, and 
fastening nut to fix the module to the FSC back plate.

\subsection{Plastic scintillator plates}

\begin{figure*}
\begin{center}
\includegraphics[width=0.7\dwidth]{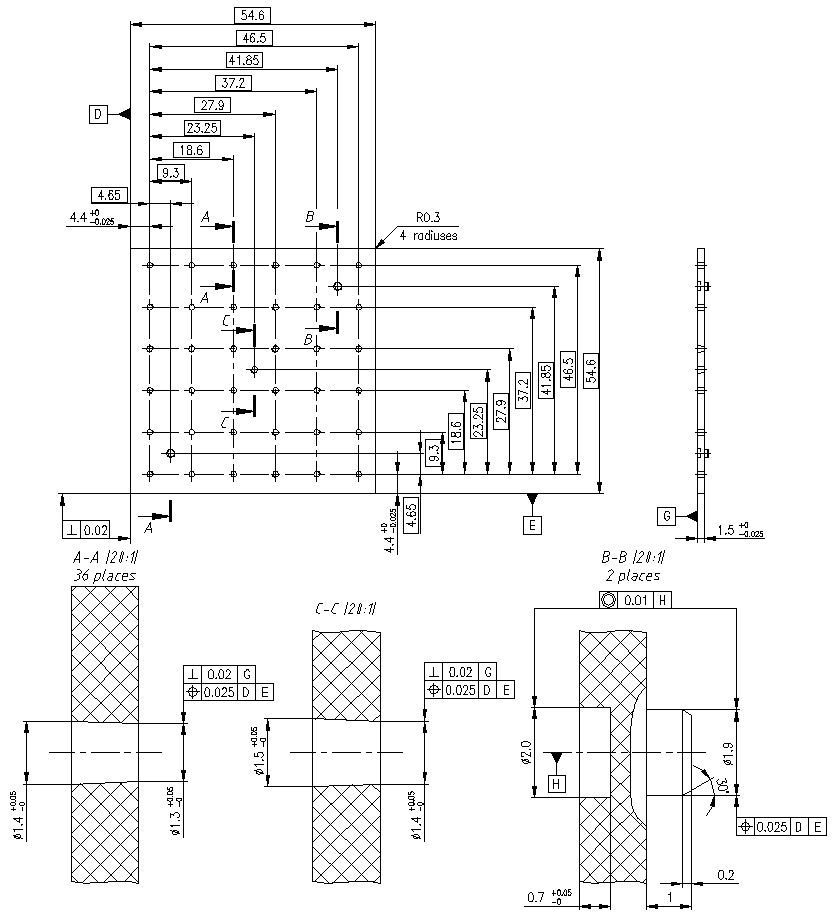}
\caption[Drawing with dimensions of scintillator tiles]
{Technical drawing of the scintillator tiles of the FSC cell.}
\label{fig:mech:scint_tile}
\end{center}
\end{figure*}

The lateral dimensions of a single cell are close to the Moli\`ere radius of 59.8 mm. In order to 
improve the light collection, 
the scintillator tiles are placed between reflector sheets of 0.15 - 0.2 mm thick
Tyvek-paper and the side faces are covered with white reflector paint. The 
scintillator material 
is made of polystyrene doped with 1.5\percent  
paraterphenyle and 0.04\percent  POPOP. 
The tiles are produced at IHEP Protvino exploiting the injection moulding technology and have 36
holes for the light collecting WLS fibres.
An earlier version of the mould had to be 
replaced by a new mould to improve the alignment by additional and 
enlarged fixing pins and to take into account the increased thickness due 
to the reflector sheets. 
Figure~\ref{fig:mech:scint_tile} shows the technical 
drawings of the scintillator tiles.

\subsection{Absorber plates, geometry, shapes and functionality}

\begin{figure*}
\begin{center}
\includegraphics[width=0.7\dwidth]{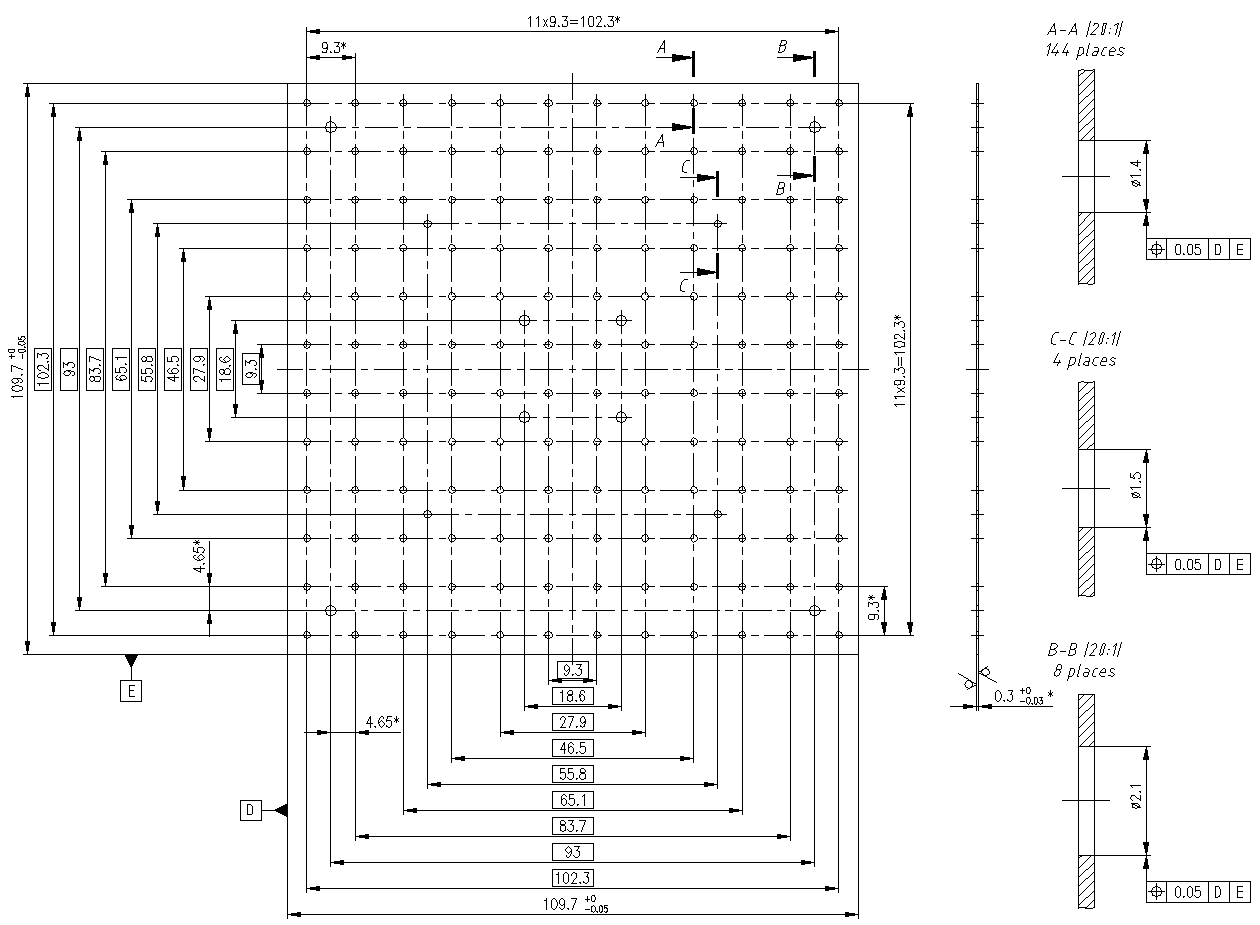}
\caption[Drawing with dimensions of lead tiles]
{Technical drawing of the lead plates of the FSC module.}
\label{fig:mech:lead_plate}
\end{center}
\end{figure*}

The quadratic lead absorber plates are common for four cells. The 0.275 mm thick lead sheets are doped with 3\percent of antimony to improve their rigidity and they provide fixing holes for four individual scintillator tiles.  
Figure \ref{fig:mech:lead_plate} shows the technical 
drawings of the lead plates.

\subsection{Light collection by WLS fibres}

The scintillation light of a single FSC cell is collected and accumulated by 36 wavelength shifting 
(WLS) multi-cladding fibres running 
axially through the sandwich structure and form 18 loops per cell  at the front 
side of the module. 
To make the loop with the required small 
radius of curvature, additional care with thermal treatment was taken during the bending of fibres.

Originally (for Type-1 and Type-2 prototypes, see definition in \Refsec{sec:types_of_proto}) the WLS 
optical fibres BCF-91A with a diameter 
of 1 mm were used in the calorimeter modules. 
However, after some beam tests, we noticed that these fibres developed 
cracks at the positions of the loop. These BCF fibres were substituted by fibres
produced by Kuraray (Type-3 prototype, see \Refsec{sec:types_of_proto}). 
In the current design, the Kuraray Y-11 (200)M 
fibres with 1mm diameter is foreseen for the mass production of FSC modules. 

\subsection {Module prototypes}
\label{sec:types_of_proto}

Three prototypes (Type-1, Type-2 and Type-3) were manufactured and tested at particle beams of miscellaneous energies.
The Type-1 and Type-2 modules were tested at the IHEP Protvino 
70 GeV proton accelerator with electrons of energy between 1 and 19 GeV. 
Subsequently, the Type-2 and Type-3 modules were tested at the electron 
accelerator at Mainz, Germany, with a tagged photon beam of energy below 1 GeV.  

The Type-1 module had a cell size  of 11$\times$11 cm$^2$.  
The Type-2 and Type-3 modules have a
cell size of 5.5$\times$5.5 cm$^2$ and one module consists of 4 cells. For all three 
prototypes, lead plates of 11$\times$11 cm$^2$ size
are used for either one cell (Type-1) or for 
four cells (Type-2 and Type-3). For the Type-2 and Type-3 modules,
scintillator plates with a size of 
5.5$\times$5.5 cm$^2$ are used. 
Type-2 scintillator tiles were produced by cutting Type-1 scintillator tiles into four quadratic parts and 
thus have only one alignment pin at the outer corner. There was no reflective material between tiles and the lead plates.
Type-3 scintillator tiles were moulded and have all four alignment pins. Besides,
for the Type-3 Tyvek sheets are placed as a reflector material between the scintillator plates and the lead plates.

For the Type-1 module the radius of the WLS fibre loop was 28 mm. 
In total 72 such looped fibres form a grid of 12$\times$12 fibres per module 
with spacings of 9.3 mm. All 144 fibre ends
are assembled into a bundle with a diameter of about 10 mm, glued, cut, 
polished and attached to the photodetector
at the downstream end of the module. No optical grease was used to 
provide an optical contact between the bundle cap
and the photodetector, thus there is a natural air gap in-between. 
For the Type-2 and Type-3 modules,
18 looped fibres are used to collect light from each cell individually.  
Presently, we employ the KURARAY optical fibre instead of the BICRON one (as it was used in Type-1 and Type-2 modules).   
The Type-2 and Type-3 modules were tested at Mainz with tagged photons of energy between 
100 and 700 MeV in 2012 and 2014.

\subsection{The photo sensor}
\label{sec:photo:PMT}

\begin{figure}
\begin{center}
\includegraphics[width=\swidth]{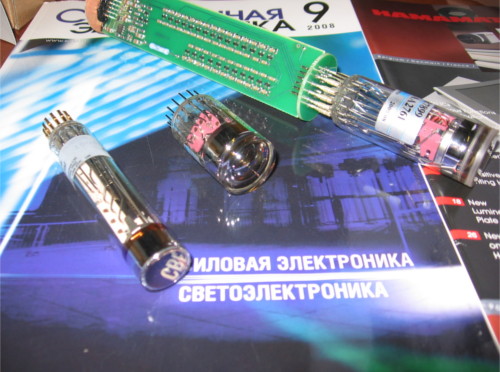}
\caption[Photograph of tested PMTs]{ Samples of PMTs to be tested.}
\label{fig:pmts}
\end{center}
\end{figure}

The \Panda Forward Spectrometer Calorimeter has to register energy depositions
in a high dynamic range with low noise at a high rate of forward-emitted photons.
Taking into account the position of the calorimeter outside of the magnetic field,
the most appropriate photo detector which can cope with the expected environment is
a photomultiplier tube (PMT). 
In this section we will describe the selected PMT, the concept
of its high-voltage base, and the performance of this photo sensor with respect to efficiency, linearity,
count rate capability, dynamic range, and noise level.
Several types of PMTs from different manufacturers (\Reffig{fig:pmts})
were selected according to their properties in the data-sheet in \Reftbl{tab:pmtsprop} and tested in the cosmic-muon test setup.

\begin{figure}
\begin{center}
\includegraphics[width=\swidth]{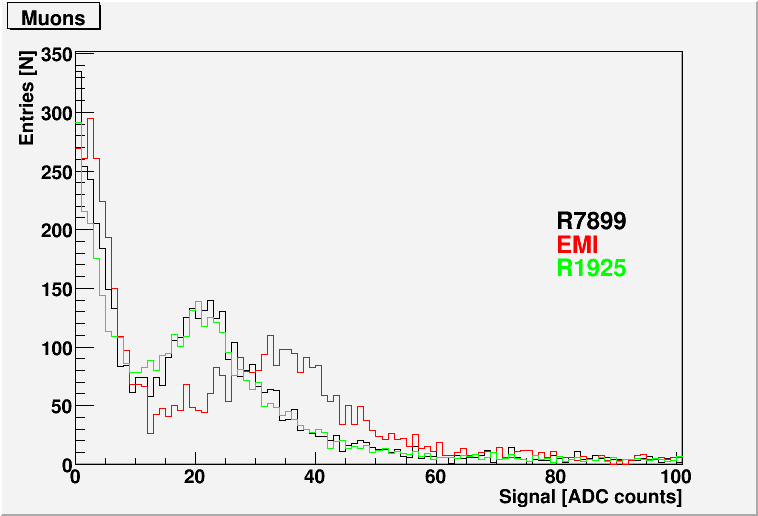}
\caption[Cosmic-muon spectra]
{Cosmic-muon spectra obtained with several types of PMT.}
\label{fig:photo:cosmics}
\end{center}
\end{figure}

In the cosmic-muon test stand the shashlyk module can be studied with muons traversing 
longitudinally or transversely. By means of blue LED the PMT gains were adjusted to the same level. 
Transversely passing muons were 
registered in the horizontal position of the module, while longitudinally passing muons were measured when
the module was turned to a vertical position. The muon rate for the transverse measurements was several times higher because of the larger detection area.
Therefore, the comparison of PMTs was based on transverse-muon data. Besides, the transverse-muon
spectrum provided a rough (upper limit) estimate of the minimal registered energy (energy threshold). The peaks in the spectra shown in
\Reffig{fig:photo:cosmics} correspond to $\approx$36 MeV.
Thus, the energy threshold for the FSC cell 
is definitely $<$36 MeV.
This figure reveals similar  noise for all three PMT types and a slightly
higher signal from
minimum-ionising particles (MIP) for the PMT from ElectronTubes (former EMI). Taking into account the size and 
(last but not least) the price, the Hamamatsu R7899 PMT was selected as the prime option for the FSC detector.

\begin{table*}
\begin{center}
\begin{tabular}{lcccccc}
\hline
PMT sample & Dark Current  & Linear range & Rise time  & QE &
Gain  & Size\\
   & nA  &  mA (type B) & ns  & @500 nm  &  $\times 10^6$  & R [mm]$\times$L [mm] \\
\hline
Photonis XP2832 & 0.5 &  & 1.8 & 15\% & 0.9 &          19$\times$61\\
ElectronTubes 9085B & 0.1 & 70 &  1.8 &  17\% & 2.4 &  19$\times$88\\
Hamamatsu R7899  &  2  &  100 &   1.6 &  14\% & 2.0 &  25$\times$68\\
Hamamatsu R1925A &  3  &  100 &   1.5 &  12\% & 2.0 &  25$\times$43\\
\hline
\end{tabular}
\caption[Compilation of PMT properties]
{Properties of tested PMTs collected from the manufacturer's data-sheets.}
\label{tab:pmtsprop}
\end{center}
\end{table*}

\subsection{High-voltage power supply and detector-control system}
\label{sec:photo:hv}

\begin{figure*}
\begin{center}
\includegraphics[width=1.1\dwidth]{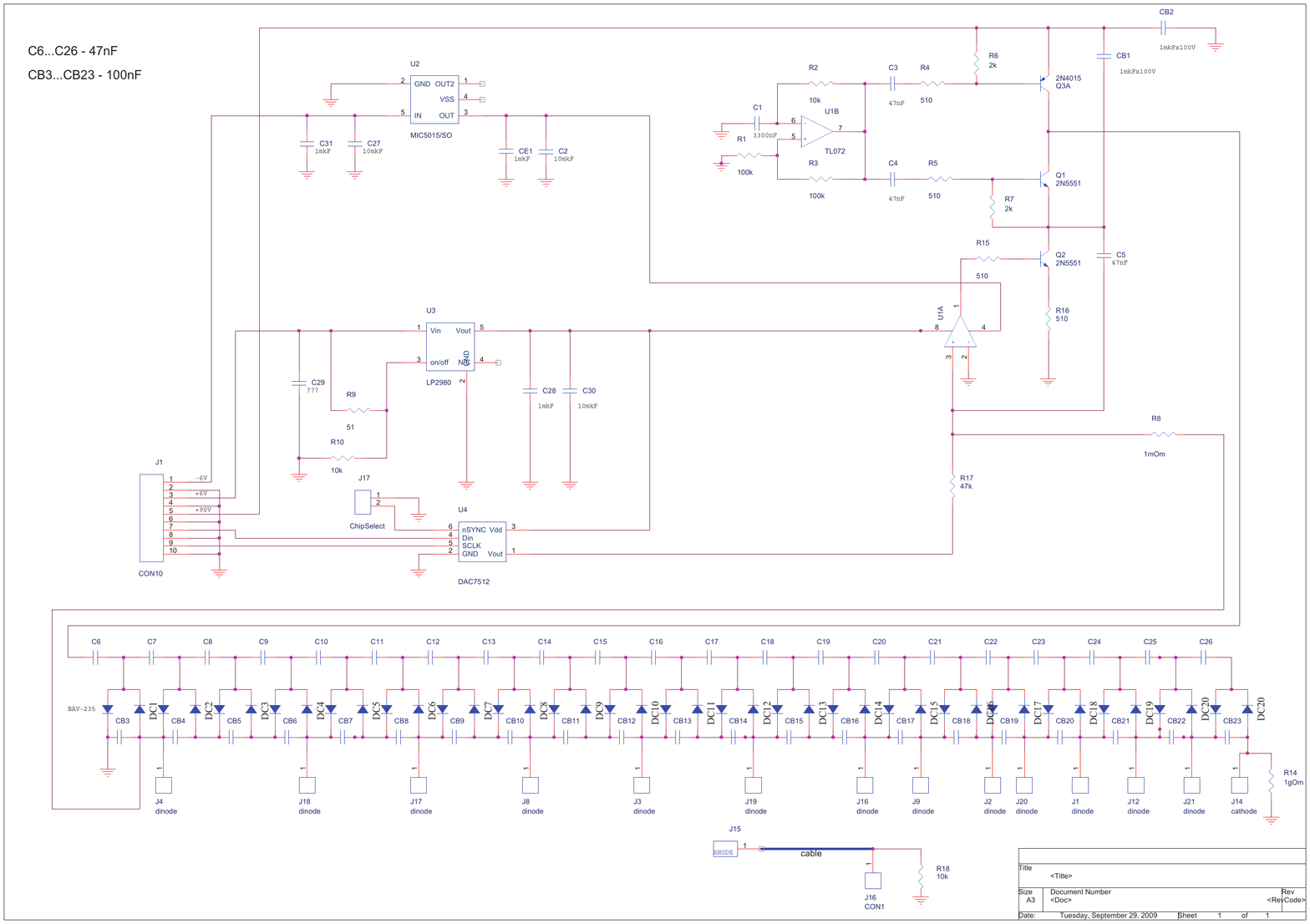}
\caption[Schematics of Cockcroft-Walton HV base]
{Schematics of Cockcroft-Walton HV base.}
\label{fig:kwbase-sch}
\end{center}
\end{figure*}

In order to provide a stable rate-independent high voltage for every dynode of the PMT, the 
high-voltage bases of Cockcroft-Walton (CW) type were designed and produced. A CW base is known 
for its high performance \cite{CW_LHCb} and relatively low cost due to the absence of long HV cables. 
The schematics of the foreseen CW base is shown in \Reffig{fig:kwbase-sch} and mainly consists of a generator, voltage multipliers, and a DAC chip to set the high voltage digitally. This base  was used for 
the test-beam measurements with Type-1 and Type-2 modules. For the Type-3 module tests, a tapered  
CW base was designed and produced, to increase the linearity range of the output current.

During the beam tests described in \Refchap{sec:perf} the FSC prototype of Type-2 comprised 64 channels and was equipped with the CW HV base.  The bases were proven to be stable and easily controllable. 
In order to control and monitor the PMT high voltage, a micro-controller-based unit (MCU) was designed, 
built and tested during the beam-test runs. Besides setting the high 
voltage and launching the user interface through the RS485 port, the MCU measures the power-supply current and provides an overload protection.

\begin{figure*}
\includegraphics[width=1.1\dwidth]{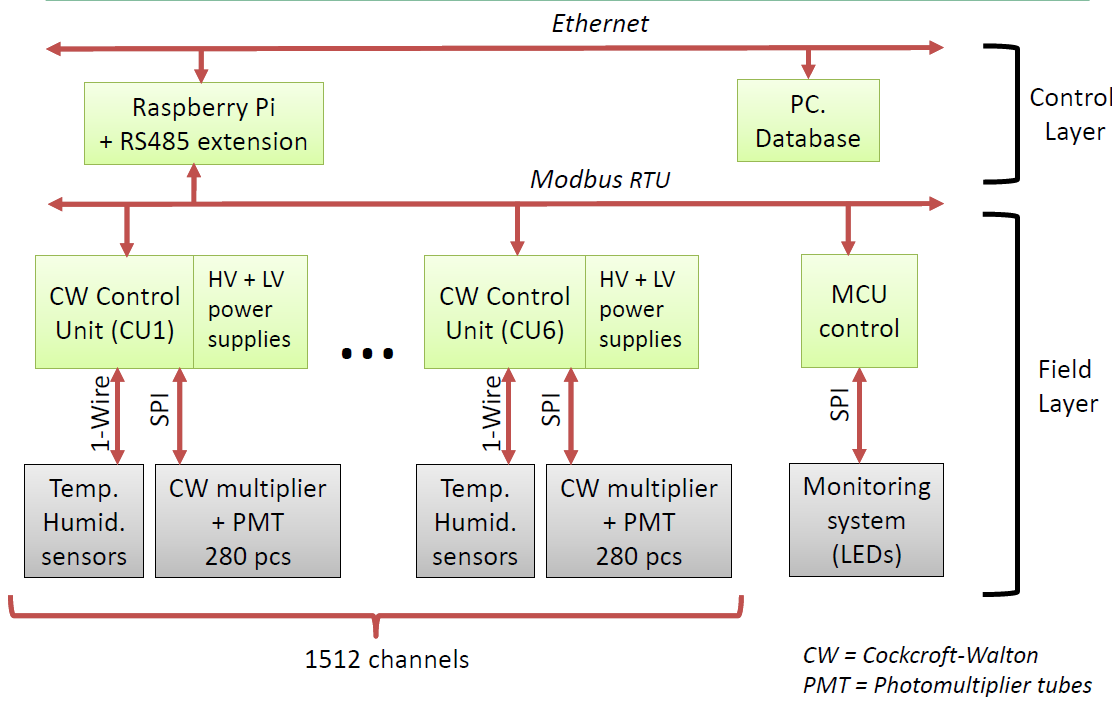}
\caption[DCS architecture for the FSC]
{DCS architecture for the FSC. The Field Layer corresponds to the devices positioned near the detector and 
consists of six control units providing HV control and temperature and humidity monitoring inside
the FSC. The link to the light-monitoring system is also envisaged.}
\label{fig:dcs}
\end{figure*}

\begin{figure*}
\includegraphics[width=\dwidth]{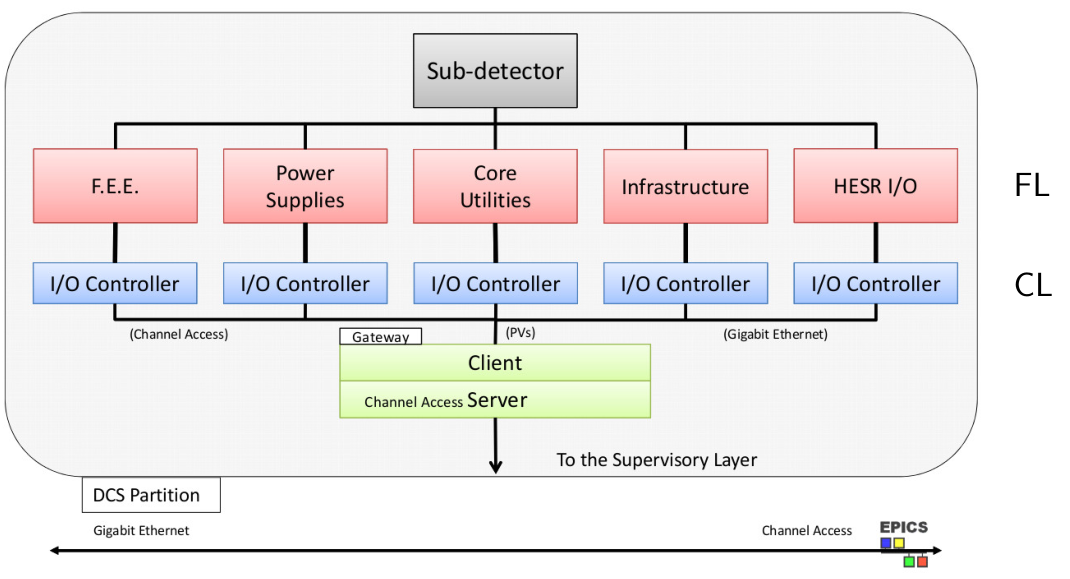}
\caption[PANDA DCS architecture]
{DCS architecture for the general PANDA subdetector.}
\label{fig:panda_dcs}
\end{figure*}

In order to provide a high-voltage control system for the FSC, we will build six control units for 280 channels 
each which will provide power and control for 10 columns $\times$ 28 lines of FSC cells.
To reduce the amount of the high-voltage cables as well as active elements in the radiation area 
and to increase the stability of the system operation, we decided to use a matrix scheme to address 
a particular channel. In this method, part of the cables run vertically for each FSC column, connecting
all channels along a column, while the other cables run horizontally for each FSC line, 
connecting channels along a row. 
By activating one column and one row one can select a unique channel to set the HV value.
All control units will be connected to the same RS485 bus and can be controlled by industry protocol 
Modbus implemented at IHEP Protvino for the control units. Thus, the whole system can be integrated 
into the \Panda Detector Control System. A set of 1-wire temperature and humidity sensors is also 
envisaged to be distributed over the FSC detector body and connected to the control units 
(\Reffig{fig:dcs}). General view if PANDA DCS based on distributed EPICS architechture is shown in the
\Reffig{fig:panda_dcs}. One can easily distinguish three layers - field layer (FL), control layer (CL) and
supervisory layer (SL) in the PANDA DCS structure. FL provides low-level access to the HV supplies, sensors, etc.
CL consists of single board computers with input-output controller executables which provide electronics control and
answers to the EPICS requests. FSC DCS is designed to be easily integrated into the general PANDA DCS and has
the same distributed and layered architechture. 

\subsection{Performance of PMT and Cockcroft-Walton high-voltage base}

\begin{figure*}
\includegraphics[width=0.95\textwidth]{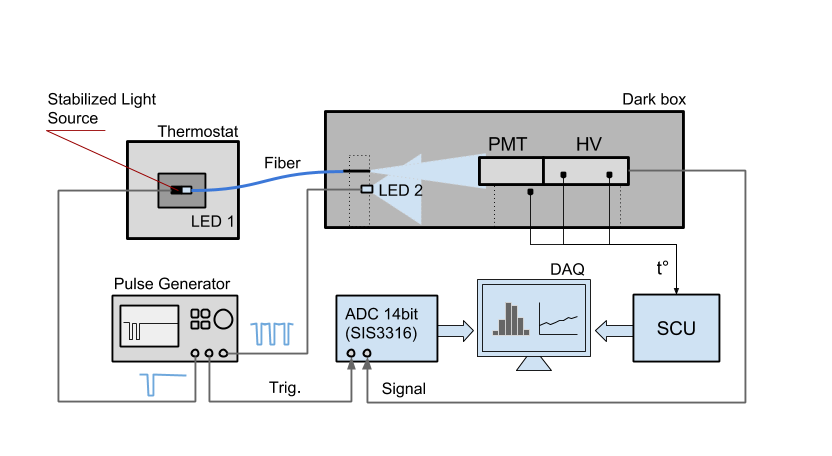}
\caption[Test setup for Cockcroft-Walton base]
{Setup for tests of CW-base parameters.}
\label{fig:bbox_PMTtest}
\end{figure*}

The \PANDA physics program requires that the FSC covers a wide range of photon energies from 10 MeV up to 15 GeV
without significant degradation of the energy resolution. To provide a linear signal from the photodetector, 
a Cockcroft-Walton base type B was designed according to Hamamatsu recommendations for the PMT R7899. This kind of 
base should provide a linear output current range up to 100 mA. To prove this and measure other 
parameters of the selected PMT and HV base, a dedicated  test setup was built at IHEP Protvino 
(see \Reffig{fig:bbox_PMTtest}).

\begin{figure*}
\begin{center}
\includegraphics[width=0.8\textwidth]{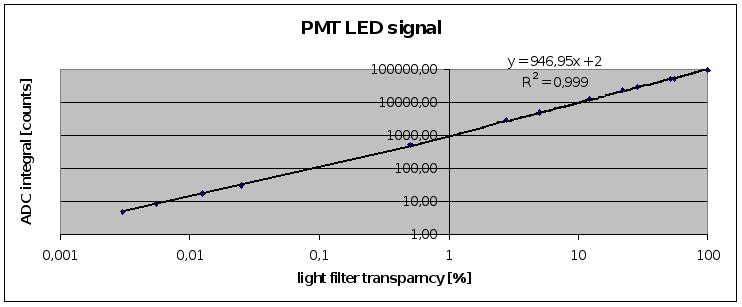}
\caption[Linearity measurement with Cockcroft-Walton base]
{Output-signal linearity for the R7899 PMT equipped with a tapered CW base (type B).}
\label{fig:CWlin}
\end{center}
\end{figure*}

As a proof of principle whether a single ADC chip would suffice to cover the full dynamic range, we provide
results of our LED measurements employing the PMT R7899, CW base type B, and the 14~bit ADC module
SIS3316 that operates at 250~M~samples/s. The LED signal was integrated over 44 samples and the base-line integral was subtracted.
The PMT high voltage was adjusted to achieve 100~mA output current, i.e., a 5~V LED signal into 50~$\Omega$ input.
For the linearity measurement the light amplitude could be reduced by a set of neutral-density filters with known optical density. The results presented in \Reffig{fig:CWlin}
show a linear response over a range of $>10^4$.

The measured root-mean-square (RMS) deviation of the base line of the Sampling ADC (SADC) was used to estimate the noise contribution of the CW base. One should note, that the RMS value in ADC counts is not the RMS of a single sample, but the RMS of the base line integrated over 44 samples equivalent to the signal integration. From
\Reftbl{tab:CWnoise} one can estimate the RMS of the noise contributed by the CW base. Assuming that noise of the CW base and the SADC are
independent, i.e.: $\sigma_{tot}^2 = \sigma_{sadc}^2 + \sigma_{CWbase}^2$, where $\sigma_{sadc}$ is the baseline RMS when 
HV and CW-base power are switched OFF, we obtain the noise contribution of  the CW base $\sigma_{CWbase}$ =7.6 ADC counts. 
With a maximum
signal of about 90.000 this give us $\approx$90~dB of signal to noise ratio.
 
\begin{table*}[h]
\begin{center}
\sf
\caption[RMS values of ADC base line]
{RMS values of the ADC base line (LED is OFF).}
\begin{tabular}{|c|c|c|}
\hline
HV status & CW power status & base line RMS (ADC counts) \\
\hline
ON & ON & 26.9   \\
\hline
OFF & ON & 26.8 \\
\hline
OFF & OFF & 25.8 \\
\hline
\end{tabular}
\label{tab:CWnoise}
\end{center} 
\end{table*}

\begin{figure}
\includegraphics[width=0.95\swidth]{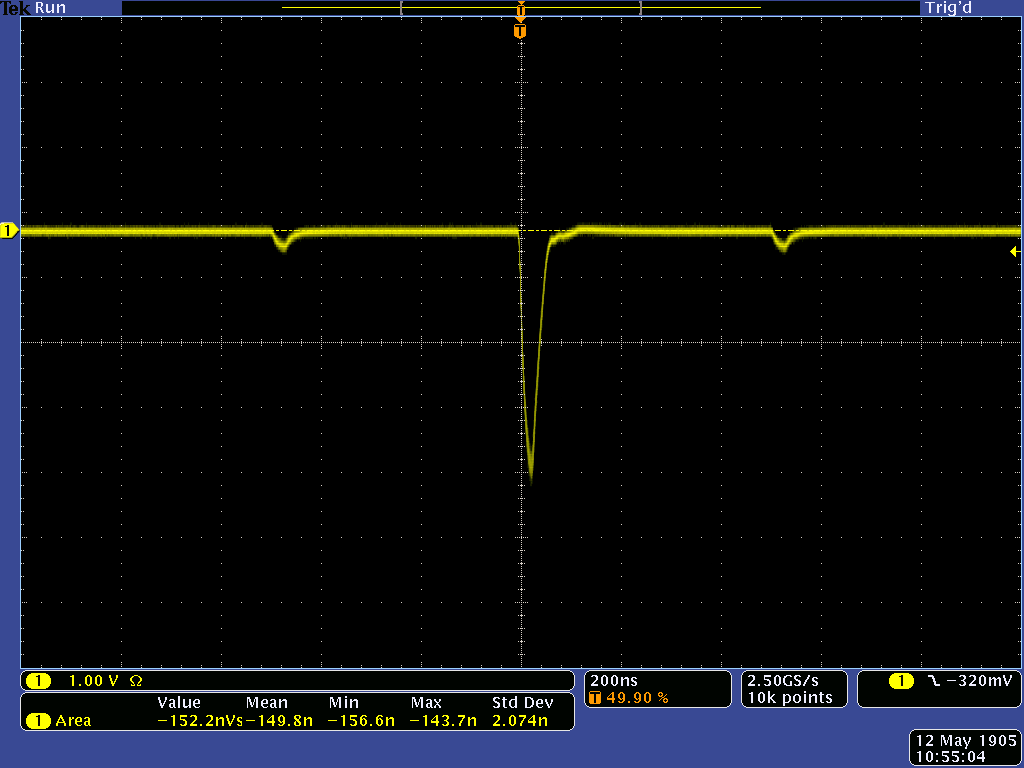}
\caption[PMT output signal for rate measurements]
{PMT output signals with load-LED frequency of 1 MHz (side signals) and the signal from the gain-LED in the centre.}
\label{fig:twoleds_ratedep}
\end{figure}

The rate performance of the CW base was studied using a dedicated setup with two LEDs. 
One LED (load-LED) emulated the background load. This was generated with variable frequency and fixed amplitude 
of approximately 300 mV, which corresponds to the most probable amplitude from the PMT during \PANDA operation, 
assuming that a 15 GeV photon generates a 5~V signal amplitude. The second LED (gain-LED) was used to measure the PMT gain and was pulsed 
by a stable-amplitude LED generator, placed in a thermo-insulated temperature-controlled box. The signal amplitude of the gain-LED was fixed to approximately 4~V, to generate an extreme-case scenario. Both LEDs were triggered by an
arbitrary function generator to control the frequency and the relative phase between them The resulting PMT output signals are shown in \Reffig{fig:twoleds_ratedep}. 

\begin{figure*}
\includegraphics[width=0.8\textwidth]{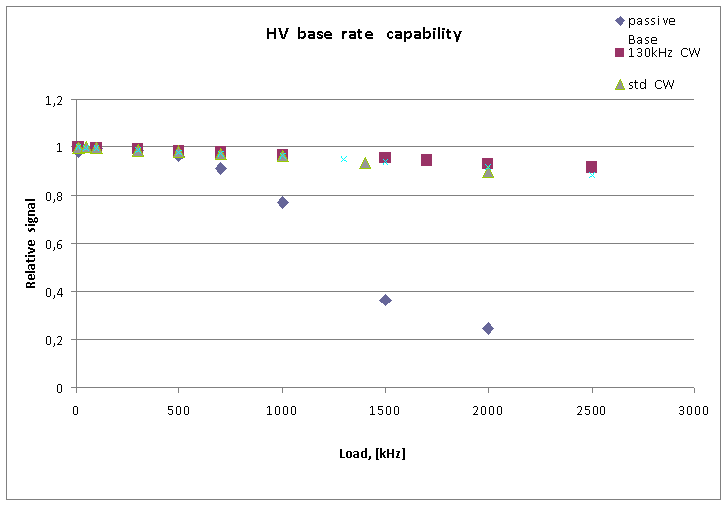}
\caption[Rate dependence of CW base]
{Rate dependence of the CW base compared to a passive HV base.}
\label{fig:ratedep_CWtest}
\end{figure*}

Three modifications of the CW base were tested. The standard base
has a conversion frequency of 80~kHz, the other two 107 and 130~kHz. Their performance was compared 
(see \Reffig{fig:ratedep_CWtest})
to that of a
passive HV base. With 2~MHz load the PMT gain dropped by 6-10\percent for the various 
modifications of CW base, while for the passive base the gain dropped by a factor of 5. As it was expected,
at higher frequency the performance of the CW base is superior.

Using the same test setup, 
the gain dependence on temperature was tested with an additional heater in the dark box. PMT and CW base were installed inside the PMT 
compartment which is a plastic tube of 40 mm diameter. This tube simulated the PMT compartment designed
for the FSC detector.

In addition, the CW-base control unit was used as a 1-wire Dallas
protocol controller with three digital thermo sensors (DS18B20) connected. The sensors provided 
the temperature measurements with a resolution of 0.05\degrees C and were installed at two points 
of the CW base inside the PMT compartment and at one point outside the PMT compartment to measure the
external temperature. The measurement provided information on
the temperature increase inside 
the PMT compartment because of the CW-base power dissipation. Even with the emulated 
most heavy load expected at the \PANDA FSC (2 MHz) the temperature increase at the hottest 
point in the PMT compartment was $<5$\degrees C relative to the external temperature.   
The measured dependence of the gain was $< 0.2$\percent/\degrees C. 

\subsection{The light-monitoring system}

Two types of a light-monitoring system (LMS) are going to be implemented in the \PANDA FSC.
Each system has specific features and will be used for different goals. 

\subsubsection{Front-side monitoring system}

\begin{figure}
\begin{center}
\includegraphics[width=0.8\swidth]{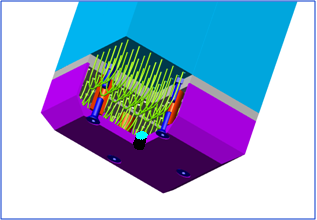}
\caption[Front-side light monitoring system]
{LED position in the front-side monitoring system. The LED is installed in the centre of the front cover of the module.}
\label{fig:front_lms}
\end{center}
\end{figure}

The front-side monitoring system consists of a set of LEDs, one for each FSC module,
and provides the possibility to emit light for each module (i.e. four cells) independently. Each LED 
is installed below the front cover of the module and illuminates all the fibre loops of the module
(see \Reffig{fig:front_lms}). Wires from the LED drivers run directly to the LED.
This simple and cheap system can be especially useful during the detector commissioning and 
maintenance.  

\subsubsection{Back-side monitoring system}
\begin{figure}
\begin{center}
\includegraphics[width=0.8\swidth]{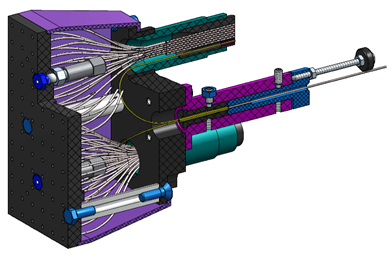}
\caption[Back-side light-monitoring system connector]
{Sketch of the 3D view of the assembled back-side region of the shashlyk module with light-monitoring system connector.}
\label{fig:mech:mb_lms}
\end{center}
\end{figure}

A more complex and precise LMS to monitor the PMT gain changes will be installed at the back 
side of the module. For the back side LMS two identical light-pulse sources will be used, one 
for each section of the FSC. A fixed fraction of the light pulse is transported to each module by means of 
quartz fibres. Each fibre connects to the optical connector of the module and is divided into four parts 
inside the module to inject light into the PMT of each cell.
A sketch of the 3D view of the assembled back-side region of the shashlyk module with the installed light-monitoring system 
connector in the centre is shown in \Reffig{fig:mech:mb_lms}.

The light-monitoring system on the back side of the module is housed in a stiff construction 
which consists of a support plate on the back side, fixed onto the module pressure plate by 
posts and screws, a centre support thread bush 
with inner collet to fix the bundle of optical fibres running through the module, an intermediate 
bush acting as light divider, and a coupling nut to hold the collet with the incoming fibre from 
the external light source. These components are placed in the back region of the FSC module. 
In this construction,
the thread bush of the centre support fulfils a double task: it serves as holder for the
optical fibres of the monitoring system and as a strong support device to fix 
the module position inside the calorimeter. 


\section{FSC module arrangement and support structure}

\begin{figure}
\begin{center}
\includegraphics[width=0.8\swidth]{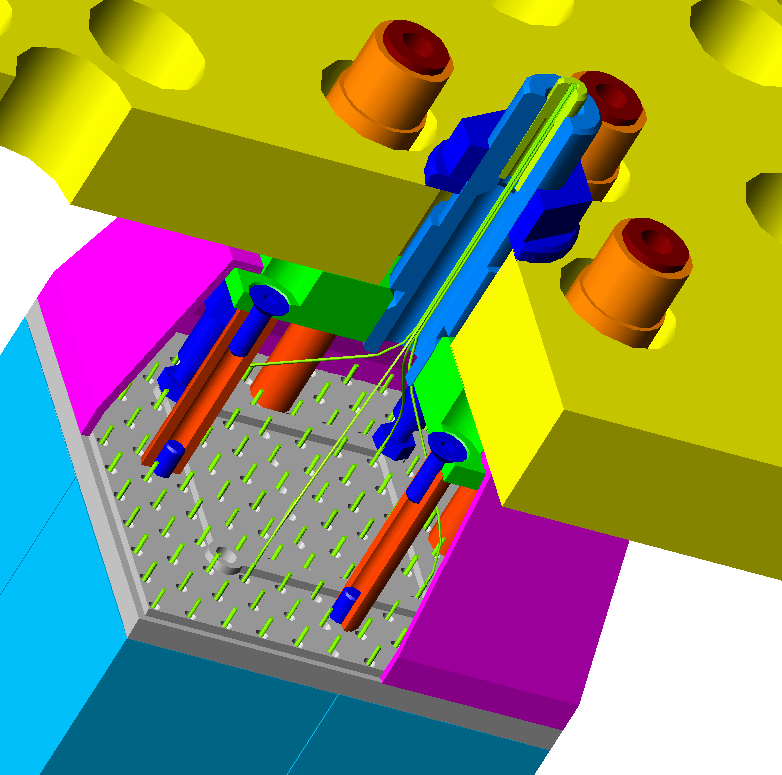}
\caption[Attachment of a module to the FSC back plate]
{Attachment of a module to the FSC back plate.}
\label{fig:mech:mod_fix}
\end{center}
\end{figure}

The modules of the completely mounted FSC are arranged in 14 horizontal layers with 27 modules in each layer.
The modules form a wall with layers placed one by one in the vertical direction. To fix the position of the modules 
in the beam direction, the FSC has a back-side vertical plate. 
Figure \ref{fig:mech:mod_fix} shows how the modules are attached and fastened to the calorimeter back plate.

The FSC is located inside the \Panda detection system in the distance range from
7800 mm to 8950 mm downstream of the interaction point with mounting gaps 
of 10 mm on the front and back side. The FSC frame should be designed such that it can be split in two sections, 
one to the left and one to the right side of the beam pipe as seen in downstream direction,
in order to avoid the disassembly of the beam pipe during installation and dismounting. The FSC 
support structure must be strong enough to carry the total weight of all modules as active detection elements, 
amounting to 4280 kg for the bigger section of the detector. The weights of the support structures and the modules, estimated from computer-aided design work, are listed in \Reftbl{tab:weight_list}.
\begin{table}[h!]
\begin{center}
\begin{tabular}{lc}
\hline
Part & Weight (tons) \\
\hline
Left support  frame &  2.9\\
Right suport frame &  3.3\\
Left set of modules  (12$\times$14 - 3) &  3.5 \\
Right set of modules (15$\times$14 - 6) &  4.3 \\
Beam pipe bracket                & 0.2 \\
Rollers, crates with support &  0.5 \\
\hline
Overall FSC weight & 14.7\\
\hline
\end{tabular}
\caption[Compilation of weight of FSC parts]
{Weight of FSC support structures and modules.}
\label{tab:weight_list}
\end{center}
\end{table}

The design has to provide two options of calorimeter movement: by crane and on rails.
The positioning tolerances and a possible module displacement due to the 
weight must be kept within the limits of 0.2 mm. In order to satisfy this requirement, 
the calorimeter needs a strong and 
stiff support structure which is able to keep the module assembly in place with high accuracy and 
to carry the total load from the heavy modules without significant flexure.

\begin{figure}
\begin{center}
\includegraphics[width=0.9\swidth]{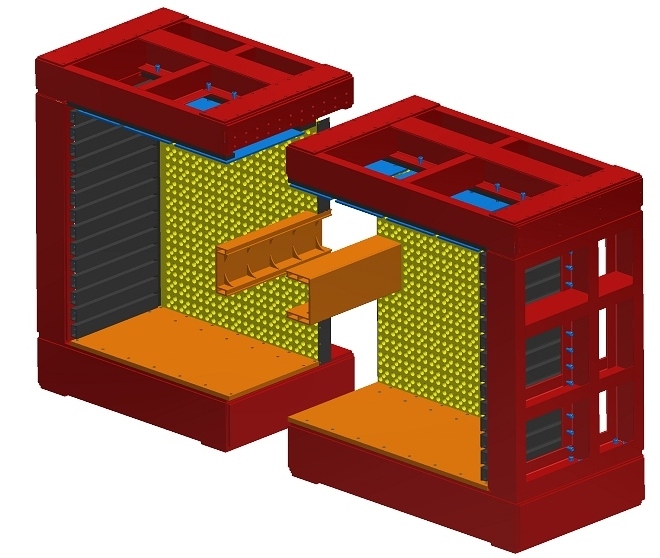}
\caption[Front-side view of FSC support frame]
{Sketch of the front-side view of the two sections of the calorimeter support frame.}
\label{fig:mech:frame}
\end{center}
\end{figure}

The FSC support-frame structure includes the following big parts: three steel frames which are 
welded of standard UPE profile and machined with high precision, the back plate to fix the position of the 
modules in axial direction, the side pressure bars for each module layer and the top pressure plates to minimise the
mounting gap between adjacent modules during the calorimeter assembly, the inner hard shield to protect
the beam pipe running through the detector from damage. All construction units are joined 
together with screws and are positioned by precision pins. The left and right sections of the calorimeter support 
frame (seen in downstream direction as in \Reffig{fig:mech:frame})
are pulled together and centred by screws and pins. 

\begin{figure}
\begin{center}
\includegraphics[width=\swidth]{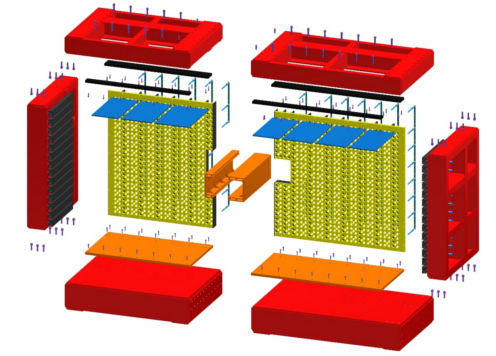}
\caption[Exploded view of FSC support frame]
{Exploded front-side view of the calorimeter support frame.}
\label{fig:mech:frame_exploded}
\end{center}
\end{figure}

\begin{figure}
\begin{center}
\includegraphics[width=0.9\swidth]{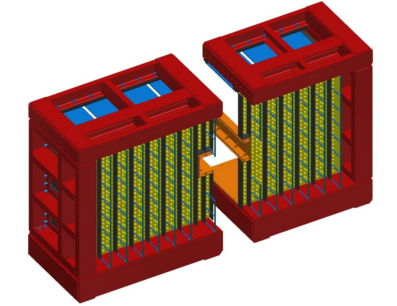}
\caption[Back-side view of FSC support frame]
{Sketch of the back-side view of the two sections of the calorimeter support frame.}
\label{fig:mech:frame_back}
\end{center}
\end{figure}

An exploded front-side view of the two sections of the support frame is shown in \Reffig{fig:mech:frame_exploded} and
reveals the details and fastening elements of the construction.
A sketch of the back-side view of the pre-assembled support 
frame as a construction unit without active elements is shown in \Reffig{fig:mech:frame_back}.

\begin{figure}
\begin{center}
\includegraphics[width=\swidth]{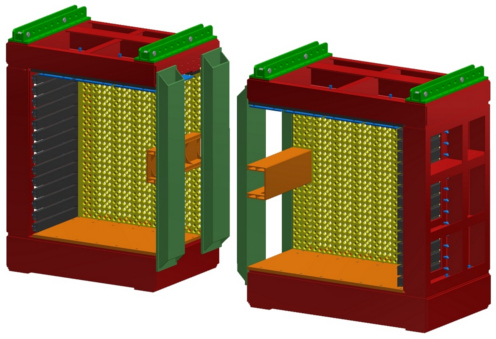}
\caption[FSC support frame with assembly and lifting tools]
{The two sections of the FSC frame with additional assembly and lifting tools shown in green colour.}
\label{fig:mech:lift_tools}
\end{center}
\end{figure}

The left and right sections of the FSC can be assembled independently in the assembly 
hall, should be lifted and positioned by crane, and inserted into the 
experiment one by one for the subsequent service operations. While mounting modules into the FSC, 
flatness deviations from the vertical plane in the region of the detector separation have to be kept as 
small as possible
in order to minimise the gap between the left and right sections, when the calorimeter will be closed. 
To meet this requirement, 
an additional stiff assembly tool has to be installed in the joint face to fix the position and to avoid 
a module displacement at the 
time of mounting the modules into the calorimeter. 
In addition, the front-side and the back-side views of the welded frames of the left and right sections, see 
\Reffig{fig:mech:frame} and \Reffig{fig:mech:frame_back}, respectively,
reveal
 a ``C'' shape with the C-opening towards the beam pipe.  Such an open support frame does not have sufficient
strength and rigidity for lifting and transportation. Therefore,  the open support frames 
should be closed temporarily during transportation by an additional strengthening structure, which has  brackets on the top for lifting by a crane. 
The required assembly and lifting tools with fastening elements are shown in \Reffig{fig:mech:lift_tools}. 
The metal structures for the FSC supports, like welded support frames, attachment elements, and 
additional assembly and installation tools, can be produced and pre-assembled 
at the production facilities of IHEP Protvino.

\subsection{Stress and deflection analysis}

The principal requirement for the detector support structure is maintaining a stable FSC geometry.
Therefore, it is important to fix the location of the active elements under different types of load during 
the detector assembly, installation and operation. 
Detailed calculations were carried out to  ensure that the support frame with the designed parameters 
can satisfy 
these requirements. A finite-element (FE) model was developed for these calculations. Results of 
the stress and deflection analysis will be presented below in \Refsec{sec:mech:stress_calc}.

\subsection{Strength and rigidity of the detector support system}

\begin{figure}
\begin{center}
\includegraphics[width=0.8\swidth]{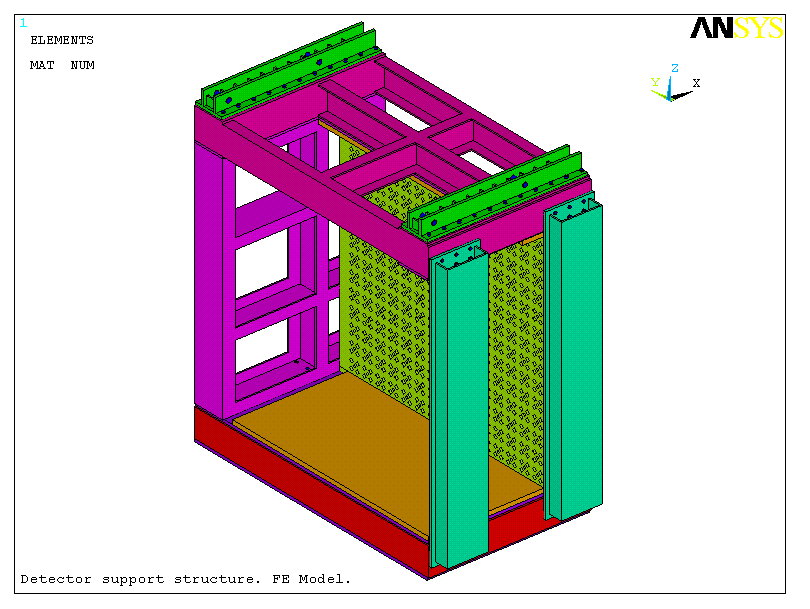}
\caption[FE model of FSC support frame]
{FE model for support-frame strength calculations with temporary reinforcement beams on the side of the frame.}
\label{fig:mech:strength_model}
\end{center}
\end{figure}

The two sections of the FSC will be assembled in the assembly hall and subsequently installed in the beam zone. 
Strength and rigidity calculations were carried out to ensure a safe transport
without damage to the FSC modules. The frames are supposed to be moved by crane using lifting 
brackets on top of the frames. 
The maximum displacement of modules when moving and lifting or lowering the detector, 
as well as the possible seismic loads must not exceed 0.2 mm. The FE model, designed according to the 3D-drawings of the detector containing 15$\times$14 modules, is shown in 
\Reffig{fig:mech:strength_model}.

The model contains all the basic elements of 
the support frame: lower, upper and side beams, the table for stacking modules on the bottom of the frame, 
the rear bearing plate (back plate), 
brackets for lifting the frame by crane, supporting temporary reinforcement beams used at 
the stages of assembly and transportation of the detector, as well as bolted connections. 
All elements of the supporting structure are made of steel. The modules itself are not included in the 
computational model. Instead, their weight is taken into account as a  constant pressure on the table.

\subsection{Loading and fixation}
\begin{figure}
\begin{center}
\includegraphics[width=0.8\swidth]{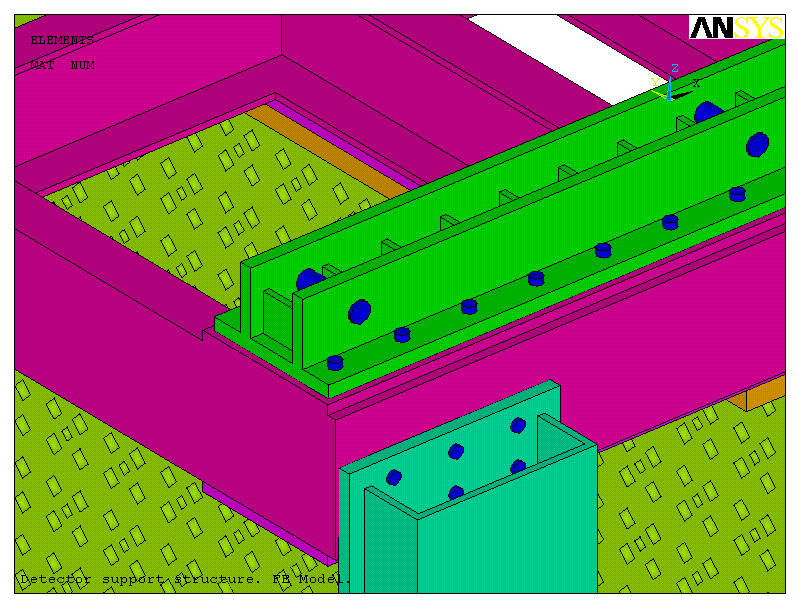}
\caption[Details of FE model for FSC frame]
{Fastening elements used in the FE model.}
\label{fig:mech:strength_model_details}
\end{center}
\end{figure}

Concerning the load on the support frame, there are two different modes of detector 
handling:
\begin{itemize}
\item The mode of detector lifting and moving (Lifting Mode);
\item The mode of the everyday detector operation (Operation Mode).
\end{itemize}
Two sets of calculations were carried out for the detector support frame. The first one corresponds to 
the raising-lowering movement by means of a crane, the second represents the operational phase, when the detector 
is laid on the rollers. In both cases, the calculations were made for a linear time-invariant elastic structure 
loaded with its own weight and the weight of modules. Differences appear in the support as well as in
the geometry of the model: In the Operation Mode there are no temporary reinforcements.
Figure
\ref{fig:mech:strength_model_details} shows the boundary conditions for the first case. 
The frame construction is suspended by crane on four hooks. The location of the hooks is chosen to 
minimise possible distortions of horizontal frame parts.
The weight of the modules is represented as a load to the table in the lower part of the support frame.

\begin{figure}
\begin{center}
\includegraphics[width=0.8\swidth]{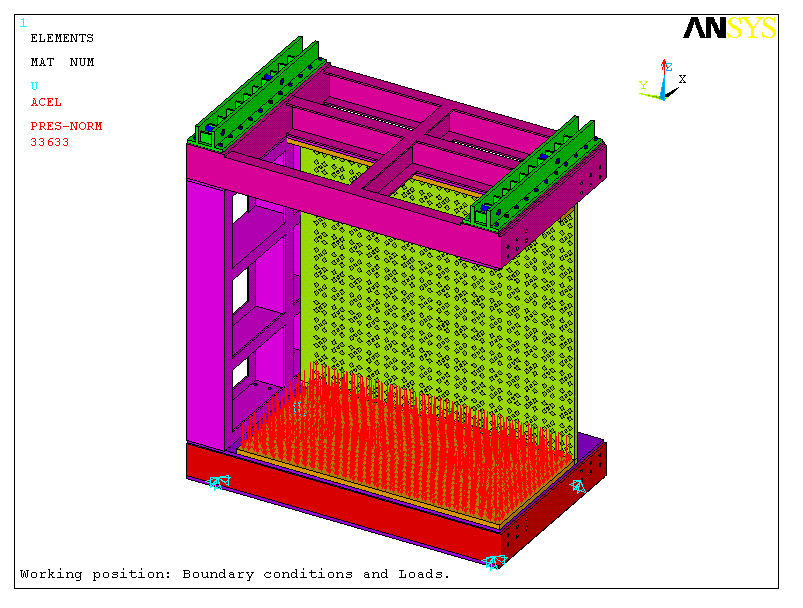}
\caption[FE Model for FSC frame in Operation Mode]
{ FE model for the strength calculation of the support frame in Operation Mode.}
\label{fig:mech:model_work}
\end{center}
\end{figure}

The model geometry and boundary conditions for the calculation of the stress distribution in the Operation 
Mode are shown in \Reffig{fig:mech:model_work}. The support frame rests on four rollers and the
temporary vertical reinforcement beams are removed, so that one corner of the upper frame is hanging freely.

\subsection{Finite-element analysis of the support frame}
\label{sec:mech:stress_calc}

\begin{figure}
\begin{center}
\includegraphics[width=0.8\swidth]{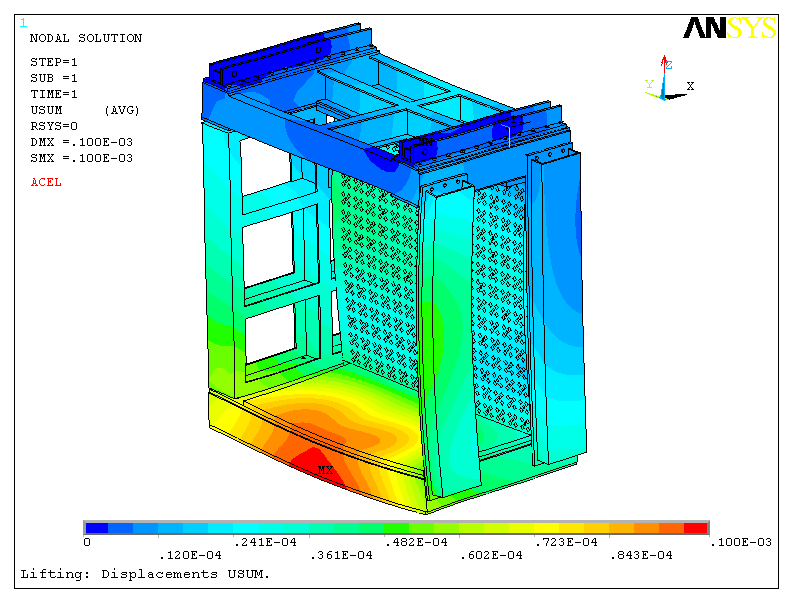}
\caption[Displacement from FE Model for FSC frame in Lifting Mode]
{FE model calculation of the displacement map for the support frame in Lifting Mode.}
\label{fig:mech:lift_res}
\end{center}
\end{figure}

\begin{figure}
\begin{center}
\includegraphics[width=0.8\swidth]{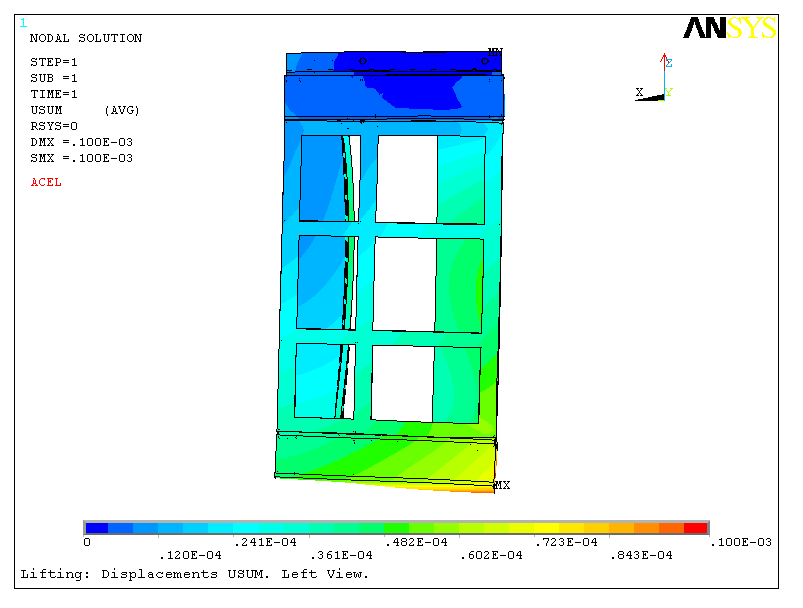}
\caption[Displacement from FE Model for FSC frame in Lifting Mode (side view)]
{FE model calculation of the displacement map for the support frame in Lifting Mode; side view.}
\label{fig:mech:lift_res_side}
\end{center}
\end{figure}

\begin{figure}
\begin{center}
\includegraphics[width=0.8\swidth]{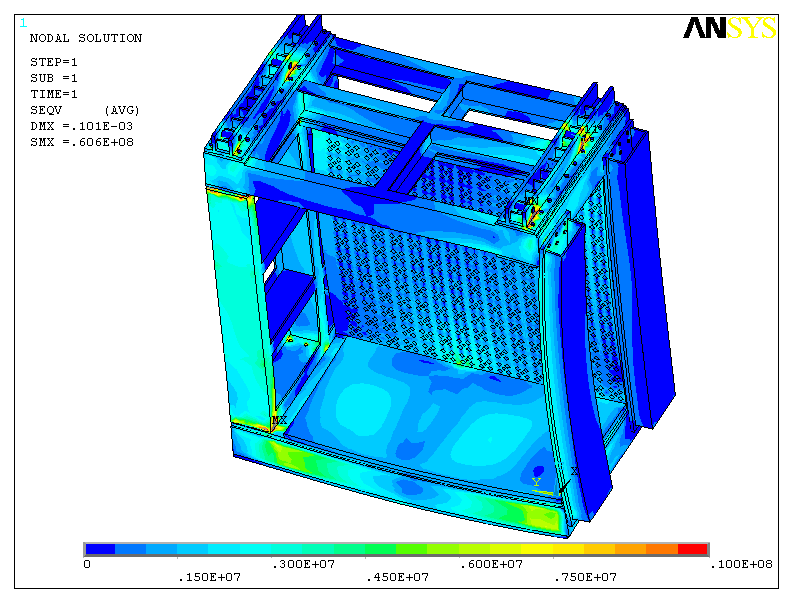}
\caption[Stress from FE Model for FSC frame in Lifting Mode]
{FE model calculation of the stress map for the support frame in Lifting Mode.}
\label{fig:mech:lift_stress}
\end{center}
\end{figure}

The results of FE model calculations for the field modulus of the displacement vector are shown in
\Reffig{fig:mech:lift_res} and \Reffig{fig:mech:lift_res_side},
and \Reffig{fig:mech:lift_stress} shows the equivalent stress field in Lifting Mode.
The maximum displacement in the support frame is 0.1~mm, which is two times smaller than the required limit. 
This result indicates a high stiffness of the design. As can be read from the figure, the stress 
at almost all points of the structure does not exceed 10 MPa.  Stress concentrations (up to 60 MPa) 
are observed only in small regions in the joints of the frames, as well as near the suspension point. 
Bearing in mind that the allowable stress of 240 MPa, the design safety factor is $>$4.

\begin{figure}
\begin{center}
\includegraphics[width=0.8\swidth]{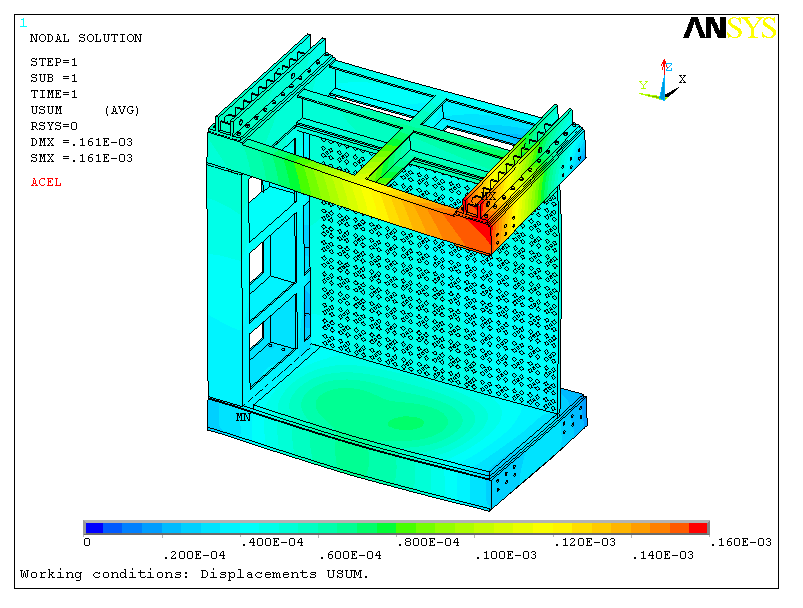}
\caption[Displacement from FE Model for FSC frame in Operation Mode]
{FE model calculation of the displacement map for the support frame in Operation Mode.}
\label{fig:mech:work_res}
\end{center}
\end{figure}

\begin{figure}
\begin{center}
\includegraphics[width=0.8\swidth]{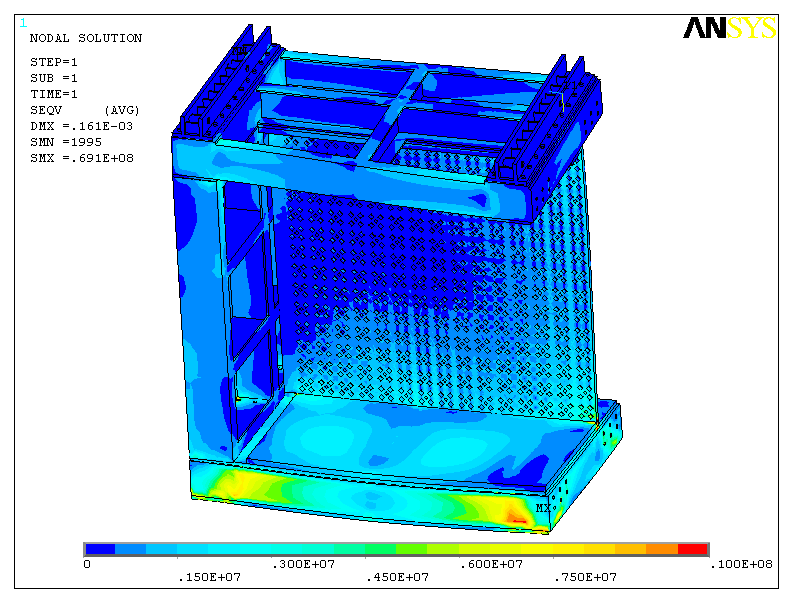}
\caption[Displacement from FE Model for FSC frame in Operation Mode]
{FE model calculation of the stress map for the support frame in Operation Mode.}
\label{fig:mech:work_res_stress}
\end{center}
\end{figure}

The results of equivalent FE model calculations for the Operation Mode at the nominal gravitational load are 
shown in \Reffig{fig:mech:work_res} and \Reffig{fig:mech:work_res_stress}. Here, as expected, the 
maximum displacement of 0.16~mm is found at the unsupported corner of the upper part of the frame. 
It is worth noting that the maximum displacement at the bottom part of the frame, loaded with the weight of the modules, 
is only 0.075 mm, which is three times lower. The maximum stress of 70 MPa is observed near 
the support points. Outside these small regions, the stress concentrations do not exceed the
value of 10 MPa. It should be noted, that the maximum stress of 70 MPa 
is due to the coarseness of the four-point support model. According to our experience in the development 
of similar  structures, the stress at the attaching points of the rollers will not exceed 20 MPa. 
Thus, for the Operation Mode a design safety factor of 12 can safely be assumed.

The stress for the bottom module can be estimated as follows.
The weight of the module is 21.5 kg and the surface of the module is flat with a
tolerance of 20 microns (provided by the injection mold tolerance). Thus the stress
is distributed uniformly over the module surface,
which results in a pressure to the bottom module of  0.046 MPa (average).
Taking into account the holes in the
tile one can calculate the effective average pressure increase up to 0.053 MPa.
Stress concentration at holes
is three times higher – 0.16 MPa. Comparing this stress with tensile strength and
ball indentation hardness
from the table of BASF-143 properties, provided by the manufacturer
(Table~\ref{tab:basf143}), we can see a safety factor of several orders of magnitude.

\begin{table}
\begin{center}
\begin{tabular}{|p{0.45\swidth}|p{0.45\swidth}|}
\hline
Material property & value \\
\hline
Density  & 1.043 g/cm3 \\
Tensile Modulus (Young Modulus) & 3.30 GPa \\
Poisson ratio & 0.22 \\
Tensile Strength & 46 MPa \\
Flexural Strength  & 72 MPa \\
Elongation at Break & 2\% \\
Ball indentation Hardness (Ball diameter 5 mm, force 358 N, time 30 seconds) & 150 MPa (indentation depth 0.167 mm) \\ 
\hline
\end{tabular}
\caption {BASF 143 strength properties}
\label{tab:basf143}
\end{center}
\end{table}

\section{Cooling system}

\label{subsec:mech:cooling}
The only heat source inside the FSC is the set of PMT bases. The average power consumption
during test-beam studies of the 64-cell FSC prototype was about 5 W, which means at most
100 mW per PMT. In order to keep the temperature inside the FSC at a constant level,
one needs to compensate the heat load of 150 W for a total of 1512 cells. 

\begin{figure}
\begin{center}
\includegraphics[width=0.8\swidth]{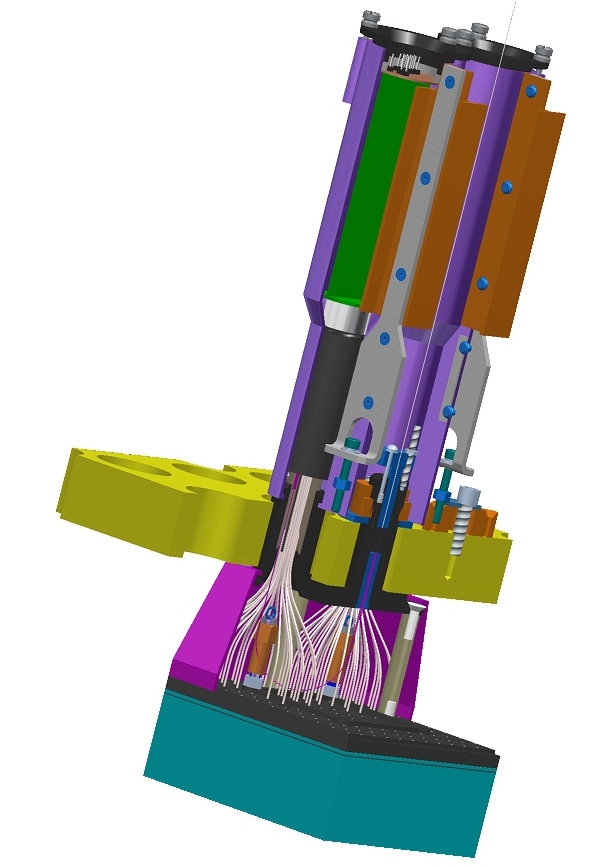}
\caption[PMT and HV-base compartment]
{PMT and HV-base compartment of the module. Orange parts are metal I-beams to transfer the heat load from the internal
volume of the plastic compartment to the outside fixation plate (yellow) through the metal mounts (grey).} 
\label{fig:pmt_heatleak}
\end{center}
\end{figure}

To keep the temperature of CW base and PMT stable, one needs to transfer heat effectively from the PMT compartment to
external structure elements and the environmental air. The original design of the PMT compartment with plastic tubes 
was modified to include metallic parts (see \Reffig{fig:pmt_heatleak}). Two metal I-beams are inserted into the plastic
tubes through slots to transfer the heat load from the internal part of the PMT and CW-base compartment.  
This design was used to calculate the  temperature distribution across the detector.

\begin{figure}
\begin{center}
\includegraphics[width=0.8\swidth]{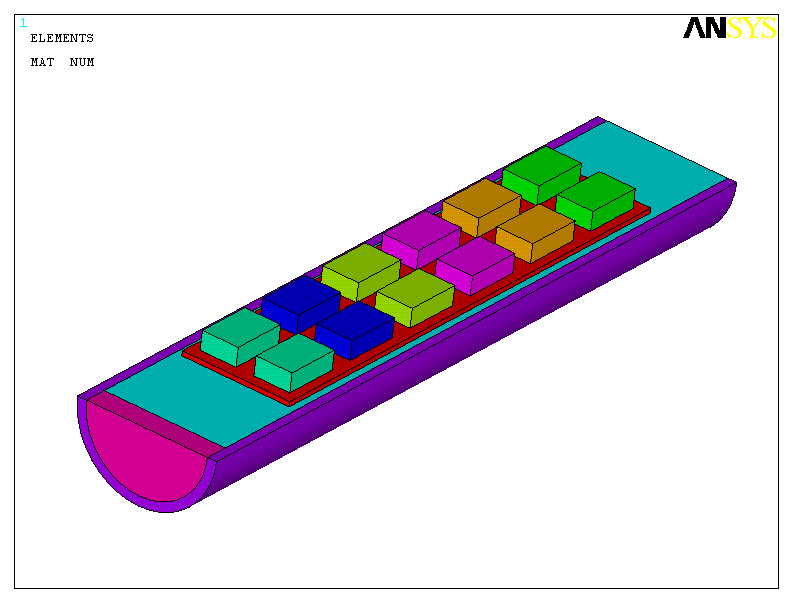}
\caption[Thermal source model of PMT]
{Model of the PMT base as a thermal source.}
\label{fig:mech:thermal_src}
\end{center}
\end{figure}

\begin{figure}
\begin{center}
\includegraphics[width=0.8\swidth]{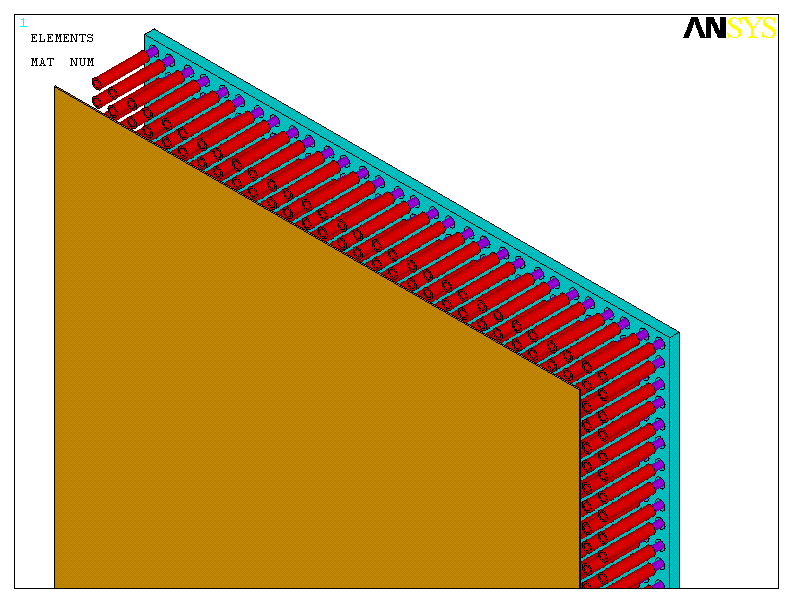}
\caption[Heat sources over the detector back-plate]
{Distribution of the heat sources over the detector back-plate.}
\label{fig:mech:thermal_det_back}
\end{center}
\end{figure}

Natural convection cooling and heat conduction through the structural elements are preferred to 
avoid vibrations of photo detectors 
and cables and instabilities caused by ageing of mechanical blowers.  
In order to check the feasibility of natural cooling, detailed simulations 
and temperature calculations were performed.  
The model of a single heat source is presented in \Reffig{fig:mech:thermal_src}. The model of the complete 
detector back-side area with photomultiplier bases, modules attached to the back-plate, and a protective cover is 
shown in \Reffig{fig:mech:thermal_det_back}. 

\begin{figure}
\begin{center}
\includegraphics[width=0.9\swidth]{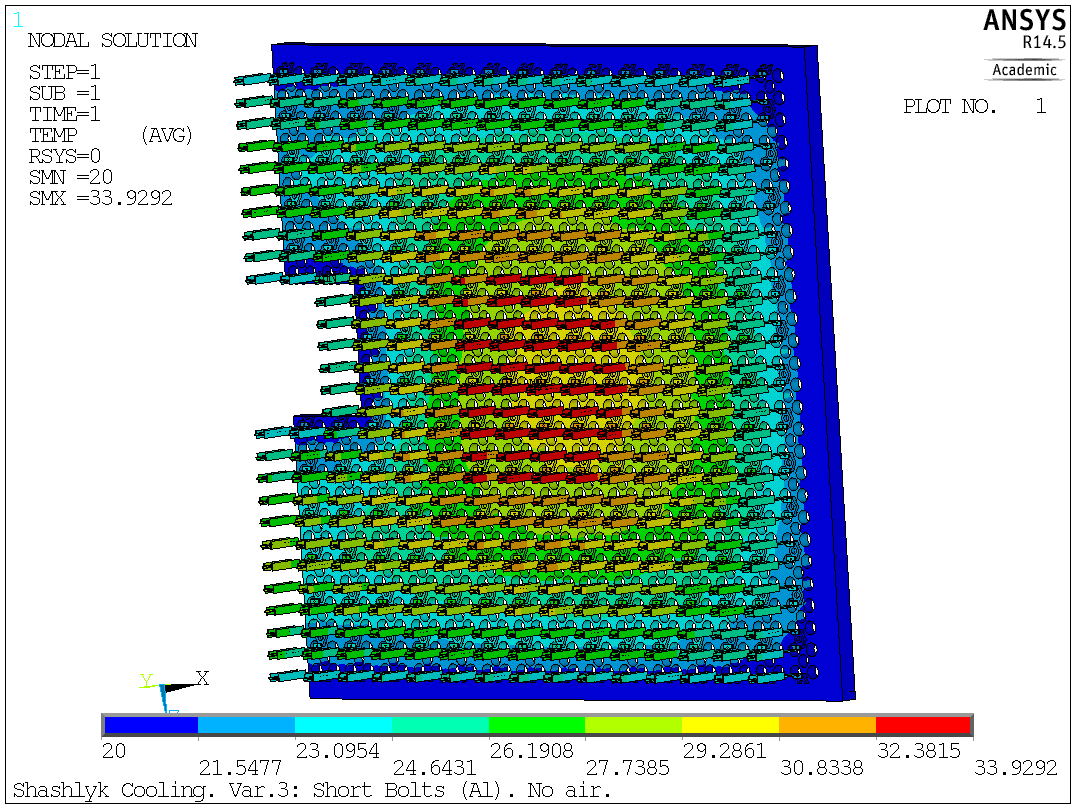}
\caption[Temperature map for the steel back plate]
{Temperature distribution over the FSC back side (steel back plate).}
\label{fig:new_temp_sim}
\end{center}
\end{figure}

The model calculations take heat conduction into account but ignore 
convection. Thus, in reality the temperatures will be lower.
The results of the simulations is a temperature distribution map 
across one half of the detector back-plate (see \Reffig{fig:new_temp_sim}).
The map shows that even without active cooling the maximum temperature 
is 32.4\degrees C assuming an external temperature 20\degrees C.



More powerful heat sources are the front-end electronics crates in AdvancedTCA (ATCA) standard, 
which will be installed at the sides of the FSC detector frame. The front-end
electronics for the FSC consists basically of high-frequency sampling ADCs with typical 
power consumption of 1-2 W per channel. The ATCA cooling system can easily
handle this heat load. The environmental system of the beam hall should be capable of 
removing this additional $\approx$2000 W of heat.

\section{Moving system}

\subsection{General requirements}
A dedicated moving system is required to install the two FSC sections in the beam position and to move them out for maintenance, 
Such a system has to include: different types of rollers to move the calorimeter, an adjustment system, rails, and a driver. 
The main function of the moving system 
is shifting the left and right sections of the FSC by approximately three meters away from the beam 
to the maintenance position and back. The rollers have to carry a total weight of $\approx$15 tons for the fully 
equipped calorimeter.

\subsection{Design of the moving system}

\begin{figure}
\begin{center}
\includegraphics[width=\swidth]{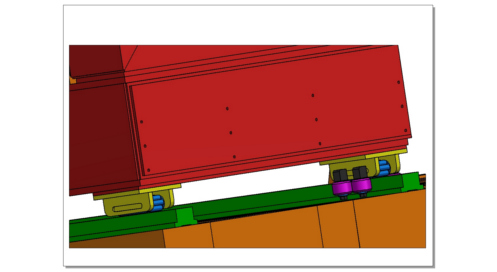}
\caption[Roller system for moving the FSC]
{Components of the FSC moving system.}
\label{fig:mech:roller_system}
\end{center}
\end{figure}

The components of the moving system are sketched in \Reffig{fig:mech:roller_system}. 
Commercial rollers were chosen as the running parts for the moving system for these reasons: rollers are widely 
used standard units which can carry a heavy load in different directions, while having small dimensions; 
the price is low; additional services (electronics, cooling or an air system) are not required; they are easily used and maintained.
Rollers are placed under the support-frame sections in the four corners and fixed 
by screws. Rails consist of fixed and removable parts: the central rails, below the beam line, are fixed to the Forward Spectrometer 
moving platform and the two side parts can be mounted, when the detector has to be opened and moved for maintenance. Standard 
profiles will be used as rails, which are attached with screws to the support platform and on additional 
temporary supports in the experimental hall. The procedure will be described in more detail in
\Refsec{sec:mech:installation} below.

According to general requirements for the moving system, we are currently considering to install two 
types of rollers for two rails: the first roller type takes only the vertical weight-load and the second type will 
carry the vertical and horizontal load to support the detector and to give a movement direction. 

Several production companies manufacture various roller systems for different objects. 
Roller blocks of INDUSTRIAL LIFTING company satisfy the needs of the FSC. The most suitable 
roller types are A-H and A-H-FR-E with a carrying capacity of 150 kN. The rollers catalog \cite{rollers}
provides a description, drawings, outer and attachment dimensions, and characteristics of the roller blocks.
The left and right sections of the detector, each weighing about 7.5 tons, have to be moved 
slowly and very accurately in order to close the calorimeter without any damage. 
Various devices can be applied as a driver for continuous or stepwise motion, e.g. cable hoist, screw or hydraulic jack. 
The final choice of the drive mechanism will be done when the complete design of the FSC has been
approved and the scenarios of the detectors installation and servicing as well as the experimental-hall equipment have been defined.

The adjustment and control system comes as an integral part of the calorimeter support structure and the moving system. 
The task of the adjustment is a precise calorimeter closing to minimise the inactive gap between the left 
and the right section 
and to place the detector back into the same position after an open-close operation. 
Small hydraulic or screw jacks and catch pins, which are placed under and between the 
support frame sections, are foreseen as adjustment units.

\section{Assembly, installation and maintenance}

The strong need to provide a calorimeter, which is split in two sections (to the left and the right of the beam), 
requires two separate frames,
which can be disconnected and connected together with high accuracy to combine into the united calorimeter. 
It is also important to design a simple and precise 
procedure for the detector assembly and installation, including the possibility to move detector sections 
by a crane or on a rail system. 
Moreover, the beam pipe passing through the centre of the calorimeter should be protected by a dedicated hard shield 
against damages at the time of detector installation and operation.

\subsection{Assembly}

FSC modules should be pre-assembled and tested at IHEP Protvino. The detector support frame can be 
manufactured by a Russian production company. Pre-assembly and testing of the detector structure is envisaged,
when all mechanical parts will be ready, to check the interfacing of components.
At the stage of the calorimeter assembly, modules have to be placed to the required position by fixing 
them to the back plate of the support frame. Thereafter, photo detectors 
and high-voltage bases can be installed. Finally, the optical 
connectors of the back-side light-monitoring system can be plugged in. For the beam pipe safety two 
protection structures will be installed around the hole for the beam pipe. The left and right sections of the FSC 
can be assembled separately in the assembly hall, tested, calibrated and transferred to the 
\Panda experiment hall.

In the beam position, the FSC detector will be placed on the Forward Spectrometer moving platform together with the other \Panda 
sub-detectors (RICH and Forward TOF wall). A special rail system should be installed for the FSC positioning 
on top of this platform. At the time of the installation and maintenance, the left and right sections of 
the FSC will be moving on two additional temporary supports and an extension of the permanent rails is needed.
The principal tasks at the time of calorimeter maintenance are:
\begin{itemize}
\item Unlocking and separating the left and right sections of the FSC.
\item Displacement of the two FSC sections from the beam position to the maintenance positions.
\item Assurance of easy access to each part of the FSC.
\end{itemize}

\begin{figure}
\begin{center}
\includegraphics[width=\swidth]{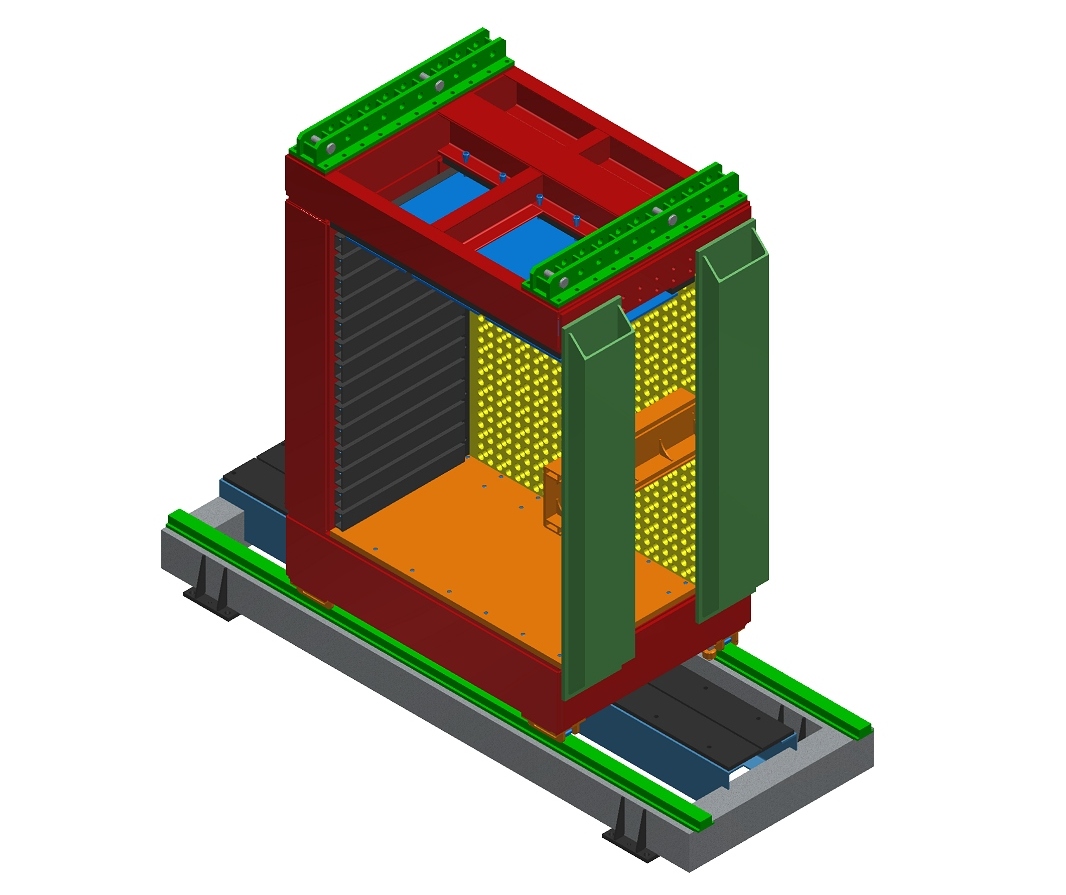}
\caption[Assembly of left section of FSC frame]
{Assembly of the left section of the FSC support frame.}
\label{fig:mech:detframe_left}
\end{center}
\end{figure}

\begin{figure}
\begin{center}
\includegraphics[width=\swidth]{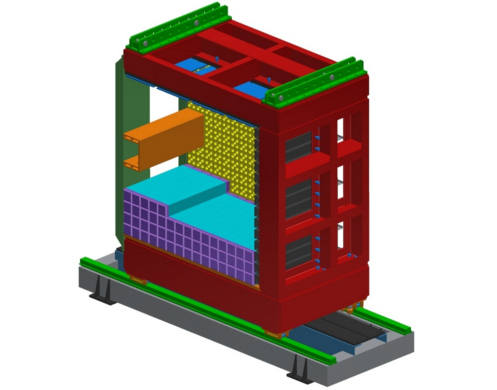}
\caption[Module installation in right section of FSC frame]
{Installation of modules in the right section of the FSC support frame.}
\label{fig:mech:modinst_right}
\end{center}
\end{figure}

\begin{figure}
\begin{center}
\includegraphics[width=\swidth]{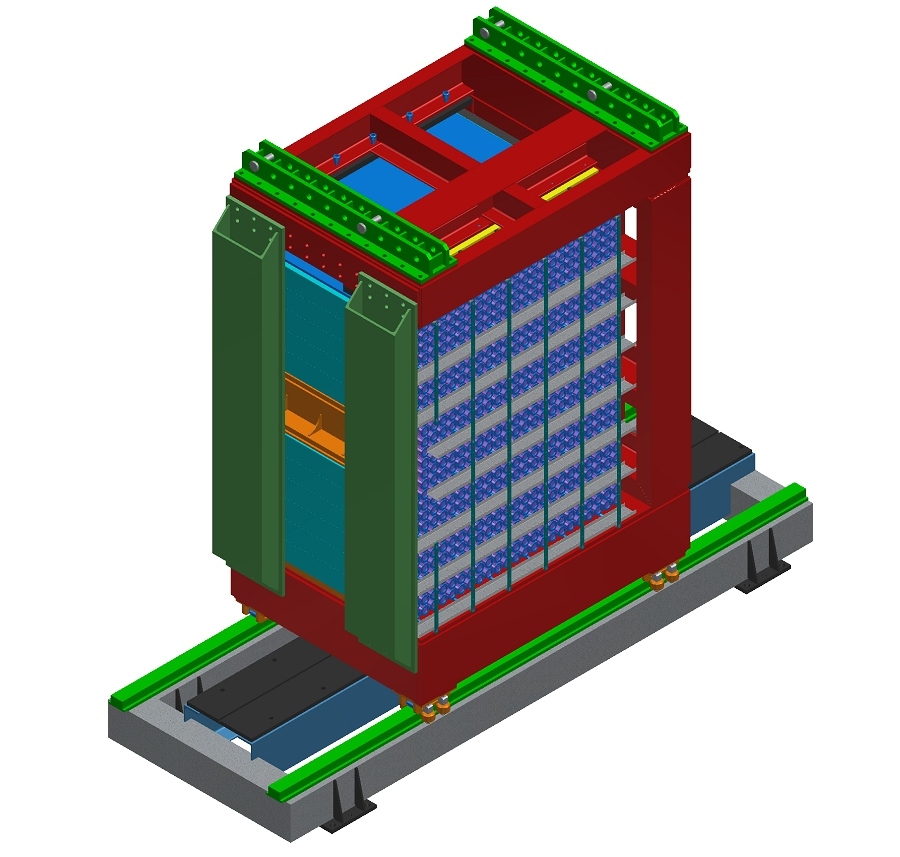}
\caption[Cable and fibre installation for the left FSC section]
{Cable and fibre installation for the left FSC section.}
\label{fig:mech:detcables_left}
\end{center}
\end{figure}

There are two possible ways of assembling the FSC. In the first way, the two FSC sections are pre-assembled 
in the assembly hall and subsequently lifted and transported to the experimental hall for the final 
installation in the beam line. In the second way, the FSC is directly assembled in the experimental 
hall close to the final position in the beam line.
Both ways of assembly are principally identical, but require different tools.
The sequence of FSC assembly in the assembly hall and the required tools are listed here:

\begin{itemize}
\item Module assembly and functional tests. Special assembly tooling is required.
\item Assembly of frames for two FSC sections, including two temporary
strengthening beams to stabilise the C-shaped frames. Temporary support rails, removable 
beams, and crane are needed (see \Reffig{fig:mech:detframe_left}).
\item Checking of all  outer dimensions of the frames, fitting of joints of left and right section and roller system.
\item Stacking of modules layer by layer inside the support frame, starting with the first layer on 
the table in the bottom part of the frame. Additional lifting and moving tools are required (see 
\Reffig{fig:mech:modinst_right}).
\item Apply pressure to each of the module layers by side pressure bars to minimise gaps between modules.
\item Fixing the location of the modules at the back plate by tightening nuts.
\item Installation of the photo-detector compartments, PMTs and high-voltage bases.
\item Installation of the back-side light-monitoring system.
\item Mounting of horizontal cable trays inside calorimeter back-side area.
\item Mounting of the front-end electronics crates on the previously fixed support bracket at the sides of 
the support frames.
\item Mounting of electronics units on top of the FSC frame.
\item Laying-out of the cables and optical fibres in the cable trays (see \Reffig{fig:mech:detcables_left})
 and connecting it to the HV-bases.
\item Functional tests of the optical monitoring system and electronics.
\item Installation of the outer environmental shields. 
\end{itemize}

Each FSC module has four high-voltage bases and a monitoring-system input. 
All parts are located behind the calorimeter support plate. Front-end electronics will be 
located aside the support frames. High-voltage control units and the monitoring-system
electronics modules will be located on top of the support frames. Inside the detector, supports are required 
for control- and power-cables for the high-voltage bases as well as 
signal output cables from the photo detectors. Also, optical fibres of the back-side monitoring 
system can be routed here. From outside, the FSC only requires a power-supply cable for the
main electric power and an optical link for the data output. 

As far as cooling is concerned, we need to remove warm air from the high-voltage bases 
(see \Refsec{subsec:mech:cooling}) and air- or water-cooling is required for the electronics crates at 
the sides of the detector.
\begin{figure*}
\begin{center}
\includegraphics[width=0.9\dwidth]{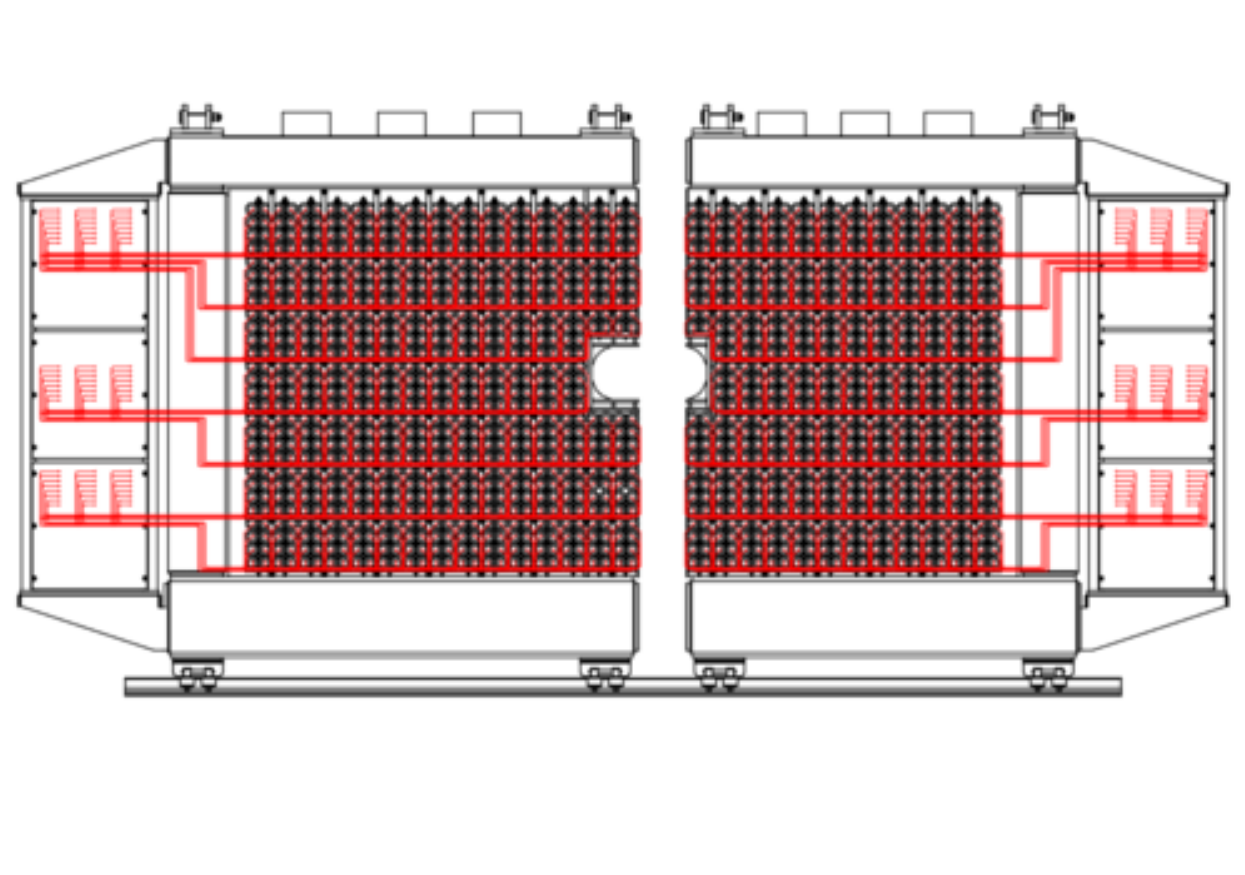}
\caption[FSC cable routing]
{Scheme of cable routing from the modules to the electronics racks aside the two FSC sections.}
\label{fig:mech:cables}
\end{center}
\end{figure*}
Electronics racks will be placed at the two sides and on top of the calorimeter. 
At the sides of the calorimeter the front-end electronics is situated, receiving all the signal cables. 
The front-end electronics consists of a set of modules placed into standard ATCA crates,
which can be positioned at the vertical shelf. 

The FSC is placed in the beam line between the TOF wall and the Muon range system and should be installed such that 
it can be easily moved out of its beam-line position and back. This requires a mounting gap of 20 mm between 
adjacent detectors to avoid damages of active elements and cables at the time of displacement. 
As additional protection for the back-side calorimeter structures we plan to use an environmental 
shield over the modules and cables. All protection shields are located inside of the calorimeter envelope lines.

\begin{figure}
\begin{center}
\includegraphics[width=\swidth]{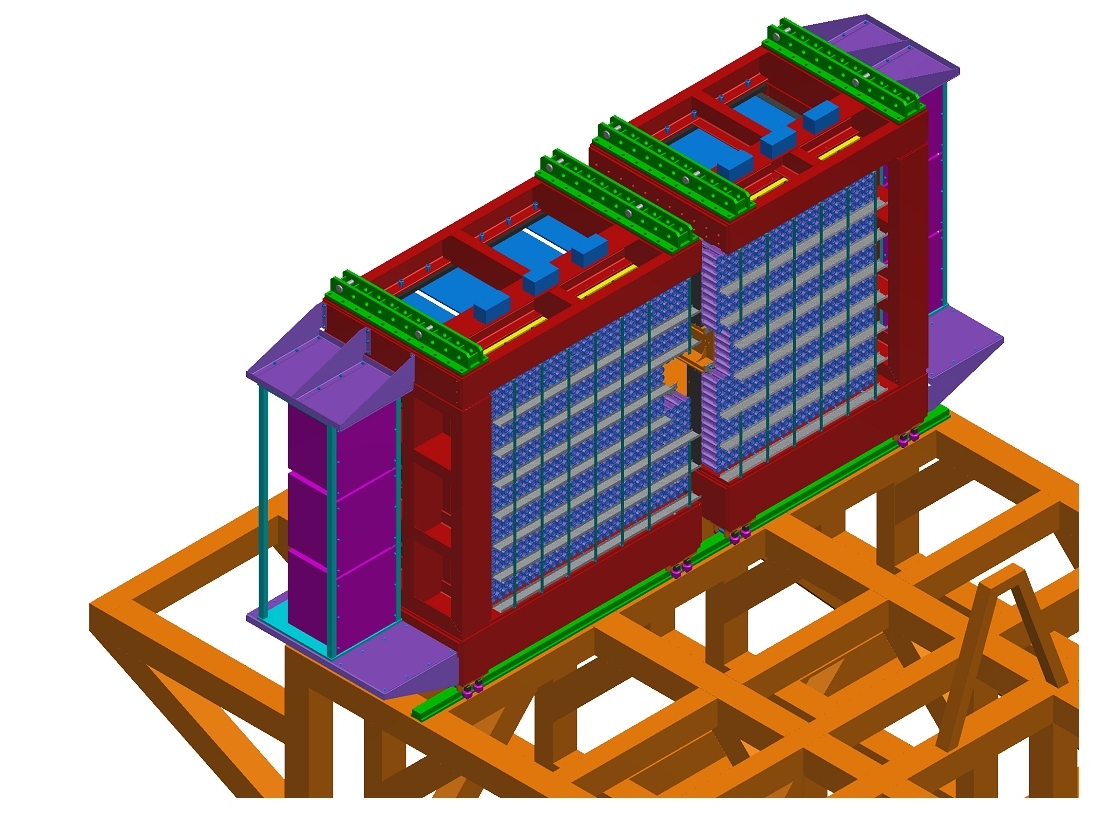}
\caption[FSC back-side view with electronics racks]
{Back-side view of the FSC detector, mounted on the Forward Spectrometer moving platform, equipped with electronics racks on both sides.}
\label{fig:mech:back_cables}
\end{center}
\end{figure}

The weight of the FSC will be supported by the
Forward Spectrometer moving platform, which also supports other parts of the Forward Spectrometer
(Forward Tracking stations, RICH, Forward TOF wall).  The FSC will be mounted on rails fixed on top of 
the platform as shown in \Reffig{fig:mech:back_cables}. 
The drive mechanism can be bolted to the platform frame.
The FSC needs additional space in the beam zone at the sides of the Forward Spectrometer moving 
platform for temporary support structures with rails to move the detector out into the maintenance
position.

During the process of calorimeter assembly one needs to use dedicated tools. The types of required 
tools are determined by the stage of the detector assembling. The set of tools is determined by the assembling scenario 
and includes: temporary support rails at the assembly hall, removable strengthening beams to lock the two
sections of the support frame, lifting and moving tools for module installation and crane in place. 
Moreover, tables or dedicated supports will be necessary for short-term placing and storing modules
before mounting. 

\subsection{Installation}
\label{sec:mech:installation}

\begin{figure}
\begin{center}
\includegraphics[width=\swidth]{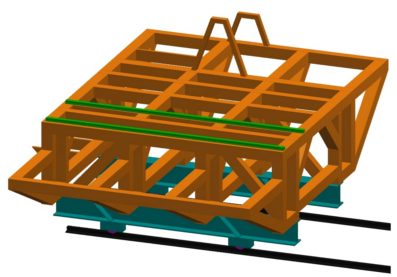}
\caption[Forward Spectrometer moving platform with rails installed]
{Forward Spectrometer moving platform with rails installed.}
\label{fig:mech:det_platform}
\end{center}
\end{figure}

\begin{figure}
\begin{center}
\includegraphics[width=\swidth]{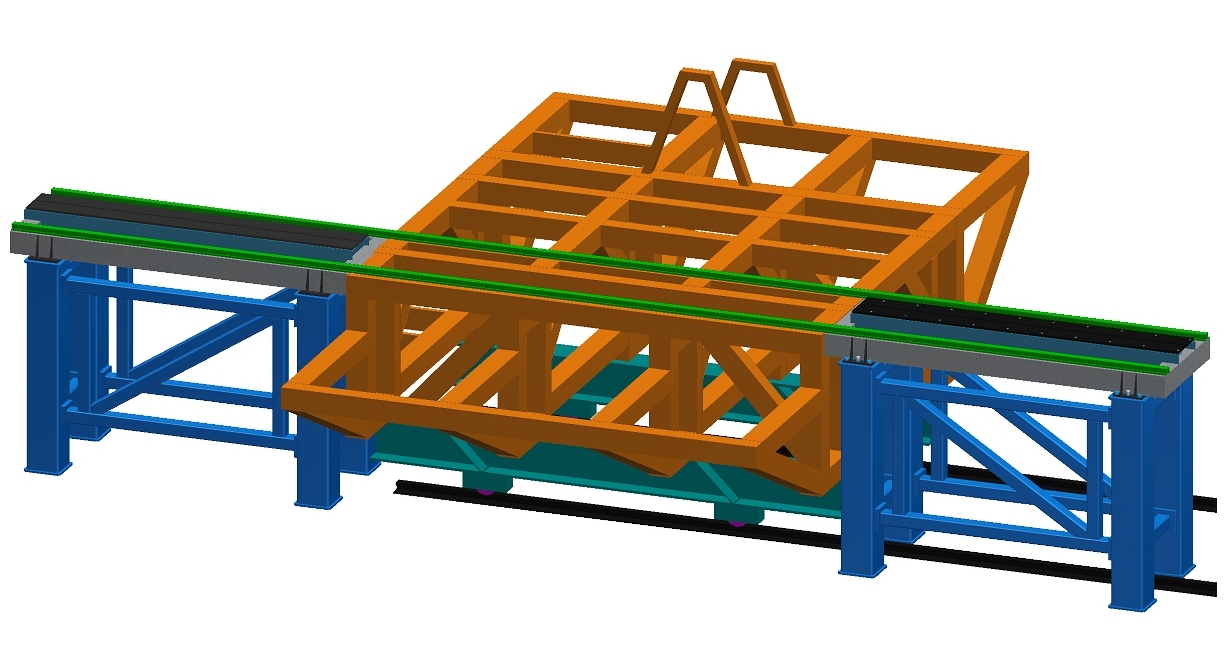}
\caption[Forward Spectrometer moving platform with extended rails]
{Forward Spectrometer moving platform and additional supports with extended rails.}
\label{fig:mech:det_rails}
\end{center}
\end{figure}

\begin{figure}
\begin{center}
\includegraphics[width=\swidth]{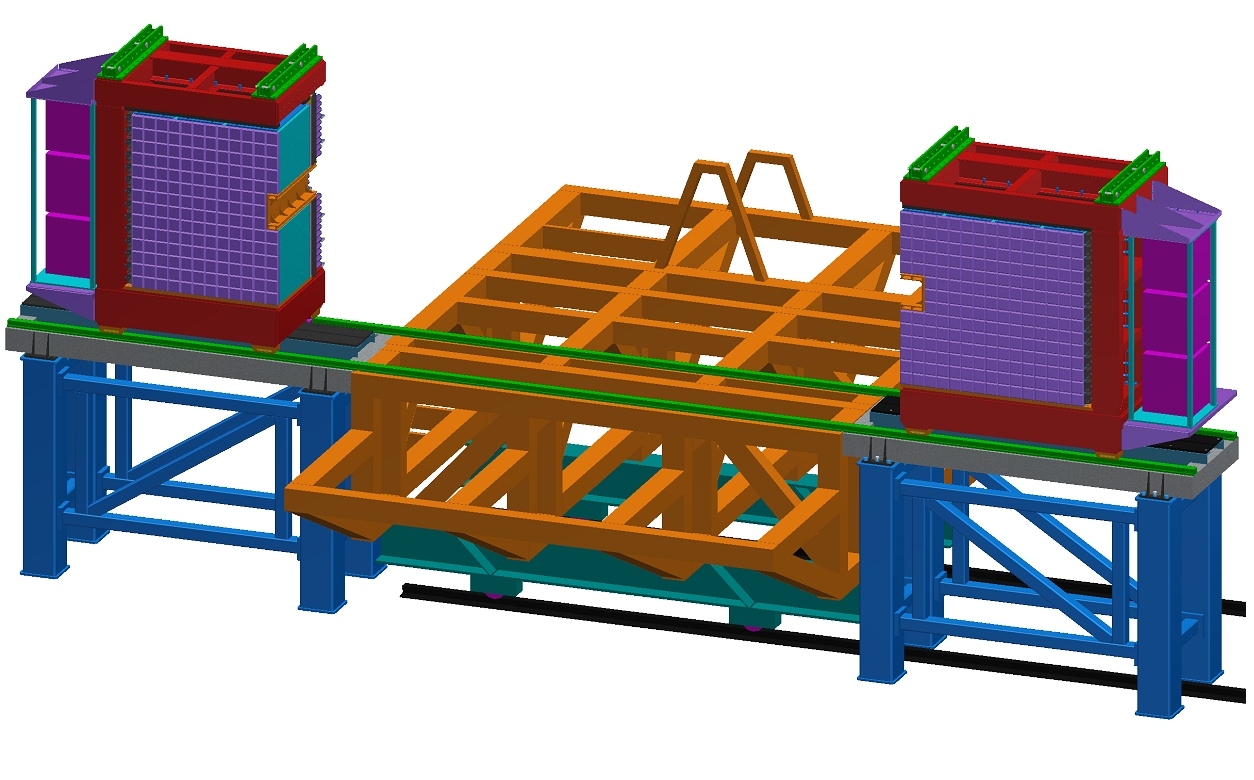}
\caption[FSC sections mounted on temporary supports]
{Left and right sections of the FSC mounted on the temporary supports.}
\label{fig:mech:det_mounting}
\end{center}
\end{figure}

\begin{figure}
\begin{center}
\includegraphics[width=\swidth]{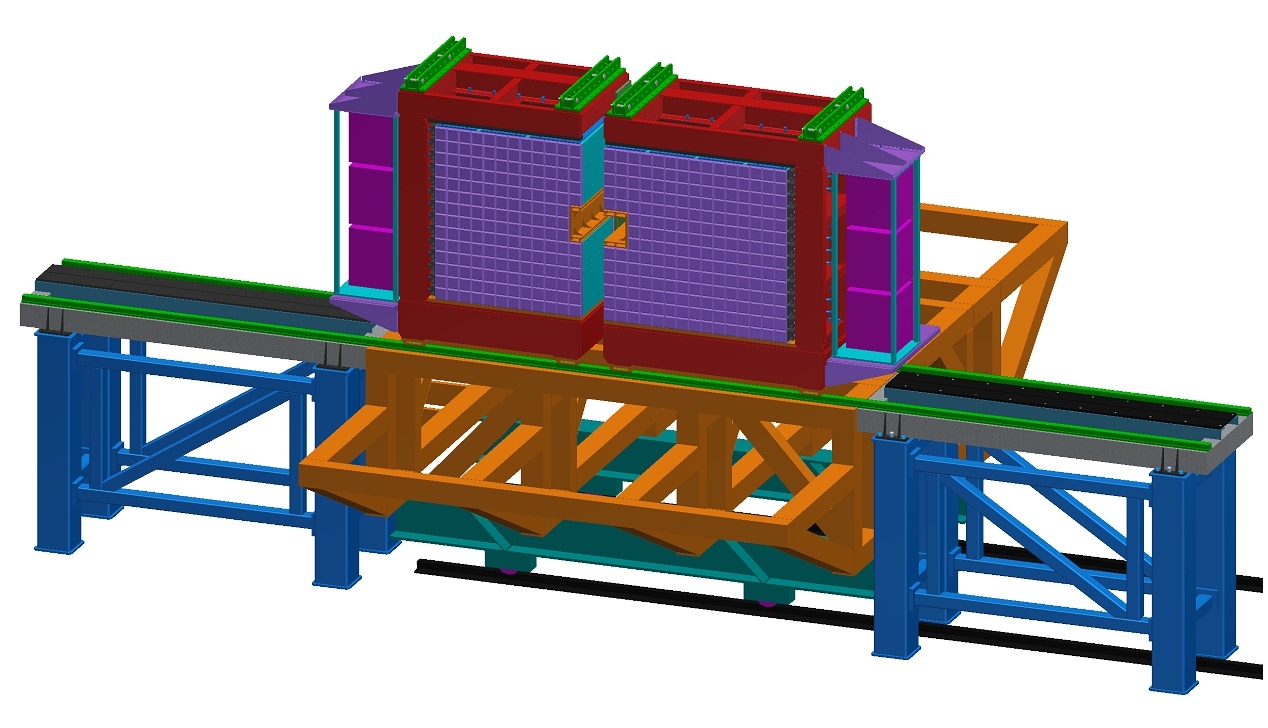}
\caption[FSC sections closed for alignment]
{Moving of the FSC sections to the beam position and relative alignment of the two sections.}
\label{fig:mech:det_closing}
\end{center}
\end{figure}

\begin{figure}
\begin{center}
\includegraphics[width=\swidth]{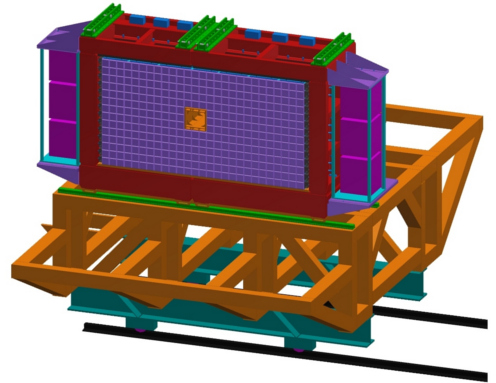}
\caption[FSC installed in beam position]
{FSC detector installed at the beam position.}
\label{fig:mech:detready_front}
\end{center}
\end{figure}

The sequence of actions for the FSC installation, when the detector was previously 
assembled at the assembly hall, includes the following steps and tools:
\begin{itemize}
\item Preparation of the Forward Spectrometer moving platform. Mounting of the rails for the FSC detector 
placement (see \Reffig{fig:mech:det_platform}).
\item  Pre-assembly of two temporary support structures with extended rails. Additional support 
frames and installation of rail system are needed (see \Reffig{fig:mech:det_rails}). It is possible 
to use only one support structure, when only the left or right section of the detector is mounted or dismounted.
\item FSC sections lifting and transportation from the assembly to the experiment hall. Crane in place 
to take $\approx$ 10 ton, special lifting beam and transport platform are required.
\item Placing of the detector sections on the side-support structures in the experiment hall 
(see \Reffig{fig:mech:det_mounting}). Crane is needed.
\item Installation of the driver system for the detector moving to the beam position.
\item Connection of power supply cables and optical fibres for data output.
\item Checking of the FSC system connections.
\item Moving of the detector sections and stop near the final position.
\item Dismounting of the temporary strengthening beams. Crane in place is needed.
\item Adjustment of relative location of the left and right sections and positioning of the calorimeter near the 
beam pipe with the help of screws or hydraulic jacks (see \Reffig{fig:mech:det_closing}).
\item Detector sections locking at the final position using a set of pins and tightening screws to join the
left and right parts into one unit with a minimum gap.
\item Fixing of the detector location at the beam position by fastening elements (see \Reffig{fig:mech:detready_front}).
\item Dismounting of the driver system and temporary support structures. Crane is needed.
\end{itemize}

\subsection{Cables routing}

In order to provide supports for cables and fibres inside the detector, the inner removable cable 
trays are fixed to the back plate by brackets and screws. The exact locations of trays depend on the routing of service connections. 
Figure \ref{fig:mech:cables} shows the scheme of the cable routing.

\begin{figure}
\begin{center}
\includegraphics[width=0.7\swidth]{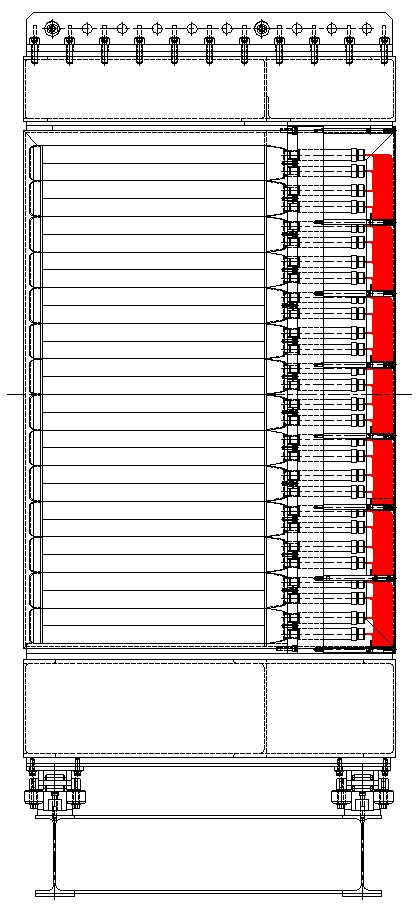}
\caption[Vertical cross section of FSC]
{Drawing of the vertical cross section of the FSC detector.}
\label{fig:mech:v_crosssection}
\end{center}
\end{figure}

\begin{figure}
\begin{center}
\includegraphics[width=\swidth]{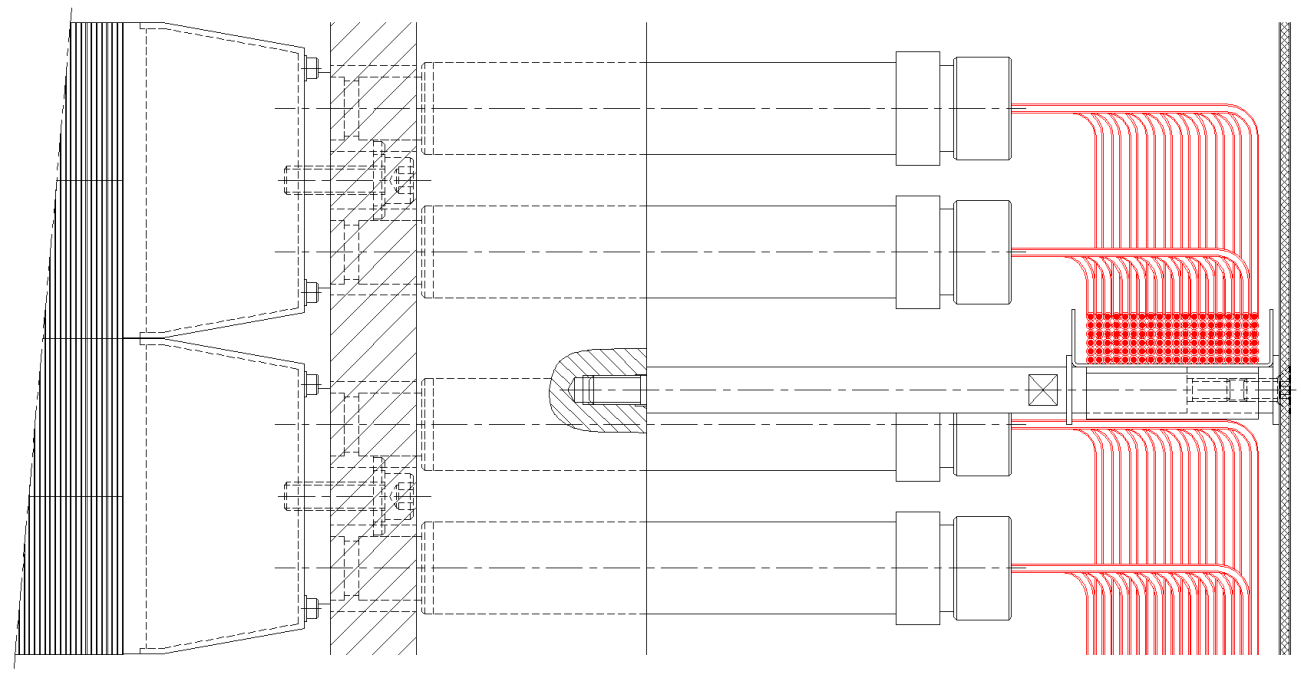}
\caption[Cable-tray of vertical cross section]
{Detail of the vertical cross section showing the cable tray area.}
\label{fig:mech:v_crossection_mod}
\end{center}
\end{figure}

Figure \ref{fig:mech:v_crosssection} shows the FSC side-view cross section revealing the vertical arrangement of cable trays. 
A detail of this cross section showing a filled tray and the support bracket is presented in 
\Reffig{fig:mech:v_crossection_mod}. Support brackets are located between the boards carrying the high-voltage base and are designed as 16 mm diameter rods with threads at both ends and screwed to the detector 
back-plate. These brackets serve a dual function:
\begin{itemize}
\item They give support for the horizontally positioned cable trays which have simple clamps for easy fixation.
\item The rod with an inner thread allows to mount the back-side environmental shield after the cables and 
fibres are laid in.
\end{itemize}

The choice of type and dimensions of cable trays are determined by: the cross section of the biggest cable 
bunch, the available gap between module HV board and back-side cover.
Commercially available parts can be used for a simple cable lay-out and fastening. 

\begin{figure}
\begin{center}
\includegraphics[width=0.9\swidth]{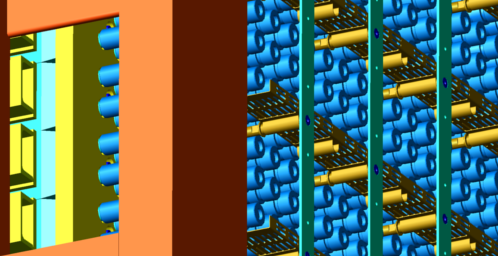}
\caption[Cable-trays in the FSC back-side area]
{Detail of the cable-tray arrangement in the FSC back-side area.}
\label{fig:mech:cable_trays}
\end{center}
\end{figure}

Figure \ref{fig:mech:back_cables} 
sketches the FSC equipped with electronics racks on both sides and mounted on the central rails, 
which are attached to the Forward Spectrometer moving platform. 
Figure \ref{fig:mech:cable_trays} sketches a detail of the cable-tray arrangement 
in the calorimeter back-side area and shows the
horizontal cable chains, support brackets, and vertical strips to install the back-side shield.

\subsection{Maintenance}

The sequence of operations and the required tools for the detector maintenance, when the detector 
is placed at the experimental hall, are listed as follows:
\begin{itemize}
\item Pre-assembly of the two temporary support structures with extended rails. Additional support 
frames and side rails are needed.
\item Installation and attachment of the driver system for the detector movement.
\item Unlocking of the left and right sections, opening and moving of the fully connected 
FSC sections with power-supply cables and optical fibres from the beam to the maintenance position.
\item Mounting a temporary scaffolding or mounting tables around the calorimeter to provide an 
easy access to different parts and systems.
\end{itemize}

\newpage

%


%% file: panda_tdr_FSC_photo.tex
%
\cleardoublepage
\chapter{FSC Readout Electronics}
\label{sec:roel}

Both \PANDA calorimeters, the Target Spectrometer EMC \cite{EMC_TDR} and the Forward Spectrometer  Calorimeter of shashlyk type, have a number of common features like the
output signal parameters and the signal treatment. Thus, the structure of the readout electronics 
will be similar for the FSC and the Target Spectrometer EMC. The main difference, however, is caused by the higher rates and shorter signals from the FSC and requires a specially designed digitiser module, which is discussed in the dedicated \Refsec{sec:fsc_digitiser}.

\begin{figure*}
\centering
\includegraphics[width=0.95\dwidth]{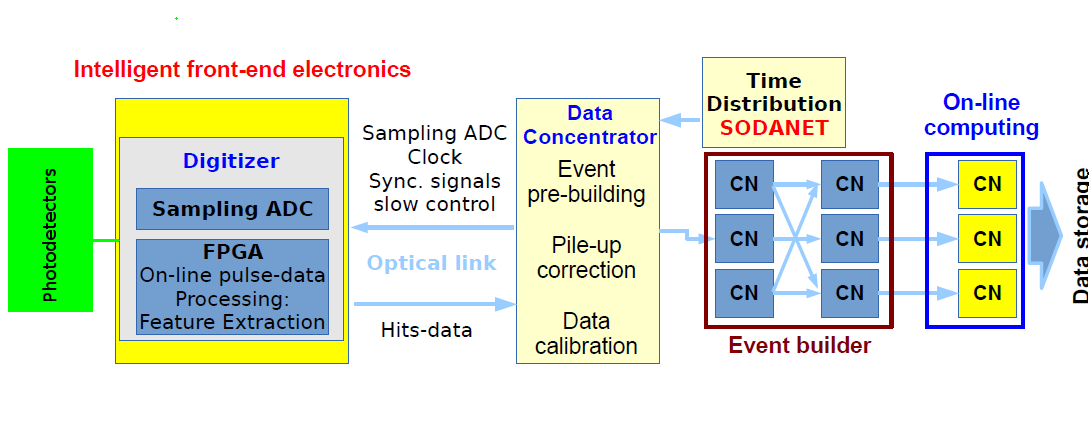}
\caption[Trigger-less readout chain of the \PANDA FSC]
{The trigger-less readout chain of the \PANDA FSC.
}
\label{fig_readout}
\end{figure*}

\section{Trigger-less readout for the \PANDA Electromagnetic Calorimeters}

A sketch of the calorimeter readout chain is shown in \Reffig{fig_readout}. 
The readout includes Digitiser, Data Concentrator (DCON) and  
Compute Node modules for On-line Computing. The digitiser module contains SADCs for 
continuous digitisation of the detector signals and a Field Programmable Gate Array (FPGA) for on-line data processing. 
The data are processed by a feature-extraction algorithm, implemented in the FPGA, which includes 
dynamic base-line compensation, hit detection, pulse pile-up detection~\cite{ref_pileup} and extraction
 of the hit information for single pulses. Using a serial optical-link connection, these reduced data 
are transferred to the DCON module, located outside the \PANDA detector. This type of data connection is 
dictated by the high data rate and mechanical constraints.

The DCON module collects data from several digitisers and performs data pre-processing, e.g. on-line 
pile-up recovery~\cite{ref_pileup}, time-ordering, and event pre-building. The pre-processed data are 
sent to the Compute Node for on-line reconstruction of the physics signatures, like shower detection and 
particle identification. The Compute Node combines calorimeter data with high-level information from all other
\PANDA sub-detectors, and only to such complete events the selection criteria are applied.

\section{Synchronisation protocol}

\begin{figure*}
\begin{center}
\includegraphics[width=0.95\dwidth]{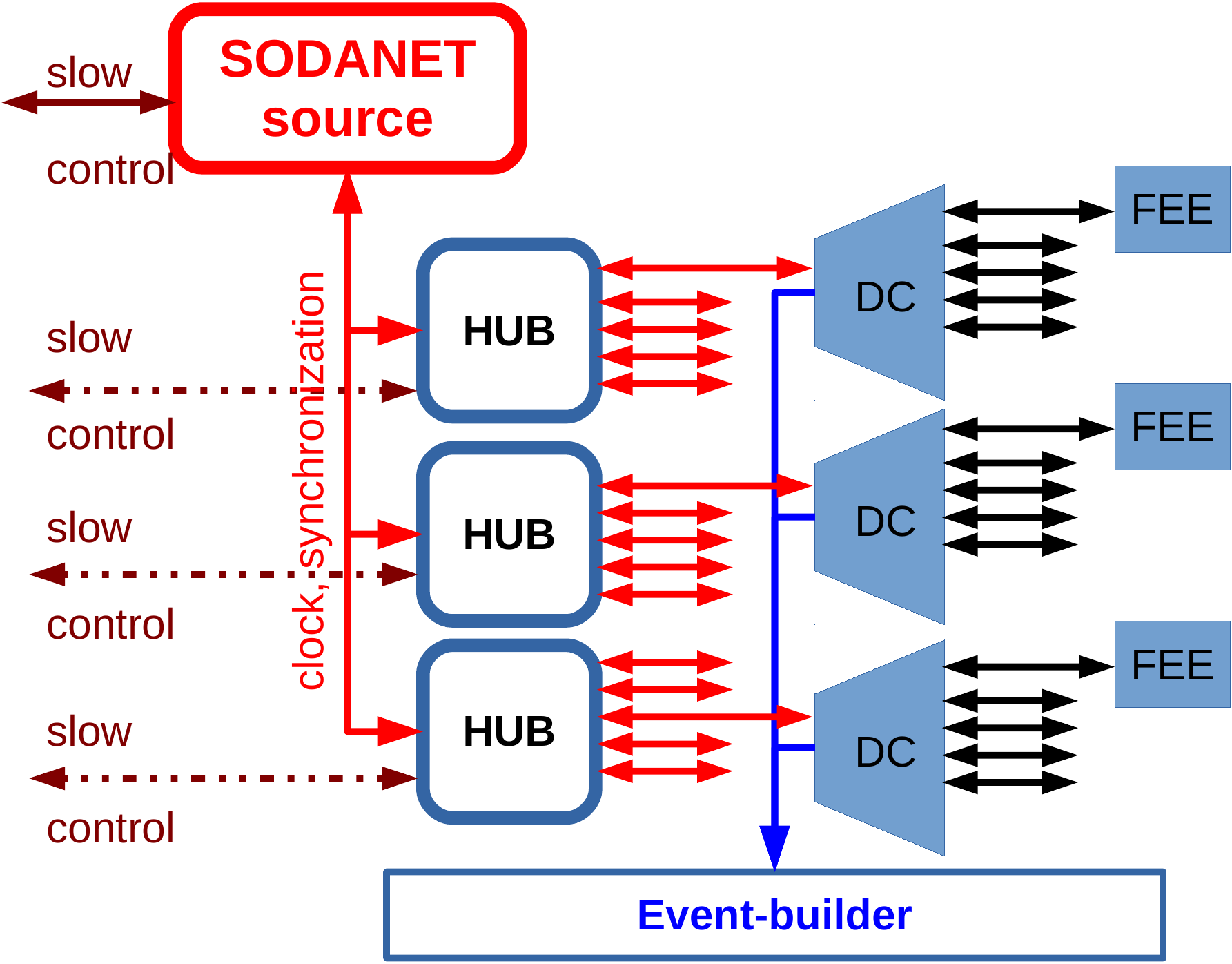}
\caption[SODANET topology]
{The topology of the SODANET synchronisation and slow-control protocol for the trigger-less readout of the \PANDA experiment.
}
\label{fig_sodanet}
\end{center}
\end{figure*}

The time-distribution system is of key importance for the trigger-less readout \cite{ref_soda}, 
since all front-end modules acquire data independently, and precise time-stamps are needed
 to correlate hits from different sub-detectors. A single clock and synchronisation source is used 
for the complete \PANDA detector. The precise clock and synchronisation commands are distributed using
 optical-fibre connections. After successful tests of the stand-alone synchronisation protocol~\cite{ref_33},
 it has been combined with the slow-control and data-transfer protocol TRBNET~\cite{ref_34}, used in the 
HADES experiment~\cite{ref_35}. The combined protocol is named SODANET~\cite{NSS2014}. The topology of the SODANET network 
for the \PANDA experiment is shown in \Reffig{fig_sodanet}. A single SODANET source supplies via hubs the 
system clock and synchronisation commands of fixed latency. 

In addition, the network is used to distribute 
slow-control (e.g. the front-end configuration, system status) data. The transmission of slow-control data can 
be interrupted at any time by a synchronisation command. The synchronisation command, a 32 bit long word, 
together with four dedicated "K-characters" (with hexadecimal value FB) is embedded into the currently 
transmitted data. At the receiving end the incoming data are continuously analysed for the presence of the 
dedicated four K-characters. Once the K-characters are detected, the synchronisation command is extracted from 
the incoming data stream and the receiving end takes the synchronisation action. The remaining slow-control data 
are not distorted. 

All SODANET links in the downstream direction, i.e. from the source to the Front-End Electronics (FEE), are operated 
in synchronous mode: all Serialiser-Deserialiser (SERDES) modules have the same phase of the serial and parallel 
clocks~\cite{ref_33}, locked on bit \#1, and no buffers in the data-path, which might change the latency of 
the data transfer. The FEE uses the recovered parallel clock as the system clock. The jitter of this recovered 
clock depends on the used hardware and is typically in the order of 20~ps. The upstream link, i.e. from the FEE 
to the SODANET source, is not synchronous. The upstream link is employed to transfer slow-control data from the 
FEE and to measure the time interval of signal propagation through the link. For such a measurement a dedicated 
synchronous command is sent by the SODANET source. Once a receiver recognises the calibration command it sends 
the feedback to the source. At the source the signal propagation-time is measured with a precision of about 10~ns. 
This precision is enough for an initial calibration of the readout system during the start-up phase. The larger 
uncertainty arises due to the lack of a synchronous upstream link. The operation of both the down- and up-stream 
links in synchronous mode would require a very long start-up time, since all transmitters and receivers should 
lock on a same phase and this would make the operation of larger systems impractical.

For the development of the SODANET protocol the Lattice ECP3 FPGA hardware is employed. At present, the network 
operates at a speed of 2.0~Gb/s while the maximum achievable synchronous speed of the link is 2.4~Gb/s. 
The synchronisation commands are periodically issued with a period chosen as a multiple of 25~ns (40~MHz clock). 
Such a combination of link speed and synchronisation frequency allows to synthesise, at the FEE side, clocks of 
40, 80, 120, 160, 200~MHz which are always in phase with the single SODANET source. At present the SODANET source 
and endpoint are implemented and perform up to specifications. The SODANET hub is currently being developed.

\section{Digitiser}
\label{sec:fsc_digitiser}

\begin{figure}
\begin{center}
\includegraphics[width=\swidth]{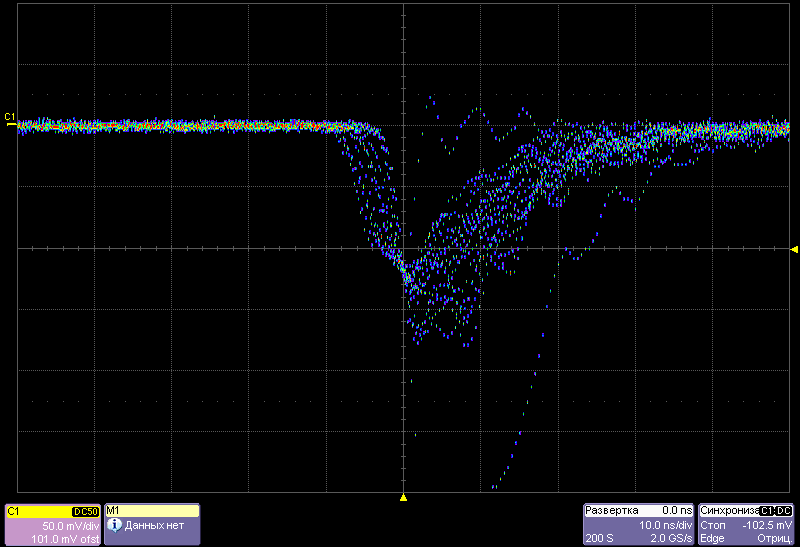}
\caption[Shashlyk signal for cosmic muons]
{Oscillogram of signals from the FSC prototype generated by cosmic muons crossing a cell.}
\label{fig:mu_signal}
\end{center}
\end{figure}

A sampling ADC with parameters required for the FSC is currently under development by the Uppsala group. 
The signal from the FSC features 10 ns rise time and  40 ns tail (\Reffig{fig:mu_signal}), 
while the average rate per cell amounts to 500 kHz for the channels near the detector outer edge 
and the rate per cell reaches 1 MHz for the channels close to the beam pipe.
In order to achieve a sub-nanosecond time resolution, the signal needs to be 
shaped by a low-pass filter to allow at least 3-4 samples on the rising edge. 
The needed 14~bit amplitude/charge resolution will be obtained by maximising 
the number of samples of the pulse, thereby increasing the signal to noise ratio. 
The input range of the ADC ($\pm$ 1V) will be adapted to a relatively high signal
 amplitude obtained from the FSC (in the order of several Volts). To further 
increase the dynamic range of the ADC, a dual-range configuration will be possible.

After the analog shaping, the signal pulses will be contained within 100 ns, being a 
compromise between the time/energy resolution and pile-up recovery efficiency. Using ADCs with a sampling rate of 240~M samples/s (MSPS), the time and energy reconstruction will be based on 24 samples of the signal pulse.
Lower ADC sampling rates are also considered in order to achieve a higher system integration factor and lower costs.
In order to minimise development cost for the ADC of the FSC and re-use the firmware algorithms, the ADC will make use of 
a double-height $\mu$TCA platform, which is currently in the design stage. 

\begin{figure*}
\begin{center}
\includegraphics[width=0.7\dwidth]{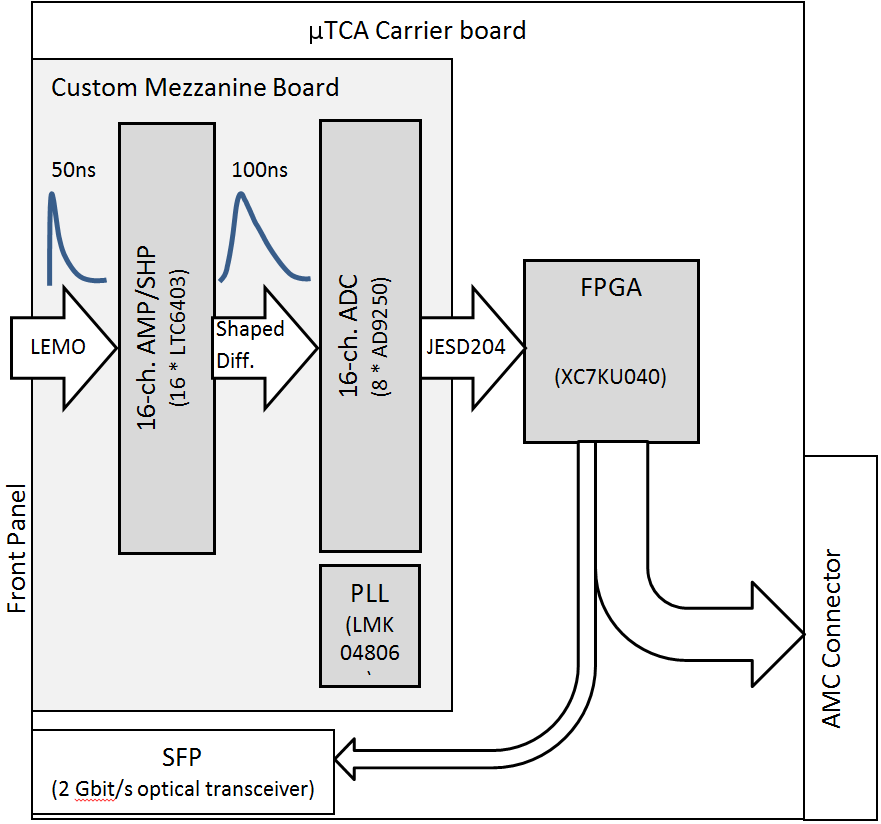}
\caption[Structure and signal flow for shashlyk digitiser]
{Structure and signal flow of the digitiser module designed for the FSC.}
\label{fig:sadc_structure}
\end{center}
\end{figure*}

The platform will be based on a XCKU040 FPGA 
from Xilinx, offering 24 multi-Gb links for communication with mezzanine devices as well as the $\mu$TCA 
backplane fabric and the front panel. The $\mu$TCA platform will feature a custom mezzanine connector, allowing 
for the attachment of a 16-channel Optical Data Concentrator module with SFP (Small Form-factor Pluggable) transceiver or a 16-channel ADC module, compatible 
with JESD204 standard, see \Reffig{fig:sadc_structure}.
The ADC mezzanine module will feature 16 input connectors of LEMO type on the front panel. 
The single-ended input signals will be processed by analog active filter/amplifiers (LTC6403), delivering 
symmetrical differential signals to the ADC. At the moment, two competing integrated dual-channel 
ADCs featuring 250 MSPS, 14~bit resolution with similar noise and power performance are considered. 
These are AD9250 from Analog Devices (National Instruments) and LTC2123 form Linear Technology.

The sampling phase will be controlled by a Phase Locked Loop (PLL) circuit LMK04806.
Signals digitised by ADCs will be transferred to the $\mu$TCA platform using JESD204 standard at a rate of 5~Gb/s 
for processing in the FPGA. Processing algorithms therein will resolve pile-up signals and extract signal parameters, 
such as arrival time, amplitude and pulse integral. Extracted parameters will be transferred to higher order
 DAQ components over the $\mu$TCA backplane or a front-panel optical link. Assuming a maximum 1 MHz pulse rate per 
channel and 16~byte parameter compression per channel, the data rate per 16-channel ADC module will amount to 2.5~Gb/s.

%
%
%

%% file: panda_tdr_FSC_cal.tex
\cleardoublepage
\chapter{FSC Calibration and Monitoring}
\label{chap:cal}
%
%
\section{Overview of the calibration and monitoring of the FSC}

The FSC should have approximately  3\percent  energy resolution (stochastic term) and  
3.5 mm position resolution (at the centre of the cell), see Table \ref{tab:req:reqs}. 
To fully utilise such a good performance one needs a monitoring system measuring variations of the PMT gain at least at the percent level in order 
to compensate gain changes. Also important for good performance are calibration procedures, which have to give us a correct conversion 
coefficient from code to energy with a precision of a few percent and with self-correctons for more long-term changes.   
This chapter describes the principles of the monitoring system
and discusses a number of procedures  for performing the FSC calibration. 

This chapter also refers to our past experience in monitoring
electromagnetic calorimeters and describes the plans for implementing this knowledge for the FSC.
  
\section{Calibration}

Each FSC cell has to be periodically calibrated.
In the intervals between calibrations the monitoring system will track the change in gain of the photomultipliers. 
On this basis, continuous adjustments  will be made  for the calibration constants. 
The monitoring and the calibration systems will  be designed based on existing  technology. 

The calibration procedures can be divided into three levels with increasing precision and, as a tradeoff, complexity and required time.
The most simple procedure is the {\it pre-calibration}. The results of the pre-calibration are not precise enough and can not be used for the final
FSC  data analysis. But it is extremely useful at the detector initial setup and commissioning. 

The {\it on-line calibration} is more precise and uses physical events from the FSC and, perhaps, other sub-detectors to calibrate the detector 
energy scale. The main goal of the on-line calibration is to give a correct energy response from the FSC during on-line data 
analysis and draw operators' attention if the detector performance deteriorates for some reason.

The most precise procedure is the  {\it off-line calibration} of the FSC and can be achieved by exploiting
physics events during full data analysis of the recorded data. At this moment, data from all sub-detectors are available and can be used 
to reconstruct events and reject background.  
Several algorithms for accurate and complete off-line calibration can be applied :
\begin{itemize}
\item[-]  Using  constraints on the $\pi^0$  and $\eta$ masses.
\item[-]  Using the E/p ratio for electrons from decays.

The \Panda Forward Spectrometer tracking system, together 
with particle identification, provides a precise momentum measurement (0.5\percent) and 
identification of electrons
and positrons.
\end{itemize} 

\subsection{ Pre-calibration }

The pre-calibration is a fast and reliable method to adjust the PMT gains,
especially  for the initial settings. This procedure does not pretend to be very 
precise. The main goal of the pre-calibration is a fast and rough determination of  detector 
parameters.

\subsubsection{ Pre-calibration with vertical cosmic-ray muons }

Vertically penetrating cosmic muons can  be used to pre-calibrate the FSC.
This  method provides an easy pre-calibration tool
because the energy loss of muons in lead and plastic is well known
and the path length through the FSC cells can be estimated.
The procedure does not require a calibration beam and neither a
vertical nor horizontal movement of the calorimeter. Furthermore, the
calorimeter is exposed approximately uniformly and the mean values of 
the signals 
from each cell are expected to be the same, which makes
this method a very effective tool, especially for low
energies. 

\begin{figure}[ht]
\centering
\includegraphics[width=\swidth]{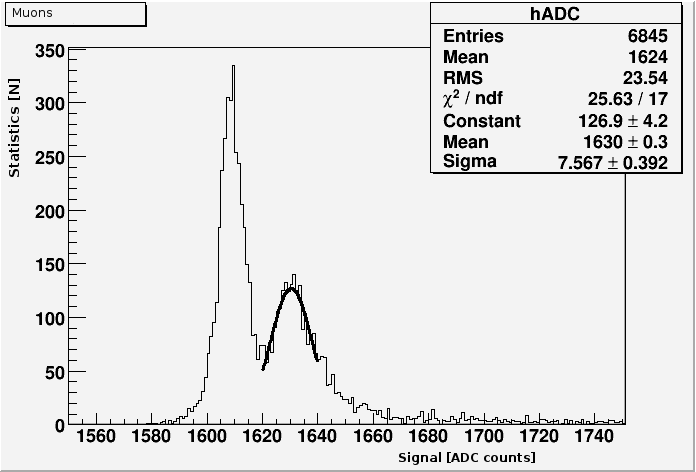}
\caption[Transverse cosmic-muon spectrum]
{Transverse cosmic muon spectrum. Vertical muons transverse the horizontally placed cell (55 mm). 
The left peak corresponds to the ADC pedestal.}
\label{fig:cal:cosmics}
\end{figure}

During testing of FSC prototypes at IHEP Protvino a dedicated test setup 
was used to check shashlyk modules and photo detectors by means of cosmic muons.
Figure~\ref{fig:cal:cosmics} shows a typical transverse 
cosmic muon spectrum obtained at this test setup. The cosmic muon from above goes thru the horizontally 
positioned FSC module in transverse direction (55~mm for each cell). 
Each event is triggered by a coincidence of signals from two scintillation counters installed 
above and below the module.  

The procedure of pre-calibration with vertical muons was applied for the 
shashlyk electromagnetic calorimeter in the E865 experiment 
at BNL \cite{E865}.
For the pre-calibration data taking a  software self-trigger was used. The
trigger required hits in at least six rows of the calorimeter prototype.
For the actual pre-calibration of a specific test-cell, cosmic ray
muons entering through the upper surface of the
cell and exiting through the lower one were
selected by requiring signals in the adjacent
cells above and below the test-cell and no signals in the left and right neighbours.
Using this setup a 4\percent accuracy was achieved.
During \PANDA operation the pre-calibration with vertical cosmic muons 
can be performed during filling of the HESR with anti-protons.

\subsubsection{ Pre-calibration with minimum-ionising particle signals }
About 30\percent of all charged hadrons with energies of several GeV reveal a minimum-ionising 
interaction in the FSC and thus are considered MIP particles. 
The position of the MIP peak is nearly independent of momentum and
particle species.
Thus, we can pre-calibrate the response of each cell by exploiting the MIP signals 
from physics events. One can easily select this kind of events on-line
or off-line by a set of software constraints like the size of the shower or the energy deposition.
The rate of MIP-like events is high, so this kind of pre-calibration can be achieved very fast
during data taking.

\subsection{The fine calibrations (on-line and off-line)}
The fine calibration of the FSC exploiting $\pi^0$ decays will apply algorithms which are
well known in high-energy physics.
Basically this is an iterative procedure which uses a set of the invariant-mass 
distributions obtained from
combinations of two photon candidates. Such distributions are constructed 
for each FSC
cell  in order to detect the $\pi^0$  invariant-mass peak. The shift of its position 
with respect to the PDG mass is
used to adjust the energy scale of the given cell. 
The method was applied successfully at the LHCb experiment which also 
employs a shashlyk-type electromagnetic calorimeter \cite{lhcb_pi0calib}.

The second algorithm exploits the E/p (energy/momentum) ratio for electrons from decays.
We concider it as a complimentary method to the main method based on  detecting $\pi^0$
peak described above.
In this method we transfer the energy scale from the
forward tracker to the FSC by measuring the
E/p ratio for isolated electrons (E from calorimeter, p from
tracker). Finally, the momentum scale is transferred to the FSC
by adjusting the E/p distribution for electrons to 1.

\subsection{Statistics and calibration time}
For a full calibration of the FSC 
a total number of  ~$5\cdot10^6$  not overlapping   $\pi^0$  are required.
With this  statistics the fine 
calibration can be performed in one to two days.
 The E/p method needs about  $3\cdot10^7$  electrons (20000 for each cell) without
bremsstrahlung photons.

\section{Light monitoring systems of the FSC}

PMT gain instabilities will deteriorate the performance of the calorimeter. 
A usual way to track the PMT gain variations is the 
use of a monitoring system employing a light pulser.
We plan to use two different systems using light emitting diodes (LED). One of these Light Monitoring 
Systems (LMS) is simple and provides access to each individual module (four cells), which is useful 
for checking each module during maintenance. The other LMS is 
envisaged to monitor PMT gain variations with very high precision over long periods
of time. Both systems are discussed below.

\subsection{Front-side monitoring system}
A relatively simple light monitoring system can be based on a set of LEDs,
one for each FSC module, and one multichannel pulser. The LED will be 
installed at the front 
of the module.  In this case the emitted light is absorbed by the optical fibre 
loops which are placed under the front cover (see~\Refchap{chap:mech}). The advantages of this 
system are simplicity and the possibility to check every four cells independently. The big 
number of independent light sources installed at the high radiation area makes the precise 
calibration of the detector questionable. One could expect LED parameter variations within 10\percent. Light capture 
efficiency by front loop fibres can differ from cell to cell by a factor of two. So this system provides only relatively
simple monitoring of the cell parameters. 
However, such a system can be very useful for a quick check of the functionality 
of the selected channel e.g. the photo detector, the high-voltage chain, and the readout electronics.
 
\subsection{Back-side monitoring system}

This system is necessary for the precise tracking of PMT gain
variations during data taking. It can be used during the HESR empty bucket.
The light pulser must provide a uniform light distribution via optical fibres to each cell of the FSC with a non-uniformity less than 10\percent.
The amplitude of the light pulses must ensure
the PMT anode response in the middle of the ADC scale (equal to those from 5 GeV electrons). The LMS does not provide the calibration of the 
electronics and PMT. Instead it monitors changes of this calibration with high precision. Calibration is done by a calibration system with physics
events.

The whole system should consist of two identical LED pulser units
coupled to a distribution network comprised of optical fibres, 
each monitoring one half of the calorimeter. 
Only one unit is planned to be fired in a given time interval. 
This solution allows to stay within the bandwidth of the data 
acquisition system while collecting monitoring data.
Each light-pulser module should include:
\begin{itemize}
\item a powerful blue LED with a driver;
\item a mixing light-guide interface to the fibre bundle;
\item reference silicon photodiode(s) to track the LED stability.
\end{itemize}

A mixing light guide of rectangular cross section ensures the required 
uniformity
of the light distribution over its output window which
will be joined to a bunch of optical fibres. In our prototype studies we could setup 
3000  fibres with 100$\mu$m diameter  for one light mixer. For the FSC we will need 
less than 1000 fibres in one bunch.    

An  essential element of the light monitoring system is a stable
reference photo detector with a good sensitivity at short wavelengths
which measures light-pulse amplitude variations in time.

\section{IHEP experience in light monitoring systems}

\begin{figure}[ht]
\centering
\includegraphics[width=\swidth]{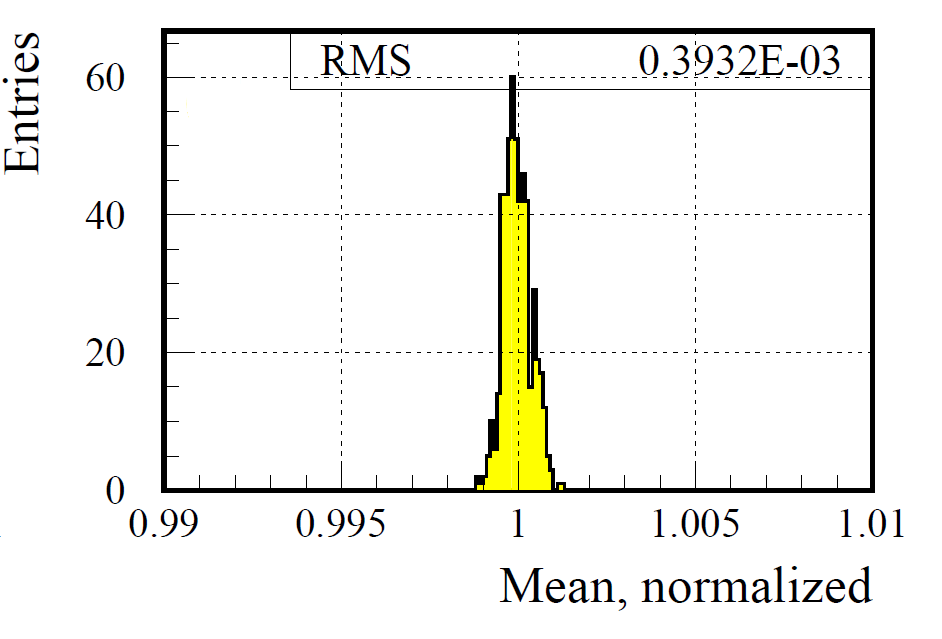}
\caption[Stability of LED light-pulse amplitude]
{LED light pulse amplitude, normalised to the initial value of the light pulse, during 160 hours of operation.
Each point is taken during 20 minutes of operation.}
\label{fig:lms_stab}
\end{figure}

IHEP has a wide experience in monitoring detectors using LEDs. 
For example, the prototype of a system with appropriate parameters 
is described in~\cite{nim8}.
A light distribution uniformity of 2\percent (RMS) was reached with this prototype.
A system stability better than 0.1\percent (RMS) has been achieved during one week of the
prototype operation, see \Reffig{fig:lms_stab}.
 
\begin{figure*}[ht]
\centering
\includegraphics[width=\dwidth]{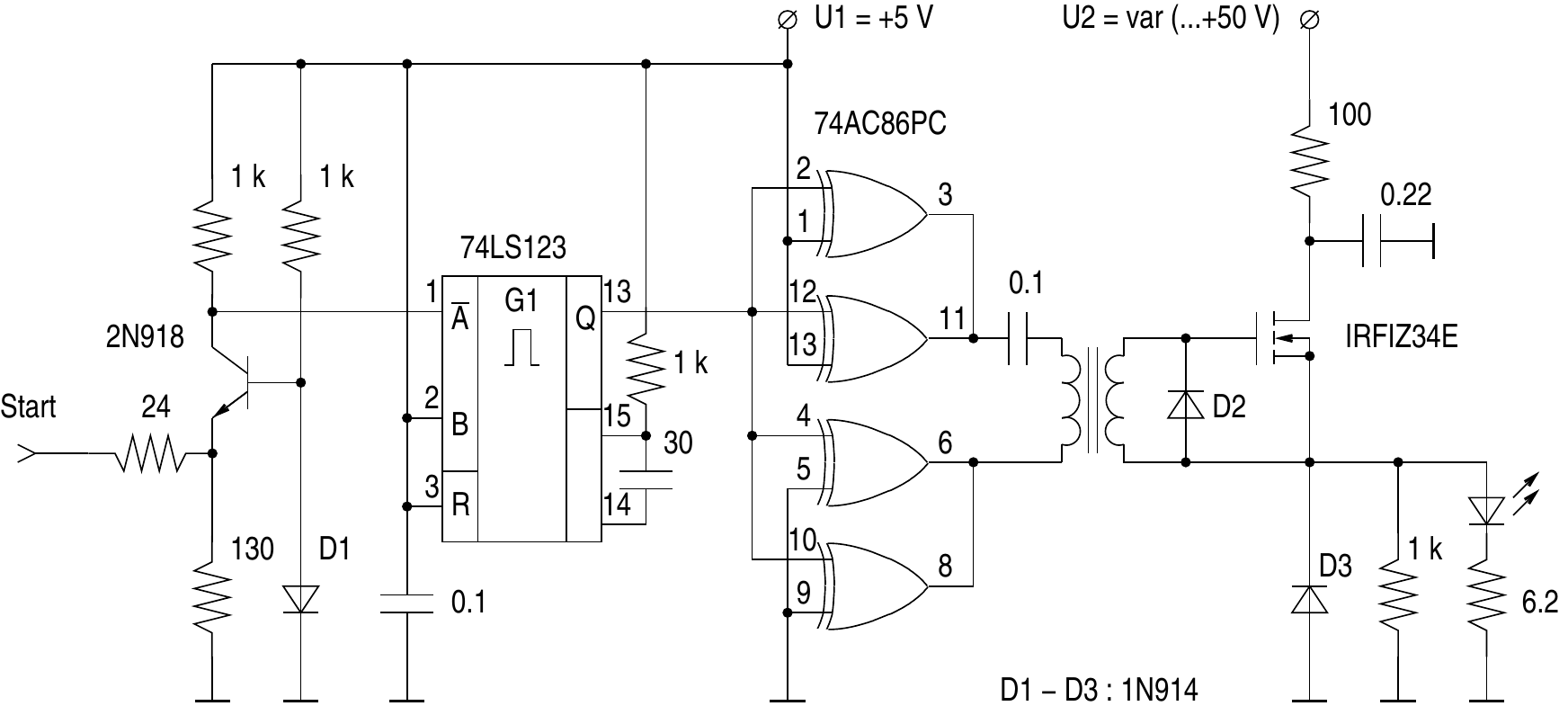}
\caption[LED driver circuit]
{LED driver circuit.}
\label{fig:circuit}
\end{figure*}

The main components which provided this performance are listed below. 
The electronic circuit for the LED driver is shown in
\Reffig{fig:circuit}. This driver
was triggered by pulses of standard NIM logic levels but the input cascade
can be modified easily to meet the requirements of the \PANDA logics.
\begin{table*}[h]
\begin{center}
\caption[Compilation of LED properties]
{The properties of LEDs used in the light pulser~\cite{luxeon,cree}.}
\begin{tabular}{|l|c|c|}
\hline
LED Property                   & LXML-PB01-0030 & XPEBLU-L1-0000-00Y01	\\
                               & (Blue )        & (Blue)		\\
\hline
Brand                          & LUXEON Rebel   & CREE Xlamp XP-E LED	\\
Min Luminous Flux @350 mA      & 30 lm          & 30.6 lm		\\
Radiation Pattern              & Lambertian     & Lambertian		\\
Viewing Angle (FWHM)           & 125\degrees    & 130\degrees		\\
Dominant Wavelength            & 460-490 nm     & 465-485 nm		\\
Forward Voltage @350 mA        & 3.03 V         & 3.2 V			\\
Maximum DC Forward Current     & 1000 mA        & 1000 mA		\\
\hline
\end{tabular}
\label{tab:leds}
\end{center}
\end{table*}
Two commercial companies, Luxeon and CREE, compete in production of high
quality LEDs.  
A comparison of the main technical parameters for the powerful Luxeon and CREE 
blue LEDs is provided in Table~\ref{tab:leds}~\cite{luxeon,cree}.
Besides the exceptional luminous fluxes, two 
features of these LEDs are very important for our purposes:
a long operating lifetime (70\percent lumen maintenance at 50,000 hours of operation
at a forward current 700 mA) and a small temperature dependence of the light output (for the blue LED 
less than 0.1\percent/\degrees C).

The LED light output and the parameters of the LED driver circuit are 
influenced by various factors like temperature and humidity. To minimise 
this influence we place the light pulser into a thermo-insulated box with a heater 
inside to keep the temperature stable. However, to obtain an even more precise light
pulser system, we used photodiodes to measure the light from the LED directly in our LMS prototype studies.

\begin{figure*}
\centering
\includegraphics[width=\dwidth]{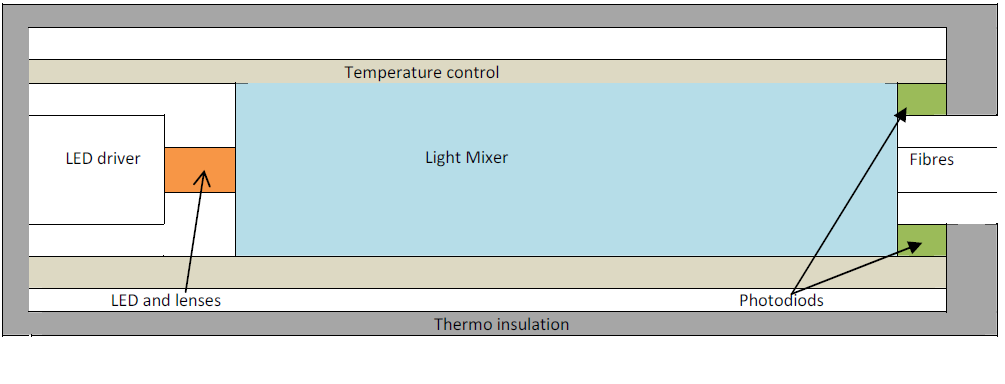}
\caption[Sketch of the stable light-pulse source]
{Sketch of the stable light-pulse source, consisting of an LED with driver, heater with controller,
light mixer, and reference photodiodes.}
\label{fig:led_tbox}
\end{figure*}

The photodiodes were installed near the light mixer inside the thermo-insulated box and measured a fixed 
fraction of the LED light pulse (see \Reffig{fig:led_tbox}). These reference measurements were used to correct the light pulser amplitude
in order to increase the stability of the monitoring system. Of course, this kind of
photodiode itself should be very stable. 
Silicon PN-photodiodes S1226-5BQ from Hamamatsu, Japan,
are well suited for this task because they have high ultraviolet and
suppressed infrared sensitivity, low dark current and a small temperature
coefficient (less than 0.1\percent/\degrees C)
in the range of wavelengths from 200 to 700~nm~\cite{ham}.
The rather large (about 6~mm$^2$) sensitive area of this photodiode allows
us to work without preamplifiers, thus improving the intrinsic stability of the reference
system.

%
%


%% file: panda_tdr_FSC_sim.tex
\cleardoublepage
\chapter{Simulations}
\label{sec:sim}

The simulation software framework to study the expected performance of the planned FSC
is described in this chapter. Software tools, reconstruction algorithms, and the digitisation procedure will be described. We focus here on the threshold dependence of energy and spatial resolution of reconstructed photons and electrons, the influence of material in front of the FSC, and the electron-hadron separation.
Since most of the physics channels have very low cross sections,
typically between pb and nb, a background rejection power up to $10^9$ has to be
achieved.
This requires an electromagnetic calorimeter which allows an accurate photon
reconstruction in the energy range from 10$\,\mev$ (20 $\,\mev$) to
15$\,\gev$, and an effective and clean electron-hadron separation.

\section{Off-line software}
The \PANDA collaboration pursues the
development of \PANDA software within the so called FairRoot framework, a GSI
project to provide a common computing structure for all experiments at
FAIR, such as \PANDA, CBM \cite{cbm} and the HADES upgrade
\cite{HADES}.
PandaRoot \cite{pandaroot} is a framework for both simulation and analysis,
and it is based on the object-oriented data analysis framework ROOT
\cite{ROOT}. It features the concept of Virtual Monte Carlo \cite{VMC}, which
allows to run different transport models such as GEANT3 \cite{GEANT3} or GEANT4
\cite{GEANT4} with exactly the same code and no need to recompile code in order to compare
results.

PandaRoot has the ambition to provide simulation software code that can easily be installed
and used by most of the collaboration members: the user should be able to install the
framework in his personal computer or laptop, without any restrictions on Linux
distribution or C++ compiler, and to run the analysis by himself. 
Through auto-configuration scripts the user can automatically install the external packages 
(as ROOT and GEANT) and the PandaRoot code without caring about configurations
and without additional manipulations. The PandaRoot computation
is divided into three main parts: event generation, transport model, and
digitisation and analysis.

First, physics events have to be generated. The generators are available in two
forms: either as external packages (since they are developed outside the collaboration)
or as direct interfaces which can be accessed internally in simulation macro code. In
the first case, the user launches the physics simulation using the original event-generator code. 
The event output file, in its original format, is used as input
for the PandaRoot simulation. In the second case, the output of the event
generator is automatically sent to the simulation stage.

The generated particles are propagated inside the detectors, and their
interactions with the spectrometer are computed by the transport model. At this
stage, the detector geometries and materials are defined together with a realistic
magnetic field map. The user has the possibility, via the Virtual Monte Carlo, to
switch between different transport models, such as GEANT3 or GEANT4, without
changing a single line of code and without recompiling, just by setting a flag in
the simulation macro code. In this way the results obtained with different models can be compared, 
cuts can be tuned, and the implemented physics can be validated by comparison to experimental data.

Finally, the simulation file is provided as input for digitisation and reconstruction tasks. Data Summary Tapes produced in this way can be used to
finally perform the physics analysis; all data in the output files are
stored in tree format, that interactively can be browsed in ROOT.

In order to perform full simulations, the first step consists in generating the
physics events of interest. In PandaRoot several event generators are
implemented, in order to comply with the many physics goals of the experiment.
Presently, the code contains the following generators:
\begin{itemize}
 \item Box generators: in order to study efficiency and acceptance, or even to check
 the response of the code, particles can be generated with uniform distributions in a given
 range, such as momentum, angular variables, and rapidity. The box generator can be
 launched directly inside the simulation macro code.
 \item EvtGen \cite{EvtGen}: EvtGen is an event generator used by many
 collaborations, such as Belle, BaBar, and CDF. It allows to
handle complex sequential decay channels, such as decays of several charmonium states, with different models or even to set angular distributions according
 to experimental results. The user can define the decay chain by himself and
 produce the event files that, through a suited interpreter, can be used as
 input for the simulation, or the user can launch the EvtGen generator directly
 from the simulation macro.
 \item Dual Parton Model \cite{DPM}: The DPM generator allows to simulate
 string fragmentation and the decay of all unstable hadrons by using the Dual
 Parton Model. In this way it is possible to generate background events for the
 main physics channels, evaluate detector occupancies and particle rates. The DPM
 generator is developed and maintained inside the \PANDA collaboration, and the
 output is provided in tree format that can be easily browsed and loaded into the
 simulation. The DPM generator can be launched directly from the simulation
 macro. Moreover it is also possible to switch the elastic and inelastic processes separately.
 \item UrQMD \cite{UrQMD1} \cite{UrQMD2}: The Ultra-relativistic Quantum
 Molecular Dynamic model is a microscopic model that can describe the
 phenomenology of hadronic interactions in nuclear collisions. The UrQMD
 generator is used to study $\bar{p}A$ collisions; the output files are read by a
 suitable interpreter and provided to the simulation procedure inside PandaRoot.
 
\end{itemize}

Moreover, an interface to the Pluto \cite{PLUTO} generator is also present. There
are ongoing activities to enlarge the number of generators that can be handled
by the framework, such as a Drell-Yan generator according to the model
\cite{DY1} \cite{DY2}, or models for Hypernuclei.
\section{Photon reconstruction}

\subsection{Reconstruction algorithm}
\label{sec:sim:recoalgo}
A photon entering one cell of the FSC develops an electromagnetic shower
which, in general, extends over several cells. A contiguous area of such
cells is called a cluster.
The energy depositions and the positions of all cell hits in a cluster allow to
determine the four-vector of the initial photon. Most of the FSC
reconstruction code used in the offline software is based on the cluster finding and bump splitting
algorithms which were developed and successfully applied by the BaBar experiment
\cite{BaBarDetector, Strother}.

The first step of the cluster reconstruction is finding a contiguous area
of cells with energy deposition. The algorithm starts at the cell exhibiting
the largest energy deposition. Subsequently, its neighbours are added to the list of cells,
if the energy deposition is above a certain threshold $E_{cell}$. The same procedure is continued on
the neighbours of newly added cells until no more cells fulfil the
threshold criterion. Finally, a cluster gets accepted if the total energy deposition
in the contiguous area is above a second threshold $\ecl$.

A cluster may be caused by more than one particle if the
distances between the particle hits on the FSC surface are small. In this case, the cluster has to be subdivided
into regions which can be associated with the individual particles. This procedure
is called the {\it bump splitting}. A bump is defined by a local maximum inside
the cluster: the energy deposition $E_{LocalMax}$ of one cell must be above an energy $\emax$,
while all neighbouring cells have smaller energies. In addition, the highest energy
$E_{NMax}$ of any of the $N$ neighbouring cells must fulfil the
following requirement:
\begin{eqnarray}
0.5\,(N-2.5) \, > \, E_{NMax} \, / \, E_{LocalMax}.
\end{eqnarray}
The total cluster energy is then shared between the bumps, taking into account the shower shape of the
cluster. For this step an iterative algorithm is used, which assigns a weight $w_i$
to each cell, so that the bump energy is defined as
$E_{bump}= \sum_{i} \, w_i \, E_i$. $E_i$ represents the energy deposition
in cell $i$ and the sum runs over all cells within the cluster.
The cell weight for each bump is calculated by
\begin{eqnarray}
w_i = \frac{E_i \, exp(-2.5 \, r_i\, / \,r_m)}
{\sum_{j} E_j \, exp(-2.5 \, r_j\, / \,r_m)},
\end{eqnarray}
with
\begin{itemize}
 \item $r_m$ = Moli\`ere radius of the cell material,
 \item $r_i$,$r_j$ = distance of cell $i$ and cell $j$ to the centre of the bump
 and
 \item index $j$ runs over all cells.
\end{itemize}
The procedure is iterated until convergence. The centre position is always
determined from the weights of the previous iteration and convergence is reached
when the bump centre stays stable within a tolerance of 1$\,\mm$.

The spatial position of a bump is calculated via a centre-of-gravity method. Due to the fact
that the radial energy distribution originating from a photon decreases mainly exponentially,
a logarithmic weighting with
\begin{eqnarray}
W_i \, = \, max(0, A(E_{bump}) \, + \, ln\,(E_i / E_{bump}))
\end{eqnarray}
was chosen, where only cells with positive weights are used. The energy
dependent factor $A(E_{bump})$ varies between 2.1 for the lowest and 3.6
for the highest photon energies.

\subsection{Digitisation of the FSC readout}
\label{sec:sim:digi}
In the \Panda experiment a fast Sampling ADC will digitise the analog response of the first amplification and
shaping stage of the FSC electronics. In order to obtain a realistically simulated detector response, the signal waveforms must be considered in the simulation.
The digitisation procedure was split in two parts: the formation of the
electronic signal shape from the transport-model hit and the conversion of the obtained shape
to an ADC-digitised signal in energy units.

To reach the first goal, the following algorithm was developed. An analytic function
of an RC-CR circuit with the following parameters has been chosen to describe
the signal shape obtained from the shashlyk cell: $T_{int} = 5$ ns integration time constant, $T_{diff} = 20$ ns differentiation time
constant, and $T_{sig} = 15$ ns signal rise time. Then the
Sampling ADC was simulated for the discrete signal shape of each hit. For this task
the class PndEmcWaveform has been used with the following parameters: $N_{samples} = 20$ is
the number of SADC counts, $SampleRate = 180$ MHz is the ADC rate, 
$N(photons/MeV) =21$ is the number of photons per $\mev$ of deposited energy (corresponds to 5 photo electrons per MeV obtained in the test beam measurement
\cite{NIMTB}), and an excess noise factor of 1.3 has been used (corresponding
to test beam measurements). Finally, an incoherent electronics noise with
Gaussian shape with 3 $\mev$ width was added to each ADC bin, and the signal was converted to an integer value in each bin.

The second stage of the digitisation procedure was implemented in the PndEmcWaveformToDigi class.
First, the maximum sample of each digitised signal shape was searched. Then, the magnitude of such
a maximum value was taken as the signal value and its position in the signal trace as the time of the signal arrival.
Moreover, two methods of  searching the maximum value were implemented, namely fitting by a parabolic function and
the convolution with a reference signal. In order to get the absolute value in
energy units, one has to correct the maximum value by a factor which is obtained when the signal for the algorithm is a $\delta$ function with energy 1 $\gev$. Finally, all
obtained digitised cell signals exceeding the threshold $E_{digiThreshold} = 3$
$\mev$ are kept for the subsequent analysis.

\subsubsection{Comparison with test-beam measurements}
\label{sec:sim:digi:compare}

In order to demonstrate that the digitisation is sufficiently well described,
the simulation was compared with the results of test beam measurements \cite{NIMTB}.
Figure \ref{fig:ECAL-eresolu} shows the measured energy resolution for Type-1
modules as a function of the incident electron energy.
Figure \ref{fig:small_eresolu} contains the experimental result for Type-2 modules.
Since the energy resolution does not depend on the transverse size of the
calorimeter cell, it is fair to compare the simulations with test-beam results achieved 
with both module types. The PandaRoot simulations contained no other \PANDA detectors but
the full FSC geometry equipped with Type-2 modules. However, the transverse non-uniformity of the light output, which was measured during the MAMI beam test (see \Refsec{ssec:non-uni-type2}) and was cured for the Type-3 modules, was not implemented.

\begin{figure}
\begin{center}
\includegraphics[width=\swidth]{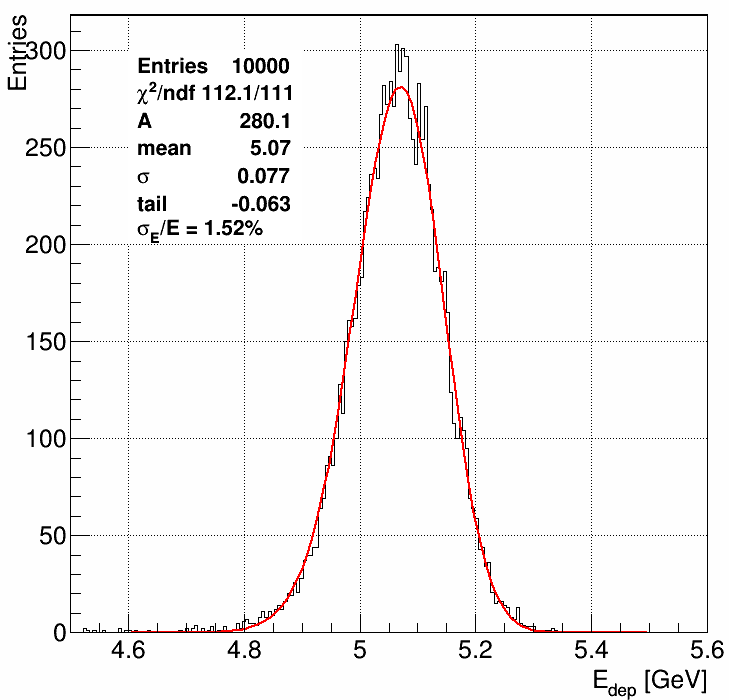}
\caption[Simulated energy deposition at 5 GeV]
{Simulated energy deposition of
FSC at the electron energy of 5$\,\gev$. The fit with a Novosibirsk function
gives an energy resolution of $\sigma_{E}/E\,=\,1.52\,$\%.}
\label{fig:sim:endep5gev}
\end{center}
\end{figure}

\begin{figure}
\begin{center}
\includegraphics[width=\swidth]{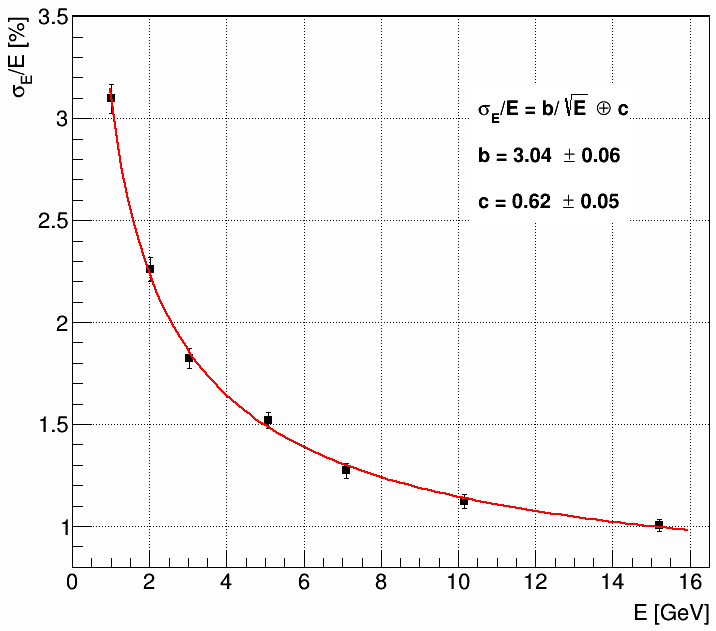}
\caption[Simulated energy resolution vs. incident electron energy]
{Simulated
energy resolution as a function of the incident electron energy between 1$\,\gev$
and 15$\,\gev$ for a digitised signal from the FSC. }
\label{fig:sim:enres}
\end{center}
\end{figure}

Electrons with discrete energies between 1$\,\gev$ and 15$\,\gev$ were directed
to the centre of the cell located far from the calorimeter edge. The resulting line shape at the
electron energy of 5$\,\gev$ is shown in \Reffig{fig:sim:endep5gev}.
A fit with a Novosibirsk function defined by
\begin{equation}
f(E) = A_S \exp ( -0.5 { \ln^2 [1 + \Lambda \tau \cdot (E - E_0) ] / \tau^2 +
\tau^2 } )
\label{eq:novofit}
\end{equation}
with
\begin{itemize}
 \item $\Lambda = \, \sinh ( \, \tau \, \sqrt{\, \ln 4} ) / ( \, \sigma \, \tau \, \sqrt{\ln 4} )$,
 \item $E_0\,$= peak position,
 \item $\sigma\,$= width,
 \item $\tau\,$= tail parameter
\end{itemize}
yields a resolution of $\sigma_{E}/E\;=\;1.52\,$\%. 
The energy resolution as a function of the electron energy (see \Reffig{fig:sim:enres})
can be reproduced very well with a fitting function represented by the equation 
\begin{equation}
  \frac{\sigma_{\rm E}}{\rm E} =
\frac{b}{\sqrt{\rm{E/GeV}}} \oplus c.
\label{eq:fitfct7}
\end{equation}

The fit to simulated data results in $b = (3.04 \pm 0.06)$\percent,
$c = (0.62 \pm 0.05)$\percent while experimental data measured with Type-1 modules 
\cite{NIMTB} give $b = (2.83 \pm 0.22)$\percent, $c = (1.30 \pm 0.04)$\percent and 
measurements with Type-2 modules yield 
$b= (3.15 \pm 0.43)$\percent, $c = (1.37 \pm 0.11)$\percent (see \Refsec{ssec:ecoor_small}).
The discrepancy in the constant term may be caused by uncertainties in the calibration coefficients
of test-beam data. Nevertheless, we can conclude that the 
digitisation procedure is accurately described in our simulation.


\subsection{Reconstruction thresholds}
\label{sec:sim:recothresh}
In order to detect low-energy photons and to
achieve a good energy resolution, the photon reconstruction thresholds should be set as
low as possible. However, the thresholds must be set sufficiently high to suppress the
mis-reconstruction of photons caused by noise of the cell readout and by
statistical fluctuations of the
electromagnetic showers. Based on the requirements for the \Panda FSC
(see \Reftbl{tab:req:reqs}) 
 the following thresholds were chosen:

\begin{itemize}
  \item $E_{cell}\,=\,3\,\mev$
  \item $\ecl\,=\,10\,\mev$
  \item $\emax\,=\,20\,\mev$.
\end{itemize}

The ability to identify photons down to
approximately $10\,\mev$ is extremely important for
\Panda. A lot of channels \--- especially in the charmonium sector (exotic and conventional) like
$\pbarp\,\to\,\eta_c\,\to \,\gamma\,\gamma$,
$\pbarp\,\to\,h_c \,\to\,\eta_c\,\gamma$ or $\pbarp\,\to\,\jpsi\,\gamma$
\-- require an accurate and clean reconstruction of isolated photons. The main background channels
in this case have the same final states with just the isolated photon being replaced by a $\piz$.
If one low-energy photon from a $\piz$ decay gets lost, the background event will
be misidentified. The cross sections for the
background channels are expected to be orders of magnitudes higher than for the channels of interest.
Therefore, an efficient identification of $\piz$ is mandatory for this important part of the
physics program of \Panda. Simulations show that roughly 1\percent of the
$\piz$ get lost for a threshold of 10$\,\mev$. However, the misidentification increases
by one order of magnitude for a scenario where photons below
30$\,\mev$ can not be detected.


\begin{figure}[!htb]
\begin{center}
\includegraphics[width=\swidth]{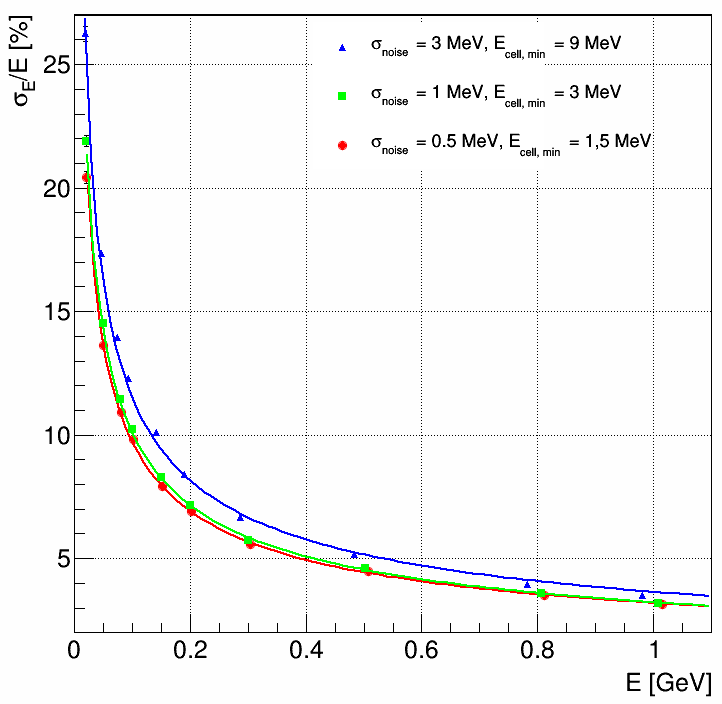}
\caption[Energy resolutions for different single-cell reconstruction thresholds]
{Comparison of the energy resolutions for three different single-cell reconstruction thresholds.
The most realistic scenario with a noise term of $\sigma_{noise} \, = \,1 \, \mev$
and a single-cell threshold of $E_{cell}=3\,\mev$ is illustrated by green rectangles, a worse case
($\sigma_{noise} \, = \, 3\, \mev$, $E_{cell} \, = \, 9 \, \mev$) by blue triangles and the better case
($\sigma_{noise} \, = \, 0.5 \, \mev$, $E_{cell} \,= \, 1.5 \, \mev$) by red circles.}
\label{fig:sim:Ethreshold}
\end{center}
\end{figure}

\subsection{Energy and spatial resolution}
\label{sec:sim:resolution}
As already described in \Refsec{sec:sim:digi} and \Refsec{sec:sim:recothresh} the choice of the
single-cell threshold $E_{cell}$, which is driven by the electronics noise term, affects
the resolution. Three different scenarios have been investigated: \Reffig{fig:sim:Ethreshold}
compares the achievable resolution for the most realistic scenario with a noise term of $\sigma$ = 1$\,\mev$
and a single-cell reconstruction threshold of $E_{cell}=3\,\mev$ with a worse case
($\sigma$ = 3 $\mev$, $E_{cell}=9$ $\mev$) and a better case ($\sigma\,= \,0.5 \,\mev$, $E_{cell}=1.5\,\mev$).
While the maximal
improvement for the better case is just 10\% for the lowest photon energies, the degradation
in the worse case increases by about 20\% . This result demonstrates
that the single-cell threshold has an influence on the energy resolution.

\begin{figure}[htb]
\begin{center}
\includegraphics[width=\swidth]{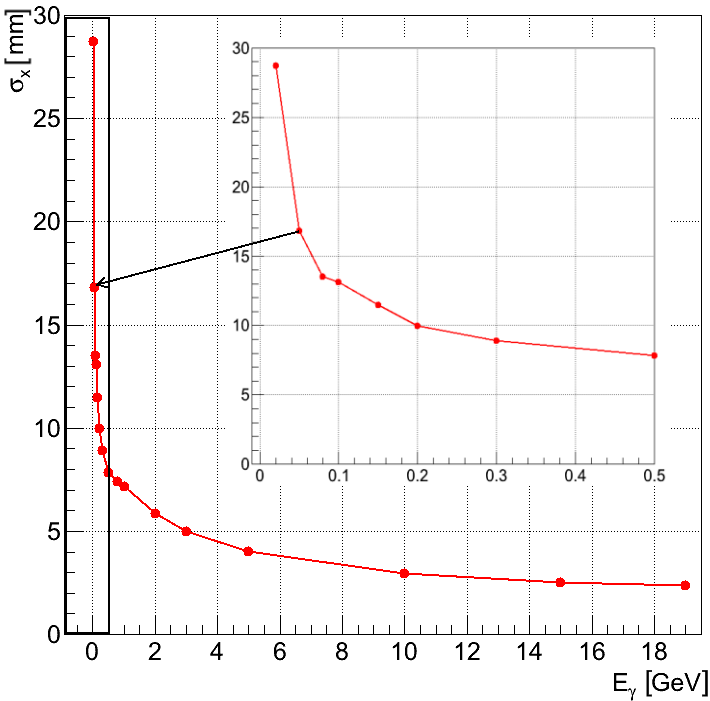}
\caption[Simulated position resolution in x-direction for photons]
{Position resolution in x-direction for photons from 20$\,\mev$ to 19$\,\gev$.}
\label{fig:sim:posResol}
\end{center}
\end{figure}

The high granularity of the planned FSC provides an excellent position reconstruction
of the detected photons. The accuracy of the spatial coordinates is
mainly determined by the dimensions of the cell with respect
to the Moli\`ere radius. Figure~\ref{fig:sim:posResol} shows the spatial
resolution in horizontal (x) direction for photons from 20$\,\mev$ up to 19$\,\gev$, when photons hit the centre of the cell (the worst case).
A resolution $\sigma_{\rm x} < 8$\,mm can be obtained
for energies above 500$\,\mev$. This corresponds to roughly 15\percent of the cell size.
For lower energies the position resolution becomes worse due to the fact that
the electromagnetic shower is contained in just a few cells. In the worst case
of only one contributing cell the x-position can be reconstructed within an
uncertainty of $\sim28$~mm (half of the cell size). The point at 19 $\,\gev$ energy is also simulated
in order to compare with test-beam data. As mentioned in \Refsec{ssec:ecoor_small}, the experimental resolution
at 19$\,\gev$ at the centre of the cell is 3~mm, while our simulation with PandaRoot gives 2.4~mm. Thus
the agreement is reasonable.

\begin{figure}[htb]
\begin{center}
\includegraphics[width=\swidth]{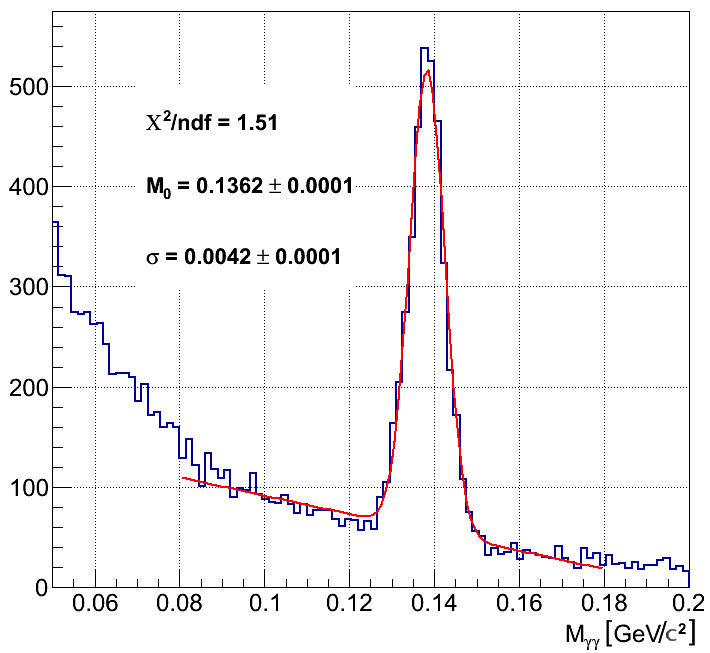}
\caption[Invariant-mass spectrum for photon pairs in FSC acceptance]
{Invariant-mass spectrum of photon pairs from decays of $\pi^0$ with energy up to 15$\,\gev$ in the FSC acceptance.}
\label{fig:sim:pi0_mass}
\end{center}
\end{figure}

The energy and position resolutions will have an impact on  the width of the reconstructed $\pi^0$
invariant mass. Figure~\ref{fig:sim:pi0_mass} shows the simulated
invariant-mass spectrum of photon pairs reconstructed in the FSC. Neutral pions were generated inside the FSC acceptance with energies from 0 to 15 $\gev$. The full \PANDA geometry was used. The resulting width of the $\pi^0$ peak was 4.2 $\mevcc$ after combinatorial background
subtraction. Thus, the requirement stated in \Refsec{sec:req:resolutions} is met. Test-beam data with a 
small-cell prototype give compatible results (see discussion in \Refsec{ssec:pi0_reco}).

\begin{figure*}[htb]
\begin{center}
\includegraphics[width=0.8\dwidth]{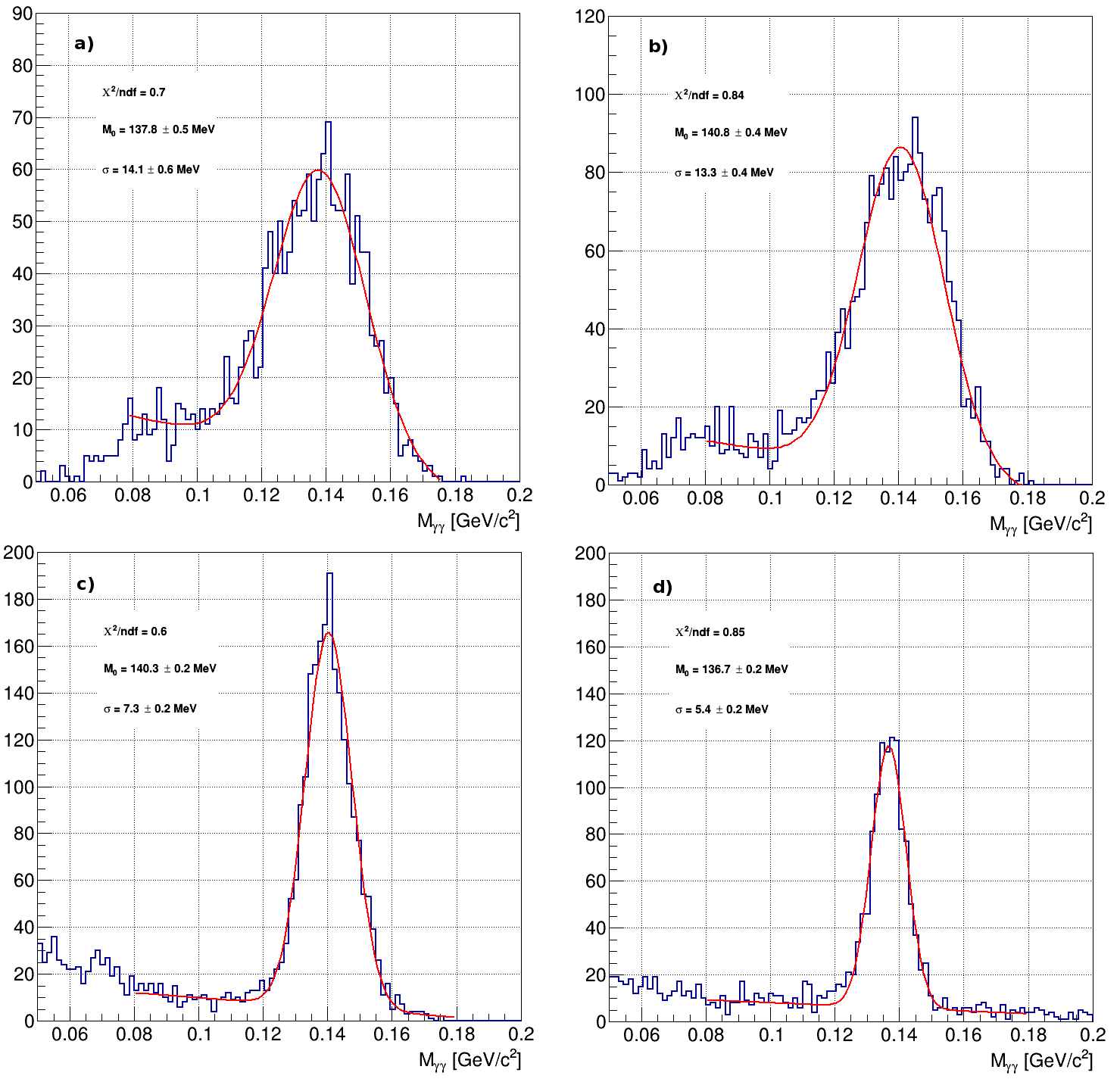}
\caption[Invariant-mass spectrum for photon pairs for several $\pi^0$ energy ranges.]
{Invariant-mass spectrum of photon pairs from decays of $\pi^0$ with energy a) 20-50$\,\mev$, b) 50-100$\,\mev$, c) 100-1000$\,\mev$
and d) 1-15$\,\gev$. The $\pi^0$ mass resolution ($\sigma$) is equal to a) 14$\,\mev$, b) 13$\,\mev$, c) 7$\,\mev$ d)  5$\,\mev$.}
\label{fig:sim:pi0_mass_4ranges}
\end{center}
\end{figure*}

In order to study the origin of the large background shown in Figure~\ref{fig:sim:pi0_mass} at low energies,
we simulated $\pi^0$ with different energy ranges. In Figure~\ref{fig:sim:pi0_mass_4ranges} one can see
the reconstruction of $\pi^0$ with energy ranges: a) 20-50 MeV, b) 50-100 MeV, c) 100-1000 MeV and d) 1-15 GeV.\
For momenta below 1 GeV/c events with at least one photon at the FSC are selected, above 1 GeV/c – with both
photons at the FSC detector. The low energy background tail comes from the large number of clusters at the FSC, wich
increases with energy. Presumably, the reason is an EM-shower formation somewhere before the FSC.
If we select the events with number of clusters below seven, the low energies tail background disappears,
as one can see in the Figure~\ref{fig:sim:pi0_mass_clucut} below. We should emphasize that for the
reconstruction here only FSC information was used (no PID data, no tracking data).

\begin{figure}[htb]
\begin{center}
\includegraphics[width=1.1\swidth]{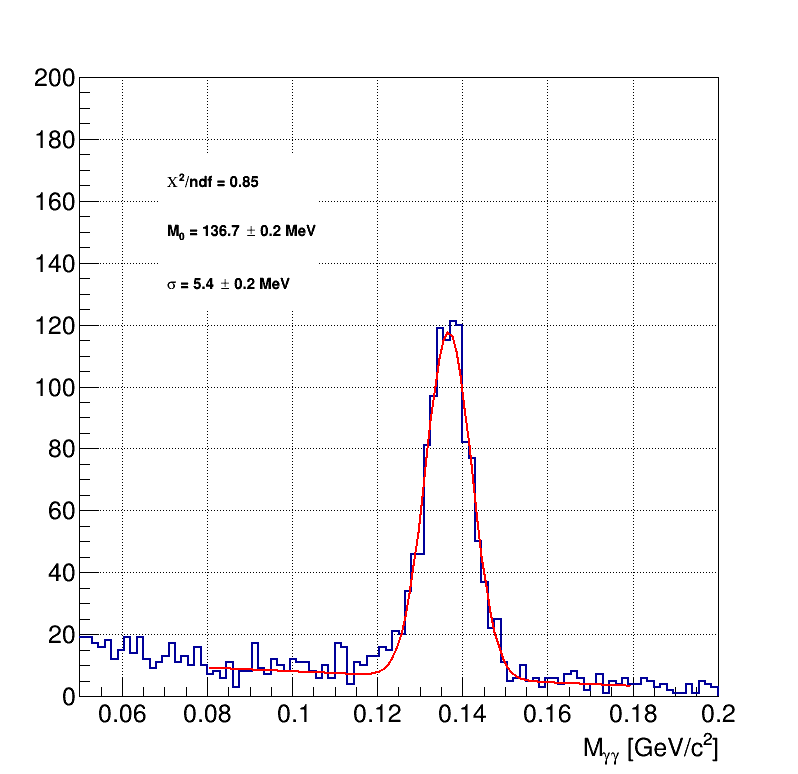}
\caption[Invariant-mass spectrum for photon pairs in FSC acceptance with number of clusters cut]
{Invariant-mass spectrum of photon pairs from decays of $\pi^0$ with energy up to 15$\,\gev$ and selection of events with number
of clusters in the FSC less than seven.}
\label{fig:sim:pi0_mass_clucut}
\end{center}
\end{figure}

\section{Electron identification}
\label{sec:sim:electronpid}
Electron identification will play an essential role for most of the physics program of
\Panda. An
accurate and clean measurement of the \jpsi decay in $e^+ \, e^-$ is needed for many
channels in the charmonium sector as well as for the study of the
$\bar{p}$ annihilation in nuclear matter like the reaction $\pbar A \, \to \, \jpsi X$.
In addition, the determination of electromagnetic form factors of the proton via
$\pbarp \, \to \,e^+ \,e^- $ requires a suppression of the main background channel
$\pbarp \, \to \, \pi^+ \, \pi^-$ in the order of $10^8$.

\begin{figure}[htb]
\begin{center}
\includegraphics[width=\swidth]{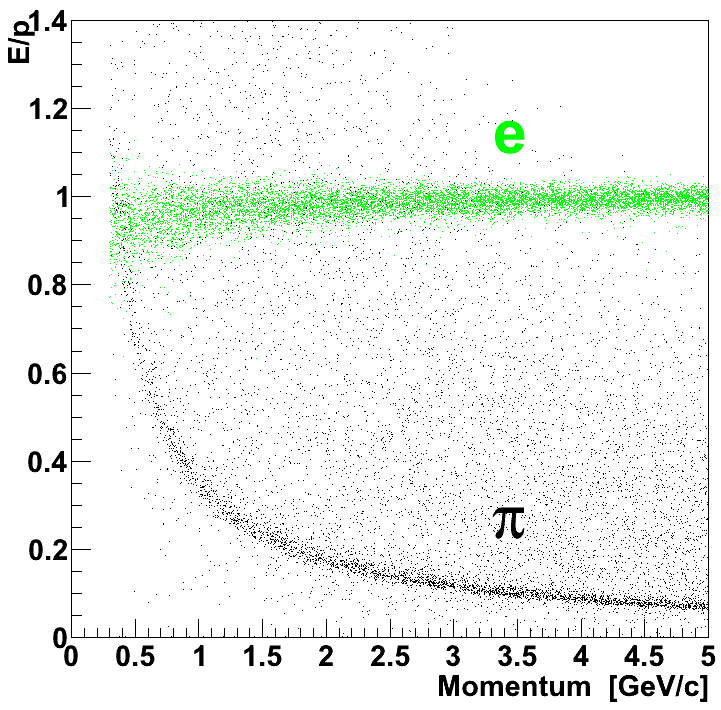}
\caption[E/p vs. track momentum for electrons and pions]
{Simulated E/p ratio as function of the track momentum for electrons (green) and pions (black) in the momentum
range between 0.3$\,\gevc$ and 5$\,\gevc$.}
\label{fig:sim:eOpElectronsPions}
\end{center}
\end{figure}

The FSC 
is designed for the detection of electromagnetic shower energy. In addition, it is also a powerful
detector for an efficient and clean discrimination of electrons from hadrons.
The character of an electromagnetic shower is distinctive for electrons, muons and hadrons. The most suitable
property is the energy deposited in the calorimeter. While muons and hadrons in
general lose only
a certain fraction of their kinetic energy by ionisation processes, electrons deposit
their complete energy in an electromagnetic shower. The
ratio E/p of the measured energy deposition E in the calorimeter over the reconstructed track momentum
p will be approximately unity. Due to the fact that hadronic
interactions within the cells
can take place, hadrons can also have a higher E/p ratio than expected from ionisation.
Figure~\ref{fig:sim:eOpElectronsPions} shows the reconstructed E/p fraction
for electrons and pions as a function of momentum.

Furthermore, the shower shape of a cluster is helpful to
distinguish between electrons, muons and hadrons. Since the chosen size of the cells
corresponds to the effective Moli\`ere radius, the largest fraction of
an electromagnetic shower originating from an electron is contained in just a few
cells. Instead, an hadronic shower with a similar energy deposition is less concentrated.
These differences are reflected in the shower shape of the cluster, which can be
characterised by the following properties:

\begin{figure}[htb]
\begin{center}
\includegraphics[width=\swidth]{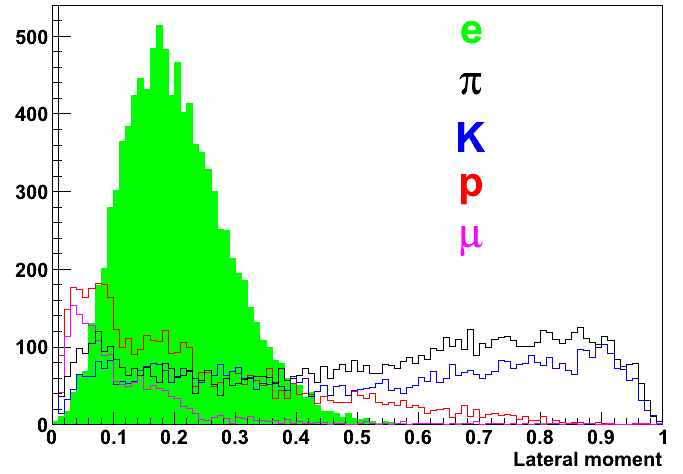}
\caption[Lateral moment of the FSC cluster for electrons, muons and hadrons]
{The lateral moment of the FSC cluster for electrons, muons and hadrons.}
\label{fig:sim:latmom}
\end{center}
\end{figure}

\begin{figure}[htb]
\begin{center}
\includegraphics[width=\swidth]{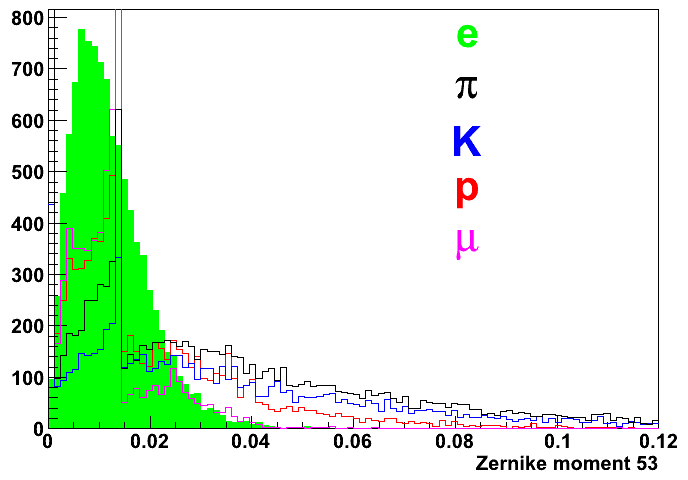}
\caption[Zernike moment $z_{53}$ for electrons, muons and hadrons]
{Zernike moment $z_{53}$ for electrons, muons and hadrons.}
\label{fig:sim::zernike53}
\end{center}
\end{figure}

\begin{itemize}
  \item $E_1/E_9$ which is the ratio of the energy $E_1$ deposited in the central cell over the energy deposition $E_9$ in
  the 3$\times$3 cells array containing the central cell and the ring of nearest cells.
  Also the ratio of $E_9$ over the energy deposition $E_{25}$ in the
  5$\times$5 cells array is useful for electron identification.
  \item The lateral moment $M_{lat}$ of the cluster defined by
\begin{equation}
 M_{lat} = \frac{\sum_{i=3}^n E_i r_i^2}{\sum_{i=3}^n E_i r_i^2 + E_1 r_0^2 +  E_2 r_0^2}
\label{eq:mlat}
\end{equation}
with
  \begin{itemize}
    \item n: number of cells associated with the shower.
    \item $E_i$:  energy deposited in cell $i$ with $E_1 \geq E_2 \geq ... \geq E_n$.
    \item $r_i$: lateral distance between the central and cell $i$.
    \item $r_0$: the average distance between two cells.
  \end{itemize}
  Figure~\ref{fig:sim:latmom} shows the simulated lateral moment for
  electrons, muons and hadrons.
  \item A set of Zernike moments \cite{zernike} which describe the energy distribution within a
  cluster by
  radial and angular dependent polynomials. An example is given in \Reffig{fig:sim::zernike53},
  where the Zernike moment $z_{53}$ is shown for all particle types.
\end{itemize}

Due to the fact that a lot of partially correlated EMC properties are suitable for electron
identification, a feedforward artificial neural network in the form of a Multilayer
Perceptron (MLP) \cite{MLP} 
can be applied. The advantage of a neural network is that it can
provide a correlation between a set of input variables and one or several output
variables without any knowledge of how the output formally depends on the input. Such
techniques are also successfully used by other HEP experiments \cite{Abramowicz, Breton}.

To demonstrate the advantages of the MLP network, we show here a result obtained for the
barrel part of the \PANDA Target Spectrometer EMC.
The training of the MLP was achieved with a data set of 850.000 single tracks for each
particle species (e, $\mu$, $\pi$, K and p) in the momentum range between 200$\,\mevc$
and 10$\,\gevc$ in such a way that the output values are constrained to be 1 for
electrons and -1 for all other particle types. In total, 10 input variables have been
used, namely E/p, the momentum p, the polar angle $\theta$ of the cluster, and 7 shower-shape 
parameters ($E_1/E_9$, $E_9/E_{25}$, the lateral moment of the shower, and 4 Zernike moments). 

\begin{figure}[htb]
\begin{center}
\includegraphics[width=\swidth]{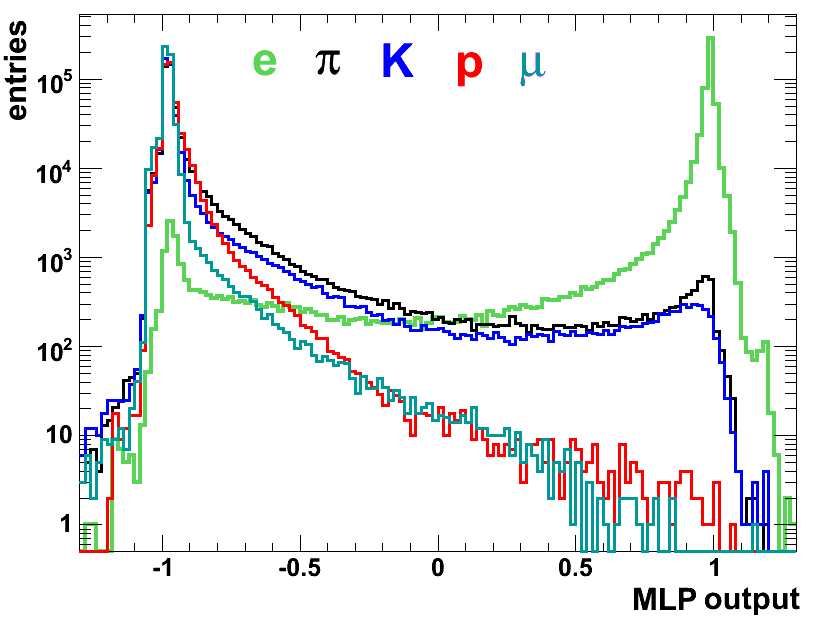}
\caption[MLP output for electrons and other particles for the barrel EMC]
{Simulation of MLP output for electrons and other particle species
in the momentum range between 300$\,\mevc$ and 5$\,\gevc$ for the barrel part of the Target Spectrometer EMC.}
\label{fig:sim:mlp}
\end{center}
\end{figure}

The
response of the trained network to a set of test data of single particles in the
momentum range between 300$\,\mevc$ and 5$\,\gevc$ is shown in \Reffig{fig:sim:mlp}.
The logarithmically scaled histogram shows that an almost clean electron recognition 
with a small contamination of muons ($<10^{-5}$) and hadrons ($0.2\%$ for
pions, $0.1\%$ for kaons and $<10^{-4}$ for protons) can be obtained by applying
a cut on the network output.
The plan is to incorporate such a method in the reconstruction algorithm.

\section{Material budget in front of FSC}
\label{sec:sim:material}

\begin{figure}[ht]
\begin{center}
\includegraphics[width=1.05\swidth]{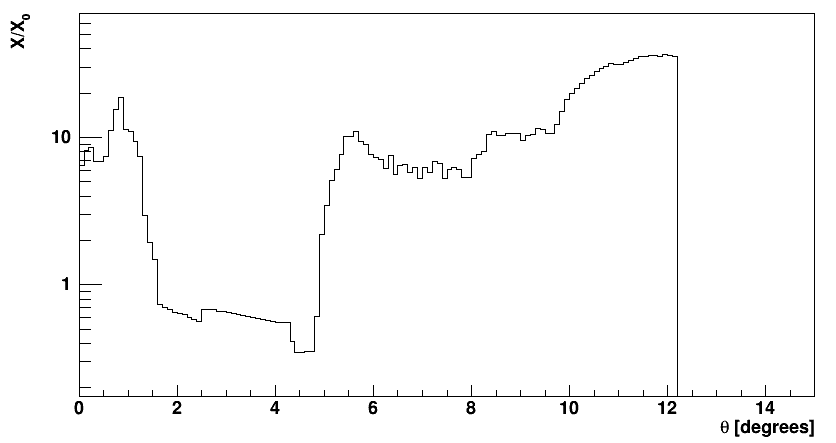}
\caption[Material budget in front of FSC versus polar angle]
{Material budget X in front of the FSC in units
of radiation length X$_0$ as a function of the polar angle $\theta$.}
\label{fig:sim:x0_vs_theta}
\end{center}
\end{figure}

\begin{figure*}[ht]
\centering
\includegraphics[width=0.85\dwidth]{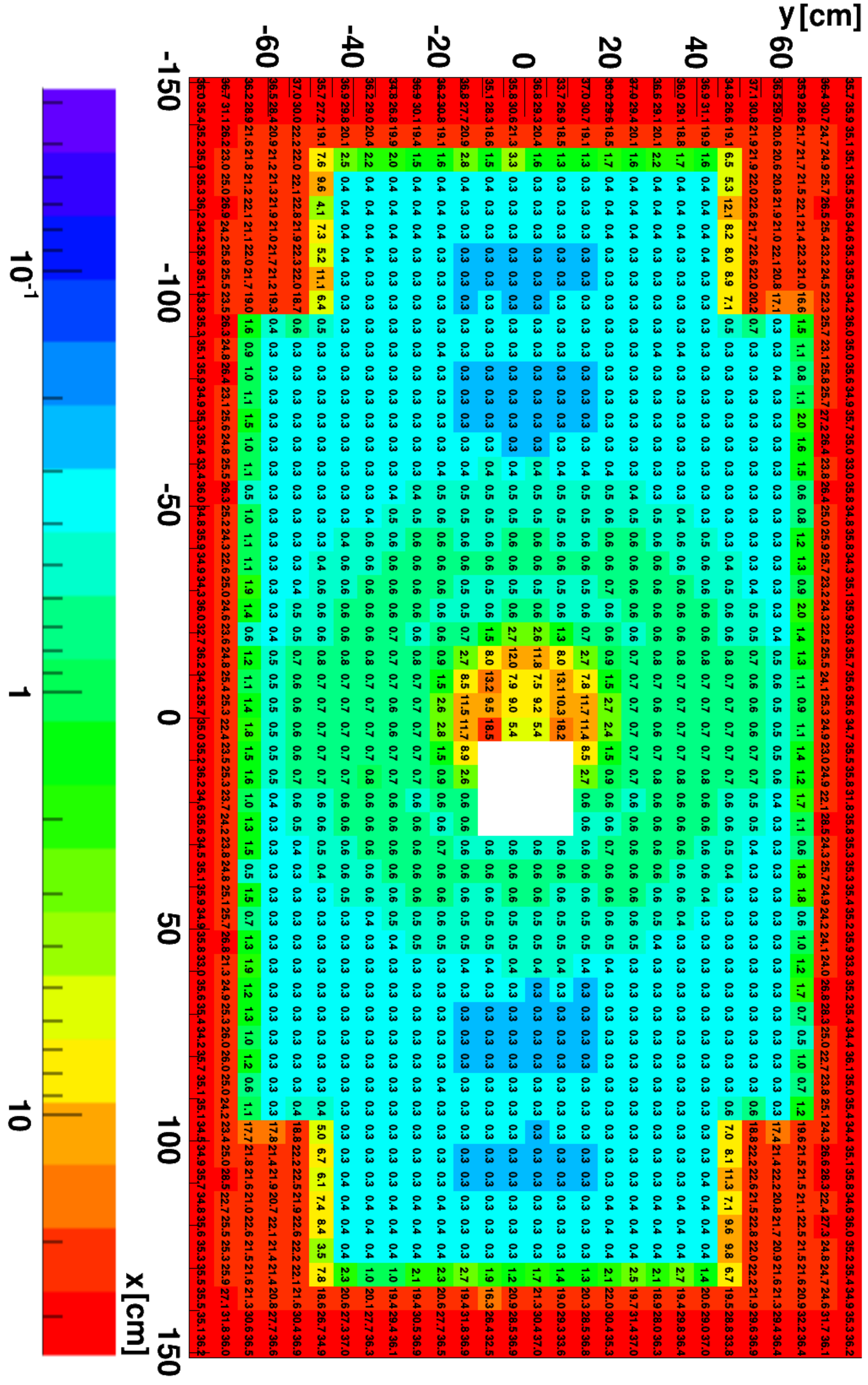}
\caption[Material-budget map in front of FSC]
{Map of the material budget X in front of the FSC in units
of radiation length X$_0$ as seen from the interaction point. Numbers in the cells and the
colour code correspond to the total material budget X in units of radiation length X$_0$.
The white rectangular area reflects the hole for beam pipe.}
\label{fig:sim:x0_vs_x_y}
\end{figure*}

The reconstruction efficiency as well as the energy and spatial resolution of the FSC
are affected by the interaction of particles with material in front of the calorimeter.
While the dominant interaction process for photons in the energy range of interest
is $e^+\,e^-$ pair production, electrons lose energy mainly via
bremsstrahlung ($e \, \to \, e \, \gamma$). The results in this section
are obtained with PandaRoot revision ``26429'' and version ``201309'' of the beam-pipe geometry
implementation. Figure \ref{fig:sim:x0_vs_theta} presents
the total material budget X in front of the FSC in units of radiation length X$_0$ as a function of the polar angle $\theta$.
The huge effect of the beam pipe is clearly seen in the region below
$\theta=1.5{\degrees}$.
In the region $\theta>5{\degrees}$ the material from the forward endcap of the
Target Spectrometer EMC and the endcap part of the Muon Detection system mainly contributes. 
Figure \ref{fig:sim:x0_vs_x_y} shows the two-dimensional map of material in front of each FSC
cell as seen from the interaction point.

The numbers in the cells and the
colour code correspond to the total material budget X in units of radiation length X$_0$
collected on the path from the interaction point (at z=0 with positive z-axis running downstream) to the respective cell. The
central red-yellow zone with values of X$\sim$10 X$_0$ reflects the effect of the beam pipe.
The beam pipe downstream of z$\sim3$ m is made of steel 
and starts to bend, in order to accommodate the beam after the dipole magnet, thus providing more
material for particles emerging from the interaction region to interact with.
The left-right asymmetry in this central zone corresponds to the bent beam pipe. It is fair to
conclude, that this zone is useless for the reconstruction of electrons and photons with a central hit in
this zone. The central zone, however, may be used to detect electrons, which were bent by the
magnetic field, or to add cells to clusters in the analysis stage, when the central cluster hit lies
close to this zone. 

Another reason to remove the central zone from the prime data analysis is the angle of incidence of detected particles.
Since the minimal angle of incidence for
particles travelling unhindered from the interaction point is $1.5{\degrees}$ (see the edge of central red-yellow zone in
\Reffig{fig:sim:x0_vs_x_y} or \Reffig{fig:sim:x0_vs_theta}), one may safely
neglect the effect of non-uniformity in the transverse light output which would
contribute at smaller angles of incidence (see \Refsec{sec:req:advandis}).

The circular zone shown in cyan with X$\sim$0.6-0.7 X$_0$ reflects the effect of the 
beam pipe made of titanium at z$<$3 m) plus material of detectors in the forward part of the Target Spectrometer and in the Forward Spectrometer: 
GEMs, DiscDIRC, straw tubes of the Forward Tracking System (FTS), Forward TOF (FTOF) and Forward RICH.
The beam pipe adds $\sim$0.3-0.4 X$_0$ to the material budget in this region. The red frame at the outer edge of the FSC acceptance with X$\sim$20-30 X$_0$
reflects the effect of material from the forward endcap of the
Target Spectrometer EMC plus the endcap part of the Muon Detection system.
This zone, too, can not be used for the direct photon reconstruction but can
help to detect electrons bent by the magnetic field. In the blue zone only detectors
of the Forward Spectrometer contribute to a material budget X$\sim$0.3-0.4 X$_0$.
\begin{figure}[ht]
\begin{center}
\includegraphics[width=1.05\swidth]{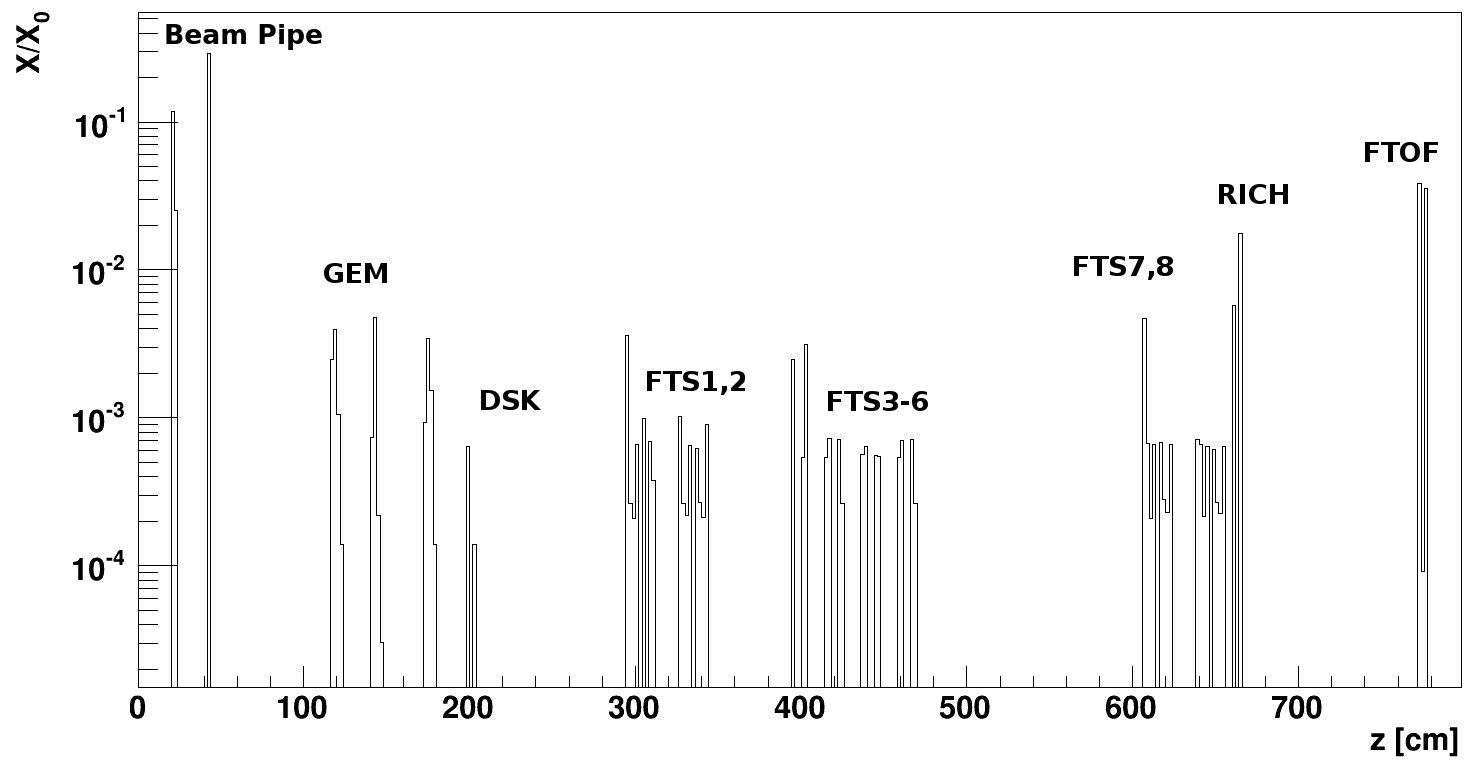}
\caption[Material budget in front of FSC vs. z axis]
{Material budget X in units of radiation length X$_0$ in front of a selected FSC cell with
coordinates (x,y) = (-7, -2)  (in cell units) as a function of the z coordinate.}
\label{fig:sim:x0_vs_z}
\end{center}
\end{figure}

Figure \ref{fig:sim:x0_vs_z} demonstrates the influence of material from the
various forward detectors in the FSC acceptance between $\theta$=1.5\degrees
and 5\degrees. The graph represents the material budget in front of a selected FSC cell
with
coordinates (x,y) = (-7, -2)  (in cell units) as a function of the z coordinate (along the beam
line). The total amount of the material budget is $\sim$0.6 X$_0$.
The main individual contribution of X$\sim$0.4 X$_0$ is caused by the
beam pipe. Thus, the most probable z coordinate of an initial
electron or photon interaction through conversion or bremsstrahlung in dead material would be
z$\sim$20-40 cm. Having the whole Forward Tracking and PID System upstream of the FSC, one
may be confident that such an event can be reconstructed. 
The second most probable z coordinate of an electron or photon interaction would be
around 7-7.5 m near the Forward RICH and FTOF (see
\Reffig{fig:sim:x0_vs_z}) with a material-budget contribution of $\sim$0.15 X$_0$. The shower initiated at such a close distance ($\sim$50 cm) from the FSC surface appears very
similar to a shower produced in the FSC without any material in front.
The material-budget contribution X$\sim$0.05 X$_0$ from all other detectors is negligible.

\newpage
\bibliographystyle{panda_tdr_lit}

%% file: panda_tdr_FSC_perf_pub.tex
\cleardoublepage
\chapter{FSC prototype test-beam studies}
\label{sec:perf} 
\begin{figure}[hb]
  \includegraphics[width=1\swidth]{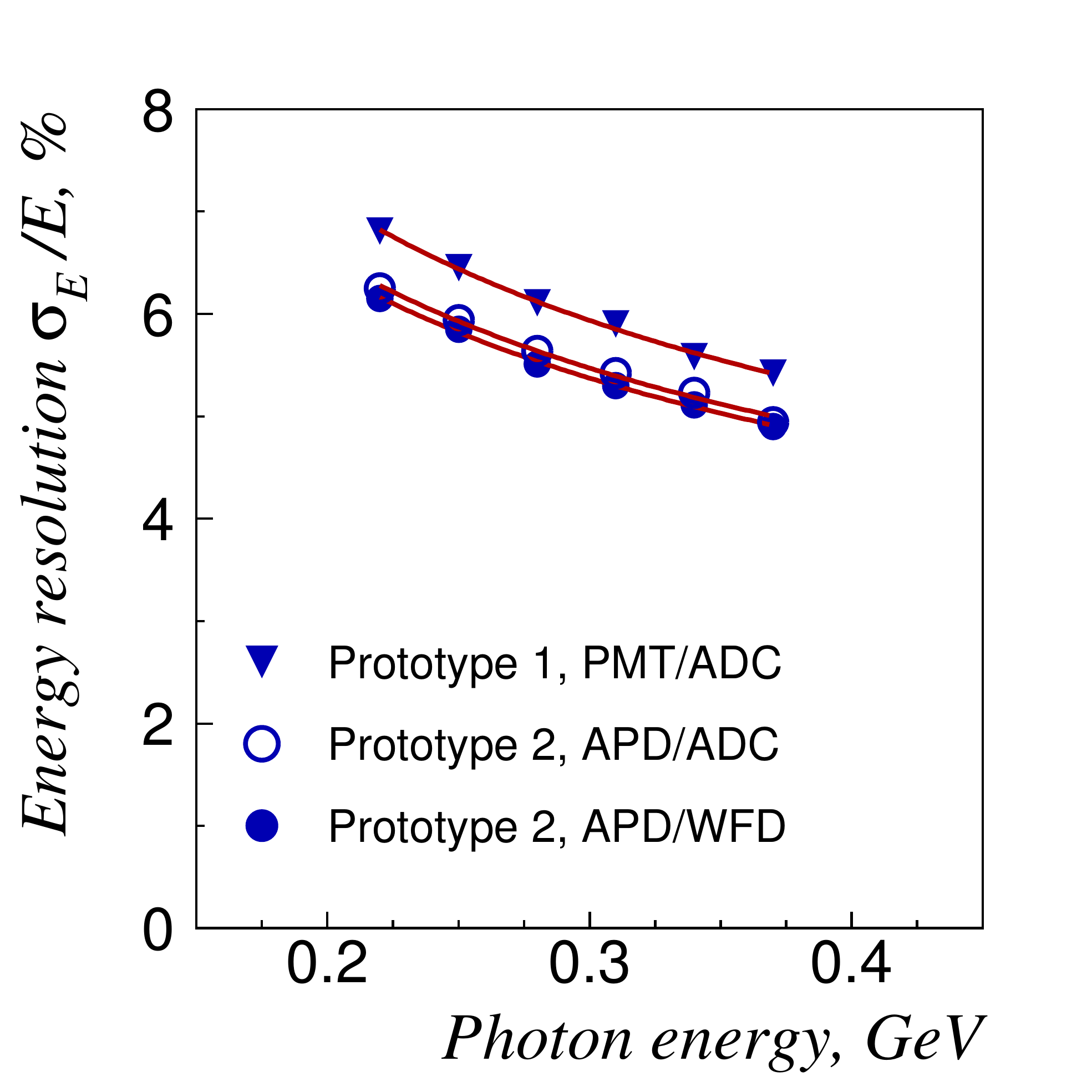}
  \caption[Energy resolution for KOPIO prototype]
{Energy resolution for KOPIO prototype calorimeters \cite{kopio2}.}
  \label{fig:test_kopio_e_resol}
\end{figure}

%
%
The design and concept of the Forward Spectrometer electromagnetic Calorimeter of the shashlyk type are based on previously
gained experience at IHEP Protvino in calorimeter development and suitable refinements to meet the requirements of \PANDA with respect to detection capabilities down to energies as low as 10 - 20~MeV.
As a start-up for the recent developments, improved versions of a finely segmented calorimeter, designed for the KOPIO experiment \cite{kopio1}, were studied. As shown in \Reffig{fig:test_kopio_e_resol}, an energy resolution \cite{kopio2} of
\begin{equation}
  \frac{\sigma_{\rm E}}{\rm E} [\% ] =
  \frac{(2.74 \pm 0.05)}{\sqrt{\rm{E/GeV}}} \oplus  (1.96 \pm 0.1)
\label{eq:fitfctdata}
\end{equation}
was reported, satisfying the requirements for the \PANDA experiment.

\section{Performance of Type-1 modules}

The first test experiments investigating energy, position and time resolution have been performed with an array of large-size modules close to the design of the KOPIO experiment.
The details of the individual modules are described in \Refchap{chap:mech}.

\subsection{The test-beam facility at IHEP Protvino}

\begin{figure}[hb]
  \includegraphics[width=1\swidth]{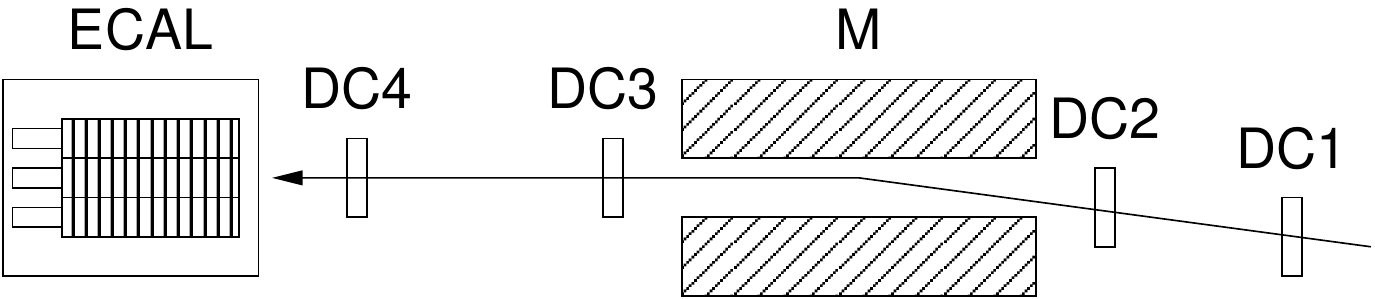}
  \caption[Protvino test-beam setup]
{Test beam-line setup at IHEP Protvino \cite{nim1}.}
  \label{fig:test_beam_line}
\end{figure}

In order to proof the achievable energy resolution of the FSC for electromagnetic probes, one needs an electron beam covering a wide range of energies with low contamination by hadrons and muons. The electron energy should be known with a precision significantly below 1-2\percent. An electron beam in the energy
range of 1 to 45 GeV satisfying the above requirements was commissioned at the U70 accelerator at Protvino, Russia.
A dedicated beam tagging setup consisting of 4 drift-chamber stations and a magnet was constructed at channel
2B (see~\Reffig{fig:test_beam_line}) of the U70 beam output building \cite{nim1}. The bending angle of the magnet was 55 mrad.

The secondary beam of negatively charged particles in the momentum range of 1 to 45~GeV/$c$ contained about 70\percent of
electrons mixed with muons and hadrons (mainly $\pi^-$ and $K^-$) at 19~GeV/$c$. However, particle identification was not available. The intrinsic momentum spread (or energy resolution) of the beam was at the level
of 1 to 5\percent at momenta from 45 to 1 GeV/$c$, respectively. However, the momentum tagging system \cite{nim1} provided final resolutions from 0.13\percent at 45 GeV/$c$, where multiple scattering was not dominating, down to a value of 2\percent at 1 GeV/$c$.
The beam spot at the detector front face covered a few cm in diameter, which could be enlarged up to 6 cm by using the so-called de-focused mode channel.
The intensity of the electron beam could be varied from 100 up to 10$^{6}$ electrons per accelerator
spill at 27~GeV energy, accompanied by low background. Alternatively, a high-intensity pion beam with up to 10$^{7}$
particles per accelerator spill allowed to study the $\pi^{0}$-meson reconstruction.

\subsection{The detector matrix}

The first FSC prototype was constructed at IHEP Protvino in 2007. The module design was based on the electromagnetic calorimeter for the KOPIO experiment with a cell size of 110$\times$110~mm$^2$.
The prototype of the electromagnetic calorimeter consisted of 9 modules assembled into a $3\times3$~matrix installed
on the remotely controlled (x,y) moving support which positioned the prototype across the beam with a precision of 0.4~mm.

The signal amplitude was measured by commercial 15~bit charge-sensitive ADC modules (LRS 2285) integrating
over a 150~ns gate with a sensitivity of 30 fC/channel. The data acquisition system was based on VME and CAMAC.
A detailed description of the DAQ system and the front-end electronics can be found in \cite{btev-nim2}.

\subsection{Calibration of the calorimeter prototype}

\begin{figure}
  \includegraphics[width=1\swidth]{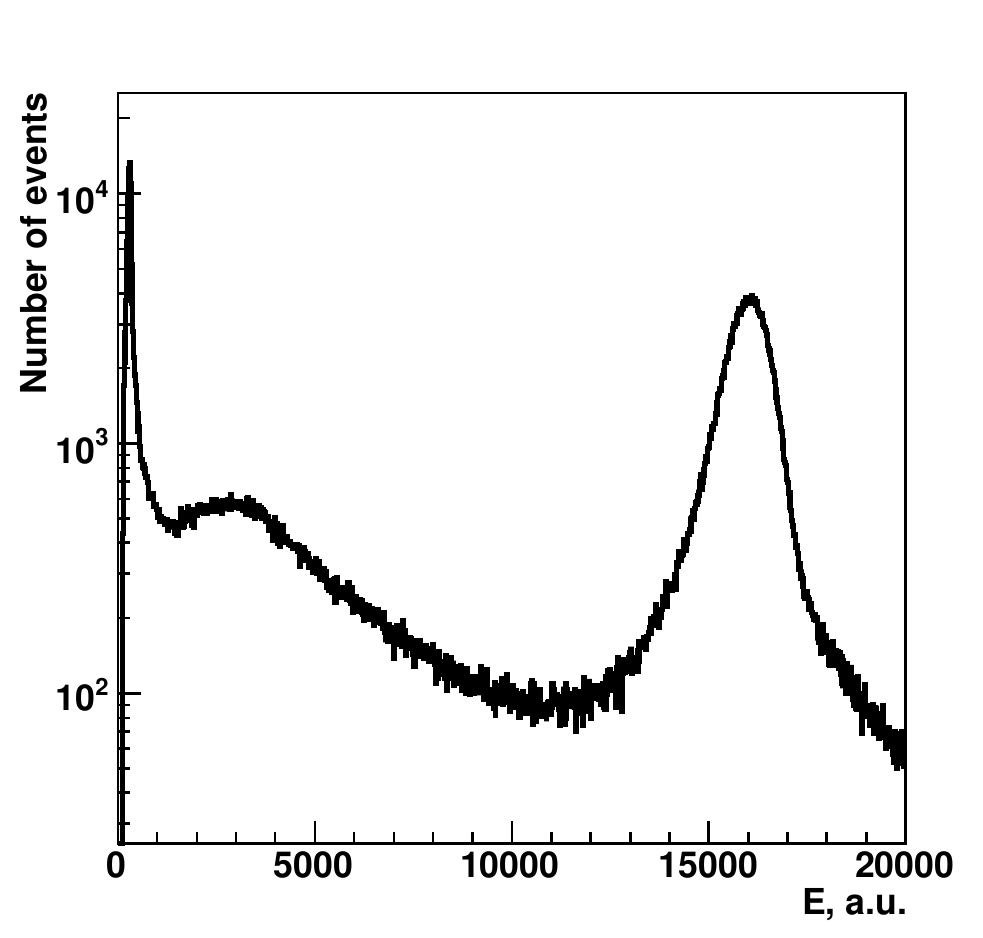}
  \caption[Energy deposition in single module]
{Energy deposited by 19 GeV/$c$ electrons in a single module \cite{btev-nim2}.}
  \label{fig:ECAL-spectrum}
\end{figure}

The modules were calibrated by exploiting the 19~GeV/$c$ electron beam. Each module was exposed directly to the beam using the (x,y) moving support. The typical energy spectrum of a single module (\Reffig{fig:ECAL-spectrum})
shows a well pronounced peak corresponding to the energy deposited by
19~GeV/$c$ electrons and another structure at low energies due to minimum ionising
(MIP) particles. The broad distribution at intermediate energies is caused by hadrons.

\begin{figure}
  \includegraphics[width=1\swidth]{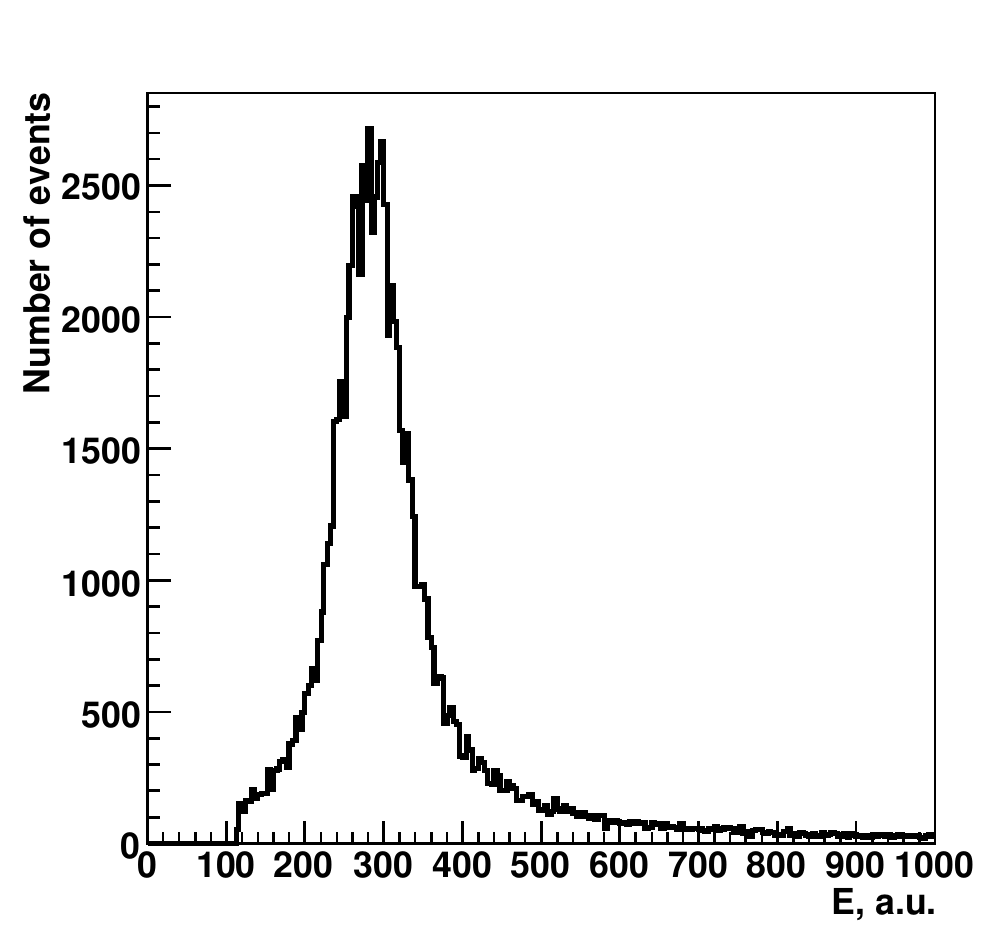}
  \caption[MIP energy distribution]
{The energy distribution around the MIP peak \cite{btev-nim2}.}
  \label{fig:ECAL-MIP}
\end{figure}

The relative energy calibration of all modules was obtained by using both the electron and the MIP signals. However, the best relative calibration coefficients were obtained by using the MIP signals only. For the MIP
calibration events were selected, when only one module triggered above a threshold of 100~MeV. A typical distribution around the MIP peak is shown in
\Reffig{fig:ECAL-MIP}. The peak with a high-energy tail was fitted by a Gaussian function
and served for the relative calibration. Finally, the absolute
calibration was obtained by setting the shower energy reconstructed in the whole
$3\times 3$ matrix to a value of 19~GeV.

%

\subsection{Energy and position resolution}

\begin{figure}[b]
  \centering
  \includegraphics[width=\swidth]{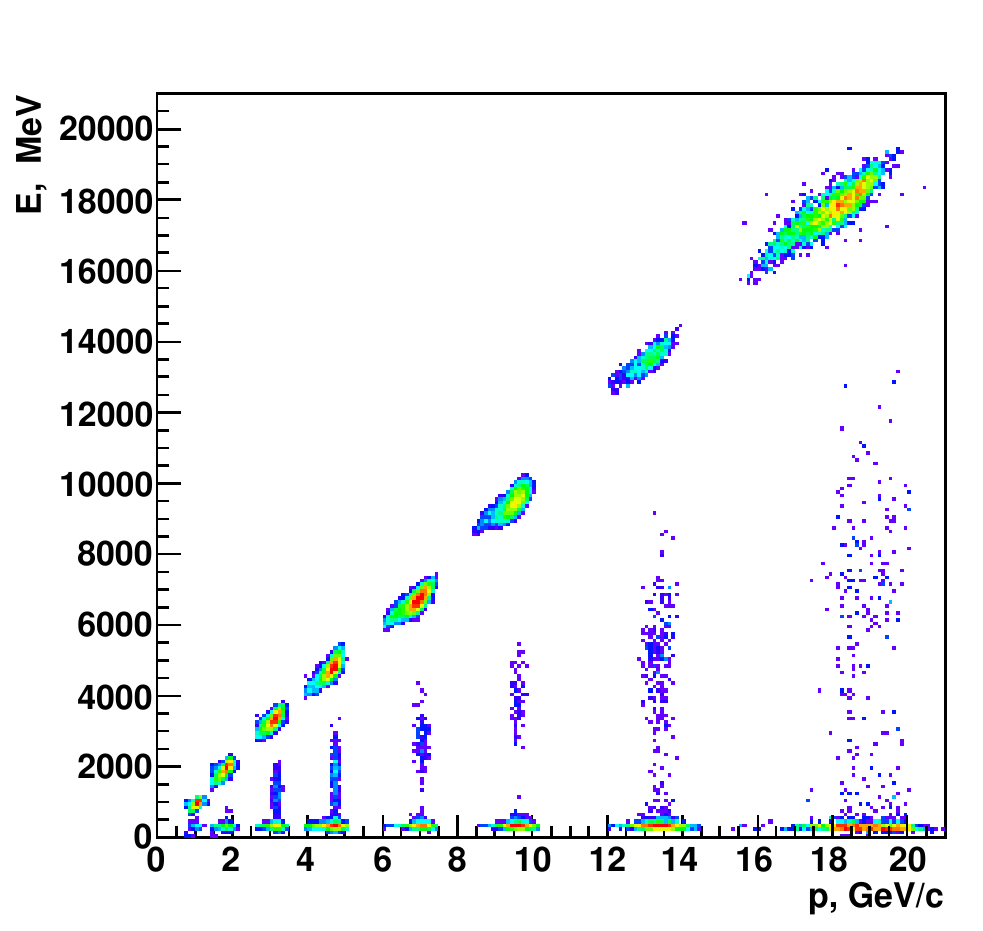}
  \caption[Energy vs. momentum correlation]
{Correlation between the energy measured in the 
calorimeter and the beam momentum measured in the magnetic 
spectrometer \cite{btev-nim2}.}
  \label{fig:E_vs_P}
\end{figure}
In order to deduce the overall resolution parameters of the prototype matrix,
the beam was directed onto the central module and beam
momenta of 1, 2, 3.5, 5, 7, 10, 14 and 19~GeV/$c$ were selected. For each setting
the magnetic field in the spectrometer magnet {\tt M}  (see~\Reffig{fig:test_beam_line})
was adjusted to provide the same bending angle of the beam. The momentum p
of the beam particle measured by the magnetic spectrometer and
the deposited energy E measured in the calorimeter prototype follow a linear correlation
as demonstrated in \Reffig{fig:E_vs_P}.

\begin{figure}[htb]
  \centering
  \includegraphics[width=\swidth]{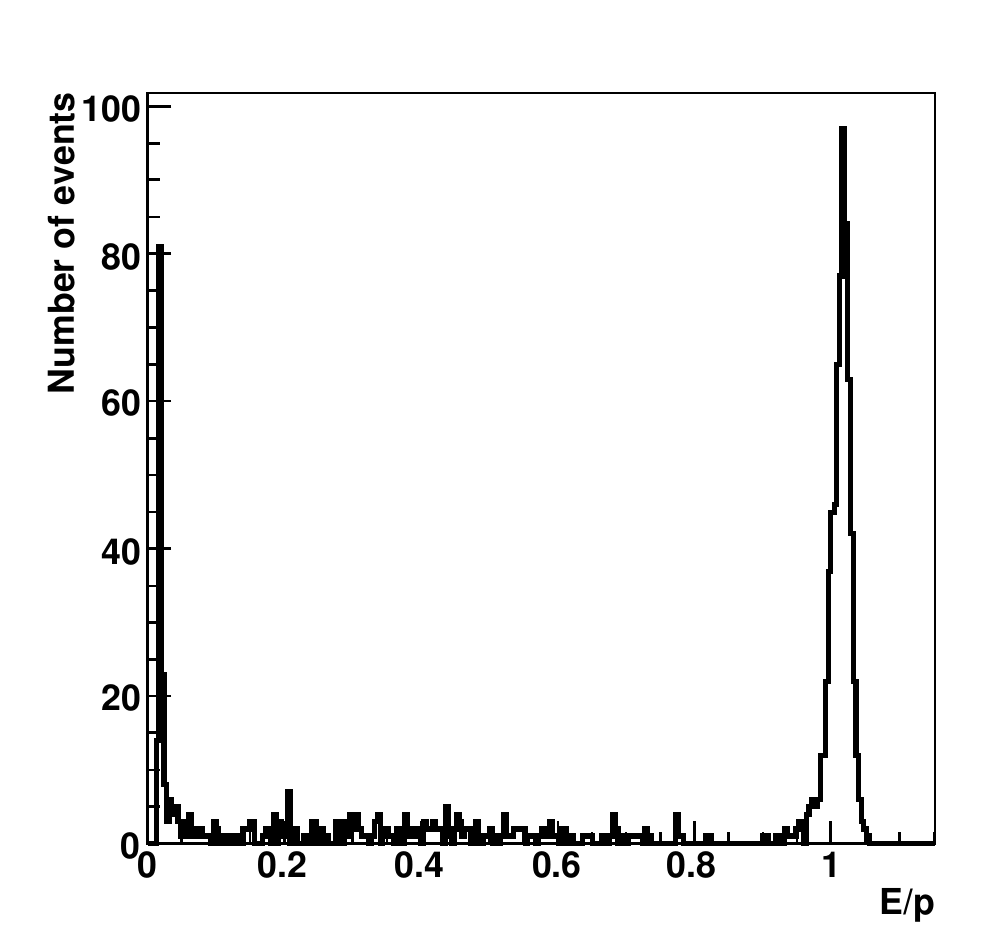}
  \caption[E/p ratio at 19 GeV in Type-1 prototype]
{Ratio of the energy E measured in the Type-1 prototype over the
    momentum p measured by the magnetic spectrometer at 19~GeV/$c$ \cite{btev-nim2}.}
  \label{fig:EoverP_19}
\end{figure}

\begin{figure}[htb]
  \centering
  \includegraphics*[width=\swidth]{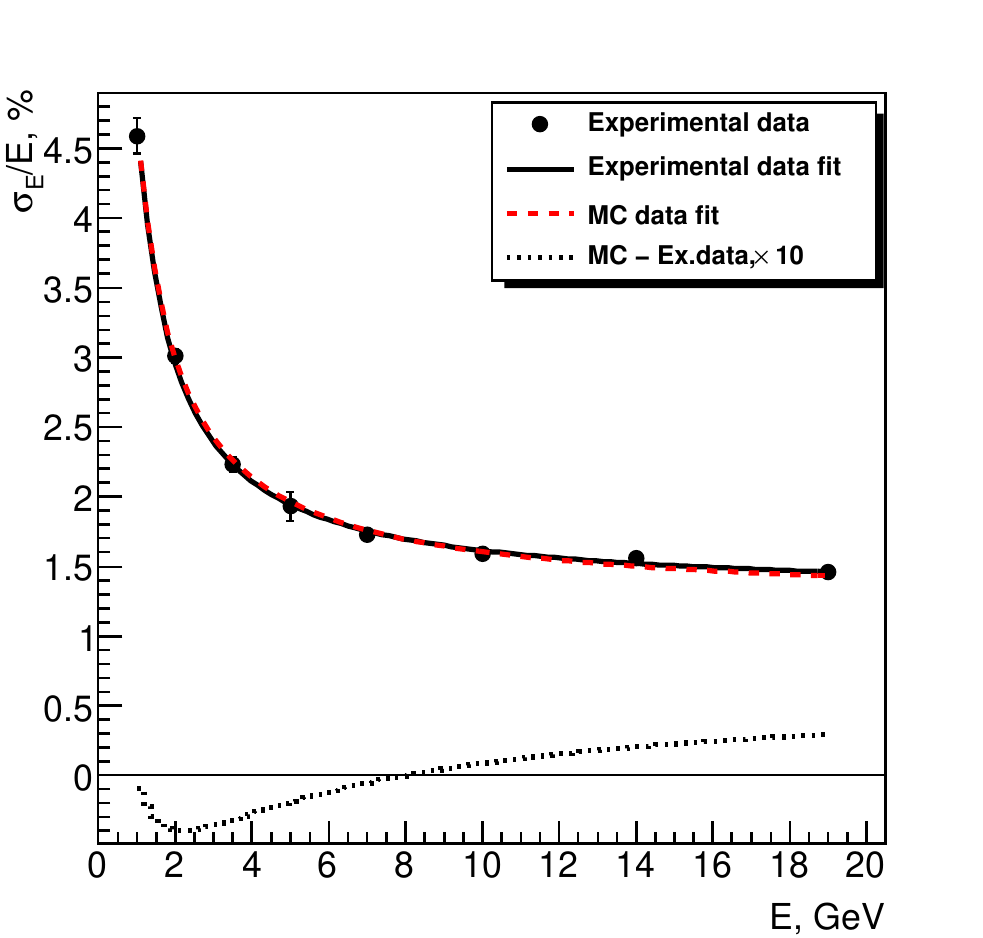}
  \caption[Measured and fitted energy resolution for Type-1 modules]
{Measured and fitted energy resolution for a 3$\times$3 matrix of Type-1 modules, compared to Monte Carlo fit. The difference between fits to experimental data and Monte Carlo is shown in the lower part \cite{btev-nim2}.}
  \label{fig:ECAL-eresolu}
\end{figure}

Therefore, in order to obtain the true intrinsic energy resolution, the measured
energy should be corrected for the beam momentum spread. Correspondingly, the energy
resolution can be represented by the width of the distribution of the
E/p ratio (\Reffig{fig:EoverP_19}).
The energy resolution is obtained from the Gaussian fit of the
peak around E/p=1. The relative energy resolution $\sigma_{E}$/E measured for
electrons at energies from 1 to 19~GeV is shown in
\Reffig{fig:ECAL-eresolu}.
Black bullets represent the measured resolution values, the solid
curve is a fit to the experimental data, and the dashed curve is a
fit of the expected resolution deduced from GEANT3 simulations. The fitting function is represented
by \Refeq{eq:fitfct8}:
\begin{equation}
  \frac{\sigma_{\rm E}}{\rm E} =
\frac{a}{\rm{E/GeV}} \oplus
\frac{b}{\sqrt{\rm{E/GeV}}} \oplus c
\label{eq:fitfct8}
\end{equation}

%
where the parameters $a$, $b$ and $c$ for the experimental and the Monte Carlo
fits are shown in Table~\ref{tab:eresolu}.
\begin{table*}[ht]
  \centering
  \begin{tabular}{|l|c|c|c|}\hline
    ~ & $a~[\%]$ & $b~[\%]$ & $c~[\%]$\, \\ \hline
    Experimental fit & $3.51 \pm 0.28$ & $2.83 \pm 0.22$ & $1.30 \pm 0.04$ \\
    Monte Carlo fit  & $3.33 \pm 0.12$ & $3.07 \pm 0.08$ & $1.24 \pm 0.02$ \\
    \hline
  \end{tabular}
  \caption[Parameters fitting energy resolution]
{Parameters of equation \ref{eq:fitfct8} fitting the energy resolution.}
  \label{tab:eresolu}
\end{table*}
The linear term $a$ of the energy resolution expansion is determined by
the beam spread rather than by the calorimeter properties. As was
shown in earlier studies performed at the beam-output channel 2B
\cite{btev-nim2}, the main contribution to this term stems from the
electronics noise and the multiple scattering of beam particles on the beam-pipe
flanges and the drift chambers. This beam momentum spread
was included in the Monte Carlo simulations as well, in order to fully reproduce
the experimental conditions. The simulated energy resolution is
shown by the red dashed line in \Reffig{fig:ECAL-eresolu}. The dotted
line in the lower part of the figure shows the difference between the fits to the experimental data
and the Monte Carlo results, multiplied by a factor 10. From this curve we conclude, that the deviation between the experimental data and the simulation result is less than 0.04\percent.

\begin{figure}
  \centering
  \includegraphics*[width=\swidth]{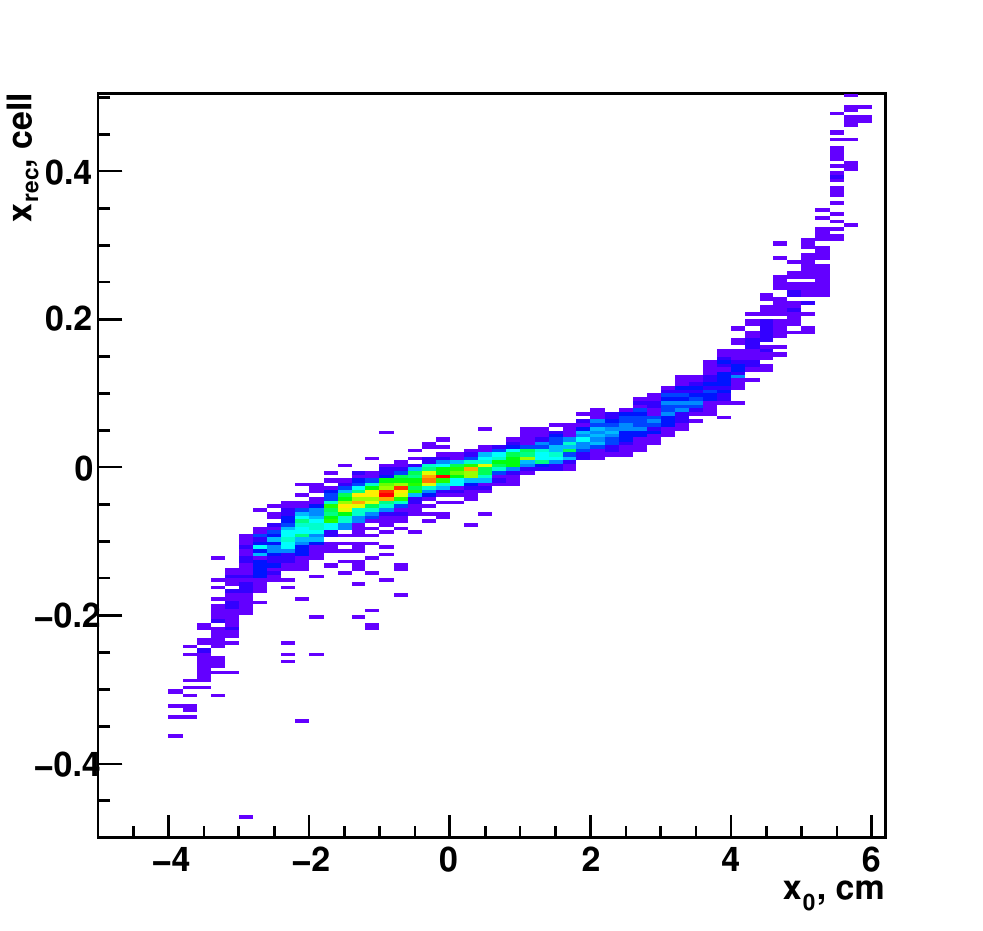}
  \caption[Reconstructed vs. impact coordinates]
{Centre of gravity x$_{\rm{rec}}$ of the electromagnetic shower as a function of the impact coordinate x$_0$ of the electron \cite{btev-nim2}.}
  \label{fig:S-curve}
\end{figure}

\begin{figure}
  \centering
  \includegraphics*[width=\swidth]{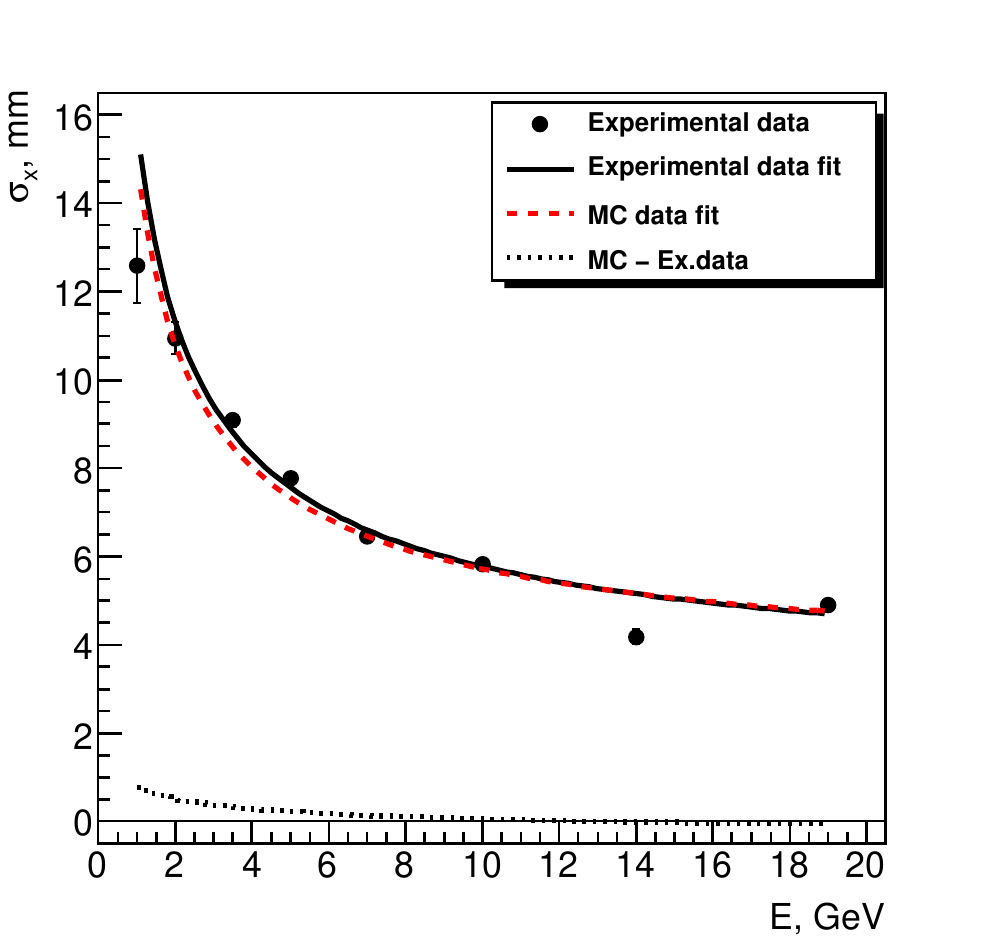}
  \caption[Measured position resolution]
{Measured position resolution. Bullets represent the
experimentally reconstructed positions. The solid curve is a fit to the
experimental data and the red dashed curve represents the Monte Carlo simulated values.
The difference between fits to experimental data and Monte Carlo is shown in the lower part \cite{btev-nim2}.}
  \label{fig:ECAL-xresolu}
\end{figure}

The position resolution has been determined from a comparison of the well-known
impact coordinate of the beam particle, measured with the last drift
chamber of the beam tagging system, and the centre of gravity of the electromagnetic
shower developed in the calorimeter prototype. Figure \ref{fig:S-curve}
shows the dependence of the measured coordinate x$_{\rm{rec}}$ on the
true coordinate x$_0$.
The deduced absolute position resolution in the centre of the matrix is shown in
 \Reffig{fig:ECAL-xresolu}.
The position-resolution values $\sigma_{\rm x}$ were fitted by the function
\begin{equation}
  \sigma_{\rm x} =
  \alpha \oplus \frac{\beta}{\sqrt{\rm{E/GeV}}},
  \label{eq:xfit}
\end{equation}
where the parameters $\alpha$ and $\beta$ are given in Table~\ref{tab:xresolu}.
\begin{table}[ht]
  \centering
  \begin{tabular}{|l|c|c|}\hline
    ~ & $\alpha~[$mm$]$ & $\beta~[$mm$]$ \\ \hline
    Experimental fit & $3.09 \pm 0.16$ & $15.4 \pm 0.3$ \\
    Monte Carlo fit  & $3.40 \pm 0.14$ & $14.5 \pm 0.3$ \\
    \hline
  \end{tabular}
  \caption[Parameters fitting position resolution]
{Parameters of equation \ref{eq:xfit} fitting the position resolution.}
  \label{tab:xresolu}
\end{table}
The dotted curve in the lower part of \Reffig{fig:ECAL-xresolu} illustrates the
deviations between both fits, which are consistent with a precision of 5\percent.

\subsection{Lateral non-uniformity}
\label{ssec:non-uniformity}

\begin{figure}[htb]
  \centering
  \includegraphics*[width=\swidth]{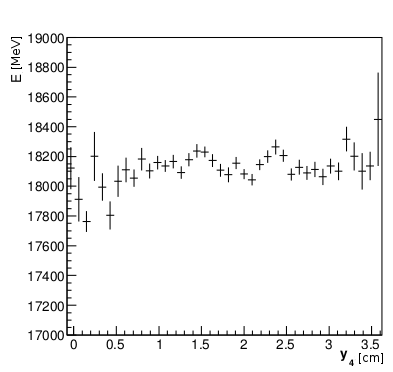}
  \caption[Profile of energy response vs. y for fixed x]
{Profile of the energy response for electron showers as function of the  y-coordinate at fixed x.}
  \label{fig:EvsY}
\end{figure}

Due to various mechanical inhomogeneities of the prototype one can
expect dependencies of the reconstructed energy E on the coordinates (x,y) of the point of impact.
Position dependencies could be caused by the positions of the WLS fibres, the steel string and
the boundaries between the modules. The lateral
uniformity of the energy response was studied with data
collected in the 19 GeV/$c$ run. The information of the last drift chamber {\tt DC4} was
used to determine the impact coordinate  of the beam particles onto the
calorimeter surface. Since the beam contained several particle species
which interact differently with the calorimeter medium (see~\Reffig{fig:ECAL-spectrum}), the mean deposited energy was measured
as a function of (x,y) for two energy intervals, E$<0.5$~GeV and
$16<\rm E<22$~GeV, corresponding to the MIP peak and that of the
electromagnetic shower, respectively. The profile of the energy response for
electron showers as function of the y-coordinate at fixed x is shown in
 \Reffig{fig:EvsY}.
The observed fluctuations stay below 1\percent. Within the statistical error no lateral non-uniformity of the
energy response is observed.

\subsection{Monte Carlo simulations of beam test with electrons}

The comparison of experimental data with Monte Carlo simulations allows to gain a good understanding of the detector performance and provides a sensitive
tool for further optimisations of the design of the calorimeter and its capability of event reconstruction.
All simulation studies were performed with GEANT3 as the Monte Carlo engine including a detailed description of the material budget and the module geometry.

The developing electromagnetic shower produces scintillation light which originates from two
different sources:
\begin{itemize}
\item scintillation light produced in the active plastic scintillator plates due to energy losses of electrons and positrons as secondary shower particles,
\item  Cherenkov radiation when charged particles pass the WLS fibres.
\end{itemize}
The simplified simulation technique consists of accounting for the energy deposition in the
active material (with some corrections to take into account the light
attenuation in the fibres) and ignoring Cherenkov radiation.
This method is very fast but cannot reproduce all details
of the calorimeter response such as the non-uniformity due to fibres and
cell borders.

For these studies, the detailed light propagation was applied taking
into account the optical properties of the materials, internal
reflections at plate borders, light capture by the fibres with double
cladding, and propagation inside fibres towards the photo sensor. An attenuation length of 70~cm in the
scintillator and 400~cm in the fibre was assumed. The refraction index of the scintillator amounts to
1.59, the total internal reflection efficiency at large scintillator faces
is 0.97, and a reflectivity of diffusion type was assumed for the side
faces of the scintillator plates with the same probability. The mean deposited
energy to generate one optical photon in the scintillator was
assumed to be 100~eV.

\subsection{Summary of the performance of Type-1 prototype}
\label{ssec:sum-large}

The measurement of the achieved energy resolution hitting a 3$\times$3 matrix right in the center has delivered a stochastic term of (2.8$\pm$0.2)\percent  which is consistent with former tests at
BNL for the KOPIO project in the energy range from 0.05 GeV to 1 GeV.
Taking into account the effect of light transmission in scintillator tiles and
WLS fibres and the photon statistics, good agreement is achieved between the measured and the Monte Carlo simulated resolution values.
The stochastic term in the dependence of position resolution on energy amounts to
(15.4$\pm$0.3) mm which is in agreement with Monte Carlo simulations as well. For 10~GeV electrons the position resolution ($\sigma$) is 6~mm in
the centre of the module and 3~mm at the boundary between two modules.
No significant non-uniformity of the energy response has been found.

\section{Performance of Type-2 modules}

\begin{figure}[b]
  \centering
  \includegraphics*[width=\swidth]{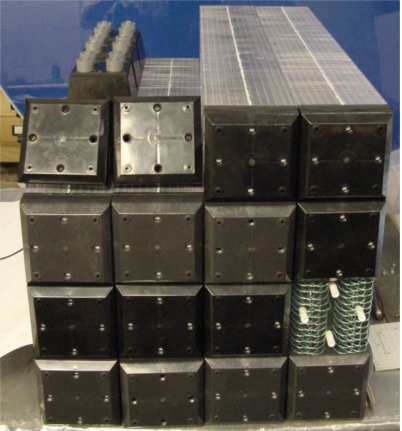}
  \caption[Small-cell Type-2 prototype]
{Small-cell prototype at the stage of assembly at the
scintillator department of IHEP Protvino.}
  \label{fig:proto64}
\end{figure}

In spite of the promising energy resolution in the high-energy region of the expected electromagnetic probes, the large geometrical cross section of the Type-1 modules will limit the achievable position resolution which becomes critical
in the forward region when dealing with the reconstruction of the invariant mass of e.g. highly energetic $\pi^0$s. In that case, the two decay photons span a small opening angle. In addition, the reconstruction of low-energy photons far below 1 GeV requires a low threshold of the individual detector modules. For such a situation, no reliable data could be provided by tests reported in the previous section.
Therefore, a second prototype module has been produced by subdividing the previous detector (Type-1) into 4 cells which are read out independently. The details of the improved geometry (Type-2) are already described in \Refchap{chap:mech}. The sampling ratio, the overall thickness, and number of layers have been kept the same as for Type-1 cells. The lead plates, however, were kept in common for the four optically isolated cells.
Two subarrays composed of 4$\times$4 or 3$\times$3 modules with a granularity of 8$\times$8 or 6$\times$6 individual cells have been prepared for test experiments at IHEP Protvino, using high-energy electrons, and at MAMI in Mainz, Germany, extending the response function for photons down to 50~MeV.

\subsection{Test with high-energy electrons at IHEP Protvino}
\label{ssec:ecoor_small}

\begin{figure}[h!]
\vspace*{-0.3cm}
  \centering
  \includegraphics*[width=0.9\swidth]{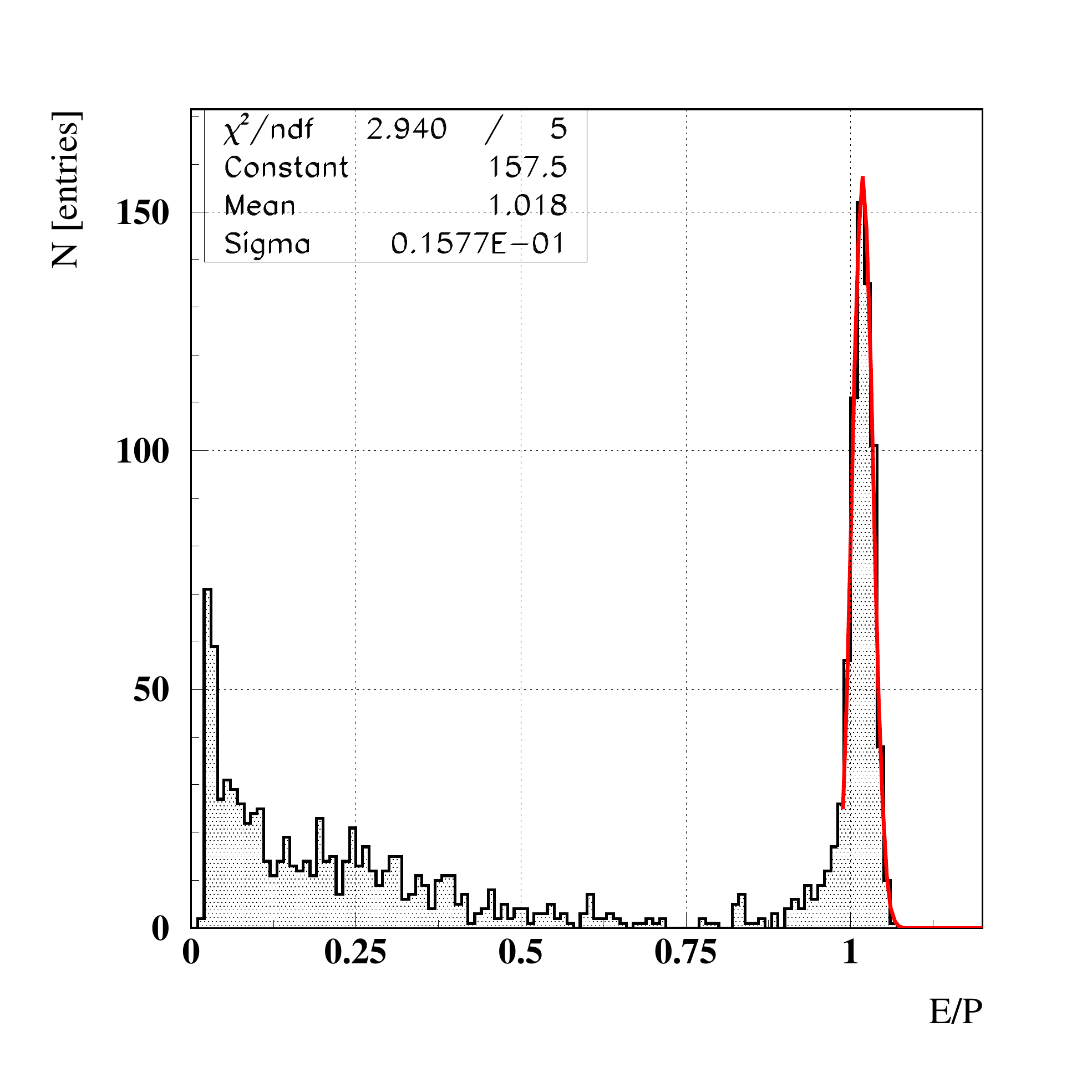}
\vspace*{-0.5cm}
  \caption[E/p ratio at 19 GeV in Type-2 prototype]
{Ratio of the energy E measured in the calorimeter over the
    momentum p measured by the magnetic spectrometer at 19~GeV/$c$.}
  \label{fig:small_19}
\end{figure}

The procedure to measure energy and position resolution for a matrix of $8\times 8$ cells (see ~\Reffig{fig:proto64}) was the same as for the Type-1 prototype described in the previous section. The measurements were performed at seven selected energies at 2.0, 3.5, 5.0, 7.0, 10.0, 14.0 and 19~GeV/$c$. The cells were pre-calibrated using a muon beam.
The spectrum of reconstructed E/p values recorded at 19~GeV is presented in \Reffig{fig:small_19}.
By quadratically subtracting the value of the beam-momentum resolution (0.62\percent) we obtain an energy resolution of 1.42\percent.

\begin{figure}[h!]
  \centering
  \includegraphics*[width=\swidth]{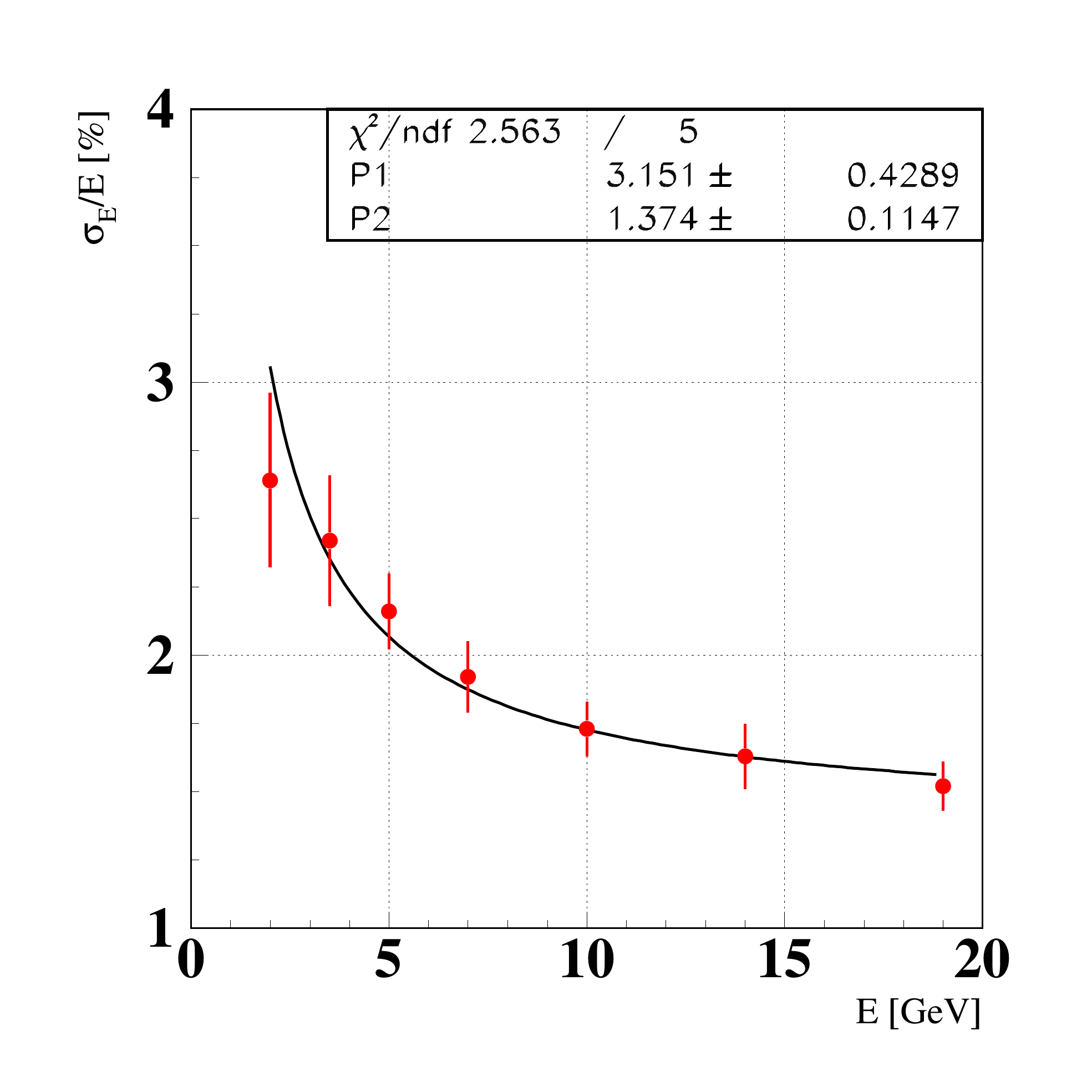}
  \caption[Energy resolution for Type-2 prototype]
{Energy resolution for a Type-2 prototype measured at IHEP Protvino.}
  \label{fig:small_eresolu}
\end{figure}

The dependence of the energy resolution on the incoming energy was fitted by the function given in \Refeq{eq:fitfct8}
with a fixed ``noise'' parameter of $a$=3.3\percent. This value for the linear term was already obtained in the test with the Type-1
prototype, since we used exactly the same beam line, magnet setting, and the same photon detection electronics.
Under these conditions, also similar values for the stochastic ($b$=3.15 $\pm$0.43 \percent) and constant ($c$=1.37 $\pm$0.11\percent) terms were obtained leading to the energy resolution shown in \Reffig{fig:small_eresolu}.


\begin{figure}[h!]
  \centering
  \includegraphics*[width=\swidth]{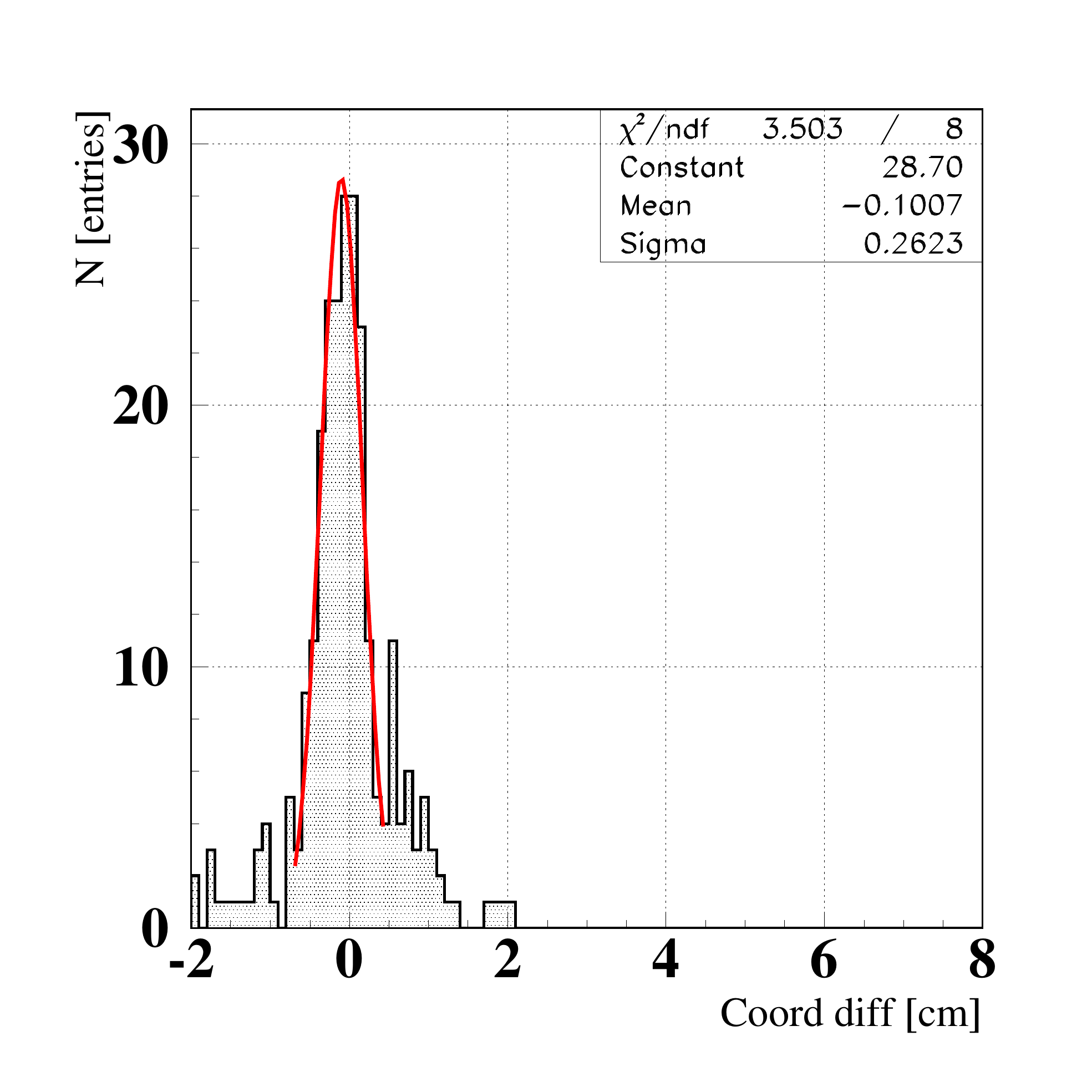}
  \caption[Position resolution for Type-2 prototype]
{Difference between 
the coordinate reconstructed in the Type-2 prototype and the track coordinate at the 
calorimeter surface.}
  \label{fig:small_coordreso}
\end{figure}

In comparison to the Type-1 prototype, which yielded about 5~mm (see \Reffig{fig:ECAL-xresolu}) position resolution at the center of the cell, 
a significantly improved resolution of 2.6~mm was reached (see \Reffig{fig:small_coordreso}), 
which is in good agreement with the MC results (\Reffig{fig:sim:posResol}). This measurement was carried out at the highest available 
test-beam energy of 19~GeV.

\subsubsection{Position dependence of the energy resolution}
\label{ssec:eres_cross}

\begin{figure}[h]
  \centering
{\hspace*{-1.3cm}
  \includegraphics*[width=1.35\swidth, height=6.5cm]{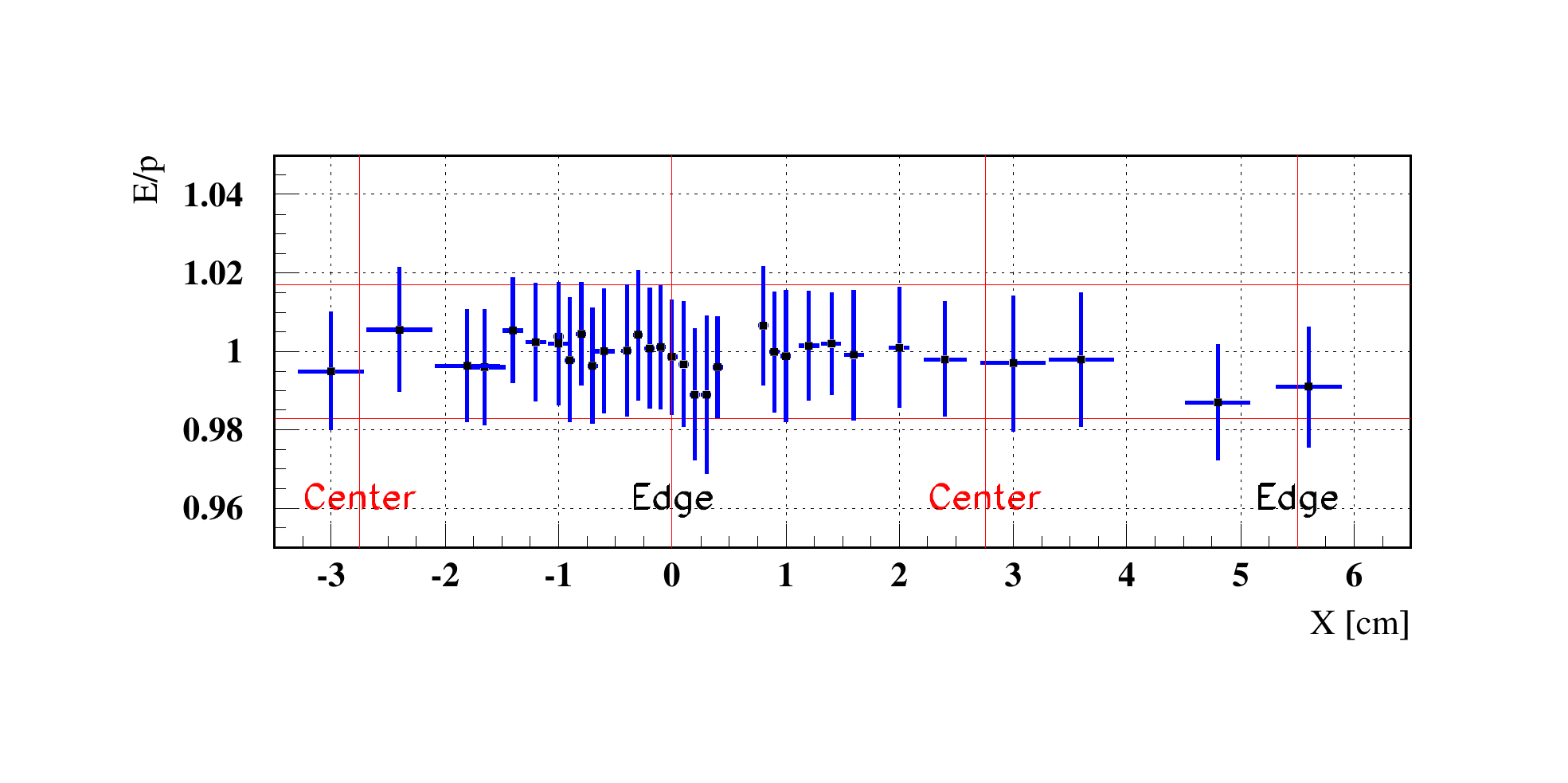}}
{\vspace*{-2.0cm}
  \caption[Dependence of E/p ratio on beam position]
{Dependence of the energy resolution, expressed as E/p ratio, on the beam position across a single-cell of a Type-2 prototype. The red data points mark the mean values of the reconstructed showers,
and the error bars indicate the corresponding ($\sigma$) resolutions.
The measurements were carried out at an electron energy of 10~GeV.}
  \label{fig:edge_eresolu}}
\end{figure}

In order to check the impact of dead material or of gaps between adjacent modules or cells, the stability of the energy resolution expressed by the E/p ratio has been investigated within a range of 6~cm in horizontal direction covering more than the width of a single cell. The results, measured at an electron energy of 10~GeV are summarised in \Reffig{fig:edge_eresolu}. The red data points mark the mean values of the reconstructed showers,
and the error bars indicate the corresponding ($\sigma$) resolutions. The horizontal red lines mark the average resolution ($\pm\sigma$) of the cell. Obviously, no significant position dependence has been observed.

\subsubsection{Reconstruction of neutral pions using a Type-2 shashlyk array}
\label{ssec:pi0_reco}

\begin{figure}
  \centering
  \includegraphics*[width=\swidth]{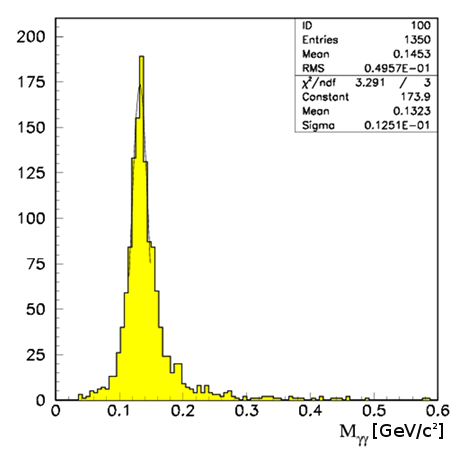}
  \caption[Reconstructed $\piz$ invariant mass]
{Spectrum of the reconstructed invariant mass of $\piz$ mesons between 1 and 2~GeV energy. The distance between the test-beam target and the FSC prototype was 1.5~m.}
  \label{fig:pizero_mass}
\end{figure}

For a combined test of position and energy performance, the prototype array composed of 8$\times$8 cells was used to detect and reconstruct neutral mesons. The setup was supplemented by an aluminium target in front of the prototype to generate
neutral mesons. Scintillator counters were positioned behind and in front of the target to provide a trigger for any interaction in the target. The prototype array was placed at a distance of 1.5~m downstream of the target. The energy of the negative pion beam impinging on the target was 28~GeV. After selecting primarily low energy $\piz$\ mesons (1-2 GeV) the obtained spectrum of the
reconstructed invariant mass is shown in \Reffig{fig:pizero_mass}.  The achieved invariant-mass resolution ($\sigma$) is 12.5 MeV/$c^2$, corresponding to 9.3\percent of the  $\piz$ PDG mass.

Considering the experimental energy and position resolutions, the contribution of the position resolution will be significantly smaller due to the large distance of $\sim$7~m of the FSC to the target in the final setup at \PANDA. One may expect an invariant-mass resolution in the order of 4~MeV/$c^2$
for 1-2~GeV $\piz$ mesons and one could reach 3 to 3.5 MeV/$c^2$ for the highest pion energies between 5-10~GeV.

\subsection{Test with high-energy photons at MAMI}

In spite of the dominance of high-energy photons in the most forward region of \PANDA, the energy response to photons with energies well below 1~GeV as well as the minimum experimental energy threshold are the most critical parameters to warrant a suitable detector design and acceptable performance.
Therefore, two different test experiments have been performed using a 3$\times$3 array composed of Type-2 shashlyk modules exploiting the beam of energy-marked photons at the electron accelerator MAMI at Mainz, Germany.

\subsubsection{The experimental setup at the tagged-photon facility at MAMI}

The tests have been performed with quasi-monochromatic photons delivered by the tagged-photon facility at the electron accelerator MAMI, exploiting the tagging of bremsstrahlung produced by a mono-energetic electron beam up to 855~MeV energy. After bremsstrahlung emission the momenta of the slowed-down electrons are analysed by the magnetic spectrometer of the Glasgow-Mainz tagger \cite{tagger-mami}, requiring a time coincidence of the detected bremsstrahlung photon with the corresponding electron identified in the focal plane. Depending on the accelerator beam energy, the typical energy width per tagging channel varies between 2.3 and 1.5~MeV for an electron beam of 855~MeV. In all experiments we have selected up to 16 photon energies to cover the investigated dynamic range. 

\begin{figure}
  \centering
  \includegraphics*[width=\swidth]{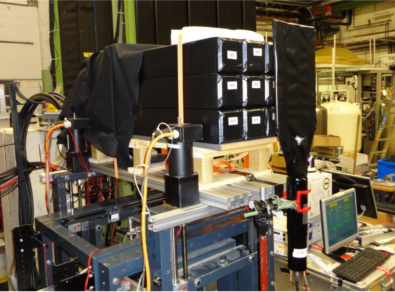}
  \caption[Test setup at MAMI tagged-photon beam]
{The experimental setup of the test matrix at the tagged-photon facility at MAMI \cite{diehl_msc}.}
  \label{fig:setup-MAMI1}
\end{figure}

The detector system was mounted on a support structure which could be moved remote-controlled in two dimensions perpendicular to the axis of the collimated beam by stepping motors. The detector array was placed typically at a distance of 12.5~m downstream of a collimator systems, which was located at a distance of 2.5~m from the radiator with a set of lead collimators of 1.5~mm diameter. The beam spot projected onto the front surface of the shashlyk matrix had a circular diameter of $\leq$10~mm. A plastic scintillator paddle in front could be used to identify leptons due to conversion of photons in air or any low-Z material in between. The mechanical setup allowed to direct the photon beam in the centre or in between two adjacent modules or cells. Figure~\ref{fig:setup-MAMI1} illustrates the experimental setup installed at the A2 tagger hall at the MAMI facility. The photon beam is hitting the detector system from the right hand side, passing the plastic veto detector in front.

\begin{figure}
  \centering
  \includegraphics*[width=\swidth]{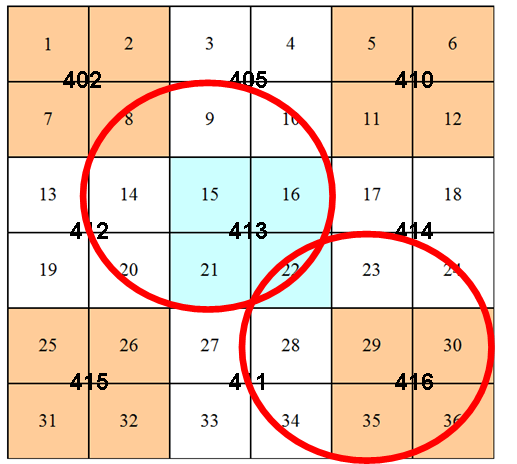}
  \caption[Labelling of 6$\times$6 detector cells at photon beam test]
{The labelling of the 6$\times$6 detector cells of the test matrix of 3$\times$3 detector modules at the tagged-photon facility at MAMI. Two main regions of impact are marked by red circles \cite{diehl_msc}.}
  \label{fig:pos-overview}
\end{figure}

\begin{figure}
  \centering
  \includegraphics*[width=\swidth]{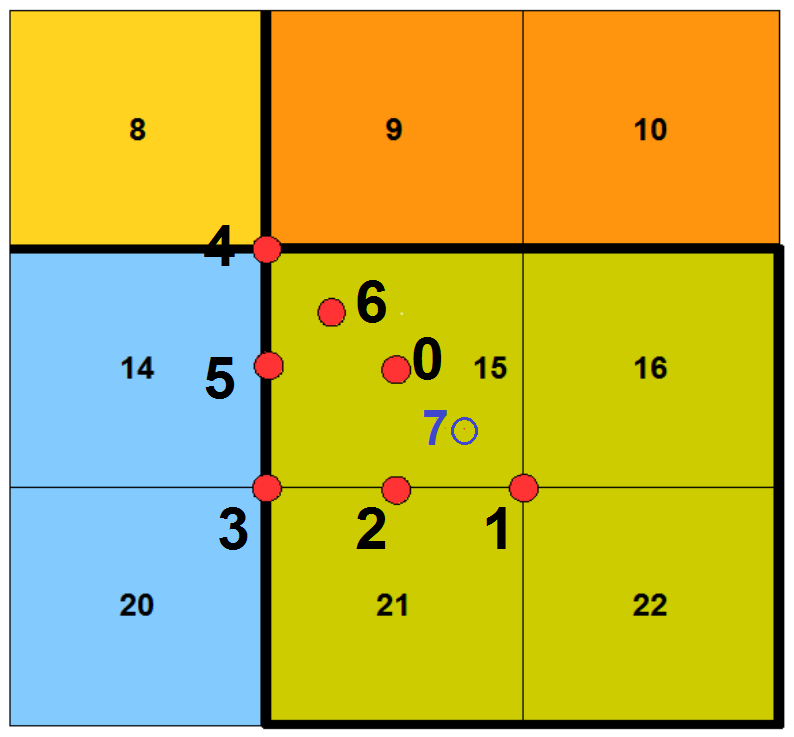}
  \caption[Beam positions on the front face of test matrix]
{The chosen beam positions (red bullets) projected onto the front face of the test matrix \cite{diehl_msc}.}
  \label{fig:pos-detail1}
\end{figure}

\begin{figure}
  \centering
  \includegraphics*[width=\swidth]{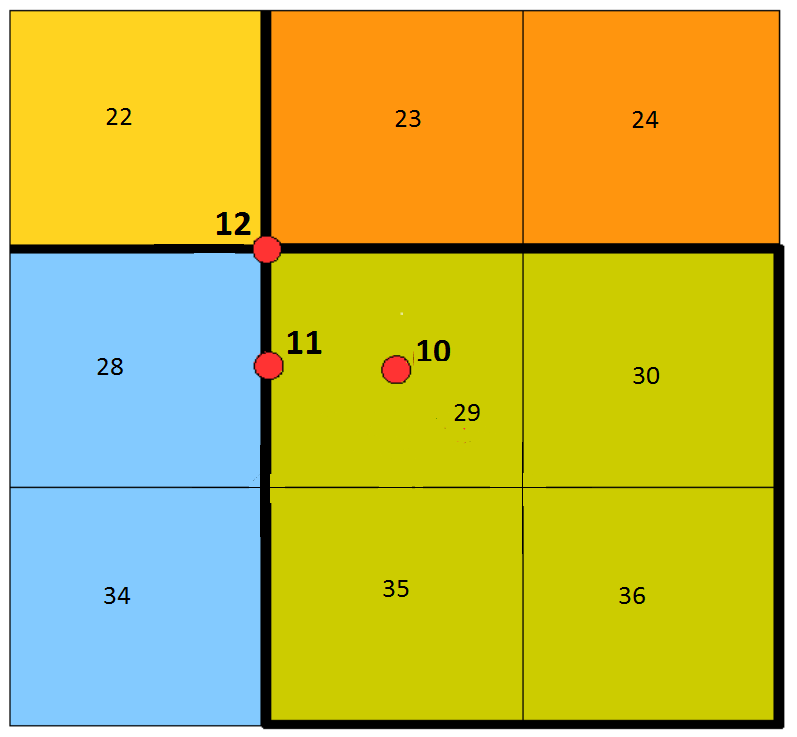}
  \caption[Beam positions on the front face of test matrix]
{The chosen beam positions (red bullets) projected onto the front face of the test matrix \cite{diehl_msc}.}
  \label{fig:pos-detail2}
\end{figure}

The specifications of the used detector modules of Type-2 are described in detail in \Refchap{chap:mech}. Two different experiments have been performed which differ in the used type of photo sensor. In the first run, photomultiplier tubes of type XP1911,1912 (Philips) operated with a commercial passive voltage divider have been used. They were replaced in the second run by R7899 (Hamamatsu) allowing a higher charge current operating with  a Cockcroft-Walton voltage divider designed at IHEP Protvino. The change of the photo sensors should exclude one possible reason for the observed strong position dependence of the detector response. To avoid non-linearities of the anode output and in the digitisation process, the setting of the bias voltage did not cover the full dynamic range of the SADC. Several measurements have been performed under identical conditions but varying the point of impact on the detector front face. Figure~\ref{fig:pos-overview}, \Reffig{fig:pos-detail1} and \Reffig{fig:pos-detail2} illustrate the labelling of the different detector cells as well as the chosen points of impact of the photon beam.

All signals of the complete matrix composed of 6$\times$6 individually read detector cells were actively split into a timing and an analog circuit to derive via a constant-fraction discriminator a fast timing signal for coincidence timing and trigger purpose. The analog signal was sampled by the 160~MHz commercial SADC (WIENER AVM16) over a range of 500~ns. The data acquisition was triggered by selected external signals from the electron-tagger detectors. The trigger information of the responding tagger channel was stored as well to select later-on one of the 16 photon-energy channels. A valid coincidence signal with the directly hit shashlyk cell was required for the readout of the complete set of 36 SADC channels. The data was analysed off-line.

\subsubsection{Signal extraction and calibration}

\begin{figure}
  \centering
  \includegraphics*[width=\swidth]{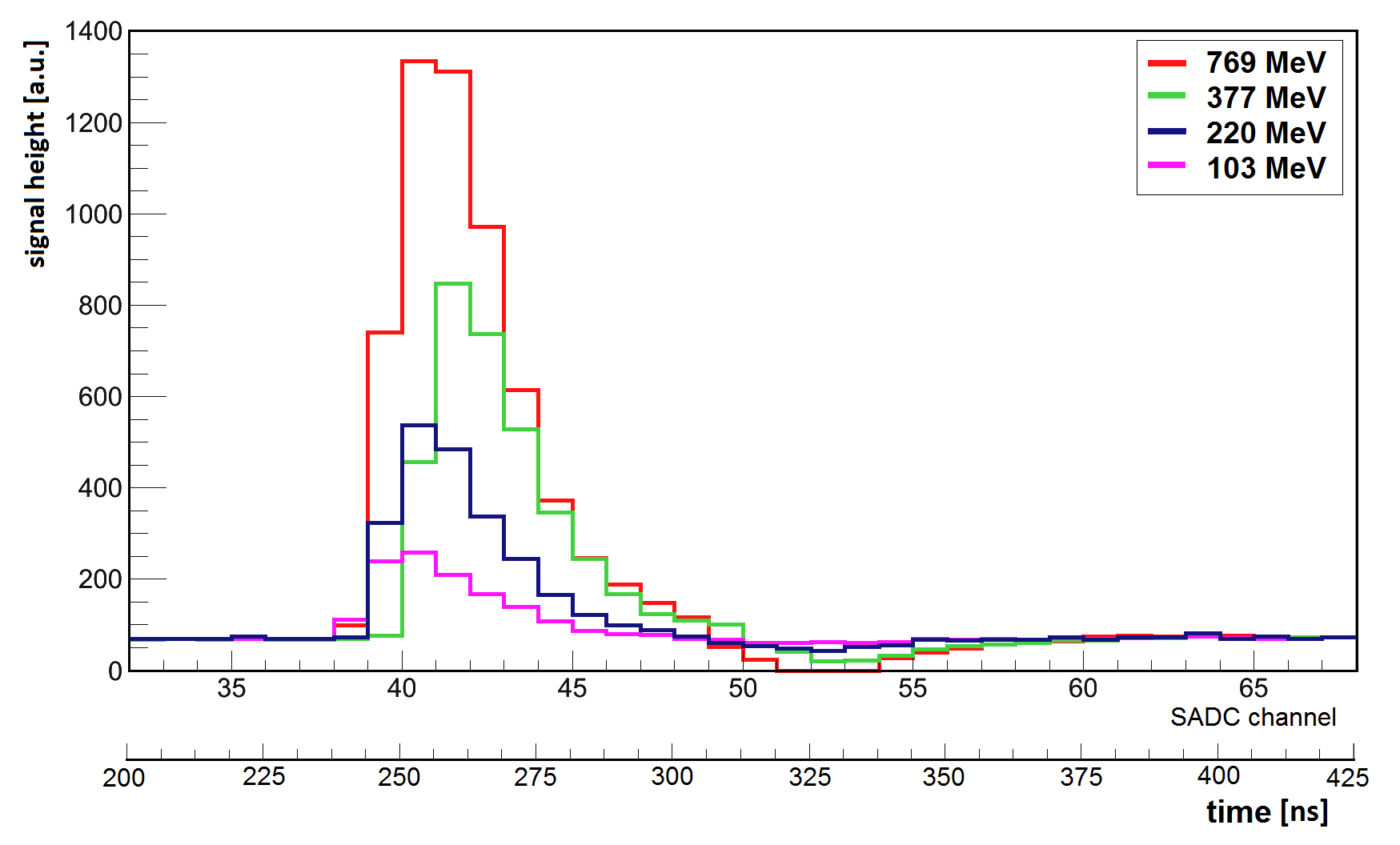}
  \caption[Signal shapes for different photon energies]
{Typical signal shapes of a responding detector cell for different photon energies recorded by the SADC \cite{diehl_msc}.}
  \label{fig:line-shape}
\end{figure}

The information on the energy deposition in the individual cells was obtained from the line shape recorded by the SADC (see~\Reffig{fig:line-shape}). The optimum resolution was achieved by integrating the signal shape over a time window of 200~ns and subtracting a base line deduced from the average of the first 32 sampling channels.
A pre-calibration for equalising the dynamical range was achieved using cosmic muons passing across the horizontally mounted modules. Thereafter, a relative calibration of all 36 cells has been performed by aiming the photon beam into each cell separately. Primarily, photon energies below 300~MeV energy have been used to confirm the linear response and to determine a relative normalisation factor. In spite of the lateral shower leakage even at low photon energies, the energy spectrum shows a nearly Gaussian shape but with a 
low-energy tail. Therefore, the peak positions have been determined by fitting the energy spectrum with the  Novosibirsk-function \cite{novo-fct}.

\subsubsection{Energy response}

The energy of the electromagnetic shower has been reconstructed by summing the energy depositions of the neighbouring responding cells. The software trigger threshold has been set to 3$\sigma$, with $\sigma$ the width of the noise distribution, which corresponds to 3.6~MeV deposited energy due to the chosen bias voltage of the photomultiplier. This threshold limit could still be reduced by a factor 2 by increasing the operating voltage and by using a dual-range readout.  

\begin{figure}
  \centering
  \includegraphics*[width=\swidth]{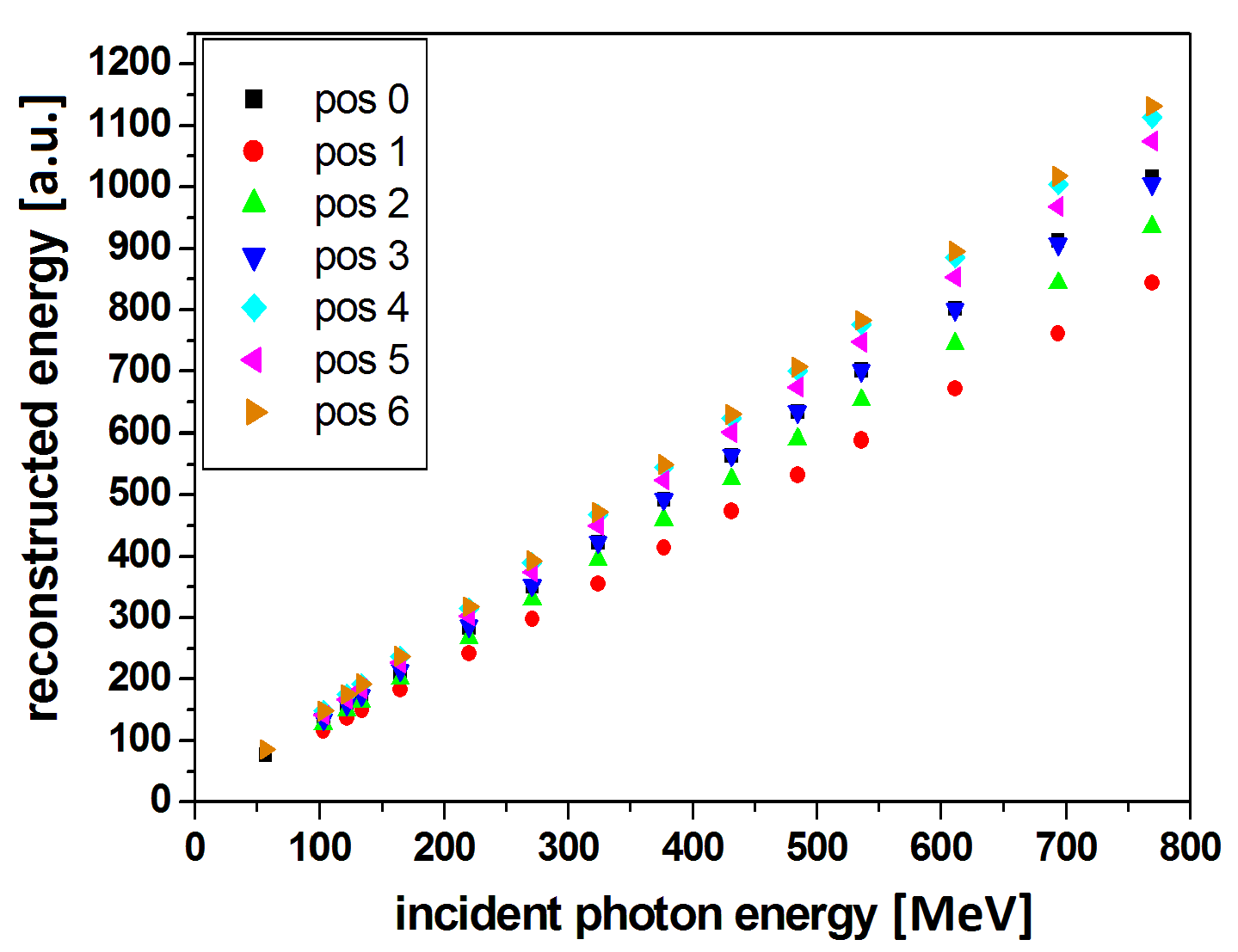}
  \caption[Linearity of energy response for different points of impact]
{Linearity of the deduced energy response over the entire range of photon energies measured for different points of impact \cite{diehl_msc}.}
  \label{fig:linearity_type2}
\end{figure}

\begin{figure}
  \centering
  \includegraphics*[width=\swidth]{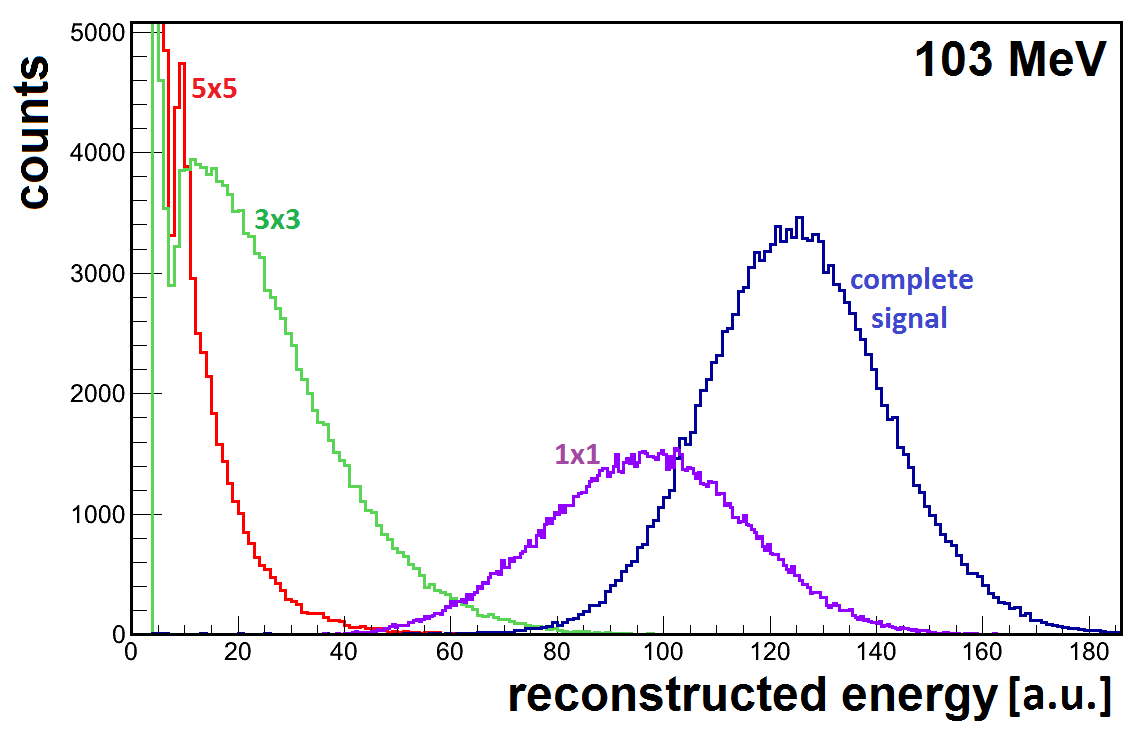}
  \caption[Energy response of Type-2 prototype for 103~MeV photons]
{Energy response of the Type-2 prototype for 103~MeV photons. The different spectra are explained in the text \cite{diehl_msc}.}
  \label{fig:ls-103}
\end{figure}

\begin{figure}
  \centering
  \includegraphics*[width=\swidth]{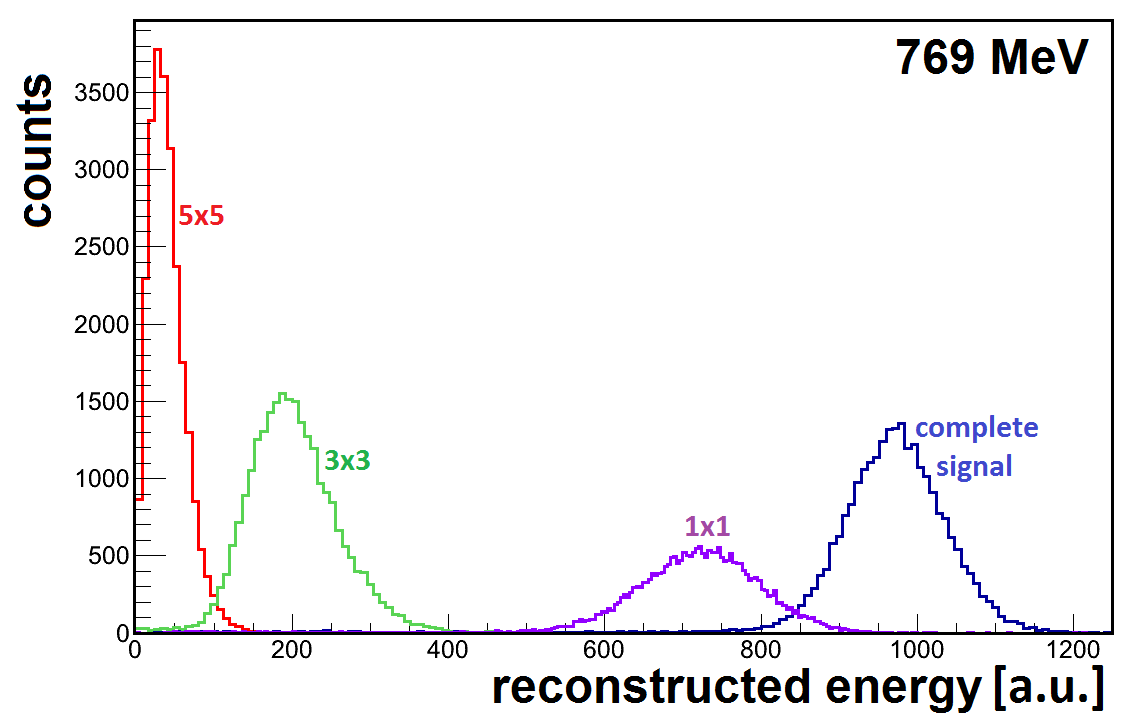}
  \caption[Energy response of Type-2 prototype for 769~MeV photons]
{Energy response of the Type-2 prototype for 769~MeV photons. The different spectra are explained in the text \cite{diehl_msc}.}
  \label{fig:ls-769}
\end{figure}

However, as illustrated in \Reffig{fig:linearity_type2}, the reconstructed shower energy is strongly depending on the impact point of the photon beam with variations up to 30\percent.
Figure~\ref{fig:ls-103} and \Reffig{fig:ls-769} illustrate the achieved line shape at two extreme photon energies of 103~MeV and 769~MeV, respectively. The figures show the energy deposition in the central module (1$\times$1), the energy sum in the first (3$\times$3) and the second (5$\times$5) ring of neighbours, and the total deposition into the complete 5$\times$5 array of cells. A nearly Gaussian shape has been obtained over the entire range of investigated photon energies.

\subsubsection{Relative energy resolution}

\begin{figure}
  \centering
  \includegraphics*[width=\swidth]{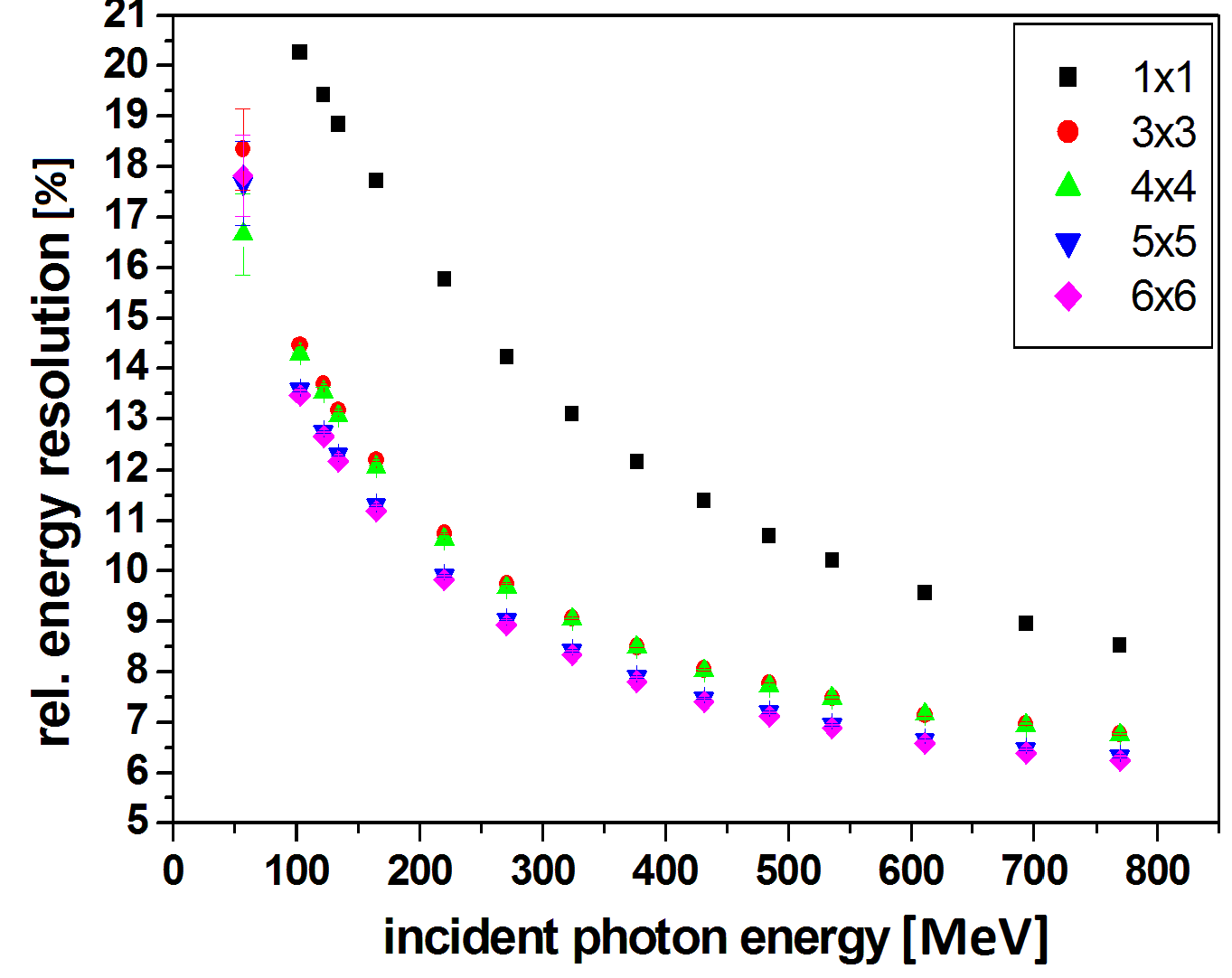}
  \caption[Energy resolution of Type-2 prototype for different cluster sizes]
{Relative energy resolution of the Type-2 prototype obtained by summing the energy deposition in different cluster sizes \cite{diehl_msc}.}
  \label{fig:res-clustersize}
\end{figure}

\begin{figure}
  \centering
  \includegraphics*[width=\swidth]{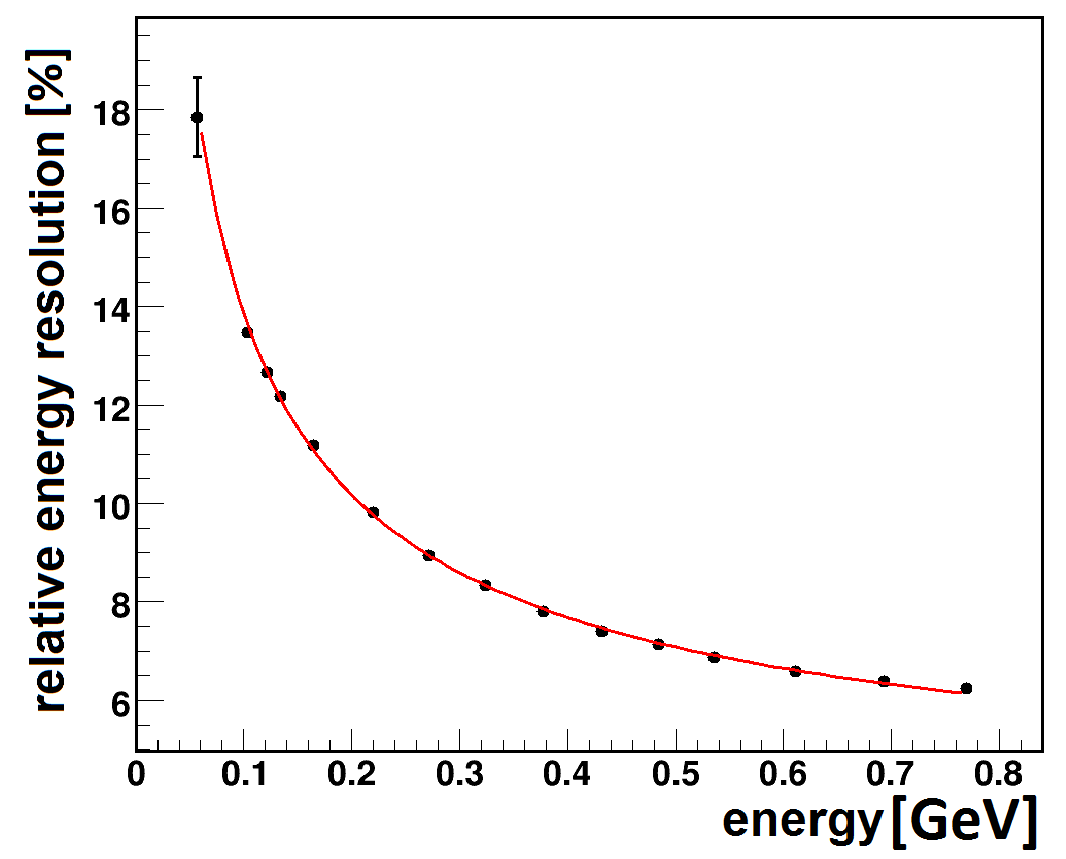}
  \caption[Energy resolution of Type-2 prototype for cluster of 6$\times$6 cells]
{Relative energy resolution of the whole matrix of the Type-2 detector composed of 6$\times$6 cells \cite{diehl_msc}.}
  \label{fig:resolution-matrix}
\end{figure}

The obtained relative energy resolution of the Type-2 prototype, using the cell numbered 15 as the central cell, depends strongly on the size of the selected matrix as illustrated in \Reffig{fig:res-clustersize}. Integrating the response of the entire matrix delivers a promising result (see~\Reffig{fig:resolution-matrix}) which can be described by \cite{diehl_msc}
\begin{equation}
  \frac{\sigma_{\rm E}}{\rm E} =
\frac{b}{\sqrt{\rm{E/GeV}}} \oplus c
\label{eq:fitfct82}
\end{equation}

with $b$=(4.21$\pm$0.01)\percent\\
 and $c$=(3.82$\pm$0.03)\percent.

\begin{figure}
  \centering
  \includegraphics*[width=\swidth]{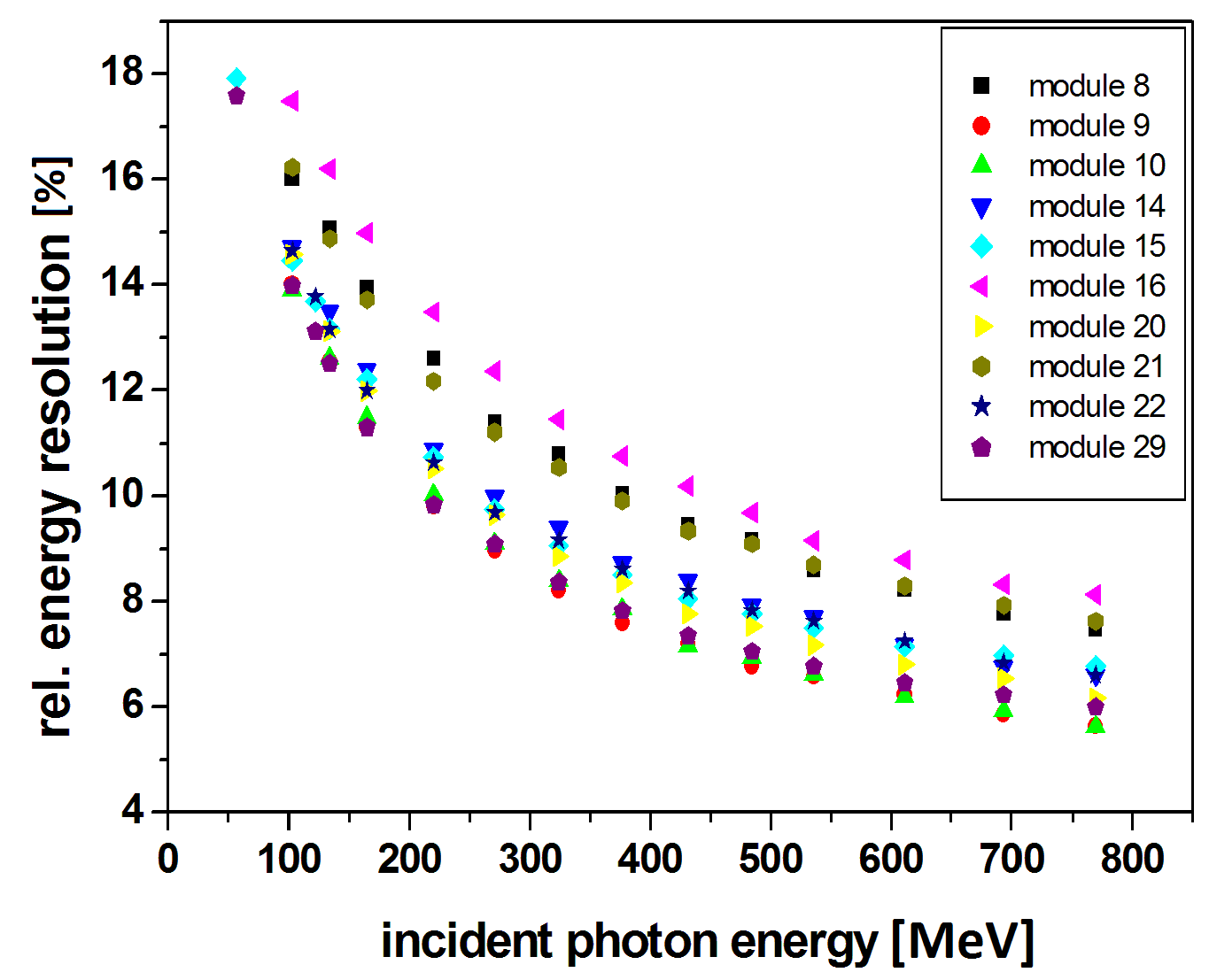}
  \caption[Energy resolution vs. photon energy for hits in cell centre]
{Relative energy resolution as function of the photon energy in the entire energy range when photons hit the centre of different cells as the central cell of a 3$\times$3 matrix of the Type-2 detector \cite{diehl_msc}.}
  \label{fig:resE-pos}
\end{figure}

When moving the point of impact of the incoming photon beam, strong position dependencies have been observed. Figure~\ref{fig:resE-pos} illustrates the unacceptable variation of the relative energy resolution of a 3$\times$3 matrix when photons hit the centre of different cells. Variations up to 4\percent are visible. The inhomogeneous response appears even more dramatic when inspecting directly the reconstructed absolute energy as a function of position. There appear differences on the level of up to 25\percent in the absolute values of the measured signal amplitude.

\subsubsection{The position resolution of Type-2 prototype}

\begin{figure}
  \centering
  \includegraphics*[width=\swidth]{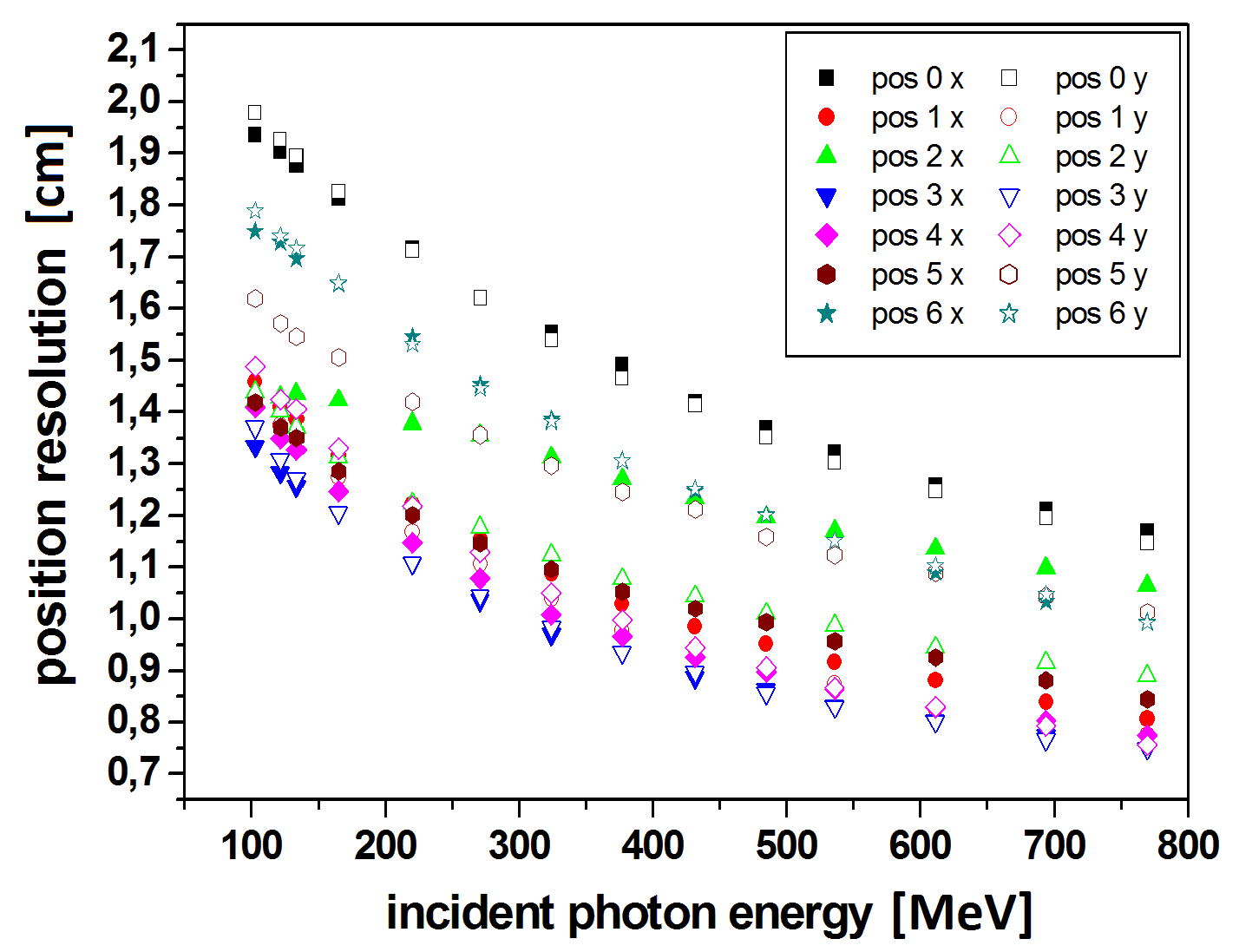}
  \caption[Energy resolution vs. photon energy for hits in marked position]
{Achieved position resolutions as a function of photon energy for different beam positions as marked in \Reffig{fig:pos-detail1} \cite{diehl_msc}.}
  \label{fig:resX-pos}
\end{figure}

The reconstruction of the point of impact of the collimated photon beam has been performed based on the lateral distribution of the detector cells. The centre of gravity in x- and y-direction perpendicular to the beam axis has been determined using a logarithmic weighting of the energy depositions with a weighting constant of W=4.6. As \Reffig{fig:resX-pos} illustrates, the achieved resolutions depend strongly on the point of impact due to the difference in energy sharing between the responding cells. The obtained values have not been corrected for the typical diameter of the beam spot of $\sim$10~mm.

\subsubsection{Time resolution of a single cell}

\begin{figure}
  \centering
  \includegraphics*[width=\swidth]{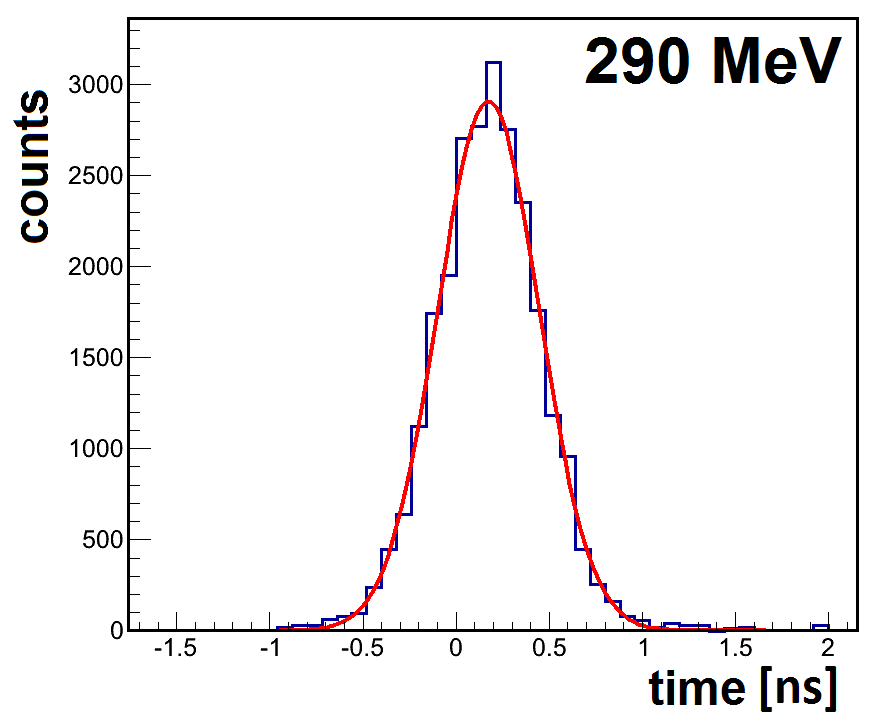}
  \caption[Coincidence timing between adjacent cells at 290~MeV energy deosition]
{Coincidence timing between two adjacent cells at a typical energy deposition of 290~MeV in both cells. The resolution is deduced by a fit with a Gaussian shape \cite{diehl_msc}.}
  \label{fig:t-res}
\end{figure}

\begin{figure}
  \centering
  \includegraphics*[width=\swidth]{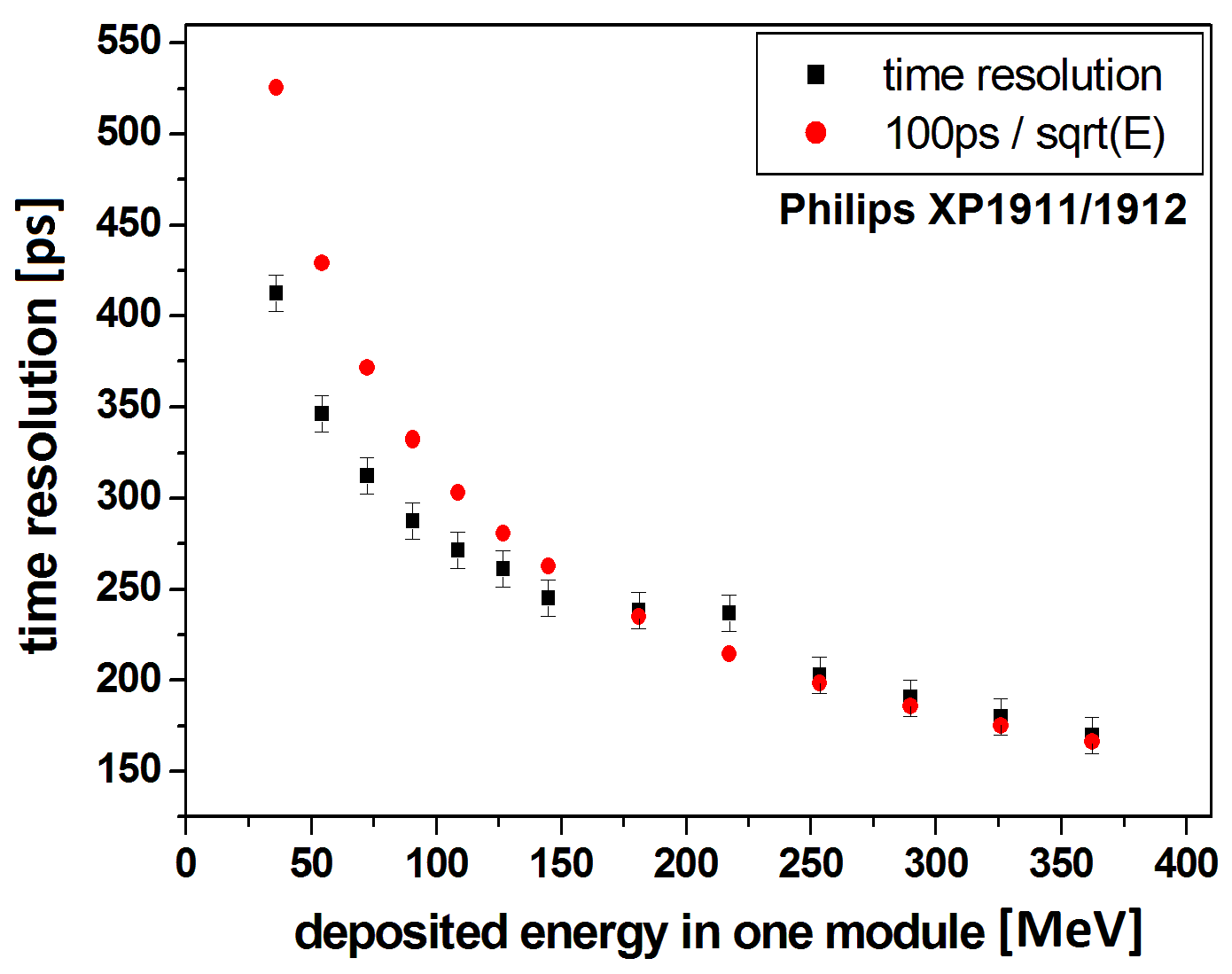}
  \caption[Time resolution of single cell vs. deposited shower energy]
{Achieved time resolution of a single cell as a function of deposited shower energy. The red dots correspond to a resolution of 100~ps/$\sqrt{\rm {E/GeV}}$ with E given in units of GeV \cite{diehl_msc}.}
  \label{fig:t-res-E}
\end{figure}

The timing performance will be a relevant parameter in reconstructing the individual events in a trigger-less data acquisition (see \Refchap{sec:roel}), as envisaged for \PANDA. A first estimate has been obtained by measuring the relative timing between two adjacent detector cells, since both obtain a very similar energy deposition if the beam hits in the middle between two cells.
This measurement at different photon energies allows to deduce the absolute time resolution as a function of energy deposition in a wide energy range. Figure~\ref{fig:t-res} shows a typical timing spectrum taken at an energy deposition of 290~MeV in one of the two considered cells.
Figure~\ref{fig:t-res-E} shows the deduced timing resolution as a function of deposited energy. A value of 100~ps can be expected at an energy deposition of 1~GeV.

\subsubsection{Unexpected non-linearities and conclusions}
\label{ssec:non-uni-type2}
The reason for the strong position dependence of the signal amplitude of the individual detector cells has been further investigated by using cosmic muons. Besides the lateral inhomogeneity, observed in the beam experiment, there has been identified as well a strong longitudinal dependence.  The recorded amplitude drops by up to 30\percent when moving from the location of the photo sensor towards the front end of the detector, where the WLS fibres take a 180\degrees turn. Detailed inspections revealed cracks in the WLS fibres at the position of the turning point. In addition, the accuracy of the alignment of the individual scintillator tiles was not sufficient due to missing fixation pins after cutting the existing scintillator tiles without a complete redesign (see \Refsec{sec:types_of_proto}).
In spite of satisfying resolution parameters, when deduced for a fixed impact point of the photon beam, the strong lateral and longitudinal inhomogeneities and the large variations from cell to cell demanded a significant improvement of the detector cell and module design.

\section{Performance of Type-3 modules}

Based on the experience gathered with prototypes of Type-2, major refinements were introduced involving modifications of the plastic scintillator tiles, the WLS fibres, reflector layers as well as the processing of the  assembly. All details can be found in \Refchap{chap:mech}. In order to quantify and certify the achieved improvement, an additional test at the tagged-photon facility at MAMI has been performed covering a similar range of photon energies to be sensitive to the limits in resolution.

\subsection{Test with high-energy photons at MAMI - setup and analysis}

\begin{figure}
  \centering
  \includegraphics*[width=\swidth]{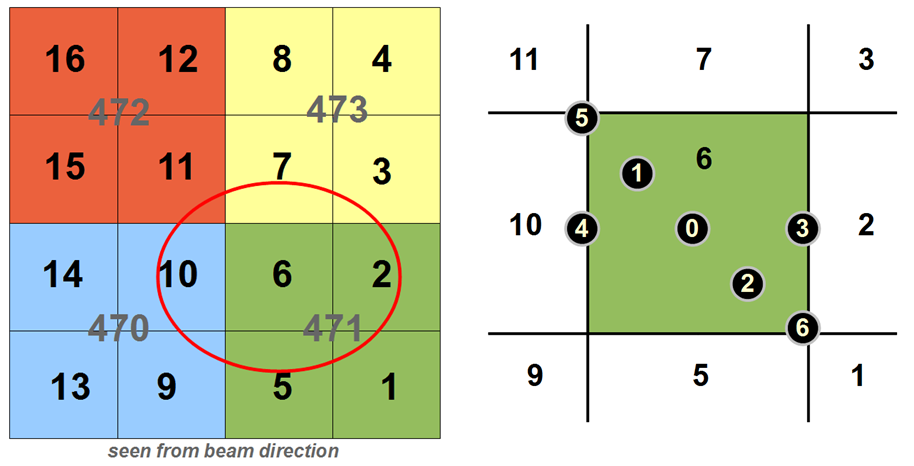}
  \caption[Scheme of selected beam positions on front face of Type-3 prototype]
{Scheme of selected positions of the incoming photon beam marked on the front face of the 2$\times$2 matrix of modules (i.e. 4$\times$4 detector cells) \cite{diehl_phd}.}
  \label{fig:3-positions}
\end{figure}

The photon energy range between 55 MeV and 650 MeV was covered by 15 selected energies with an intrinsic energy width
of 1-2~MeV. The diameter of the spot of the collimated beam at the detector front face was again in the order of 1~cm. The prototype consisted of a 2$\times$2 matrix of modules composed of 16 cells with individual readout and was positioned on a remotely controlled (x,y) moving support which could vary the point of impact of the beam to study the position dependence of the response. Figure~\ref{fig:3-positions} defines schematically the selected positions of impact of the incoming photon beam. The photomultiplier signals were digitised using a VME-based commercial sampling ADC (WIENER AVM16). In case of the largest energy depositions the input stage of the SADC caused saturation of the signal, which, however, did not spoil the final results and conclusions.

The relative calibration of the 16 cells has been performed in a first iteration using the response to cosmic muons in order to adjust the gain to the SADC range. For the final calibration the photon beam has been directed into the centre of each cell and the energy deposition for the lowest tagger energies was used as reference. In spite of the lateral shower distribution, the signal amplitudes have been correlated absolutely with the incoming photon energy. Therefore, the true energy deposition in a single cell scales according to the Moli\`ere radius with a typical factor of 0.7 when reading the given energy scales.

Several software procedures have been applied to deduce the optimum information on the deposited energy. Finally, the signal shape has been integrated over a fixed time window of 200~ns. In addition, a constant-fraction algorithm was applied to deduce the optimum timing information.

\subsubsection{Energy response to high-energy photons}

\begin{figure}
  \centering
  \includegraphics*[width=\swidth]{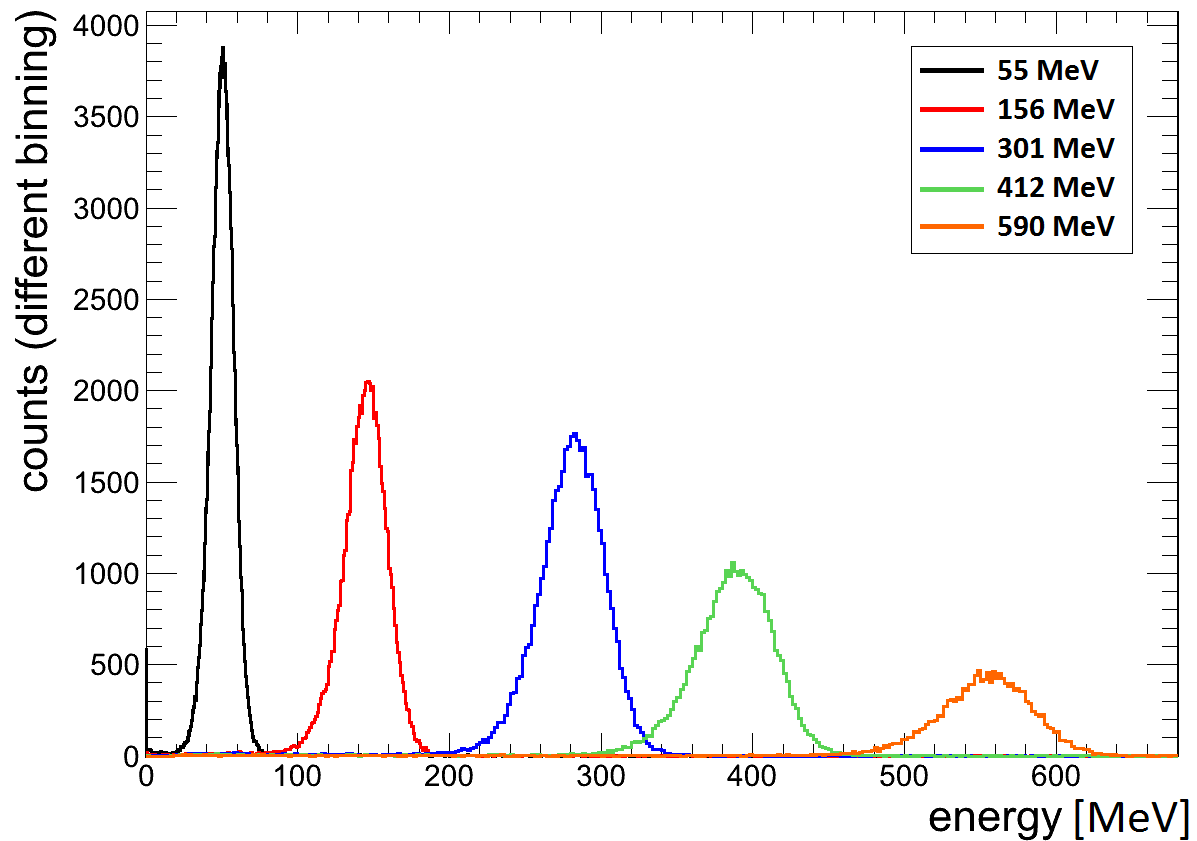}
  \caption[Response function of Type-3 prototype for tagged photons]
{Response function of a 3$\times$3 matrix of cells of the Type-3 prototype for tagged photons. The results are not corrected for the intrinsic width of the selected tagger channels \cite{diehl_phd}.}
  \label{fig:3-lineshape}
\end{figure}

After the relative calibration and the off-line selection of the different energies of the tagged photons, the total energy of the electromagnetic showers was reconstructed by summing the response of a 3$\times$3 matrix of cells. A software threshold corresponding to ~ 1 MeV and a coincidence window of 50~ns were applied.
Figure~\ref{fig:3-lineshape} demonstrates the obtained response functions for 5 selected energies. At the highest energies the slightly asymmetric shape and the low-energy tails indicate the onset of lateral shower leakage due to the restriction to a 3$\times$3 matrix.

\begin{figure}
  \centering
  \includegraphics*[width=\swidth]{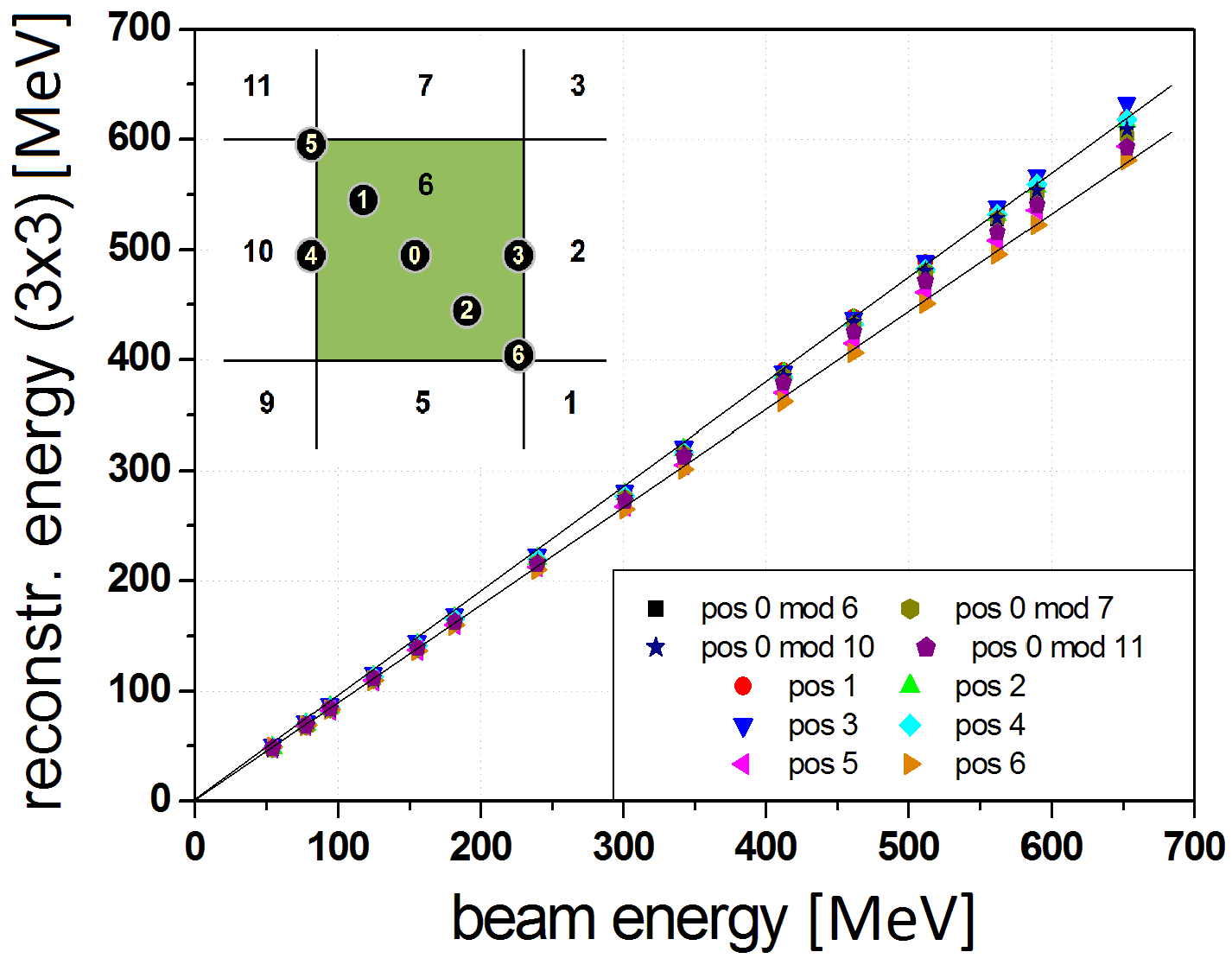}
  \caption[Linearity of energy response for Type-3 prototype for different points of impact]
{Linearity of the deduced energy response over the entire range of photon energies measured with Type-3 prototype for different points of impact \cite{diehl_phd}.}
  \label{fig:3-linearity}
\end{figure}

The response over the entire energy range is linear with a slightly different slope depending on the point of impact and the possible leakage at the highest energies. The overall behavior is depicted in \Reffig{fig:3-linearity}. In contrast to the former test experiment with a significantly different design and less accurate assembly of the sampling sandwich (see~\Reffig{fig:linearity_type2}), the observed variation of the reconstructed energy for different points of impact on the central cell remains below 3-4\percent \cite{diehl_phd}; this value is fully acceptable.

\subsubsection{Relative energy resolution for photons}

\begin{figure}
  \centering
  \includegraphics*[width=\swidth]{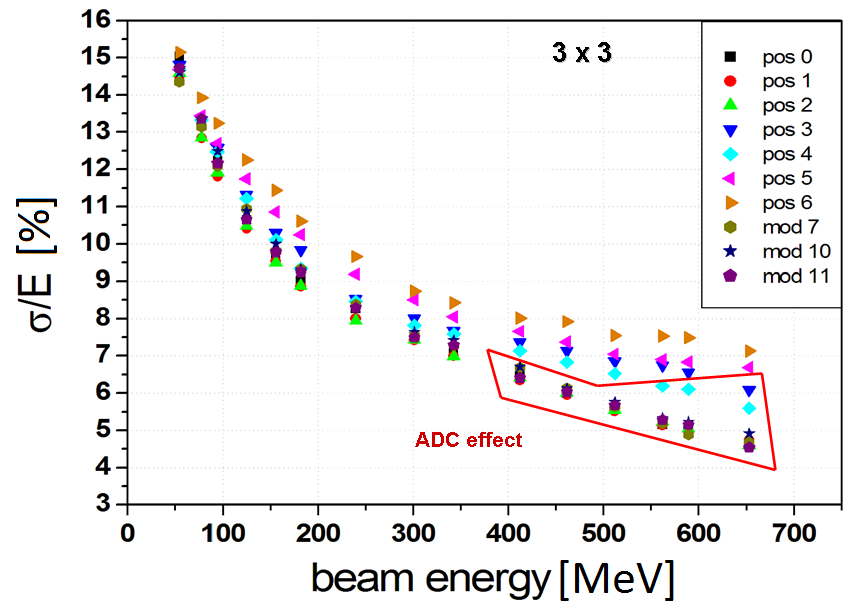}
  \caption[Energy resolution for Type-3 prototype for different points of impact]
{Relative energy resolution obtained for a 3$\times$3 matrix of cells over the entire range of photon energies. The points of impact of the photon beam on the central cell are noted and the data spoiled by the non-linear response of the ADC are marked by the red frame \cite{diehl_phd}.}
  \label{fig:3-ener-res}
\end{figure}

Due to the limited performance of the commercial SADC with respect to a linear response, a saturation effect has been observed at photon energies above 300~MeV, when the beam was hitting the centre of a cell and depositing the maximum fraction of the energy. Since the test experiment was focusing on the performance towards the lowest energies, the deteriorated results at the highest energies had no severe impact on the overall conclusion. Figure~\ref{fig:3-ener-res} summarise the obtained relative energy resolutions for different positions of impact. The results spoiled by the saturation of the input stage of the SADC are marked. The differences in resolution are to some extent caused by differences in the available lateral volume due to the limited size of the used Type-3 prototype array composed of 4$\times$4 cells.

\subsubsection{Timing response}

\begin{figure}
  \centering
  \includegraphics*[width=\swidth]{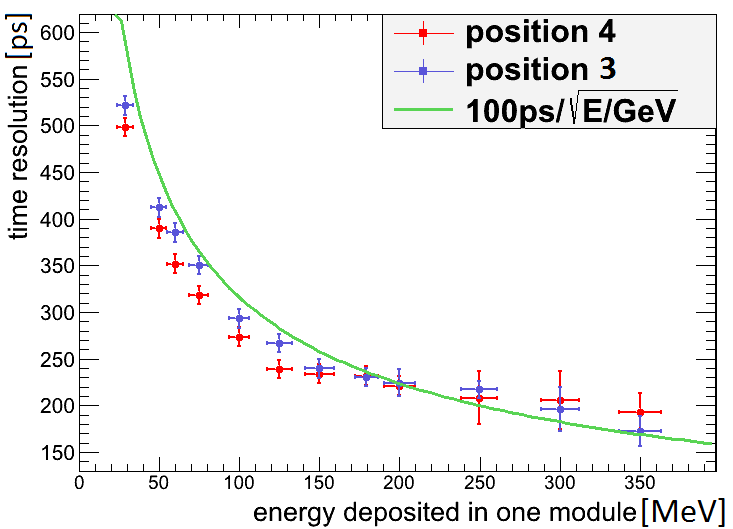}
  \caption[Time resolution vs. deposited energy for two impact points between adjacent cells]
{Measured time resolution as a function of the deposited energy for two impact points between two adjacent cells. The green line corresponds to a resolution of 100~ps/$\sqrt{\rm {E/GeV}}$ \cite{diehl_phd}.}
  \label{fig:3-time}
\end{figure}

The time resolution of an individual module has been estimated by directing the photon beam between two adjacent cells.  As a consequence, the shower is symmetrically shared. Under the assumption that both cells have a similar timing performance the time resolution can be deduced as a function of deposited energy from the time difference between both timing signals. The obtained values are shown in \Reffig{fig:3-time} for two impact points and compare reasonably to a parametrisation of 100~ps/$\sqrt{\rm {E/GeV}}$.

\subsubsection{The position reconstruction}

\begin{figure}
  \centering
  \includegraphics*[width=\swidth]{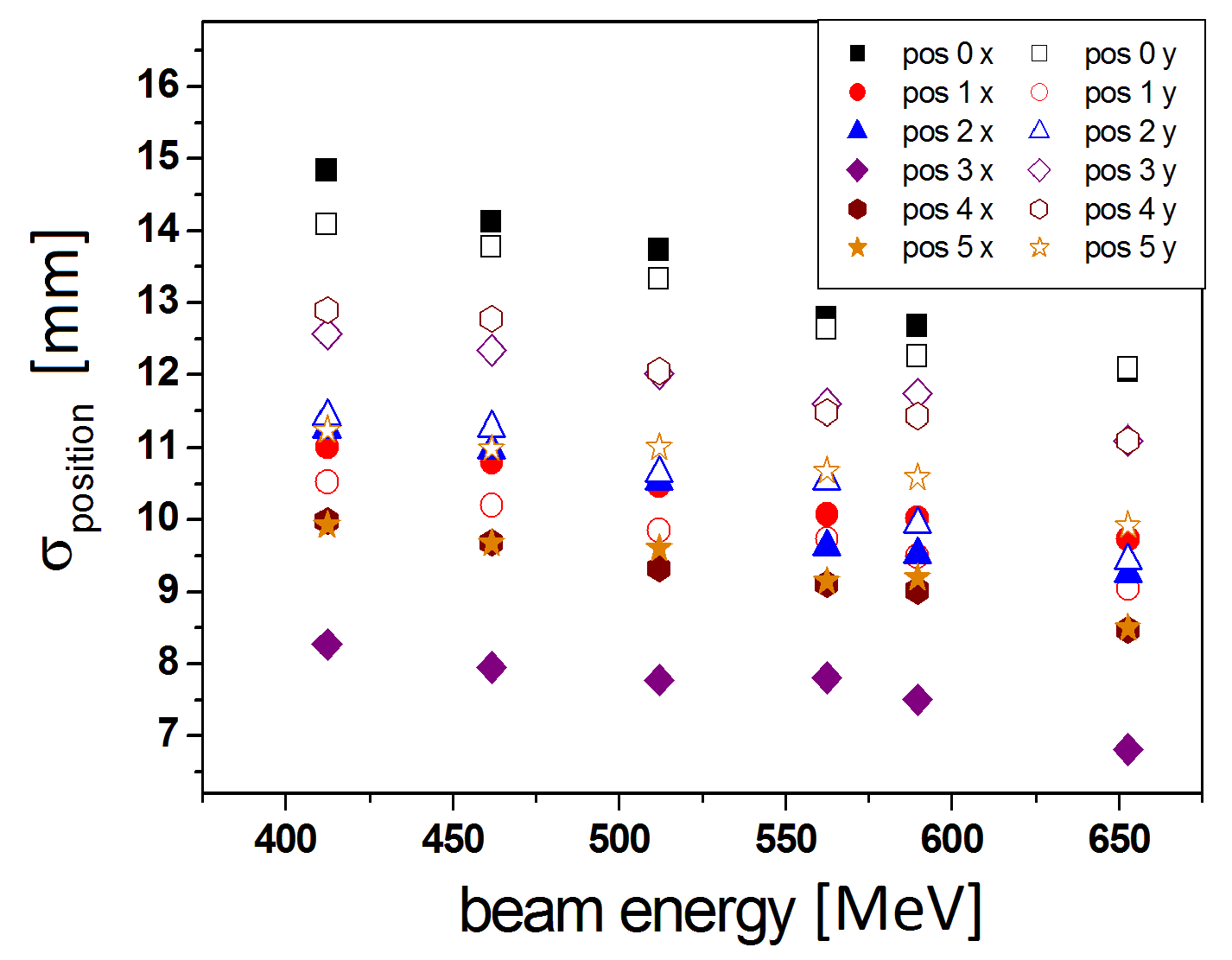}
  \caption[Position resolution vs. photon energy for different beam positions]
{Experimental resolution of the reconstructed position of the impinging tagged photons as function of the photon energy for different beam positions \cite{diehl_phd}.}
  \label{fig:3-position}
\end{figure}

The measurement of the distribution of the electromagnetic shower over the responding cells of the 3$\times$3 matrix allows to reconstruct the most probable point of impact of the tagged photon. The impact point has been deduced using a logarithmic weighting of the energy distribution in the neighbouring cells and typical resolution values are shown in \Reffig{fig:3-position} for different beam positions and energies. In this case the large variation of the obtained resolution reflects the limited size of the prototype matrix. The results are not corrected for the typical radius of the beam spot of $\approx$5.1~mm.

\subsubsection{Comparison to GEANT4 simulations}

\begin{figure}
  \centering
  \includegraphics*[width=\swidth]{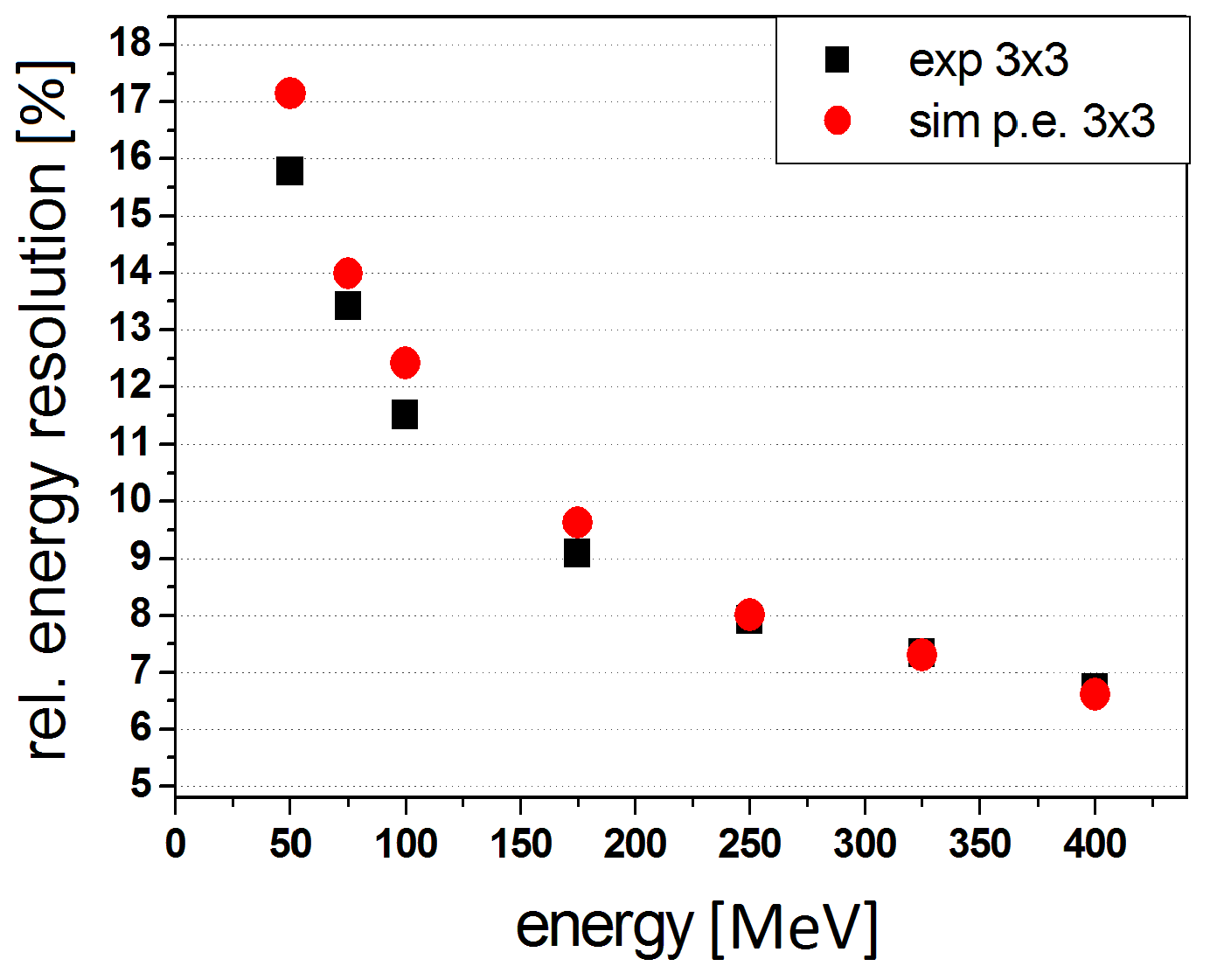}
  \caption[Energy resolution vs. photon energy compared to GEANT4]
{Experimental relative energy resolutions as function of the tagged-photon energy in comparison to GEANT4 simulations which take into account the number of photo electrons (p.e.) per MeV energy deposition \cite{diehl_phd}.}
  \label{fig:3-GEANT}
\end{figure}

The obtained experimental results can be well understood and reproduced by a standard simulation based on GEANT4. The simulated energy deposition in the active scintillator elements is statistically smeared taking into account the determined number of photo electrons per MeV energy deposition. As an example, \Reffig{fig:3-GEANT} shows the comparison of experimental and simulated energy resolutions for the lower photon energies and reveals a good agreement.

\subsubsection{Achieved performance of Type-3 prototypes}

The Type-3 shashlyk cells and modules represent a significant improvement of the overall performance with respect to energy, position and time information necessary for shower reconstruction. The significantly better reproducibility and homogeneity of the modules and individual cells should guarantee to reach the necessary performance even at photon energies as low as 10-20~MeV. The implementation of additional reflector material and WLS fibres of better quality and light collection have almost doubled the recorded light yield which now amounts to $\sim$2.8$\pm$0.3 photo electrons per MeV deposited energy \cite{diehl_phd}. 

A similar test procedure using cosmic muons has confirmed a significantly lower longitudinal inhomogeneity below 15\percent with nearly identical values for all available detector cells \cite{diehl_phd}. Such a value is expected due to the light attenuation within the WLS fibres.  Therefore, the overall detector design appears to be well suited for the \PANDA application also with respect to the lateral and longitudinal dimensions to provide the required granularity and to minimise shower leakage up to the highest photon energies. However, the described tests were restricted to a narrow energy range using commercial electronics and have not yet exploited the expected capability of the envisaged and dedicated readout electronics to cover the entire dynamic range over 4 orders of magnitude in energy.

%
%
\newpage

%

%% file: panda_tdr_FSC_org_pub.tex
\cleardoublepage
\chapter{Project Management}
\label{sec:org}
%
%

Until recently, the only participating institution in the \PANDA FSC project 
was the Institute of High Energy Physics (IHEP) Protvino, 
Russian Federation. IHEP Protvino is responsible for all parts of the FSC,
except the front-end electronics. The \Panda group from the University of Uppsala, 
Sweden, involved in the EMC front-end electronics development, now also 
participates in the FSC project. In this chapter, the provisions to be taken
for quality control of components upon delivery from the vendors 
will be discussed. Aspects of safety, risk assessment and schedule will be considered as well.

\section{Quality control and assembly}
\subsection{Production logistics and Quality Control procedures}

The FSC is a modular detector. It consists of a set of independent objects
that can be produced and tested independently before the final installation.
The realisation of the major components of the FSC is 
split into several parts, which also require significantly different 
logistics. 

The detector components comprise the 1512 cells as active elements
and 1512 photomultipliers (PMT). PMTs can safely be used since the 
FSC is located outside the dipole magnetic field. Each cell consists 
of a sandwich of scintillator tiles and lead plates, as described in \Refchap{chap:mech}.
All these components, as well as fibres, photomultiplier tubes and the Cockcroft-Walton 
type high-voltage bases, 
require a sophisticated program of production by manufacturers, as well as 
quality control and assurance. 


A special test-stand to measure the light output of the scintillator plates has been built at
the IHEP Scintillator Workshop. The raw material with the necessary additives is loaded into the
injection moulding machines before the production work. For each batch of the raw
material several scintillator plates will be studied selectively at this stand. If these
pieces do not pass the quality control (QC) test, the production of the scintillator plates
with this raw material is stopped, and the raw material is removed from the injection 
moulding machines to be substituted by a new portion of raw material with additives. 

During the mass production of the tiles the quality control is performed at several levels
\begin{itemize}
\item Visual inspection of the tiles color and transparency. Each stopping of the moulding machines
requires a few dummy cycles of tiles moulding (production of 12-16 tiles) to get a good optical quality tiles.
The tiles produced during the first cycles contain air bubbles inside or “silver” near the surface. Also
rough errors of the dopants percentage can be easily detected by change of the tile color in comparison
with the reference sample.       
\item Testing of the light output at the dedicated setup. There is a dedicated test setup with a
radioactive source (Sr-90) mentioned above to test four tiles contained in one layer of the module.
Light is collected by the same WLS optical fiber, which is used in the module.
The light output from each tile is measured and compared to the reference sample.
The measurements are preformed at the beginning and at the end of each working day
and with each new batch of raw materials. The same setup can measure both the level as well as
the uniformity of light output.
\item The final quality control of each assembled module consisting of four cells will be
performed at the cosmic stand at IHEP where the mean amplitude of the MIP signal created 
in the cells by cosmic muons must be greater than some threshold value at the same 
fixed high-voltage value applied to the photomultiplier. If one or more out of four cells in 
the module do not pass such a quality control test, this module will be set aside,
be rebuilt and again be sent to the MIP quality control test until it successfully passes this test.
\end{itemize}

This kind of QC resulted in rather low losses of raw material (up to 5\%). The main point of the
losses are setup cycles of moulding machines and adjustment of the molding parameters.
The quality of the molded tiles is very stable after a few “dummy cycles” at the beginning of the moulding machine
operation. So far we encountered no reasons for the rejection of tiles. As for the modules, there is a rare chance
to make a mistake during the module assembly (too high pressure, optical fiber break during the assembly).
In this case the module can be reassembled.

From our experience BASF-143E and BASF-124N polystyrene types, which are used for the FSC tiles production,
have rather stable properties from batch to batch. Thus, it can be purchased for the complete production at once.
The same is true for lead. One load of the moulding machine requires about 15 kg of raw material.
Assuming 2.9 ton of material for the complete set of tiles for the FSC,
one needs 200 loads. Scintillating dopants (PTP and POPOP) usually are bought in several stages because
their production is a lengthy process. The time schedule is defined by the manufacturer.
For the optimal production schedule the dopants should come once per quarter.  

\begin{figure*}[h]
\begin{center}
\includegraphics[width=0.8\dwidth]{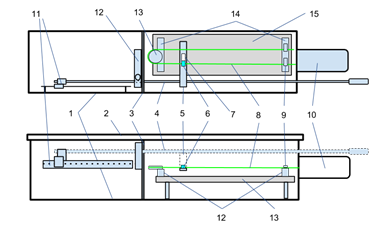}
\caption[Optical fibers QC]
{Dedicated test setuo to check the quality of the optical fiber bending procedure. 1 - light tight box, 2 - box cover,
3 - light tight division, 4 - guide, 5 - moving head, 7 - LED, 8 - loop of optical fiber,  9 - clamps, 10 - PMT,
11,12 - position sensor, 13,14 - fiber supports, 15 - table}
\label{fig:fiber_qc}
\end{center}
\end{figure*}

There was no problem with the stability of the light attenuation length of optical fibers
during the production of prototypes.
Nevetherless, it was measured at the bent loop with a dedicated setup to confirm the bending
procedure parameters
(\Reffig{fig:fiber_qc}). The same setup can be used for QC during the mass production.
Also cosmic muons will be used to check the longitudinal uniformity of the light output for the complete module. 

Lead tiles quality is estimated first by visual control of the surface. The main parameter to measure
is thickness which is controlled by micrometer and digital caliper.

We plan to purchase the photomultipliers for the whole FSC outside the Russian Federation 
and send them directly to  FZ J\"ulich where the assembly will take place. 
The results of the quality control of all photomultipliers and all modules will be presented 
at J\"ulich by IHEP personnel.
Before the final assembly 
of the FSC, a partial assembly (pre-assembly) of several FSC modules will be done including the required 
cabling, to test modules and PMTs with cosmic muons after the shipping.   
Before assembling the results of the quality control of all photomultipliers and all modules will be presented 
at J\"ulich by IHEP personnel.
Only one institute, namely IHEP Protvino, is 
involved in the mass production of the FSC components, except the digitiser 
electronics, which is under the responsibility of Uppsala University.

\subsection{Detector module assembly}

First, a mould for producing quadratic plastic tiles with a side length 
of 5.5 cm, holes for fibres, pins for relative tiles position fixation
will be manufactured at a Russian company close to Protvino (PlastProduct from Kaluga or Maral-2000 
from Vladimir). 
The same company can produce a high precision stamp to cut and punch quadratic lead plates with a side 
length of 11 cm. Then, both the scintillator tiles (using the injection moulding 
technology) and the lead plates will be manufactured at the IHEP Protvino 
scintillator workshop. 
The mechanical support structures of a module will be fabricated at the IHEP Protvino central 
mechanical workshop. Thereafter, the module assembly, including the bending of the fibre loops and pulling the fibres through the holes in the tiles and the lead plates, will be accomplished at IHEP Protvino as well. 
All 36 fibre ends for each cell will be assembled into a bundle with a diameter of
about 5 mm, glued, cut, polished, and prepared to be attached later
to the photo detector at the downstream end of the cell.   

\subsection{Final assembly, pre-calibration and implementation into \PANDA}
In order to avoid delays in FSC assembling and testing caused by a possible delay in the
FAIR civil construction, the  
final assembling of the FSC including the intermediate
storage is anticipated to take place at FZ J\"ulich, Germany. 
The separately assembled modules 
(arrays of 2$\times$2 cells) as well as all mechanical support structures will
be sent from Protvino to J\"ulich. The photomultipliers will be sent
by Hamamatsu directly to Germany. The fibres will be attached to the
photomultipliers at J\"ulich.  Instead of optical grease a silicon rubber cookie is going to be used 
to provide an optical contact between the fibre-bundle cap and the photo detector. The final assembly of
the FSC will be completed by the implementation of the optical
fibres for the monitoring system, the electronics for digitisation
(will be delivered to FZ J\"ulich by some other institute in \PANDA, not
by IHEP Protvino, but supervised by IHEP Protvino), 
the cables for signal transfer and low- and high-voltage support, the sensors 
and lines for slow control.       

The basic functionality of the assembled units will be tested with the help of the
monitoring system in order to validate the electronics chain including 
the photo sensors. Knowing the gain of the photomultipliers, a good estimate of the expected sensitivity of each
detector module is obtained from the measured light, resulting in a pre-calibration with an accuracy of about 20 \%. 
Exploiting minimum ionising cosmic muons, a more sensitive relative and absolute calibration of the calorimeter modules 
will be obtained. This procedure will deliver pre-calibration parameters
with an accuracy of about 5 \%. 
Later, the final in situ calibration will be achieved with a sub-percent 
accuracy as mentioned in \Refchap{sec:req}. 

\subsection{Integration in \PANDA}

The design of the FSC is fully compatible
with the overall layout of the whole \PANDA detector. The
components including service connections and mounting spaces fit
within the fiducial volume defined by the forward dipole magnet
and the other sub-detector components. It is foreseen that fully assembled 
and tested modules are stored in an air-conditioned area with
stabilised temperature and low humidity. 

\section{Safety}

The components of the FSC including
the infrastructure for the operation will be built according to the safety
requirements of FAIR and the European and German safety regulations. 
Assembly and installation ask for detailed procedures to avoid 
interference with and take into account concomitant assembly and 
movement of other detectors. Radiation safety aspects during FSC commissioning with
a beam will be assured by the common accelerator safety interlock system.

A FSC risk analysis has to take into account mechanical 
and electrical parts as well as potentially hazardous materials. 
Some important aspects of this risk analysis are discussed in the following:
The mechanical design of the support structures of the FSC has been 
checked by a FEM analysis. This analysis revealed a safety factor of 12 (the ratio of 
the material strength to the design load, see \Refsec{sec:mech:stress_calc} for more details),  
which is quite comfortable for the FSC mechanics.
Cables, pipes and optical fibres as well as the other components 
of the FSC are made of non-flammable halogen-free materials. 

In addition, the materials are chosen to be radiation tolerant 
at the radiation level expected in the \PANDA environment. All supplies
have appropriate safety circuits and fuses against shorts and the power 
channels have to be equipped with over-current and over-voltage control 
circuits. Safety interlocks on the electrical parts have to be planned 
to prevent accidents induced by cooling leakage from other detectors.

\subsection{Mechanics}

The strength of the calorimeter support structure has been computed 
with physical models in the course of the design process. Details of a 
finite-element analysis are shown in \Refsec{sec:mech:stress_calc}. 
Each mechanical component will undergo a quality-acceptance examination 
including stress and loading tests for weight bearing parts. Spare samples
 may also be tested up to the breaking point. A detailed material map of 
the entire apparatus showing location and abundance of all materials used 
in the construction will be created. For structural components radiation 
resistance levels will be taken into account in the selection process 
and quoted in the material map. 

\subsection{Electrical equipment}

All electrical equipment in \PANDA will comply to the legally required
safety code and concur to standards for large scientific installations 
following guidelines worked out at CERN to ensure the protection of all
personnel working at or close to the components of the \PANDA system.
Power supplies will have safe mountings independent of large mechanical 
loads. Hazardous voltage supplies and lines will be marked visibly and 
protected from damage by near-by forces, like pulling or squeezing. All
supplies will be protected against over-current and over-voltage and have 
appropriate safety circuits and fuses against shorts. All cabling and
optical-fibre connections will be executed with nonflammable halogen-free
materials according to up-to-date standards and will be dimensioned with
proper safety margins to prevent overheating. A safe ground scheme will be
employed throughout all electrical installations of the experiment. 
Smoke detectors will be mounted in all appropriate locations.

High output LEDs will be employed in the monitoring system and their
light is distributed throughout the calorimeter system. For these 
devices all necessary precautions like safe housings, colour coded 
protection pipes, interlocks, proper warnings and instructions as well 
as training of the personnel with access to these components will be taken.
  
 \subsection{Radiation aspects}

The FSC modules can become radioactive due to two
main processes, which induce radioactivity in materials,
neutron activation and inelastic hadronic interactions at high energy.
Tentative  simulations based on the computer code MARS deliver, in case 
of a maximum luminosity of 2$\cdot$10$^{32}$ cm$^{-2}$ s$^{-1}$ 
and for an operation period of 30 days, a dose rate of 30 $\mu$Sv/h
at the calorimeter surface closest to the beam axis.
The estimated dose rate is several orders of magnitude lower
than in case of CMS at LHC. Shielding, operation
and maintenance will be planned according to European
and German safety regulations to ensure the
proper protection of all personnel.

%
%
%
%
\newpage
\bibliographystyle{panda_tdr_lit}
\bibliography{./lit_emc}

%% file: panda_tdr_FSC_end.tex
\cleardoublepage
\onecolumn
\input{./main/thanks}
%
\cleardoublepage
\twocolumn
\input{./main/acronyms}
\cleardoublepage
\addcontentsline{toc}{chapter}{List of Figures}
\listoffigures
\cleardoublepage
\addcontentsline{toc}{chapter}{List of Tables}
\listoftables
%

%% file: main/thanks.tex
%
\begin{center}
\vspace*{2cm}
{\Large\bf Acknowledgments}\addcontentsline{toc}{chapter}{Acknowledgements}
\vskip 2cm
\begin{minipage}[t]{8cm}
\sloppy\large
We acknowledge the dedicated financial support from the Russian State 
Corporation ``ROSATOM'' over the years 2008-2013. In addition, this work has 
been partially supported by the IHEP Protvino budget.  

We gratefully acknowledge the generous support by the A2 Collaboration in providing the tagged-photon beam time at MAMI, Mainz, Germany.
\end{minipage}
\end{center}
\vfill
%
%

%% file: main/acronyms.tex
%
\addcontentsline{toc}{chapter}{List of Acronyms}
\begin{acronym}
\acro{ADC}{Analog to Digital Converter}
\acro{ATCA}{Advanced Telecommunications Computing Architecture}
\acro{APD}{Avalanche Photo Diode}
\acro{BaBar}{B and B-bar}
\acro{BELLE}{B detector at KEK in Japan}
\acro{BES}{BEijing Spectrometer}
\acro{BNL}{Brookhaven National Laboratory}
\acro{CAMAC}{Computer Aided Measurement And Control}
\acro{CBM}{Compressed Baryonic Matter}
\acro{CDF}{Collider Detector at Fermilab}
\acro{CERN}{Conseil Europ\'een pour la Recherche Nucl\'eaire}
\acro{CLEO}{Detector at Cornell's CESR accelerator}
\acro{c.m.} centre of mass
\acro{CMS}{Compact Muon Solenoid}
\acro{CN}{Compute Node}
\acro{CR}{Collector Ring}
\acro{CW}{Cockcroft-Walton generator}
\acro{D0}{detector named for location on the Tevatron Ring at Fermilab}
\acro{DAC}{Digital-to-Analog Converter }
\acro{DAQ}{Data Acquisition}
\acro{DCON}{Data Concentrator Board}
\acro{DCS}{Detector Control System}
\acro{DESY}{Deutsches ElektronenSYnchrotron}
\acro{DIRC}{Detector for Internally Reflected Cherenkov Light}
\acro{DPM}{Dual Parton Model}
\acro{EMC}{Electromagnetic Calorimeter}
\acro{EPICS}{Experimental Physics and Industrial Control System}
\acro{EvtGen}{Event Generator}
\acro{FAIR}{Facility for Antiproton and Ion Research}
\acro{FairRoot}{Root-based computing framework for FAIR experiments}
\acro{FE}{Finite Element}
\acro{FEM}{Finite Element Model}
\acro{FEE}{Front-End Electronics}
\acro{FF} {Form Factor}
\acro{FNAL}{Fermi National Accelerator Laboratory}
\acro{FPGA}{Field Programmable Gate Array}
\acro{FSC}{Forward Spectrometer Calorimeter}
\acro{FZ}{Forschungszentrum}
\acro{FZJ}{Forschungszentrum J\"ulich}
\acro{GEANT}{GEometry ANd Tracking}
\acro{GEM}{Gas Electron Multiplier}
\acro{GSI}{Gesellschaft f\"ur Schwerionenforschung}
\acro{HADES}{High Acceptance DiElectron Spectrometer}
\acro{HESR}{High Energy Storage Ring}
\acro{HV}{High Voltage}
\acro{IHEP}{Institute for High Energy Physics}
\acro{KOPIO}{K0-to-PI0 experiment proposal at BNL}
\acro{LEAR}{Low Energy Antiproton Ring}
\acro{LED}{Light Emitting Diode}
\acro{LHC}{Large Hadron Collider}
\acro{LHCb}{Large Hadron Collider beauty experiment}
\acro{LMS}{Light Monitoring System} 
\acro{microTCA}{TCA for extremely high bandwidth}
\acro{MAMI}{Mainz Microtron}
\acro{MDT}{Mini Drift Tubes}
\acro{MIP}{Minimum Ionising Particle}
\acro{MLP}{Multilayer Perceptron neural network}
\acro{Modbus} {a serial communication protocol published originally by Modicon}
\acro{MSPS}{MegaSamples per Second}
\acro{MSV}{Modularised Start Version}
\acro{MVD}{Micro-Vertex Detector}
\acro{PANDA}{AntiProton ANnihilation at DArmstadt}
\acro{PandaRoot}{Root-based computing framework for PANDA}
\acro{PCB}{Printed Circuit Board}
\acro{PDG}{Particle Data Group}
\acro{PID}{Particle Identification}
\acro{p-LINAC}{Proton Linear Accelerator}
\acro{PLL}{Phase Locked Loop}
\acro{PMT}{Photomultiplier}
\acro{POPOP}{1,4-bis(5-phenyloxazol-2-yl) benzene}
\acro{PWA}{Partial Wave}
\acro{PWO}{Lead Tungstate}
\acro{PWOII}{improved Lead Tungstate version II.}
\acro{QCD}{Quantum ChromoDynamics}
\acro{QED}{Quantum ElectroDynamics}
\acro{RECTO}{Non-Relativistic QCD}
\acro{RESR}{Recuperated Experimental Storage Ring}
\acro{RF}{Radio Frequency}
\acro{RICH}{Ring Imaging Cherenkov Counter}
\acro{RMS}{Root Mean Square}
\acro{ROOT}{Object-oriented data analysis framework} 
\acro{SADC}{Sampling Analog-to-Digital Converter}
\acro{SciTil}{Scintillator Tile hodoscope}
\acro{SCU}{Slow Control Unit}
\acro{SERDES} {Serialiser-Deserialiser}
\acro{SFP}{Small Form-factor Pluggable}
\acro{SIS-100}{Superconducting fast cycling Synchrotron}
\acro{SIS-18}{Schwerionen-Synchrotron}
\acro{SODA}{Synchronisation Of Data Acquisition}
\acro{SODANET}{SODA and TRBNET combined}
\acro{STT}{Straw Tube Tracker}
\acro{TCA}{Telecommunications Computing Architecture}
\acro{TDC}{Time-to-Digital Converter}
\acro{TDR}{Technical Design Report}
\acro{TOF}{Time-of-Flight Detector}
\acro{TRBNET}{slow-control and data-transfer protocol}
\acro{TS}{Target Spectrometer}
\acro{UNILAC}{Universal Linear Accelerator}
\acro{UPE}{European standard U}
\acro{UrQMD}{Ultra-relativistic Quantum MolecularDynamics}
\acro{VMC}{Virtual Monte Carlo}
\acro{VME}{Versa Module Europa bus}
\acro{WACS}{Wide Angle Compton Scattering}
\acro{WASA}{Wide Angle Shower Apparatus}
\acro{WLS}{WaveLength Shifting fibre}
\end{acronym}
\vfill
%
%